\documentclass[12pt,tightenlines,eqsecnum,floats,aps,amsmath,amssymb,nofootinbib,prd,shownopacs]{revtex4}

\usepackage{color}
\usepackage{graphicx}

\usepackage{setspace}
\usepackage{amsmath,amssymb,amsfonts,amsthm}
\usepackage{enumerate} 
\usepackage{colordvi} 
\usepackage{subfigure}

\def\be{\nopagebreak[3]\begin{equation}}
\def\ee{\end{equation}}
\def\ba{\nopagebreak[3]\begin{eqnarray}}
\def\ea{\end{eqnarray}}
\def\f{\frac}

\def\ul{\underline}
\def\t{\tilde}
\def\h{\hat}
\def\sint{\textstyle{\int}}
\def\dd{{\rm d}}
\def\d{{\rm d}}
\def\g{{\rm grav}}

\def\b{{\rm b}}
\def\v{{\rm v}}
\def\B{{}_{\rm B}}
\def\la{\lambda}

\def\a{\alpha}
\def\ep{\varepsilon}

\def\L{{\cal L}}
\def\pt{\tilde{p}_{(a)}}
\def\p{\partial}

\def\C{\mathcal{C}}
\def\R{\mathbb{R}}

\def\WDW{\rm WDW\,\,}

\def\H{{\cal H}}
\def\Hkg{\H_{\rm kin}^{\rm grav}}
\def\Hk{\H_{\rm kin}^{\rm total}}
\def\Hkwdw{\H_{\rm kin}^{\rm wdw}}
\def\Ha{\H_{\rm aux}}

\def\Hp{\H_{\rm phy}}

\def\Hpwdw{\H_{\rm phy}^{\rm wdw}}

\def\N{\mathcal{N}}

\def\C{\mathcal{C}}
\def\E{\mathcal{E}}

\def\SU{{\rm SU}}

\def\lp{{\ell}_{\rm Pl}}
\def\mpl{m_{\rm Pl}}

\def\rcr{\rho_{\rm max}}

\def\R{\mathbb{R}}
\def\Z{\mathbb{Z}}
\def\e{\mathring{e}}
\def\o{\mathring{\omega}}
\def\q{\mathring{q}}

\newcommand{\ip}[2]{{\langle#1\,|\,#2\rangle}}

\def\pphi{p_{(\phi)}}

\def\gpi{12 \pi G}
\def\chiwb{\underline\chi(\b,\phi)}
\def\chiwy{\underline \chi(y,\phi)}

\def\chilb{\chi(\b,\phi)}
\def\chilx{\chi(x,\phi)}
\def\lam{\la}

\def\ep{\epsilon}

\def\Dz{\Delta z}
\def\lu{l_1}
\def\ld{l_2}
\def\lt{l_3}
\def\lvp{(\lu,\ld,v;\phi)}
\def\vfu{\f{v \pm 4}{v \pm 2}}
\def\vfd{\f{v \pm 2}{v}}
\def\vft{\f{v}{v \pm 2}}

\def\Dz{\Delta z}
\def\lu{l_1}
\def\ld{l_2}
\def\lt{l_3}
\def\lvp{(\lu,\ld,v;\phi)}
\def\vfu{\f{v \pm 4}{v \pm 2}}
\def\vfd{\f{v \pm 2}{v}}
\def\vft{\f{v}{v \pm 2}}
\def\ox{\mathring{\xi}}

\begin{document}
\preprint{\vbox{\baselineskip=12pt \rightline{IGC-11/03-??} }}
\title{Loop Quantum Cosmology: A Status Report}
\author{Abhay Ashtekar}
\email{ashtekar@gravity.psu.edu} \affiliation{Institute for
Gravitation and the Cosmos \& Physics Department, The Pennsylvania
State University, University Park PA 16802, USA}
\author{Parampreet Singh}
\email{psingh@phys.lsu.edu} \affiliation{Department of
Physics and Astronomy, Louisiana State University, Baton Rouge, LA
70803, USA }
\begin{abstract}

Loop quantum cosmology (LQC) is the result of applying principles of
loop quantum gravity (LQG) to cosmological settings. The
distinguishing feature of LQC is the prominent role played by the
\emph{quantum geometry effects} of LQG. In particular, quantum
geometry creates a brand new repulsive force which is totally
negligible at low space-time curvature but rises very rapidly in the
Planck regime, overwhelming the classical gravitational attraction.
In cosmological models, while Einstein's equations hold to an
excellent degree of approximation at low curvature, they undergo
major modifications in the Planck regime: For matter satisfying the
usual energy conditions any time a curvature invariant grows to the
Planck scale, quantum geometry effects dilute it, thereby resolving
singularities of general relativity. Quantum geometry corrections
become more sophisticated as the models become richer. In
particular, in anisotropic models there are significant changes in
the dynamics of shear potentials which tame their singular behavior
in striking contrast to older results on anisotropies in bouncing
models. Once singularities are resolved, the conceptual paradigm of
cosmology changes and one has to revisit many of the standard issues
---e.g., the `horizon problem'--- from a new perspective.
Such conceptual issues as well as potential observational
consequences of the new Planck scale physics are being explored,
especially within the inflationary paradigm. These considerations
have given rise to a burst of activity in LQC in recent years, with
contributions from quantum gravity experts, mathematical physicists
and cosmologists.

The goal of this article is to provide an overview of the current
state of the art in LQC for three sets of audiences: young
researchers interested in entering this area; the quantum gravity
community in general; and, cosmologists who wish to apply LQC to
probe modifications in the standard paradigm of the early universe.
An effort has been made to streamline the material so that each of
these communities can read only the sections they are most
interested in, without a loss of continuity.

\end{abstract}

\maketitle

\vfill\break

\tableofcontents

\vfill\break

\section{Introduction} \label{s1}

This section is divided into five parts. In the first, we provide a
broad overview of how cosmological paradigms have evolved over time
and why we need quantum cosmology. In the second, we first discuss
potential limitations of restricting quantum gravity considerations
to cosmological contexts and explain why quantum cosmology is
nonetheless an essential frontier of quantum gravity. In the third,
we list some of the most important questions any quantum cosmology
theory should address and explain why this challenge has proved to
be so non-trivial. In the fourth we introduce the reader to loop
quantum cosmology (LQC) and in the fifth we provide an outline of
how the review is organized to best serve primary interests of three
research communities.

\subsection{Cosmological paradigms}
\label{s1.1}

As recorded history shows, cosmological paradigms have evolved
considerably over time as notions of space and time themselves
matured. It is illuminating to begin with a broad historical
perspective by recalling paradigms that seemed obvious and most
natural for centuries only to be superseded by radical shifts.

Treatise on Time, the Beginning and the End date back at least
twenty five centuries and it is quite striking that some of the
fundamental questions were posed and addressed already in the early
literature. Does the flow of time have an objective, universal
meaning beyond human perception? Or, is it only a convenient and
perhaps merely psychological notion? If it does have an objective
meaning, did the physical universe have a finite beginning or has it
been evolving eternally? Leading thinkers across cultures meditated
on these issues and arrived at definite but strikingly different
answers, often rooted in theology. Eastern and Greek traditions
generally held that the universe is eternal or cyclic with no
beginning or end while the western religions promoted the idea of a
finite beginning. A notable variation is St. Augustine who argued in
the fourth century CE that time itself started with the world.

Although founding fathers of modern Science, including Galileo and
Newton, continued to use theology for motivation and justifications,
they nonetheless developed a much more successful paradigm, marked
by precision. Before Newton, boundaries between the absolute and the
relative and the mathematical and the common were blurry. Through
precise axioms stated in the \emph{Pricipia}, Newton isolated time
from the psychological \emph{and} the material world, making it
objective and absolute. It now ran uniformly from the infinite past
to the infinite future, indifferent to matter and forces. This
paradigm became the dogma over centuries. The universe came to be
identified with matter. Space and time provided an \emph{eternal}
background or a stage on which the drama of dynamics unfolds.
Philosophers often used this clear distinction to argue that the
universe itself \emph{had} to be eternal. For, as Immanuel Kant
emphasized, otherwise one could ask ``what was there before?''

As we know, general relativity toppled this paradigm in one fell
swoop. Now the gravitational field was encoded in the very geometry
of space-time. Geometry became a dynamical, physical entity and it
was now perfectly feasible for the universe to have had a finite
beginning ---the big-bang--- at which not only matter but
\emph{space-time itself} is born. In this respect, general
relativity took us back to St. Augustine's paradigm but in a
detailed, specific and mathematically precise form. In books and
semi-popular articles relativists now like to emphasize that the
question ``what was there before?'' is rendered meaningless because
the notion of `before' requires a pre-existing space-time geometry.
We now have a new paradigm: \emph{In the Beginning there was the Big
Bang.}

However, the very fusion of gravity with geometry now gives rise to
a new tension. In Newtonian (or Minkowskian) physics, if a given
physical field becomes singular at a space-time point it can not be
unambiguously evolved to the future but this singularity has no
effect on the global arena: since the space-time geometry is
unaffected by matter, it remains intact. Other fields can be evolved
indefinitely; trouble is limited to the one field which became ill
behaved. However, because gravity is geometry in general relativity,
when the gravitational field becomes singular, the continuum tears
and the space-time itself ends. There is no more an arena for other
fields to live in. All of physics, as we know it, comes to an abrupt
halt. Physical observables associated with both matter and geometry
simply diverge, signalling a fundamental flaw in our description of
Nature.

This problem arises because the reasoning assumes that general
relativity ---with its representation of space-time as a smooth
continuum--- provides an accurate description of Nature
arbitrarily close to the singularity. But general relativity
completely ignores quantum physics and over the last century we
have learned that quantum effects become important at high
energies. Indeed, they should in fact be \emph{dominant} in
parts of the universe where matter densities become enormous.
Thus \emph{the occurrence of the big-bang and other
singularities are predictions of general relativity precisely
in a regime where it is inapplicable!} Classical physics of
general relativity does come to a  halt at the big-bang and the
big-crunch. But this is not an indication of what really
happens because the use of general relativity near
singularities is an extrapolation which has no physical
justification whatsoever. We need a theory that incorporates
not only the dynamical nature of geometry but also the
ramifications of quantum physics. We need a
quantum theory of gravity, a new paradigm.%
\footnote{It is sometimes argued that the new paradigm need not
involve quantum mechanics or $\hbar$; new classical field
equations that do not break down at the big-bang should suffice
(see e.g. \cite{rp-ccc}). But well established physics tells us
that quantum theory is essential to the description of matter
\emph{much before} one reaches the Planck density, and $\hbar$
features prominently in this description. Stress energy of this
quantum matter must couple to gravity. So it is hard to imagine
that a description of space-time that does not refer to $\hbar$
would be viable in the early universe.}
Indeed, cosmological singularities where the space-time
continuum of general relativity simply ends are among the most
promising gates to physics beyond Einstein.

In quantum cosmology, then, one seeks a `completion' of general
relativity, as well as known quantum physics, in the hope that it
will provide the next paradigm shift in our overall understanding of
the universe. A focus on cosmology serves three purposes. First, the
underlying large scale symmetries of cosmological space-times
simplify technical issues related to functional analysis. Therefore
it is possible to build mathematically complete and consistent
models and systematically explore their physical consequences.
Second, the setting is well suited to address the deep conceptual
issues in quantum gravity, discussed in subsequent sections, such as
the problem of time, extraction of dynamics from a `frozen'
formalism, and the problem of constructing Dirac observables in a
background independent theory. These problems become manageable in
quantum cosmology and their solutions pave the way to quantum
gravity beyond the S-matrix theory that background dependent
approaches are wedded to. Finally, the last decade has seen
impressive advances in the observational cosmology of the very early
universe. As a result, quantum cosmology offers the best avenue
available today to confront quantum gravity theories with
observations.

\subsection{Quantum cosmology: Limitations?}
\label{s1.2}

The first point we just listed to highlight the benefits of focusing
quantum gravity considerations to cosmology also brings out a
fundamental limitation of this strategy. Symmetry reduction used in
the descent from full quantum gravity is severe because it entails
ignoring infinitely many degrees of freedom (even in the
`midi-superspaces').
So, a natural question arises: Why should we trust predictions of
quantum cosmology? Will results from full quantum gravity resemble
anything like what quantum cosmology predicts? There is an early
example \cite{kr} in which a mini-superspace A was embedded in a
larger mini-superspace B and it was argued that quantization of A by
itself is inequivalent to the sector of the quantum theory of B that
corresponds to A. However, to unravel the relation between the two
quantum theories, one should `integrate out' the extra degrees of
freedom in B rather than `freezing them out'. As an example, let A
be the k=0 Friedmann-LeMa\^itre-Robertson-Walker (FLRW) model with a
massless scalar field and let B be the Bianchi I model with the same
matter source. Then, if one first constructs the quantum theory of
the Bianchi I model and \emph{integrates out} the anisotropies in a
precise fashion, one does recover the quantum theory of the FLRW
model \cite{awe2}. Thus, a comparison between quantum theories of
the larger and the smaller systems has to be carried out with due
care. The question is: Will the quantum theory of the smaller system
capture the \emph{relevant qualitative features} of the quantum
theory of the larger system? We would like to give three arguments
which suggest that the answer is likely to be in the affirmative,
\textit{provided} quantum cosmology is so constructed that the
procedure captures the essential features of the full quantum
gravity theory.

First, consider an analogy with electrodynamics. Suppose,
hypothetically, that we had full QED but somehow did not have a good
description of the hydrogen atom. (Indeed, it is difficult to have a
complete control on this bound state problem in the framework of
full QED!) Suppose that Dirac came along at this juncture and said:
let us first impose spherical symmetry, describe the proton and
electron as particles, and then quantize the system. In this
framework, all radiative modes of the electromagnetic field would be
frozen and we would have quantum mechanics: the Dirac theory of
hydrogen atom. One's first reaction would again have been that the
simplification involved is so drastic that there is no reason to
expect this theory to capture the essential features of the physical
problem. Yet we know it does. Quantum cosmology may well be the
analog of the hydrogen atom in quantum gravity.

Second, recall the history of singularities in classical
general relativity. They were first discovered in highly
symmetric models. The general wisdom derived from the detailed
analysis of the school led by Khalatnikov, Lifshitz and others
was that these singularities were artifacts of the high
symmetry and a generic solution of Einstein's equations with
physically reasonable matter would be singularity free. But
then singularity theorems of Penrose, Hawking, Geroch and
others shattered this paradigm. We learned that lessons derived
from symmetry reduced models were in fact much more general
than anyone would have suspected. LQC results on the resolution
of the big-bang in Gowdy models which have an infinite number
of degrees of freedom
\cite{hybrid1,hybrid2,hybrid3,hybrid4,hybrid5}, as well as
\emph{all} strong curvature singularities in the
homogeneous-isotropic context \cite{ps} may be hints that the
situation would be similar with respect to singularity
\emph{resolution} in LQC.

Finally, the Belinskii-Khalatnikov-Lifshitz (BKL) conjecture in
classical general relativity says that as one approaches space-like
singularities in general relativity, terms in the Einstein equations
containing `spatial derivatives' of basic fields become negligible
relative to those containing `time derivatives' (see, e.g.,
\cite{bkl1,bkl-ar}). Specifically, the dynamics of each spatial
point follow the `Mixmaster' behavior
---a sequence of Bianchi I solutions bridged by Bianchi II
transitions. By now there is considerable support for this
conjecture both from rigorous mathematical and numerical
investigations \cite{ar,berger-rev,dg1,zurich}. This provides some
support for the idea that lessons on the quantum nature of the
big-bang (and big-crunch) singularities in Bianchi models may be
valid much more generally.

Of course none of these arguments shows conclusively that the
qualitative features of LQC will remain intact in the full
theory. But they do suggest that one should not a priori
dismiss quantum cosmology as being too simple. If quantum
cosmology is constructed by paying due attention to the key
features of a full quantum gravity theory, it is likely to
capture qualitative features of dynamics of the appropriate
\emph{coarse-grained macroscopic variables}, such as the mean
density, the mean anisotropic shears, etc.

\subsection{Quantum cosmology: Some key questions}
\label{s1.3}

Many of the key questions that any approach to quantum gravity
should address in the cosmological context were already raised
in the seventies by DeWitt, Misner, Wheeler and others. More
recent developments in inflationary and cyclic models raise
additional issues. In this section, we will present a prototype
list. It is far from being complete but provides an approach
independent gauge to compare the status of various programs.

\begin{itemize}
\item How close to the big-bang does a smooth space-time of
    general relativity make sense?  Inflationary scenarios,
    for example, are based on a space-time continuum. Can
    one show from `first principles' that this is a safe
    approximation already at the onset of inflation?

\item Is the big-bang singularity naturally resolved by
    quantum gravity? It is this tantalizing possibility
    that led to the development of the field of quantum
    cosmology in the late 1960s. The basic idea can be
    illustrated using an analogy to the theory of the
    hydrogen atom. In classical electrodynamics the ground
    state energy of this system is unbounded below. Quantum
    physics intervenes and, thanks to a non-zero Planck's
    constant, the ground state energy is lifted to a finite
    value, $-me^4/2\hbar^2 \approx - 13.6{\rm eV}$. Since
    it is the Heisenberg uncertainly principle that lies at
    the heart of this resolution and since the principle is
    fundamental also to quantum gravity, one is led to ask:
    Can a similar mechanism resolve the big-bang and big
    crunch singularities of general relativity?

\item Is a new principle/ boundary condition at the
    big-bang or the big-crunch essential? The most well
    known example of such a boundary condition is the `no
    boundary proposal' of Hartle and Hawking \cite{hh}. Or,
    do quantum Einstein equations suffice by themselves
    even at the classical singularities?

\item Do quantum dynamical equations remain well-behaved even at
    these singularities? If so, do they continue to provide a
    (mathematically) deterministic evolution? The idea that
    there was a pre-big-bang branch to our universe has been
    advocated in several approaches (see Ref. \cite{novello} for
    a review), most notably by the pre-big-bang scenario in
    string theory \cite{pbb}, ekpyrotic and cyclic models
    \cite{ekp1,ekp2} inspired by the brane world ideas and in
    theories with high order curvature terms in the action (see
    eg. \cite{bran1,bran2}). However, these are perturbative
    treatments which require a smooth continuum in the
    background. Therefore, their dynamical equations break down
    at the singularity whence, without additional input, the
    pre-big-bang branch is not joined to the current
    post-big-bang branch by deterministic equations. Can one
    improve on this situation?

\item  If there is a deterministic evolution, what is on
    the `other side'? Is there just a quantum foam from
    which the current post-big-bang branch is born, say a
    `Planck time after the putative big-bang'? Or, was
    there another classical universe as in the pre-big-bang
    and cyclic scenarios, joined to ours by deterministic
    equations?

\item In bouncing scenarios the universe has a contraction
    phase before the bounce. In general relativity, this
    immediately gives rise to the problem of growth of
    anisotropy because the anisotropic shears dominate  in
    Einstein's equations unless one introduces by hand
    super-stiff matter (see, e.g., \cite{bb}). Can this
    limitation be naturally overcome by quantum
    modifications of Einstein's equations?

\end{itemize}

Clearly, to answer such questions we cannot start by assuming
that there is a smooth space-time in the background. But
already in the classical theory, it took physicists several
decades to truly appreciate the dynamical nature of geometry
and to learn to do physics without recourse to a background
space-time. In quantum gravity, this
issue becomes even more vexing.%
\footnote{There is a significant body of literature on this issue;
see e.g., \cite{as-book,hartle,crbook,rgjp1,rgjp2} and references
therein. These difficulties are now being discussed also in the
string theory literature in the context of the AdS/CFT conjecture.}

For simple systems (including Minkowskian field theories) the
Hamiltonian formulation generally serves as a `royal road' to
quantum theory. It was therefore adopted for quantum gravity by
Dirac, Bergmann, Wheeler and others. But absence of a background
metric implies that the Hamiltonian dynamics is generated by
constraints \cite{kk}. In the quantum theory, physical states are
solutions to quantum constraints. All of physics, including the
dynamical content of the theory, has to be extracted from these
solutions. But there is no external time to phrase questions about
evolution. Therefore we are led to ask:

\begin{itemize}

\item Can we extract, from the arguments of the wave
    function, one variable which can serve as
    \emph{emergent time} with respect to which the other
    arguments `evolve'? Such an internal or emergent time
    is not essential to obtain a complete, self-contained
    theory. But its availability makes the physical meaning
    of dynamics transparent and one can extract the
    phenomenological predictions more easily. In a
    pioneering work, DeWitt proposed that the determinant
    of the 3-metric can be used as internal time
    \cite{bdw1}. Consequently, in much of the literature on
    the Wheeler-DeWitt (\WDW) approach to quantum
    cosmology, the scale factor is assumed to play the role
    of time, although sometimes only implicitly. However,
    in closed models the scale factor fails to be monotonic
    due to classical recollapse and cannot serve as a
    global time variable already in the classical theory.
    Are there better alternatives at least in the simple
    setting of quantum cosmology?

\end{itemize}

Finally there is an important ultraviolet-infrared tension
\cite{gu}:

\begin{itemize}

\item Can one construct a framework that cures the
    short-distance limitations of classical general relativity
    near singularities, while maintaining an agreement with it
    at large scales?

\end{itemize}

By their very construction, perturbative and effective
descriptions have no problem with the low energy limit.
However, physically their implications can not be trusted at
the Planck scale and mathematically they generally fail to
provide a deterministic evolution across the putative
singularity. Since non-perturbative approaches often start from
deeper ideas, they have the potential to modify classical
dynamics in such a way that the big-bang singularity is
resolved. But once unleashed, do these new quantum effects
naturally `turn-off' sufficiently fast, away from the Planck
regime? The universe has had some \emph{14 billion years} to
evolve since the putative big-bang and even minutest quantum
corrections could accumulate over this huge time period leading
to observable departures from dynamics predicted by general
relativity. Thus, the challenge to quantum gravity theories is
to first create huge quantum effects that are capable of
overwhelming the extreme gravitational attraction produced by
matter densities of some $10^{94}\, {\rm gms/cc}$ near the
big-bang, and then switching them off with extreme rapidity as
the matter density falls below this Planck scale. This is a
huge burden!

In sections \ref{s2} -- \ref{s6} we will see that all these
issues have been satisfactorily addressed in LQC.

\subsection{Loop quantum cosmology}
\label{s1.4}

Wheeler's geometrodynamics program led to concrete ideas to
extract physics from the Dirac-Bergmann approach to canonical
quantum gravity already in the seventies \cite{bdw1,kk,jw1}.
However, mathematically the program still continues to remain
rather formal, with little control on the functional analysis
that is necessary to adequately deal with the underlying
infinite dimensional spaces, operators and equations.
Therefore, the older Wheeler-DeWitt (WDW) quantum cosmology did
not have guidance from a more complete theory. Rather, since
the cosmological symmetry reduction yields a system with only a
finite number of degrees of freedom, quantum kinematics was
built simply by following the standard Schr\"odinger theory
\cite{cwm}. Then, as we will see in section \ref{s3}, the
big-bang singularity generically persists in the quantum
theory.

The situation is quite different in LQG. In contrast to the
\WDW theory, a well established, rigorous kinematical framework
\emph{is} available in full LQG
\cite{almmt,alrev,crbook,ttbook}. If one mimics it in symmetry
reduced models, one is led to a quantum theory which is
\emph{inequivalent} to the \WDW theory already at the kinematic
level. Quantum dynamics built in this new arena agrees with the
\WDW theory in `tame' situations but differs dramatically in
the Planck regime, leading to new physics. This, in turn, leads
to a natural resolution of the big-bang singularity.

These developments occurred in three stages, each of which involved
major advances that overcame limitations of the previous one. As a
consequence, the viewpoint and the level of technical discussions
has evolved quite a bit and some of the statements made in the
literature have become outdated. Occasionally, then, there is an
apparent tension between statements made at different stages of this
evolution. Since this can be confusing to non-experts, we will
briefly summarize how the subject evolved. Readers who are not
familiar with the loop quantum cosmology literature can skip the
rest of this sub-section in the first reading without loss of
continuity.

The first and seminal contribution was Bojowald's result \cite{mb1}
that, in the FLRW model, the quantum Hamiltonian constraint of LQC
does not break down when the scale factor vanishes and the classical
singularity occurs. Since this was a major shift that overcame the
impasse of the \WDW theory, it naturally led to a flurry of activity
and the subject began to develop rapidly (see eg.
\cite{mb_isotropic,mb_inverse,mb_inflation,mb_kevin,mb_homo,bdv,tm}).
This success naturally drew scrutiny. Soon it became clear that
these fascinating results came at a cost: it was implicitly assumed
that $K$, the trace of the extrinsic curvature (or the Hubble
parameter, $\dot{a}/a$), is periodic, i.e. takes values on a circle
rather than the real line. Since this assumption has no physical
basis, at a 2002 workshop at Schr\"odinger Institute, doubts arose
as to whether the unexpectedly good behavior of the quantum
Hamiltonian constraint was an artifact of this assumption.

However, thanks to key input from Klaus Fredenhagen at the same
workshop, it was soon realized \cite{abl} that if one mimics the
procedure used in full LQG even more closely, the periodicity
assumption becomes unnecessary. In the full theory, the requirement
of diffeomorphism covariance leads to a \emph{unique} representation
of the algebra of fundamental operators \cite{lost,cf}. Following
the procedure in the full theory, in LQC one finds $K$ naturally
takes values on the real line as one would want physically. But, as
mentioned above, the resulting quantum kinematics is
\emph{inequivalent} to that of the \WDW theory. On this new arena,
one can still construct a well-defined quantum Hamiltonian
constraint, but now without having to assume the periodicity in $K$.
This new kinematical framework ushered in the second stage of LQC. A
number of early papers based on periodicity of $K$ cannot be taken
at their face value but results of \cite{abl} suggested how they
could be reworked in the new kinematical framework. This led to
another flurry of activity in which more general models were
considered. However, at this stage the framework was analogous to
the older \WDW theory in one respect: the models did not yet include
a physical Hilbert space or well-defined Dirac observables. While
there is a general method to introduce the physical inner product on
the space of solutions to the quantum constraints
\cite{dm,abc,almmt}, it could not be applied directly because often
the Hamiltonian constraint failed to be self-adjoint in these
models. Consequently, new questions arose. In particular, Brunnemann
and Thiemann \cite{bt} were led to ask: What is the precise sense in
which the physical singularity is resolved?

To address these key physical questions, one needs a physical
Hilbert space and a complete family of Dirac observables at least
some of which diverge at the singularity in the classical theory.
Examples are matter density, anisotropic shears and curvature
invariants (all evaluated at an instant of a suitably chosen
relational time). The question then is: Do the corresponding
operators all remain bounded on the \emph{physical} Hilbert space
even in the deep Planck regime? If so, one can say that the
singularity is resolved in the quantum theory. In the \WDW theory,
for example, generically these observables fail to remain bounded
whence the singularity is not resolved. What is the situation in
LQC?

The third stage of evolution of LQC began with the detailed
construction of a mathematical framework to address these issues
\cite{aps1,aps2,aps3}. The physical Hilbert space was constructed
using a massless scalar field $\phi$ as internal time. It was found
\cite{aps2} that the self-adjoint version of the Hamiltonian
constraint introduced in the second stage \cite{abl} ---called the
$\mu_o$ scheme in the literature--- does lead to singularity
resolution in the precise sense mentioned above. Since the detailed
theory could be constructed, the Hamiltonian constraint could be
solved numerically to extract physics in the Planck regime. But this
detailed analysis also brought out some glaring limitations of the
theory which had remained unnoticed because the physical sector of
the theory had not been constructed. (For details see, e.g.,
Appendix 2 of \cite{aps3}, and \cite{cs1}.) In a nutshell, while the
singularity was resolved in a well-defined sense, the theory
predicted large deviations from general relativity in the low
curvature regime: in terms of the key questions raised in section
\ref{s1.3}, it had infrared problems.

Fortunately, the problem could be traced back to the fact that
quantization of the Hamiltonian constraint had ignored a conceptual
subtlety. Roughly, at a key step in the procedure, the Hamiltonian
constraint operator of \cite{abl} implicitly used a kinematic
3-metric $\q_{ab}$ defined by the co-moving coordinates rather than
the physical metric $q_{ab}= a^2 \q_{ab}$ (where $a$ is the scale
factor). When this is corrected, the new, improved Hamiltonian
constraint again \emph{resolves the singularity and, at the same
time, is free from all three drawbacks of the $\mu_o$ scheme}. This
is an excellent example of the deep interplay between physics and
mathematics. The improved procedure is referred to as the `$\bar\mu$
scheme' in the literature. The resulting quantum dynamics has been
analyzed in detail and has provided a number of insights on the
nature of physics in the Planck regime. $\bar\mu$ dynamics has been
successfully implemented in the case of a non-zero cosmological
constant \cite{aps3,bp,ap,kp1}, the k=1, spatially compact case
\cite{apsv,warsaw1}, and to the Bianchi models
\cite{awe2,madrid-bianchi,awe3,we}. In the k=1 model, Green and
Unruh \cite{gu} had laid out more stringent tests that LQC has to
meet to ensure that it has good infrared behavior. These were met
successfully. Because of these advances, the $\bar\mu$ strategy has
received considerable attention also from a mathematical physics
perspective \cite{warsaw2,warsaw3,warsaw4,warsaw5}. This work uses a
combination of analytic and numerical techniques to enhance rigor to
a level that is unprecedented in quantum cosmology.

Over the last 4 years or so LQC has embarked the fourth stage
where two directions are being pursued. In the first, the
emphasis is on extending the framework to more and more general
situations (see in particular
\cite{ps,awe2,awe3,we,hybrid1,hybrid2,hybrid3,hybrid4,
hybrid5,kp1,aps4}). Already in the spatially homogeneous
situations, the transition from $\mu_o$ to $\bar\mu$ scheme
taught us that great care is needed in the construction of the
quantum Hamiltonian constraint to ensure that the resulting
theory is satisfactory both in the ultraviolet \emph{and}
infrared. The analysis of Bianchi models \cite{awe2,awe3,we}
has reinforced the importance of this requirement as a valuable
guide. The hope is that these generalizations will guide us in
narrowing down choices in the definition of the constraint
operator of full LQG. The second important direction is LQC
phenomenology. Various LQC effects have been incorporated in
the analysis of the observed properties of CMB particularly by
cosmologists (see, e.g.,
\cite{copeland1,copeland2,barrau1,barrau2,barrau3,barrau4,joao1,joao2}).
These investigations explored a wide range of issues,
including: i) effects of the quantum-geometry driven
super-inflation just after the big-bounce, predicted by LQC;
ii) production of gravitational waves near the big bounce and
LQC corrections to the spectrum of tensor modes; and iii)
possible chirality violations. They combine very diverse ideas
and are therefore important. However, in terms of heuristics
versus precision, there are large variations in the existing
literature and the subject is still evolving. As we will see in
sections \ref{s5} and \ref{s6}, over the last year or so this
frontier has begun to mature. It is likely to become the most
active forefront of LQC in the coming years.

\subsection{Organization} \label{s1.5}

In section \ref{s2} we will introduce the reader to the main issues
of quantum cosmology through the k=0 FLRW model
\cite{aps1,aps2,aps3}. We begin with the \WDW theory, discuss its
limitations and then introduce LQC which is constructed by paying
due attention to the Riemannian quantum geometry underlying full
LQG. In both cases we explain how one can use a relational time
variable to extract dynamics from the otherwise `frozen-formalism'
of the canonical theory. It turns out that if one uses the
relational time variable already in the classical Hamiltonian theory
\emph{prior} to quantization, the model becomes exactly soluble also
in LQC \cite{acs}. Section \ref{s3} is devoted to this soluble
model. Because the quantization procedure still mimics full LQG and
yet the model is solvable analytically, it leads to a direct and
more detailed physical understanding of singularity resolution. One
can also obtain precise results on similarities and differences
between LQC and the \WDW theory. We conclude section \ref{s3} with a
path integral formulation of the model. This discussion clarifies an
important conceptual question: How do quantum gravity corrections
manage to be dominant near the singularity in spite of the fact that
the classical action is large? As we will see, the origin of this
phenomenon lies in quantum geometry \cite{ach3}.

Section \ref{s4} is devoted to generalizations. We summarize results
that have been obtained in a number of models beyond the k=0, FLRW one:
the closed k=1 model, models with cosmological constant with either
sign, models with inflationary potentials and the Bianchi models
that admit anisotropies \cite{apsv,bp,ap,kp1,awe2,awe3,we}. In each
of these generalizations, new conceptual and mathematical issues
arise that initially appear to be major obstacles in carrying out
the program followed in the k=0 FLRW case. We explain these issues
and provide a succinct summary of how the apparent difficulties are
overcome. Although all these models are homogeneous, the
increasingly sophisticated mathematical tools that had to be
introduced to arrive at a satisfactory LQC provide useful guidance
for full LQG.

One of the most interesting outcomes of the detailed analysis of
several of the homogeneous models is the power of effective
equations \cite{jw,vt,psvt}. They involve only the phase space
variables without any reference to Hilbert spaces and operators.
Their structure is similar to the constraint and evolution equations
in classical general relativity; the quantum corrections manifest
themselves only through additional terms that explicitly depend on
$\hbar$. As in the classical theory, their solutions provide a
smooth space-time metric and smooth matter fields. Yet, in all cases
where the detailed evolution of \emph{quantum} states has been
carried out, effective equations have provided excellent
approximations to the full quantum evolution of LQC even in the deep
Planck regime, provided the states are semi-classical initially in
the low curvature regime \cite{aps3,apsv,bp,ap}. Therefore, section
\ref{s5} is devoted to this effective dynamics and its consequences
\cite{ps,cs1,cs2}. It brings out the richness of the Planck scale
physics associated with the singularity resolution and also sheds
new light on inflationary scenarios \cite{as2,as3,ck-inflation}.

Section \ref{s6} summarizes the research that goes beyond
homogeneity. We begin with a discussion of the one polarization
Gowdy models that admit infinitely many degrees of freedom,
representing gravitational waves. These models have been
analyzed in detail using a `hybrid' quantization scheme
\cite{hybrid1,hybrid2,hybrid3,hybrid4} in which LQC is used to
handle the homogeneous modes that capture the essential
non-trivial features of geometry, including the intrinsic time
variable, and the familiar Fock theory is used for other modes
that represent gravitational waves. Rather surprisingly, this
already suffices for singularity resolution; a full LQG
treatment of all modes can only improve the situation because
of the ultraviolet finiteness that is built into LQG. The
current treatment of this model is conceptually important
because it brings out the minimal features of quantum geometry
that are relevant to the singularity resolution. We then
summarize a framework to study general inhomogeneous
perturbations in an inflationary paradigm \cite{aan}. It
encompasses the Planck regime near the bounce where one must
use quantum field theory on cosmological \emph{quantum}
space-times \cite{akl}. This analysis has provided a step by
step procedure to pass from this more general theory to the
familiar quantum field theory on curved, classical space-times
that is widely used in cosmological phenomenology. Finally,
through a few specific examples we illustrate the ideas that
are being pursued to find observational consequences of LQC
and, reciprocally, to constrain LQG through observations
\cite{nucleo,barrau1,barrau2,barrau3,barrau4,bct,aan}.

In section \ref{s7} we provide illustrations of the lessons we have
learned from LQC for full LQG. These include guidance
\cite{aps3,apsv,warsaw1,awe2} for narrowing down ambiguities in the
choice of the Hamiltonian constraint in LQG and a viewpoint
\cite{awe1} towards entropy bounds that are sometimes evoked as
constraints that any satisfactory quantum gravity theory should
satisfy \cite{bousso1,bousso2}. A program to complete the
Hamiltonian theory was launched recently \cite{warsaw-full} based on
ideas introduced in \cite{rs-ham,kr2}. We provide a brief summary
because this program was motivated in part by the developments in
LQC and the construction of a satisfactory Hamiltonian constraint in
LQC is likely to provide further concrete hints to complete this
program. Next, we summarize the insights that LQC has provided into
spin foams \cite{perezrev,crbook} and group field theory
\cite{bou,gft1,gft2}. In broad terms, these are sum-over-histories
formulations of LQG where one integrates over \emph{quantum}
geometries rather than smooth metrics. Over the last three years,
there have been significant advances in the spin foam program
\cite{eprl,fk,newlook}. LQC provides an arena to test these ideas in
a simple setting. Detailed investigations \cite{ach1,ach2,chn,hrvw}
have provided concrete support for the paradigm that underlies these
programs and the program has also been applied to cosmology
\cite{rv,bkrv1,bkrv2}. Finally, the consistent histories framework
provides a generalization of the `Copenhagen' quantum mechanics that
was developed specifically to face the novel conceptual difficulties
of non-perturbative quantum gravity \cite{hartle-halliwell,hartle}.
Quantum cosmology offers a concrete and perhaps the most important
context where these ideas can be applied. We conclude section
\ref{s7} with an illustration of this application
\cite{consistent1,consistent3}.

Our conventions are as follows. We set $c=1$ but generally
retain $G$ and $\hbar$ explicitly to bring out the conceptual
roles they play in the Planck regime and to make the role of
quantum geometry more transparent. We will use Planck units,
setting $\lp^2 = G\hbar$ and $\mpl^2 = \hbar/G$ (rather than
the reduced Planck units often used in cosmology). The
space-time metric has signature - + + +. Lower case indices in
the beginning of the alphabet, $a,b,c, \ldots$ refer to
space-time (and usually just spatial) indices while
$i,j,k,\ldots$ are `internal' ${\rm SU(2)}$ indices. Basis in
the ${\rm su(2)}$ Lie algebra is given by the $2\times 2$
matrices $\tau^i$ satisfying $\tau_i\tau_j =
\f{1}{2}\epsilon_{ijk}\tau^k$. We regard the metric (and hence
the scale factor) as dimensionless so the indices can be raised
and lowered without changing physical dimensions. Most of the
plots are taken from original papers and have not been updated.

As mentioned in the abstract, this review is addressed to three sets
of readers. Cosmologists who are primarily interested in the basic
structure of LQC and its potential role in confronting theory with
observations may skip Section II, IV, VI.A and VII without a loss of
continuity. Similarly, mathematically inclined quantum gravity
readers can skip sections V and VI.D. Young researchers may want to
enter quantum gravity through quantum cosmology. They can focus on
sections II, III.A -- III.C, IV.A, IV.B, VA, VB, VII.A and VII.B in
the first reading and then return to other sections for a deeper
understanding. There are several other complementary reviews in the
literature. Details, particularly on the early developments, can be
found in \cite{mb-livrev}, a short summary for cosmologists can be
found in \cite{aa-paris}, for general relativists in \cite{aa-gr}
and for beginning researchers in \cite{aa-cimento,aa-badhonef}.
Because we have attempted to make this report self-contained, there
is some inevitable overlap with some of these previous reviews.

\section{k=0 FLRW Cosmology: Role of Quantum Geometry}
\label{s2}

The goal of this section is to introduce the reader to LQC.
Therefore will discuss in some detail the simplest cosmological
space-time, the k=0, $\Lambda=0$ FLRW model with a massless
scalar field. We will proceed step by step, starting with the
classical Hamiltonian framework and explain the conceptual
issues, such as the problem of time in quantum cosmology. We
will then carry out the \WDW quantization of the model. While
it has a good infrared behavior, it fails to resolve the
big-bang singularity. We will then turn to LQC. We now have the
advantage that, thanks to the uniqueness theorems of
\cite{lost,cf}, we have a well-defined kinematic framework for
full LQG. Therefore we can mimic its construction step by step
to arrive at a specific quantum kinematics for the FLRW model.
As mentioned in section \ref{s1.4}, because of quantum geometry
that underlies LQG, the LQC kinematics differs from the
Schr\"odinger theory used in the \WDW theory. The \WDW quantum
constraint fails to be well-defined in the new arena and we are
led to carry out a new quantization of the Hamiltonian
constraint that is tailored to the new kinematics. The ensuing
quantum dynamics
---and its relation to the \WDW theory--- is discussed in section
\ref{s3}. Since the material covered in this section lies at the
foundation of LQC, as it is currently practiced, discussion is
deliberately more detailed than it will be in the subsequent
sections.

\subsection{The Hamiltonian framework}
\label{s2.1}

This sub-section is divided into two parts. In the first we recall
the phase space formulation that underlies the \WDW theory and in
the second we recast it using canonical variables that are used in
LQG.

\subsubsection{Geometrodynamics}
\label{s2.1.1}

In the canonical approach, the first step toward quantization
is a Hamiltonian formulation of the theory. In the k=0 models,
the space-time metric is given by:
\be \label{g-proper} g_{ab}\,\dd x^a \dd x^b\, = \, -\dd t^2 +
q_{ab}\,\dd x^a \dd x^b \, \equiv \, -\dd t^2 + a^2 (\dd x_1^2 + \dd
x_2^2 + \dd x_3^2) \ee
where $q_{ab}$ is the physical spatial metric and $a$ is the scale
factor. Here the coordinate $t$ is the proper time along the world
lines of observers moving orthogonal to the homogeneous slices. In
the quantum theory, physical and mathematical considerations lead
one to use instead relational time, generally associated with
physical fields. In this section we will use a massless scalar field
$\phi$ as the matter source and it will serve as a physical clock.
Since $\phi$ satisfies the wave equation with respect to $g_{ab}$,
in LQC it is most natural to introduce a \emph{harmonic time
coordinate} $\tau$ satisfying $\Box \tau =0$. Then the space-time
metric assumes the form
\be \label{g-harmonic} g_{ab}\,\dd x^a \dd x^b \,\, =\,\, -a^6\, \dd
\tau^2 + q_{ab}\,\dd x^a \dd x^b \,\, \equiv\,\, -a^6\, \dd\tau^2 +
a^2 (\dd x_1^2 + \dd x_2^2 + \dd x_3^2) \ee
since the lapse function $N_\tau$, defined by $N_\tau \dd\tau = dt$,
is given by $a^3$. This form will be useful later in the analysis.

In the k=0 FLRW  models now under considerations, the spatial
topology can either be that of a 3-torus $\mathbb{T}^3$ or $\R^3$.
In non-compact homogeneous models, spatial integrals in the
expressions of the Lagrangian, Hamiltonian and the symplectic
structure all diverge. Therefore due care is needed in the
construction of a Hamiltonian framework \cite{as-bianchi}. Let us
therefore begin with the $\mathbb{T}^3$ topology. Then, the
co-moving coordinates define a non-dynamical, fiducial metric
$\q_{ab}$ via
\be \q_{ab} \dd x^{a} \dd x^{b} = \dd x_1^2 + \dd x_2^2 + \dd x_3^2
\qquad {\rm where}\,\, x^a \in [0,\ell_o] \,\, \hbox{for some
fixed}\,\, \ell_o \, .\ee
We will set $V_o = \ell_o^3$; this is the volume of $\mathbb{T}^3$
with respect to $\q_{ab}$. The physical 3-metric $q_{ab}$ will be
written as $q_{ab} = a^2 \q_{ab}$. Since we have fixed the fiducial
metric with a well-defined gauge choice, (unlike in the case with
$\R^3$ topology) the scale factor $a$ has direct physical meaning:
$V := a^3 V_o$ is the \emph{physical} volume of $\mathbb{T}^3$.

We can now start with the Hamiltonian framework for general
relativity coupled with a massless scalar field and systematically
arrive at the following framework for the symmetry reduced FLRW
model. The canonical variables are $a$ and $\pt = -a\dot{a}$
for geometry and $\phi$ and $\pphi= V\, \dot\phi$ for the
scalar field. \emph{Here and in what follows, `dot' denotes
derivative with respect to proper time $t$.} The non-vanishing
Poisson brackets are given by:
\be \label{pb1} \{a,\, \pt\}\, = \,\f{4\pi G}{3V_o}, \quad\quad
\{\phi, \pphi\} = 1\, .\ee
Because of symmetries, the (vector or the) diffeomorphism constraint
is automatically satisfied and the (scalar or the) Hamiltonian
constraint (i.e. the Friedmann equation) is given by
\be \label{hc1} C_H = \f{\pphi^2}{2V} - \f{3}{8\pi G}\,\f{\pt^2
V}{a^4} \, \approx 0\, .\ee

Next, let us now consider the $\R^3$ spatial topology. Now, we
cannot eliminate the freedom to rescale the Cartesian
coordinates $x^a$ and hence that of rescaling the fiducial
metric $\q_{ab}$. Therefore the scale factor no longer has a
direct physical meaning; only ratios of scale factors do. Also,
as mentioned above, volume integrals in the expressions of the
action, the Hamiltonian and the symplectic structure diverge. A
natural viable strategy is to introduce a fiducial cell $\C$
and restrict all integrals to it \cite{as-bianchi}. Because of
the symmetries of the k=0 model, we can let the cell be cubical
with respect to every physical metric $q_{ab}$ under
consideration on $\R^3$. The cell serves as an infrared
regulator which has to be removed to extract physical results
by taking the limit $\C \to \R^3$ at the end. We will find that
many of the results are insensitive to the choice of the cell;
in these cases, the removal of the limit is trivial.

Given $\C$, the phase space is again spanned by the quadruplet
$a,\pt;\, \phi,\pphi$; the fundamental non-vanishing Poisson
bracket are again given by (\ref{pb1}) and the Hamiltonian
constraint by (\ref{hc1}). However, now $V_o$ and $V$ refer to
the \emph{volume of the cell} $\C$ with respect to the fiducial
metric $\q_{ab}$ and the physical metric $q_{ab}$ respectively.
Thus, we now have the possibility of performing two rescalings
under which physics should not change:
\be \label{rescaling1} \q_{ab} \to \alpha^2 \q_{ab} \qquad \C \to
\beta^3 \C \, .\ee
One may be tempted to just fix the cell by demanding that its
fiducial (i.e. coordinate) volume be unit, thereby setting $\beta$
to $1$. But because there is no natural unit of length in classical
general relativity, for conceptual clarity (and for manifest
dimensional consistency in equations), it is best not to tie the
two. Under these recalings we have the following transformation
properties:
\ba \label{rescaling2}
    a &\to& \alpha^{-1}a,\qquad \pt \to\alpha^{-2}\pt,\qquad
    \phi\to \phi,\nonumber\\
    \pphi &\to& \beta^3\pphi,\qquad\,  V \to \beta^3 V, \qquad
    \,\,\,\,\, V_o \to \alpha^3\beta^3 V_o \, .\ea
Next, let us consider the Poisson brackets and the Hamiltonian
constraint. Since the Poisson brackets can be expressed in terms of
the symplectic structure on the phase space as $\{f,g\} =
\Omega^{\mu\nu}\,\p_\mu f\, \p_\nu g$ we have:
\be \label{rescaling3} \Omega^{\mu\nu} \to  \beta^{-3}
\Omega^{\mu\nu},\quad{\rm and} \quad C_H \to \beta^3 C_H \ee
Consequently, the Hamiltonian vector field $X_H^\mu =
\Omega^{\mu\nu}\p C_H$ is left invariant under both rescalings.
Thus, as one would hope, although elements of the Hamiltonian
formulation do make an essential use of the fiducial metric
$\q_{ab}$ and the cell $\C$, the final equations of motion are
insensitive to these choices. By explicitly taking the Poisson
brackets, it is easy to verify that we have:
\be \f{\ddot{a}}{a}\, =\,  -2\f{\dot{a}^2}{a^2}\, \equiv \,  -
\f{16\pi G}{3}\, \rho, \qquad {\rm and}\qquad  \ddot\phi =0  \ee
Cosmologists may at first find the introduction of a cell somewhat
strange because the classical general relativity makes no reference
to it. However, in passage to the quantum theory we need more than
just the classical field equations: We need either a well-defined
Hamiltonian theory (for canonical quantization) or a well-defined
action (for path integrals) which descends from full general
relativity. It is here that a cell enters. In the classical theory,
we know from the start that equations of motion do not require a
cell; cell-independence of the final physical results is  priori
guaranteed. But since elements that enter the very construction of
the quantum theory require the introduction of a cell $\C$, a priori
cell dependence can permeate the scalar product and definitions of
operators. The theory can be viable only if the final physical
results are well defined in the limit $\C \to \R^3$.

\subsubsection{Connection-dynamics}
\label{s2.1.2}

The basic strategy underlying LQG is to cast general relativity in a
form that is close to gauge theories so that: i) we have a unified
kinematic arena for describing all four fundamental forces of
Nature; and, ii) we can build quantum gravity by incorporating in it
the highly successful non-perturbative techniques based on Wilson
loops (i.e. holonomies of connections) \cite{aa-newvar}. Therefore,
as in gauge theories, the configuration variable is a gravitational
spin connection $A_a^i$ on a Cauchy surface $M$ and its conjugate
momentum is the electric field $E^a_i$ ---a Lie-algebra valued
vector field of density weight one on $M$. A key difference from
Yang-Mills theories is that the gauge group $\SU(2)$ does not refer
to rotations in some abstract internal space, but is in fact the
double cover of the rotation group ${\rm SO(3)}$ in the tangent
space of each point of $M$ (where the double cover is taken because
LQG has to accommodate fermions). Because of this `soldering' of the
gauge group to spatial geometry, the electric fields now have a
direct geometrical meaning: they represent orthonormal triads of
density weight 1. Thus, the contravariant physical metric on $M$ is
given by $ q^{ab} = q^{-1} E^a_i E^b_j\, \q^{ij}$ where $q^{-1}$ is
the inverse of the determinant $q$ of the covariant metric $q_{ab}$,
and $\q^{ij}$ the Cartan-Killing metric on the Lie algebra ${\rm
su(2)}$. To summarize, the canonical pair $(q_{ab}, p^{ab})$ of
geometrodynamics is now replaced by the pair $(A_a^i, E^a_i)$.
Because we deal with triads rather than metrics, there is now a new
gauge freedom, that of triad rotations. In the Hamiltonian theory
these are generated by a new Gauss constraint.

Let us now focus on the k=0 FLRW model with $\R^3$ spatial topology.
Again, a systematic derivation of the Hamiltonian framework requires
one to introduce a fiducial cell $\C$, which we again take to be
cubical. (As in geometrodynamics, for the $\mathbb{T}^3$ spatial
topology this is unnecessary.) As before, let us fix a fiducial
metric $\q_{ab}$ of signature +,+,+ and let $\e^a_i$ and $\o_a^i$ be
the orthonormal frames and co-frames associated with its Cartesian
coordinates $x^a$. The symmetries underlying FLRW space-times imply
that from each equivalence class of gauge related homogeneous,
isotropic pairs $(A_a^i, E^a_i)$ we can select one such that
\cite{abl}
\be \label{AE} A_a^i = \t{c}\, \o_a^i \qquad {\rm and} \qquad E^a_i
= \t{p}\, (\q)^{\f{1}{2}}\, \e^a_i \, . \ee
(In the literature, one often uses the notation $\delta^a_i$ for
$\e^a_i$ and $\delta_a^i$ for $\o_a^i$.) Thus, as one would expect,
the gauge invariant information in the canonical pair is again
captured in just two functions $(\t{c}, \t{p})$ of time. They are
related to the geometrodynamic variables via:
\be \t{c}= \gamma \dot{a} \quad {\rm and} \quad \t{p} = a^2 \ee
where $\gamma>0$ is the so-called Barbero-Immirzi parameter of LQG.
Whenever a numerical value is needed, we will set $\gamma \approx
0.2375$, as suggested by the black hole entropy calculations (see,
e.g., \cite{alrev}). It turns out that the equations of connection
dynamics in full general relativity are meaningful even when the
triad becomes degenerate. Therefore, the phase space of connection
dynamics is larger than that of geometrodynamics. In the FLRW
models, then, we are also led to enlarge the phase space by allowing
physical triads to have both orientations and, in addition, to be
possibly degenerate. On this full space, $\t{p} \in \R$, and $\t{p}
>0$ if $E^a_i$ and $\e^a_i$ have the same orientation, $\t{p} <0$ if
the orientations are opposite, and $p=0$ if $E^a_i$ is degenerate.

The LQC phase space is then topologically $\R^4$, naturally
coordinatized by the quadruplet $(\t{c},\t{p};\, \phi,\pphi)$. The
non-zero Poisson brackets are given by
\be \label{pb2} \{\t{c},\, \t{p}\} = \f{8\pi G\gamma}{3V_o} \qquad
{\rm and} \qquad \{\phi, \pphi\} =1 \ee
where as before $V_o$ is the volume of $\C$ with respect to the
fiducial metric $\q_{ab}$. As in geometrodynamics, the basic
canonical pair depends on the choice of the fiducial metric: under
the rescaling (\ref{rescaling1}) we have
\be \t{c} \to \alpha^{-1} \t{c}, \qquad {\rm and}\qquad \t{p} \to
\alpha^{-2} \t{p}  \, ,\ee
and the symplectic structure carries a cell dependence. Following
\cite{abl}, it is mathematically convenient to rescale the canonical
variables as follows
\be \label{pb3} {\rm set}\,\,\, c := V_o^{\f{1}{3}}\, \t{c},\,\,\,\,
p := V_o^{\f{2}{3}}\, \t{p},\quad {\hbox{\rm so that}}\quad
 \{c,\, p\} = \f{8\pi G\gamma}{3} \, . \ee
Then $c,p$ are insensitive to the choice of $\q_{ab}$ and the
Poisson bracket between them does not refer to $\q_{ab}$ \emph{or
to} the cell $\C$. Again because of the underlying symmetries (and
our gauge fixing) only the Hamiltonian constraint remains. It is now
given by:
\be \label{hc2} C_H = \f{\pphi^2}{2|p|^{\f{3}{2}}} - \f{3}{8\pi
G\,\gamma^2}\, |p|^{\f{1}{2}}\, c^2\, \approx 0 \ee
As before, $C_H$ and the symplectic structure are insensitive to the
choice of $\q_{ab}$ but they do depend on the choice of the fiducial
cell $\C$:
\be \label{rescaling4} \Omega^{\mu\nu} \to  \beta^{-3}
\Omega^{\mu\nu},\qquad{\rm and} \qquad C_H \to \beta^3 C_H \, . \ee
Since there is no $V_o$ on the right side of the Poisson bracket
(\ref{pb3}) it may seem surprising that the symplectic structure
still carries a cell dependence. But note that $\{c,\, p\} =
\Omega^{\mu\nu}\p_\mu c \p_\nu p$, and since $c,p$ transform as
\be \label{rescaling5} c \to \beta\,c \quad {\rm and}\quad  p\to
\beta^{2}\, p\ee
but the Poisson bracket does not change, it follows that
$\Omega^{\mu\nu}$ must transform via (\ref{rescaling4}).

So the situation with cell dependence is exactly the same as in
geometrodynamics: While the classical equations of motion and the
physics that follows from them are insensitive to the initial choice
of the cell $\C$ used in the construction of the Hamiltonian (or
Lagrangian) framework, a priori there is no guarantee that the final
physical predictions of the quantum theory will also enjoy this
property. That they must be well-defined in the limit $\C \to \R^3$
is an important requirement on the viability of the quantum theory.

The gravitational variables $c,p$ are directly related to the basic
canonical pair $(A_a^i,\, E^a_i)$ in full LQG and will enable us to
introduce a quantization procedure in LQC that closely mimics LQG.
However, we will find that quantum dynamics of the FLRW model is
significantly simplified in terms of a slightly different pair of
canonically conjugate variables, $(\b,\v)$:
\be \label{bv} \b := \f{c}{|p|^{\f{1}{2}}} 
,\,\,\,\, \v := \f{|p|^{\f{3}{2}}}{2\pi G}\, {\rm sgn}\,p \quad\quad
{\hbox{\rm so that}}\quad \{\b,\, \v\} = 2\gamma \ee
where ${\rm sgn}\, p$ is the sign of $p$ ($1$ if the physical triad
$e^a_i$ has the same orientation as the fiducial $\e^a_i$ and $-1$
if the orientation is opposite). In terms of this pair, the
Hamiltonian constraint becomes
\be \label{hc3} C_H = \f{\pphi^2}{4\pi G |\v|} - \f{3}{4\gamma^2}\b^2
|\v|\,\, \approx 0 \, .\ee
As with (\ref{hc1}) and (\ref{hc2}), canonical transformations
generated by this Hamiltonian constraint correspond to time
evolution in \emph{proper} time. As mentioned in the beginning of
this sub-section, it is often desirable to use other time
parameters, say $\tau$. The constraint generating evolution in
$\tau$ is $N_\tau\,C_H$ where $N_\tau = \dd t/\dd\tau$. Of
particular interest is the harmonic time that results if the scalar
field is used as an internal clock, for which we can set $N =a^3
\propto |\v|$.

We conclude  with a remark on triad orientations. Since we do not
have any spinor fields in the theory, physics is completely
insensitive to the orientation of the triad. Under this orientation
reversal we have $p \rightarrow -p$ and $\v \rightarrow -\v$. In the
Hamiltonian framework, constraints generate gauge transformations in
the connected component of the gauge group and these have been
gauge-fixed through our representation of $A_a^i, E^a_i$ by $c,p$.
The orientation flip, on the other hand, is a `large gauge
transformation' and has to be handled separately. We will return to
this point in section \ref{s2.6} in the context of the quantum
theory.

\subsection{The WDW theory}
\label{s2.2}

As remarked in section \ref{s1.4}, mathematically, full quantum
geometrodynamics continues to remain formal even at the kinematical
level. Therefore, in quantum cosmology, the strategy was to analyze
the symmetry reduced models in their own right \cite{cwm,swh1,ck}
without seeking guidance from the more complete theory. Since the
reduced FLRW system has only 2 configuration space degrees of
freedom, field theoretical difficulties are avoided from the start.
It appeared natural to follow procedures used in standard quantum
mechanics and use the familiar Schr\"odinger representation of the
canonical commutation relations that emerge from the Poisson
brackets (\ref{pb1}). But there is still a small subtlety. Because
$a>0$, its conjugate momentum cannot be a self-adjoint operator on
the Hilbert space $\H = L^2(\R^+, \dd a)$ of square integrable
functions $\Psi(a)$: If it did, its exponential would act as an
unitary displacement operator $\Psi(a) \to \Psi(a+\lambda)$ forcing
the resulting wave function to have support on negative values of
$a$ for some choice of $\lambda$. This difficulty can be easily
avoided by working with $z = \ln a^3$ and its conjugate momentum
(where we introduced the power of 3 to make comparison with LQC more
transparent in section \ref{s3}). Then, we have:
\be z = 3\ln a,\,\,\,\, \t{p}_{(z)} = \f{a}{3}\,\pt,\quad {\hbox{\rm
so that}}\,\, \{z,\, \t{p}_{(z)}\} = \f{4\pi G}{3V_o} \quad {\rm
and}\,\,\, C_H= \f{\pphi^2}{2V} - \f{27 V_o^2}{8\pi
G}\,\f{\t{p}_{(z)}^2 }{V} \, \approx 0 \ee

Now it is straightforward to carry out the Schr\"odinger
quantization. One begins with a kinematic Hilbert space $\Hkwdw =
L^2(\R^2, \dd z\dd\phi)$ spanned by wave functions $\Psi(z,\phi)$.
The operators $\h{z},\h\phi$ act by multiplication while their
conjugate momenta act by ($-i\hbar$ times) differentiation. The
Dirac program for quantization of constrained systems tells us that
the physical states are to be constructed from solutions to the
quantum constraint equation. Since $1/V$ is a common factor in the
expression of $C_H$, it is simplest to multiply the equation by $V$
before passing to quantum theory. We then have
\be \label{qhc1} \h{C}_H\, \ul{\Psi}(z,\phi)=0 \qquad {\rm
i.e.}\quad \p_\phi^2 \ul{\Psi}(z,\phi) = 12\pi G\, \partial_z^2
\ul{\Psi}(z,\phi)\,  \ee
where underbars will serve as reminders that the symbols refer to
the \WDW theory. The factor ordering we used in this constraint is
in fact independent of the choice of coordinates on the phase space;
it `covariant' in the following precise sense. The classical
constraint is quadratic in momenta, of the form $G^{AB}p_Ap_B$ where
$G^{AB}$ is the Wheeler DeWitt metric on the mini-superspace and the
quantum constraint operator is of the form $G^{AB}\nabla_A \nabla_B$
where $\nabla_A$ is the covariant derivative associated with
$G^{AB}$ \cite{ag}.

Note that this procedure is equivalent to using a lapse function
$N=a^3$. As explained in the beginning of section \ref{s2.1},
the resulting constraint generates evolution in harmonic time
in the classical theory. Since the scalar field $\phi$
satisfies the wave equation in space-time, it is natural to
regard $\phi$ as a relational time variable with respect to
which the scale factor $a$ (or its logarithm $z$) evolves. In
more general models, the configuration space is richer and the
remaining variables ---e.g., anisotropies, other matter fields,
density, shears and curvature invariants--- can be all regarded
as evolving with respect to this relational time variable.
However, this is a switch from the traditional procedure
adopted in the \WDW theory where, since the pioneering work of
DeWitt \cite{bdw1}, $a$ has been regarded as the internal time
variable. But in the closed, k=1 model, since $a$ is
double-valued on any dynamical trajectory, it cannot serve as a
global time parameter. The scalar field $\phi$ on the other
hand is single valued also in the k=1 case. So, it is
better suited to serve as a global clock.%
\footnote{$\phi$ shares one drawback with $a$: it also does not have
the physical dimensions of time ($[\phi] = [M/L]^{\f{1}{2}}$). But
in both cases, one can rescale the variable with suitable multiples
of fundamental constants to obtain a genuine harmonic time $\tau$.}

In the Dirac program, the quantum Hamiltonian constraint
(\ref{qhc1}) simply serves to single out \emph{physical} quantum
states. But none of them are normalizable on the kinematical Hilbert
space $\Hkwdw$ because the quantum constraint has the form of a
Klein-Gordon equation on the $(z,\phi)$ space and the wave operator
has a continuous spectrum on $\Hkwdw$. Therefore our first task is
to introduce a physical inner product on the space of solutions to
(\ref{qhc1}). The original Dirac program did not provide a concrete
strategy for this task but several are available \cite{aa-book,at}.
The most systematic of them is the `group averaging method'
\cite{dm,almmt,abc}. Since the quantum constraint (\ref{qhc1}) has
the form of a Klein-Gordon equation, as one might expect, the
application of this procedure yields, as physical states, solutions
to the positive (or, negative) frequency square root of the
constraint \cite{aps2},
\be \label{qhc2} -i\,\p_\phi \ul\Psi(z,\phi) =
\sqrt{\ul\Theta}\,\,\, \ul\Psi (z,\phi)\qquad {\rm with} \,\,\,
\ul\Theta = -12\pi G \p_z^2\, , \ee
where $\sqrt{\ul \Theta}$ is the square-root of the positive definite
operator $\Theta$ on $\Hkwdw$ defined, as usual, using a Fourier
transform. The physical scalar product is given by
\be \label{ip1} \ip{\ul\Psi_1}{\ul\Psi_2} = \int_{\phi=\phi_o}\!\!
\dd z\,\, {\ul{\bar\Psi}}_1(z,\phi)\,\ul{\Psi}_2(z,\phi) \ee
where the constant $\phi_o$ can be chosen arbitrarily because the
integral on the right side is conserved. The physical Hilbert space
$\Hpwdw$ is the space of normalizable solutions to (\ref{qhc2}) with
respect to the inner product (\ref{ip1}).

Let us summarize. By regarding the scalar field as an internal
clock and completing the Dirac program, one can interpret the
quantum constraint as providing a Schr\"odinger evolution
(\ref{qhc2}) of the \emph{physical} state $\ul\Psi$ with
respect to the internal time $\phi$. Conceptual difficulties
associated with the frozen formalism \cite{kk} in the
Bergmann-Dirac program are thus neatly bypassed. This is an
example of the \emph{deparametrization procedure} which enables
one to reinterpret the quantum constraint as an evolution
equation with respect to a relational time variable.

Next, we can introduce Dirac observables as self-adjoint operators
on $\Hpwdw$. Since $\h{p}_{(\phi)}$ is a constant of motion, it is
clearly a Dirac observable. Other useful observables are relational.
For example, the observable $\h{V}|_{\phi_o}$ that defines the
physical volume at a fixed instant $\phi_o$ of internal time $\phi$
is given by
\be \label{vol1} \h{V}|_{\phi_o}\, \ul\Psi(z,\phi) =
e^{i\sqrt{\ul\Theta}\,(\phi-\phi_o)}\,\, (e^{\h{z}}V_o)\,\,
e^{-i\sqrt{\ul\Theta}\,(\phi-\phi_o)}\, \ul\Psi(z,\phi)\, .\ee
Since classically $V = a^3V_o = e^z\, V_o$, conceptually the
definition (\ref{vol1}) simply corresponds to evolving the physical
state $\ul\Psi(z,\phi)$ back to the time $\phi=\phi_o$, operating on
it by the volume operator, and evolving the new wave function at
$\phi=\phi_o$ to all times $\phi$. Therefore, the framework enables
us to ask and answer physical questions such as: How do the
expectation value of (or fluctuations in) the volume or matter
density operator evolve with $\phi$?

These questions were first analyzed in detail by starting with a
unit lapse in the classical theory (so that the evolution is in
proper time), quantizing the resulting Hamiltonian constraint
operator, and finally re-interpreting the quantum constraint as
evolution in the scalar field time \cite{aps2,aps3}. Then the
constraint one obtains has a more complicated factor ordering;
although it is analogous to the Klein Gordon equation in
(\ref{qhc1}), $\ul\Theta$ is a rather complicated positive definite
operator whose positive square-root cannot be computed simply by
performing a Fourier transform. Therefore, the quantum evolution was
carried out using numerics. However, as we have seen, considerable
simplification occurs if one uses harmonic time already in the
classical theory. Then, one obtains the simple constraint
(\ref{qhc1}) so that the quantum model can be studied analytically.%
\footnote{Although setting $N=1$ and using proper time in the
classical theory makes quantum theory more complicated, it has
the advantage that the strategy is directly meaningful also in
the full theory. The choice $N=a^3$, by contrast does not have
a direct analog in the full theory.}

As one would hope, change of factor ordering only affects the
details; qualitative features of results are the same and can be
summarized as follows. Consider an expanding classical dynamical
trajectory in the $a,\phi$ space and fix a point $a_o,\phi_o$ on it
at which the matter density (and hence curvature) is very low
compared to the Planck scale. At $\phi=\phi_o$, construct a
`Gaussian' physical state $\ul\Psi(z,\phi_o)$ which is sharply
peaked at this point and evolve
it backward and forward in the internal time.%
\footnote{Because of the positive frequency requirement, due
care is necessary in constructing these `Gaussian' states. As
was pointed out in \cite{mv1}, an approximate analytical
argument following Eq. (3.19) in \cite{aps2} ignored this
subtlety. However, the exact numerical simulation did not; the
analytical argument should be regarded only as providing
intuition for explaining the general physical mechanism behind
the numerical result.}
Does this wave packet remain sharply peaked at the classical
trajectory? The answer is in the affirmative. This implies that
the \WDW theory constructed here has the correct infrared
behavior. This is an interesting and non-trivial feature
because even in simple quantum systems, dispersions in physical
observables tend to spread rapidly. However, the peakedness
also means that in the ultraviolet limit it is as bad as the
classical theory: matter density grows unboundedly in the past.
In the consistent histories framework, this translates to the
statement that the probability for the Wheeler-DeWitt quantum
universes to encounter a singularity is unity, independently of
the choice of state \cite{consistent2}. \emph{In this precise
sense, the big-bang singularity is not avoided by the WDW
theory.} The analytical calculation leading to this
conclusion is summarized in section \ref{s3.1}.\\

\textbf{Remarks:}

1) In the literature on the \WDW theory, the issue of
singularity resolution has often been treated rather loosely.
For example, it is sometimes argued that the wave function
vanishes at $a=0$ \cite{kks}. By itself this does not imply
singularity resolution because there is no a priori guarantee
that the physical inner product would be local in $a$ and, more
importantly, matter density and curvature may still grow
unboundedly at early times. Indeed, even in the classical
theory, if one uses harmonic (or conformal) time, the big-bang
is generally pushed back to the infinite past and so $a$ never
vanishes! Also, sometimes it is argued that the singularity is
resolved because the wave function becomes highly non-classical
(see, e.g., \cite{kks,np}). Again, this by itself does not mean
that the singularity is resolved; one needs to show, e.g., that
the expectation values of classical observables which diverge
in the classical theory remain finite there. In much of the
early literature, the physical inner product on the space of
solutions to the constraints was not spelled out whence one
could not even begin to systematically analyze of the behavior
of physical observables.

2) Notable exceptions are \cite{lr,lemos} where in the k=1 model
with radiation fluid, the physical sector is constructed using a
matter variable $T$ as relational time. The Hamiltonian constraint
takes the form $(\h{p}_{(T)} - \h{H}) \psi(a,T) =0$ where $\h{H}$ is
just the Hamiltonian of the harmonic oscillator. To use the standard
results from quantum mechanics of harmonic oscillators, the range of
$a$ is extended to the entire real line by assuming that the wave
function is \emph{antisymmetric} in $a$. It is then argued that the
expectation value of the volume ---$\h{V}= |\h{a}|^3V_o$ in our
terminology--- never vanishes. However, since by definition this
quantity could vanish only if the wave function has support just at
$a=0$, this property does not imply singularity resolution. As noted
above, one should show that operators corresponding to physical
observables that diverge in the classical theory are bounded above
in the quantum theory. This, to our knowledge, was not shown.

\subsection{Bypassing von Neumann's uniqueness}
 \label{s2.3}

In quantum mechanics of systems with a finite number of degrees
of freedom, a theorem due to von Neumann uniquely leads us to
the standard Schr\"odinger representation of the canonical
commutation relations. This is precisely the representation
used in the \WDW theory. The remaining freedom ---that of
factor ordering the Hamiltonian constraint--- is rather
limited. Therefore, the fact that the big-bang is not naturally
resolved in the \WDW theory was long taken to be an indication
that, due to symmetry reduction, quantum cosmology is simply
not rich enough to handle the ultraviolet issues near
cosmological singularities.

However, like any theorem, the von Neumann uniqueness result is
based on some assumptions. Let us examine them to see if they are
essential in quantum cosmology. Let us begin with non-relativistic
quantum mechanics and, to avoid unnecessary complications associated
with domains of unbounded operators, state the theorem using
exponentials $U(\sigma)$ and $V(\zeta)$ of the Heisenberg operators
$\hat{q},\hat{p}$. The algebra generated by these exponentials is
often referred to as the Weyl algebra.
\begin{quote}
Let $\mathfrak{W}$ be the algebra generated by 1-parameter groups of
(abstractly defined) operators $U(\sigma), V(\zeta)$ satisfying
relations:
\ba \label{weyl} U(\sigma_1)U(\sigma_2) &=&
U(\sigma_1+\sigma_2), \qquad V(\zeta_1) V(\zeta_2) = V(\zeta_1+\zeta_2)\nonumber\\
&& U(\sigma) V(\zeta) = e^{i\sigma\zeta} V(\zeta)U(\sigma) \ea
Then, up to unitary equivalence, there is an unique irreducible
representation of  $\mathfrak{W}$ on a Hilbert space $\H$ in which
$U(\sigma),V(\zeta)$ are unitary \emph{and weakly continuous} in the
parameters $\sigma,\zeta$. Furthermore, this representation is
unitarily equivalent to the Schr\"odinger representation with
$U(\sigma) = \exp i\sigma \h{q}$ and $V(\zeta) = \exp i \zeta \h{p}$
where $\h{q},\h{p}$ satisfy the Heisenberg commutation relations.
(For a proof, see, e.g., \cite{emch}.)
\end{quote}
The Weyl relations (\ref{weyl}) are just the exponentiated
Heisenberg commutation relations and the requirements of unitarity
and irreducibility are clearly natural. The requirement of weak
continuity is also well motivated in quantum mechanics because it is
necessary to have well defined self-adjoint operators
$\h{q},\,\h{p}$, representing the position and momentum observables.
What about the \WDW theory? Since a fundamental assumption of
geometrodynamics is that the 3-metric and its momentum be well
defined operators on $\Hkwdw$, again, the weak continuity is well
motivated. Therefore, in the \WDW quantum cosmology the theorem does
indeed lead uniquely to the standard Schr\"odinger representation
discussed in section \ref{s2.2}.

However, the situation is quite different in the connection dynamics
formulation of full general relativity. Here $\mathfrak{W}$ is
replaced by a specific $\star$-algebra $\mathfrak{a}_{\rm hf}$
(called the holonomy-flux algebra \cite{ai,al5,alrev}). It is
generated by holonomies of the gravitational connections $A_a^i$
along 1-dimensional curves and fluxes of the conjugate electric
fields $E^a_i$ ---which serve as orthonormal triads with density
weight $1$--- across 2-surfaces. Furthermore, even though the system
has infinitely many degrees of freedom, thanks to the background
independence one demands in LQG, theorems due to Lewandowski,
Okolow, Sahlmann and Thiemann \cite{lost} again led to a unique
representation of $\mathfrak{a}_{\rm hf}$! A little later
Fleischhack \cite{cf} established uniqueness using a $C^\star$
algebra $\mathfrak{W}_{\rm F}$ generated by holonomies and the
exponentials of the electric flux operators, which is analogous to
the Weyl algebra $\mathfrak{W}$ of quantum mechanics. As was the
case with von-Neumann uniqueness, this representation had already
been constructed in the 90's and used to construct the kinematics of
LQG \cite{alrev}. The uniqueness results provided a solid foundation
for this framework.

However, there is a major departure from the von Neumann uniqueness.
A fundamental property of the LQG representation is that while the
holonomy operators $\h{h}$ are well defined, there is no local
operator $\h{A}_a^i(x)$ corresponding to the gravitational
connection. Since classically the holonomies are (path ordered)
exponentials of the connection, uniqueness theorems of
\cite{lost,cf} imply that the analog of the weak continuity with
respect to $\sigma$ in $U(\sigma)$ \emph{cannot} be met in a
background independent dynamical theory of connections. Thus, a key
assumption of the von Neumann theorem is violated. As a consequence
even after symmetry reduction, we are led to new representations of
the commutation relations which replace the Schr\"odinger theory.
\emph{LQC is based on this new quantum kinematics.}

To summarize, since we have a fully rigorous kinematical framework
in LQG, we can mimic it in the symmetry reduced systems. As we will
see in the next sub-section, this procedure forces us to a theory
which is inequivalent to the \WDW theory already at the kinematical
level.

\subsection{LQC: kinematics of the gravitational sector}
\label{s2.4}

Let us begin by spelling out the `elementary' functions on the
classical phase space which are to have well-defined analogs in
quantum theory (without any factor ordering or regularization
ambiguities). We will focus on the geometrical variables $(c,p)$
because the treatment of matter variables is the same as in the \WDW
theory. As in full LQG, the elementary variables are given by
holonomies and fluxes. However, because of homogeneity and isotropy,
it now suffices to consider holonomies only along edges of the
fiducial cell $\C$ and fluxes across faces of $\C$; these functions
form a `complete set' (in the sense that they suffice to separate
points in the LQC phase space). Let $\ell$ be a line segment
parallel to the $k$th edge of $\C$ and let its length w.r.t.
\emph{any} homogeneous, isotropic metric $q_{ab}$ on $M$ be $\mu$
times the length of the $k$th edge of $\C$ w.r.t. the \emph{same}
$q_{ab}$. Then, a straightforward calculation shows that the
holonomy of the connection $A_a^i= \t{c}\,\o_a^i$ along $\ell$ is
given by:
\be \label{hol} h_\ell\, =\, \exp\, [\t{c}\,\mu V_o^{\f{1}{3}}\,
\tau_k]\,\, =\, \cos \f{\mu c}{2} \mathbb{I} + 2\sin \f{\mu c}{2}\,
\tau_k\, . \ee
where $\tau_k$ is a basis in su(2) introduced in section
\ref{s1.5} and $\mathbb{I}$ is the identity $2\times 2$ matrix.
Hence it suffices to restrict ourselves only to $\N_\mu (c):=
e^{i\mu c/2}$ as elementary functions of the configuration
variable $c$. Next, the flux of the electric field, say\,
$E^a_3$,\, is non-zero only across the 1-2 face of $\C$ and is
given simply by $p$. Therefore, following full LQG, we will let
the triad-flux variable be simply $p$. These are the elementary
variables which are to have unambiguous quantum analogs,
$\h{\N}_{(\mu)}$ and $\h{p}$. Their commutation relations are
dictated by their Poisson brackets:
\be \label{ccr1} \{\N_{(\mu)},\, p\} = \f{8\pi i\gamma G}{3}\,
\f{\mu}{2}\, \N_{(\mu)} \quad \Rightarrow\quad  [\h{\N}_{(\mu)}, \,
\h{p} ] = - \f{8\pi \gamma G \hbar}{3}\, \f{\mu}{2}\, \h{\N}_{(\mu)}
\ee
The holonomy flux algebra $\mathfrak{a}$ of the FLRW model is (the
free $\star$ algebra) generated by $\h{\N}_{(\mu)}, \h{p}$ subject
to the commutation relations given in (\ref{ccr1}). The idea is to
find the representation of $\mathfrak{a}$ which is the analog of the
canonical representation of the full holonomy-flux algebra
$\mathfrak{a}_{\rm hf}$ \cite{ai,al5,acz,alrev} singled out by the
uniqueness theorem of \cite{lost}. Now, in LQG the canonical
representation of $\mathfrak{a}_{\rm hf}$ can be constructed using a
standard procedure due to Gel'fand, Naimark and Segal \cite{gns1},
and is determined by a specific expectation value (i.e., a positive
linear) functional on $\mathfrak{a}_{\rm hf}$. In the FLRW model,
one uses the `same' expectation value functional, now on
$\mathfrak{a}$. As one would expect, the resulting quantum theory
inherits the salient features of the full LQG kinematics.

We can now summarize this representation of $\mathfrak{a}$. Quantum
states are represented by \emph{almost periodic} functions $\Psi(c)$
of the connection $c$ ---i.e., countable linear combinations
$\sum_n\, \alpha_n \, \exp i\mu_n (c/2)$ of plane waves, where
$\alpha_n \in \mathbb{C}$ and $\mu_n \in \R$.%
\footnote{These functions are called \emph{almost periodic} because
the coefficients $\mu_n$ are allowed to be arbitrary real numbers,
rather than integral multiples of a fixed number. It is interesting
to note that this theory of almost periodic functions was developed
by the mathematician Harold Bohr, Niels' brother.}
The (gravitational) kinematic Hilbert space $\Hkg$ is spanned by
these $\Psi(c)$ with a finite norm:
\be ||\Psi||^2 \,=\, \lim_{D\to \infty}\,
\frac{1}{2D}\,\int_{-D}^{D} \bar\Psi(c)\, \Psi(c)\,\, \dd c \,\,\,
=\, \sum_n |\alpha_n|^2 \, . \ee
Note that normalizable states are \emph{not} integrals $\int
\dd\mu\, \alpha(\mu)\, e^{i\mu (c/2)}$ but \emph{discrete} sum of
plane waves $e^{i\mu(c/2)}$. Consequently, the intersection between
$\Hkg$ and the more familiar Hilbert space $L^2(\R, \dd c)$ of
quantum mechanics (or of the \WDW theory) consists only of the zero
function. Thus, already at the kinematic level, the LQC Hilbert
space is \emph{very} different from that used in the \WDW theory. An
orthonormal basis in $\Hkg$ is given by the almost periodic
functions of the connection, $\N_{(\mu)}(c) := e^{i\mu c/2}$. (The
$N_{(\mu)} (c)$ are in fact the LQC analogs of the spin network
functions in full LQG \cite{rs2,jb2}). They satisfy the relation
\be \label{basis} \langle \N_{(\mu)}|\N_{(\mu^\prime)}\rangle \,
\equiv \, \langle e^{\f{i\mu c}{2}}|e^{\f{i\mu^\prime c}{2}} \rangle
\,\, = \,\, \delta_{\mu, \mu^\prime}\, . \ee
Again, note that although the basis is of the plane wave type, the
right side has a Kronecker delta, rather than the Dirac
distribution.

The action of the fundamental operators, however, is the familiar
one. The configuration operator acts by multiplication and the
momentum by differentiation:
\be \label{ops1} \hat{\N}_{(\sigma)}\Psi (c)= \exp \f{i\sigma c}{2}
\Psi(c), \quad {\rm and} \quad \hat{p}\, \Psi (c) \, =\,  -i
\f{8\pi\gamma \lp^2}{3}\, \f{\dd\Psi}{\dd c} \ee
where, as usual, $\lp^2 =G\hbar$. The first of these provides a
1-parameter family of unitary operators on $\Hkg$ while the second
is self-adjoint. Finally, although the action of $\h{\N}_{(\sigma)}$
is \emph{exactly} the same as in the standard Schr\"odinger theory,
there is a key difference in their properties because the underlying
Hilbert spaces are distinct: Its matrix elements fail to be
continuous in $\sigma$ at $\sigma=0$. (For example, the expectation
value $\langle \N_{(\mu)}(c)|\,\h{\N}_{(\sigma)}|\,
\N_{(\mu)}(c)\rangle$ is zero if $\sigma\not= 0$ but is $1$ at
$\sigma=0$.) This is equivalent to saying that $\h{\N}_{(\sigma)}$
fails to be weakly continuous in $\sigma$; this is the assumption of
the von Neumann uniqueness theorem that is violated. \emph{As a
result we cannot introduce a connection operator $\h{c}$ by taking
the derivative of $\h{\N}_{(\mu)}$} as we do in the Schr\"odinger
(and the \WDW ) theory. (For further discussion, see
\cite{afw,cvz2}.)

Since $\h{p}$ is self-adjoint, it is often more convenient to use
the representation in which it is diagonal. Then quantum states are
functions $\Psi(\mu)$ of $\mu$ with support only on a
\emph{countable} number of points. The scalar product is given by
\be \label{ip2} \ip{\Psi_1}{\Psi_2} = \sum_n\, \bar{\Psi}_1(\mu_n)\,
\Psi_2(\mu_n)\, ; \ee
and the action of the basic operators by
\be \label{ops2} \h{\N}_{(\alpha)}\, \Psi(\mu) =
\Psi(\mu+\alpha)\qquad {\rm and} \qquad
 \h{p}\,\Psi(\mu) = \f{8\pi \gamma \lp^2}{6}\, \mu \Psi(\mu)\, . \ee
Finally, there is one conceptual subtlety that we need to address.
In the \WDW theory, $\mu \sim a^2$ is positive while in LQC because
of the freedom in the orientation of triads, it takes values on the
entire real line. However, as discussed at the end of section
\ref{s2.1.2}, orientation reversal of triads, i.e. map $\mu \to
-\mu$, corresponds to large gauge transformations which are not
generated by constraints and therefore not eliminated by gauge
fixing. Nonetheless, in absence of fermions, they represent gauge
because they do not change physics. In the quantum theory, these
large gauge transformations are induced by the `parity' operator,
$\hat\Pi$, with the action $\hat\Pi\, \Psi (\mu) = \Psi(-\mu)$.
There is a well-established procedure to incorporate large gauge
transformations in the Yang-Mills theory: Decompose the state space
into irreducible representations of this group and discuss each of
them separately. In the present case, since $\h{\Pi}^2=1$, we are
led to consider wave functions which are either symmetric or
anti-symmetric under $\hat\Pi$. There is no qualitative difference
in the physics of these sectors \cite{bp}. However, since
anti-symmetric wave functions vanish at $\mu=0$, to avoid the wrong
impression that the singularity resolution is `put by hand', in LQC
it is customary to work with the symmetric representation. Thus,
from now, we will restrict ourselves to states satisfying $\Psi(\mu)
= \Psi(-\mu)$.

To summarize, by faithfully mimicking the procedure used in
full LQG, we arrive at a kinematical framework for LQG which is
inequivalent to the Schr\"odinger representation underlying
\WDW theory. We are now led to revisit the issue of singularity
resolution in this new quantum arena.

\subsection{LQC: Gravitational part of the Hamiltonian constraint}
\label{s2.5}

In this sub-section we will construct the quantum Hamiltonian
constraint on the gravitational part of the kinematic Hilbert space
$\Hkg$. Right at the beginning we encounter a problem: While the
expression (\ref{hc2}) of the classical constraint has a factor
$c^2$, there is no operator $\h{c}$ on $\Hkg$. To compare the
situation with the \WDW theory, let us use the representation in
which $\mu$ (or, the scale factor) is diagonal and $c$ is analogous
to the momentum $p_z$. Now, in the \WDW theory, $\h{p}_z$ acts by
differentiation: $\h{p}_z \psi(z) \sim -i\, \dd\Psi/\dd z$. But
since $\Psi(\mu)$ has support on a \emph{countable} number of points
in LQC, we cannot define $\h{c}\,\Psi(c)$ using $-i\, \dd\Psi/\dd
\mu$. Thus, because the kinematic arena of LQC is qualitatively
different from that of the \WDW theory, we need a new strategy.

For this, we return to the key idea that drives LQC: Rather than
introducing ab-initio constructions, one continually seeks guidance
from full LQG. There, the gravitational part of the Hamiltonian
constraint is given by
\ba \label{cgrav} C_{\mathrm{grav}} &=& - \gamma^{-2}\,
\textstyle{\int}_{\cal C}\,\, \dd^3 x\,\big[N ({\det
q})^{-\f{1}{2}}\,\epsilon^{ij}{}_k E^a_i
E^b_j\big]\,\, F_{ab}{}^k\nonumber\\
&=& - \gamma^{-2} V_o^{-\f{1}{3}} \, \epsilon^{ij}{}_k
\,\e^a_i\e^b_j\,\, |p|^2 F_{ab}{}^k \, \ea
where, in the second step we have evaluated the integral in the FLRW
model and, as in the \WDW theory, geared the calculation to harmonic
time by setting $N=a^3$. The non-trivial term is $F_{ab}{}^k$ ($\sim
c^2$ when expanded out using the form (\ref{AE}) of the connection).
Now, while holonomies are well defined operators on the kinematic
Hilbert space of full LQG, there is no local operator corresponding
to the field strength. Therefore, as in gauge theories, the idea is
to expresses $F_{ab}{}^k$ in terms of holonomies.

\subsubsection{The non-local curvature operator}
\label{s2.5.1}

Recall first from differential geometry that the $a$-$b$ component
of $F_{ab}{}^k$ can be written in terms of holonomies around a
plaquette in the $a$-$b$ plane:
\be \label{F1} F_{ab}{}^k = 2\lim_{Ar_\Box\rightarrow 0}\,\,\,
\mathrm{Tr}\,\,\left(\f{h_{\Box_{ij}}- \mathbb{I}} {Ar_\Box}
\:\tau^k\right)\, \o_a^i\, \o_b^j, \ee
where $Ar_\Box$ is the area of the plaquette $\Box$. In the FLRW
model, because of spatial homogeneity and isotropy, it suffices to
compute $F_{ab}{}^k$ at any one point and use square plaquettes that
lie in the faces of $\C$ with edges that are parallel to those of
$\C$. Then the holonomy $h_{\Box_{ij}}$ around the plaquette
$\Box_{ij}$ is given by
\be \label{holonomy} h_{\Box_{ij}}= {h_j^{(\bar\mu)}}^{-1}\,
{h_i^{(\bar\mu)}}^{-1}\, h_j^{(\bar\mu)}\, h_i^{(\bar\mu)}\, . \ee
where there is no summation over $i,j$ and $\bar\mu$ is the (metric
independent) ratio of the length of any edge of the plaquette with
the edge-length of the cell $\C$. Prior to taking the limit, the
expression on the right side of (\ref{F1}) can be easily promoted to
a quantum operator in LQG or LQC. However, the limit does not exist
precisely because the weak continuity with respect to the edge
length $\bar\mu$ fails. Now, the uniqueness theorems \cite{lost,cf}
underlying LQG kinematics imply that this absence of weak continuity
is a direct consequence of background independence \cite{aa-je}.
furthermore, it is directly responsible for the fact that the
eigenvalues of geometric operators ---in particular, the area
operator--- are purely discrete. Therefore, in LQC the viewpoint is
that the non-existence of the limit $Ar_\Box \rightarrow 0$ in
quantum theory is not accidental: quantum geometry is simply telling
us that we should shrink the plaquette not till the area it encloses
goes to zero, but only till it equals the minimum non-zero
eigenvalue $\Delta \lp^2$ of the area operator. That is, the action
$\hat{F}_{ab}^k \Psi(\mu)$ is to be determined essentially by the
holonomy operator $\h{h}_{\Box_{ij}} \Psi(\mu)$ where the area
enclosed by the plaquette $\Box$ is $\Delta \lp^2$ \emph{in the
quantum geometry determined by} $\Psi(\mu)$. The resulting
$\h{F}_{ab}{}^k$ would be non-local at Planck scale and the local
curvature used in the classical theory will arise only upon
neglecting quantum geometry at the Planck scale, e.g. by
coarse-graining of suitable semi-classical states.

To implement this idea it only remains to specify the plaquette,
i.e., calculate the value of $\bar\mu$ in (\ref{holonomy}) that
yields the desired plaquette $\Box$.

\subsubsection{Determining $\bar\mu$}
\label{s2.5.2}

The strategy is to use a heuristic but well-motivated correspondence
between kinematic states in LQG and those in LQC. Fix an eigenstate
$\Psi_\alpha(\mu) = \delta_{\mu,\alpha}$ of geometry in the LQC
Hilbert space $\Hkg$. It represents quantum geometry in which each
face of $\C$ has area $|\alpha|\,(4\pi/3) \gamma \lp^2 $. How would
this quantum geometry be represented in full LQG? First, the
macroscopic geometry must be spatially homogeneous and, through our
initial gauge fixing, we have singled out three axes with respect to
which our metrics $q_{ab}$ are diagonal (and to which the cubical
cell $\C$ is adapted.) Therefore, semi-heuristic considerations
suggest that the corresponding LQG quantum geometry state should be
represented by a spin network consisting of edges parallel to the
three axes (see Fig. 1(a)). Microscopically this state is not
exactly homogeneous. But the \emph{coarse grained} geometry should
be homogeneous. To achieve the best possible coarse grained
homogeneity, the edges should be `packed as tightly as is possible'
in the desired quantum geometry. That is, each edge should carry the
smallest non-zero label possible, namely $j =1/2$.

\begin{figure}[tb]
  \begin{center}
  \subfigure[]
      {\includegraphics[width=2.5in]{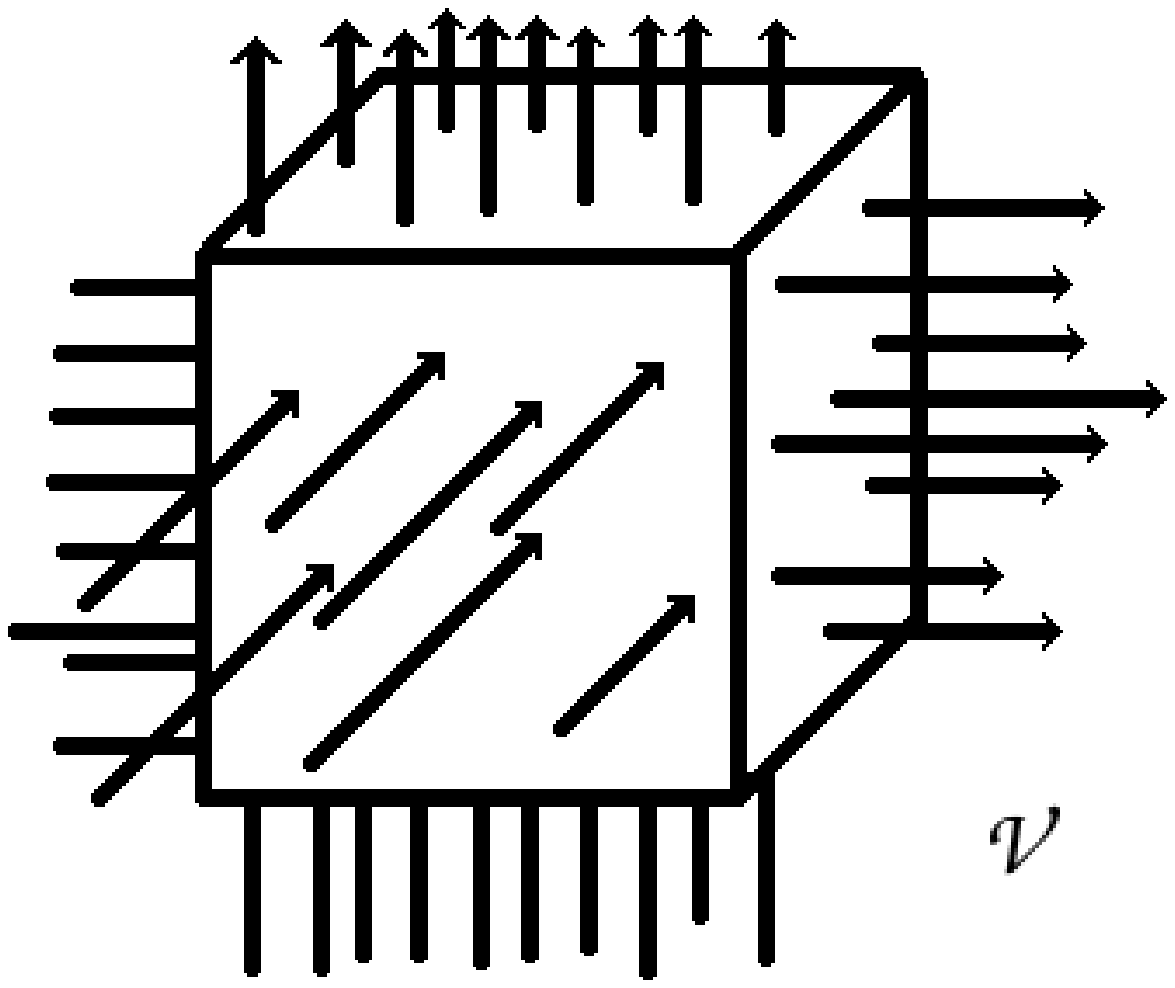}}
    \subfigure[]
      {\includegraphics[width=3.5in]{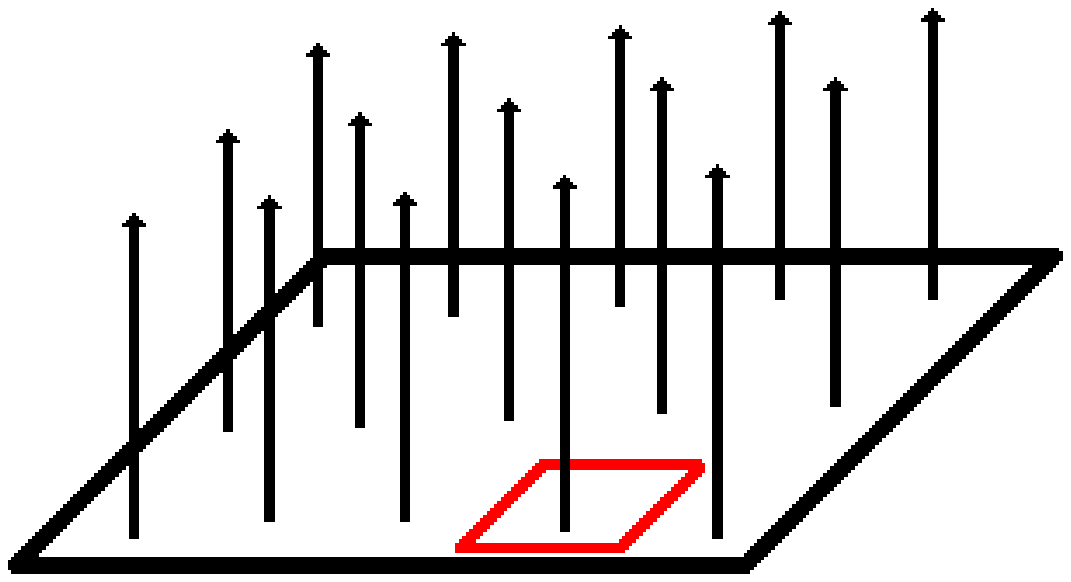}}
    \caption{Depiction of the LQG quantum geometry state corresponding to
    the LQC state $\Psi_\alpha$. The LQG spin-network has edges parallel
    to the edges of the cell $\C$, each carrying a spin label $j=1/2$.\,\,
    (a) Edges of the spin network traversing through the fiducial cell $\C$.
    (b) Edges of the spin network traversing the 1-2 face of $\C$ and
        an elementary plaquette associated with a single flux line. This plaquette
        encloses the smallest quantum, $\Delta\, \lp^2$, of area. The curvature
        operator $\hat{F}_{12}{}{}^k$ is defined by the holonomy around such a
        plaquette.}
  \label{fig2}
 \end{center}
\end{figure}

For definiteness, let us consider the 1-2 face $S_{12}$ of the
fiducial cell $\C$ which is orthogonal to the $x_3$ axis (see Fig.
1(b)). Quantum geometry of LQG tells us that at each intersection of
any one of its edges with $S_{12}$, the spin network contributes a
quantum of area $\Delta\, \lp^2$ on this surface, where $\Delta =
4\sqrt{3}\pi \gamma$ \cite{al5}. For this LQG state to reproduce the
LQC state $\Psi_\alpha(\mu)$ under consideration, $S_{12}$ must be
pierced by $N$ edges of the LQG spin network, where $N$ is given by
$$N\, \Delta\, \lp^2 = |p| \equiv \f{4\pi \gamma\lp^2}{3} |\alpha|\, .$$
Thus, we can divide $S_{12}$ into $N$ identical squares each of
which is pierced by exactly one edge of the LQG state, as in Fig.
1(b). Any one of these elementary squares encloses an area $\Delta
\lp^2$ and provides us the required plaquette $\Box_{12}$. Recall
that the dimensionless length of each edge of the plaquette is
$\bar\mu$. Therefore their length with respect to the fiducial
metric $\q_{ab}$ is $\bar{\mu}\,V_o^{1/3}$. Since the area of
$S_{12}$ with respect to $\q_{ab}$ is $V_o^{2/3}$, we have
$$ N\,\, (\bar\mu V_o^{\f{1}{3}})^2\, = V_o^{\f{2}{3}}. $$
Equating the expressions of $N$ from the last two equations, we
obtain
\be \bar\mu^2 = \f{\Delta\, \lp^2}{|p|}\,\, \equiv \,\,
\f{3\sqrt{3}}{|\alpha|}\, .  \ee
Thus, on a general state $\Psi(\mu)$ the value of $\bar\mu$ we
should use in (\ref{holonomy}) is given by
\be \label{mubar} \bar\mu =
\big(\f{3\sqrt{3}}{|\mu|}\big)^{\f{1}{2}}. \ee

To summarize, by exploiting the FLRW symmetries and using a simple
but well-motivated correspondence between LQG and LQC states we have
determined the unknown parameter $\bar\mu$ and hence the required
elementary plaquettes enclosing an area $\Delta\,\lp^2$ on each of
the three faces of the cell $\C$.

\subsubsection{The final expression}
\label{s2.5.3}

It is now straightforward to compute the product of holonomies
in (\ref{holonomy}) using (\ref{hol}) and arrive at the
following expression of the field strength operator:
\be \label{F2} \h{F}_{ab}^k\, \Psi(\mu) = \epsilon_{ij}{}^k\,
V_o^{-\f{2}{3}}\,\o_a^i\,\o_b^j\,
\widehat{\big(\f{\sin^2\bar{\mu}c}{\bar{\mu}^2}\big)}\,
 \Psi(\mu)  \ee
where, for the moment, we have postponed the factor ordering issue.
From now on, for notational simplicity we will generally drop hats
over trignometric operators. To evaluate the right side of
(\ref{F2}) explicitly, we still need to find the action of the
operator ${\exp [i\bar\mu\, (c/2)]}$ on $\Hk$. This is not
straightforward because $\bar\mu$ is \emph{not} a constant but a
function of $\mu$. However recall that $e^{i\mu_o (c/2)}$ is a
displacement operator: ${\exp [i\mu_o (c/2)]}\, \Psi(\mu) =
\Psi(\mu+\mu_o)$. That is, the operator just drags the wave function
a unit affine parameter distance along the vector field
$(\mu_o)\dd/\dd\mu$. A geometrical argument tells us that the action
of ${\exp {i\bar\mu\, (c/2)}}$ is completely analogous: it drags the
wave function a unit affine parameter distance along the vector
field $(\bar\mu)\, \dd/\dd\mu$. The action on wave functions
$\Psi(\mu)$ has been spelled out in \cite{aps3} but is rather
complicated. But it was also shown in \cite{aps3} that the action
simplifies greatly if we exploit the fact that the affine parameter
is proportional to $|\mu|^{3/2}\sim \v$ (see (\ref{bv})).

The idea then is to make a trivial transition to the volume
representation. For later convenience, let us rescale $\v$ by
setting
\be \label{nu} \nu = \f{\v}{\gamma\hbar}  \ee
and regard states as functions of $\nu$ rather than $\mu$. This
change to the `volume representation' is trivial because of the
simple form (\ref{ip2}) of inner product on $\Hkg$ of LQC. Since the
volume representation is widely used in LQC, let us summarize its
basic features. The kinematical Hilbert space $\Hkg$  consists of
complex valued functions $\Psi(\nu)$ of $\nu\in \R$ with support
only on a countable number of points, and with a finite norm:
\be ||\Psi||^2 = \sum_n\, |\Psi(\nu_n) |^2 \,  \ee
where $\nu_n$ runs over the support of $\Psi$. In this
representation, the basic operators are $\h{V}$, the volume of the
fiducial cell $\C$, and its conjugate $\exp {i \sigma\b}$ where
$\sigma$ is a parameter (of dimensions of length) and $\b$ is
defined in (\ref{bv}). The volume operator acts simply by
multiplication:
\be \label{eq:V_nu}
\h{V}\Psi(\nu) = 2\pi \gamma \lp^2 |\nu|\, \Psi(\nu)\, ,     \ee
while the conjugate operator acts via displacement:
\be \label{displacement} [\exp i\sigma \b]\, \Psi (\nu) = \Psi(\nu -
2\sigma) \, ,\ee
where $\sigma$ is a parameter. ($\b$ has dimension $[L]^{-1}$ and
$\nu$ has dimension $[L]$. For details, see \cite{aps2,awe2}.) In
terms of the original variables $c,\mu$ we have $\bar\mu c = \lambda
\b$ where $\lambda$ is a constant, the square-root of the area gap:
\be \lambda^2 = \Delta\, \lp^2 = 4\sqrt{3}\pi\gamma\lp^2\, , \ee
whence in the systematic procedure from the $\mu$ to the $\nu$
representation, the operator $\exp\, i{\bar\mu}c$ becomes just $\exp
i \lambda\b$.

Substituting (\ref{F2}) in the expression of the gravitational part
(\ref{cgrav}) of the Hamiltonian constraint, we obtain
\be \label{qhc3} \h{C}_{\rm grav}\, \Psi (\nu,\phi)
= -24\pi^2 G^2\gamma^2\hbar^2\,\, |\nu| \f{\sin{\lambda\b}}{\lambda}
|\nu| \f{\sin \lambda \b}{\lambda}\, \Psi(\nu, \phi) \ee
where we have chosen a factor ordering analogous to the covariant
ordering in the \WDW theory.

We will conclude with a discussion of the important features of
this procedure and properties of $\h{C}_{\rm grav}$.

1. Since the operator $\h{F}_{ab}{}^k$ is defined in terms of
holonomies and the loop has not been shrunk to zero area, at a
fundamental level curvature is non-local in LQC. This
non-locality is governed by the area gap, and is therefore at a
Planck scale. In the classical limit, one recovers the familiar
local expression. The size of the loop, i.e., $\bar\mu$ was
arrived using a semi-heuristic correspondence between LQG and
LQC states. This procedure parachutes the area gap from full
LQG into LQC; in LQC proper there is no area gap. It is
somewhat analogous to the way Bohr arrived at his model of the
hydrogen atom by postulating quantization of angular momentum
in a model in which there was no a priori basis for this
quantization. Recall that although the Bohr model captures some
of the essential features of the full, quantum mechanical
hydrogen atom, there are also some important differences. In
particular, the correct eigenvalues of angular momentum
operators are $\sqrt{j(j+1)}\hbar$ rather than $n\hbar$ used by
Bohr. Similarly, it is likely that, in the final theory, the
correct correspondence between full LQG and LQC will require us
to use not the `pure' area gap used here but a more
sophisticated coarse grained version thereof, and that will
change the numerical coefficients in front of $\bar\mu$ and the
numerical values of various physical quantities such as the
maximum density we report in this review. So, specific numbers
used in this review should not be taken too literally; they
only serve to provide reasonable estimates, help fix parameters
in numerical simulations, etc.

2. The functional dependence of $\bar\mu$ on $\mu$ on the other hand
is robust. Under the rescaling of the fiducial cell, $\C \to
\beta^3\C$, we have $\bar\mu c \to \bar\mu c$, or equivalently $\b
\to \b$, whence $\sin\lambda\b$ does not change and the
gravitational part of the constraint simply acquires an overall
rescaling as in the classical theory. Had the functional dependence
been different, e.g. if we had used $\bar\mu= \mu_o$, a constant
\cite{abl}, or $\bar\mu \sim |\mu|^{3/2}$ \cite{bck}, we would have
found that $\b \to \beta^n \b$ with $n\not=0$ whence we would have
$\sin \lambda \b \to \sin\beta^n\lambda\b$ and the constraint would
not have simply rescaled. Consequently, the quantum Hamiltonian
constraint would have acquired a non-trivial cell dependence and
even in the effective theory (discussed in section \ref{s4})
physical predictions would have depended on the choice of $\C$. This
would have made quantum theory physically inadmissible.

3. A quick way to arrive at the constraint (\ref{qhc3}) is to write
the classical constraint in terms of the canonical pair $(\b,\v)$
(see (\ref{hc3})), and then simply replace $\b$ by
$(\sin\,\lambda\b) /\lambda$. While this so-called `polymerization
method' \cite{cvz2,other} yields the correct final result, it is not
directly related to procedures used in LQG because $\b$ has no
natural analog in LQG. In particular, since it is not a connection
component, it would not have been possible to use holonomies to
define the curvature operator. For a plausible relation to LQG, one
has to start with the canonical pair $(A_a^i, E^a_i)$, i.e.,
$(c,p)$, mimic the procedure used in LQG as much as possible and
then pass to the $\b$ representation as was done here. Without this
anchor, there is no a priori justification for using
$\sin\lambda\b/\lambda$; even if one could argue that $\b$ should be
replaced by a trignometric function, there are many other candidates
with the same behavior in the $\lambda \to 0$ limit, e.g. $\tan
\lambda \b/\lambda$. However, \emph{a posteriori} it is possible,
and indeed often very useful, to use shortcut $\b \rightarrow
(\sin\lambda\,\b)/\lambda$.

\subsection{LQC: The full Hamiltonian constraint}
\label{s2.6}

We can now add the matter Hamiltonian to obtain the total quantum
Hamiltonian constraint using $C_H = C_{\rm grav} + 16\pi G C_{\rm
matt}$ from the classical phase space formulation \cite{alrev}. The
quantum constraint $\hat{C}_H \Psi (\v,\phi) =0$ then yields:
\ba \label{qhc4} \p_\phi ^2 \Psi(\nu,\phi) &=& 3\pi G\gamma^2\,\,
\nu\, \f{\sin\lambda\b}{\lambda}\,
\nu\,\f{\sin\lambda\b}{\lambda}\, \Psi(\nu,\phi)\nonumber\\
&=& \f{3\pi G}{4\lambda^2}\, \nu\, \big[(\nu+2\lambda)
\Psi(\nu+4\lambda) - 2\nu \Psi(\nu,\phi) + (\nu
-2\lambda)\Psi(\nu-4\lambda)\big]\nonumber\\
&=:& - \Theta\, \Psi(\nu,\phi) \ea
where, in the second step, we have used (\ref{displacement}). Thus,
the second order \WDW \emph{differential} equation (\ref{qhc1}) is
replaced by a second order \emph{difference} equation, where the
step-size $4\lambda$ is dictated by the area gap (which is
$\lambda^2$). Nonetheless, there is a precise sense in which the
\WDW equation (\ref{qhc1}) ---with its `covariant' factor
ordering--- emerges from (\ref{qhc4}) in the limit in which the area
gap goes to zero.

Let us begin by setting:
\be \label{chi} \chi(\nu) := \nu\,\, \big(\Psi(\nu+2\lambda) -
\Psi(\nu-2\lambda)\big)\, , \ee
so that (\ref{qhc4}) can be rewritten as
\be \p_\phi^2 \Psi(\nu,\phi) = \f{3\pi G}{4\lambda^2}\,\, \nu\,
[\chi(\nu+2\lambda) -\chi(\nu-2\lambda)]\, .\ee
To obtain the \WDW limit, let us assume $\Psi(\nu)$ is smooth. Then,
we have:
\ba \p_\phi^2 \Psi(\nu,\phi) &=& {3\pi G}\, \nu\, \p_\nu\,
\chi(\nu) + O \left(\f{(4\lambda)^n}{n!} \nu \p^{n}_\nu\,
\chi\right)\nonumber\\ &=& 12\pi G \nu\, \p_\nu \,\, \nu \p_\nu\,
\Psi + O\left(\f{(4\lambda)^{m+n}}{m!n!}\,\, \nu\, \p^m_\nu\,\,
\nu\, \p^n_\nu \Psi(\nu,\phi)\right) \ea
where $m,n\, \ge 2$. Thus, if we restrict ourselves to wave
functions $\Psi$ which are slowly varying in the sense that the
`error terms' under $O$ can be neglected, we obtain the \WDW limit
of the Hamiltonian constraint (\ref{qhc4}):
\be \label{wdw} \p_\phi^2\Psi(\nu,\phi) = 12\pi G\,\, \nu\p_\nu\,
\nu\p_\nu \Psi(\nu,\phi)\, . \ee
This procedure has several noteworthy features. First, in the
reduction, in addition to the $\lambda \to 0$ limit, we had to
assume that $\Psi(\nu)$ is smooth and slowly varying in $\nu$.
Therefore, it cannot be in the LQC Hilbert space $\Hkg$.
Second, the final form (\ref{wdw}) of the \WDW limit is exactly
the same as Eq (\ref{qhc1}), including the `covariant' factor
ordering. Third, this approximation is not uniform because the
terms which are neglected depend on the state $\Psi$. However,
these assumptions are realized at late times on semi-classical
states of interest. In this precise sense LQC dynamics is well
approximated by the \WDW theory at late times.

Let us return to LQC. We can now construct the physical Hilbert
$\Hp$ by applying a group averaging procedure \cite{dm,almmt,abc}.
This requires the introduction of an auxiliary Hilbert space $\Ha$
with respect to which the constraint operator is self-adjoint, which
can be achieved simply by a slight modification of the inner product
on $\Hkg$ to $\langle \Psi_1|\Psi_2\rangle = \sum_{\nu}\, |\nu|^{-1}
\bar{\Psi}_1(\nu)\, \Psi_2(\nu)$. (The analogous modification was
carried out implicitly in the \WDW theory when we considered wave
functions which are square integrable with respect to the measure
$\dd z = \dd a/a$ in place of $\dd a$.) Then, physical states are
again those solutions to the `positive frequency' part of the
Hamiltonian constraint,
\be \label{qhc5} -i\hbar\p_\phi \Psi(\nu,\phi) = \sqrt\Theta\,
\Psi(\nu,\phi) \ee
which are symmetric under $\nu \rightarrow -\nu$ and have finite
norms under the inner product
\be \label{ip3} \ip{\Psi_1}{\Psi_2} = \sum_{\nu}\,\,
{\bar\Psi}_1(\nu,\phi_o) \,\,|\nu|^{-1}\,\, \Psi_2(\nu,\phi_o)\, .
\ee
Here, the constant $\phi_o$ can be chosen arbitrarily because the
integral on the right side is conserved. As in the \WDW theory, we
can again introduce Dirac observables: $\hat{p}_{(\phi)} \equiv
\sqrt\Theta$, representing the scalar field momentum and, using
(\ref{vol1}), the operator $\h{V}|_{\phi}$, representing the volume
at internal time $\phi$. This completes the specification of the
mathematical framework in LQC. One can now explore physical
consequences and compare them with those of the \WDW theory.

\begin{figure}[]
  \begin{center}
    $a)$\hspace{8cm}$b)$
    \includegraphics[width=3.2in,angle=0]{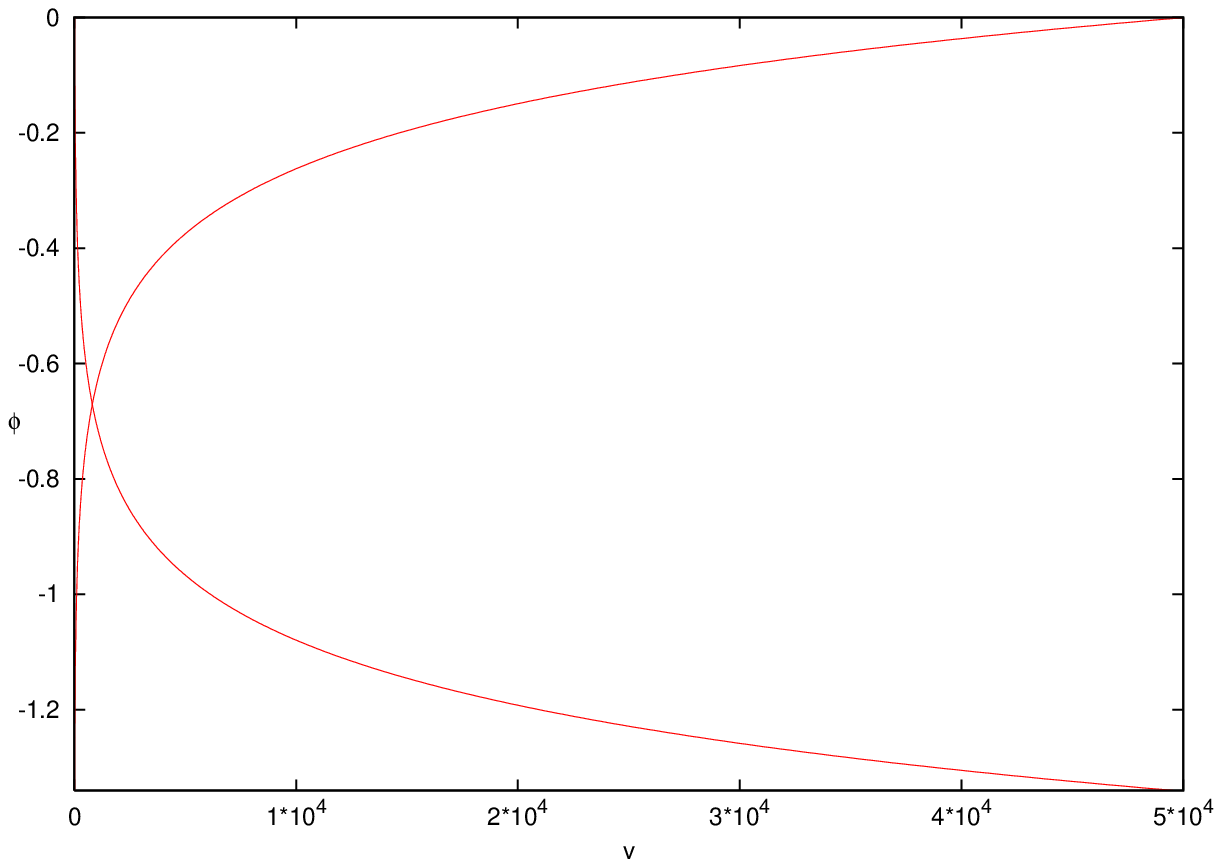}
    \includegraphics[width=3.2in,angle=0]{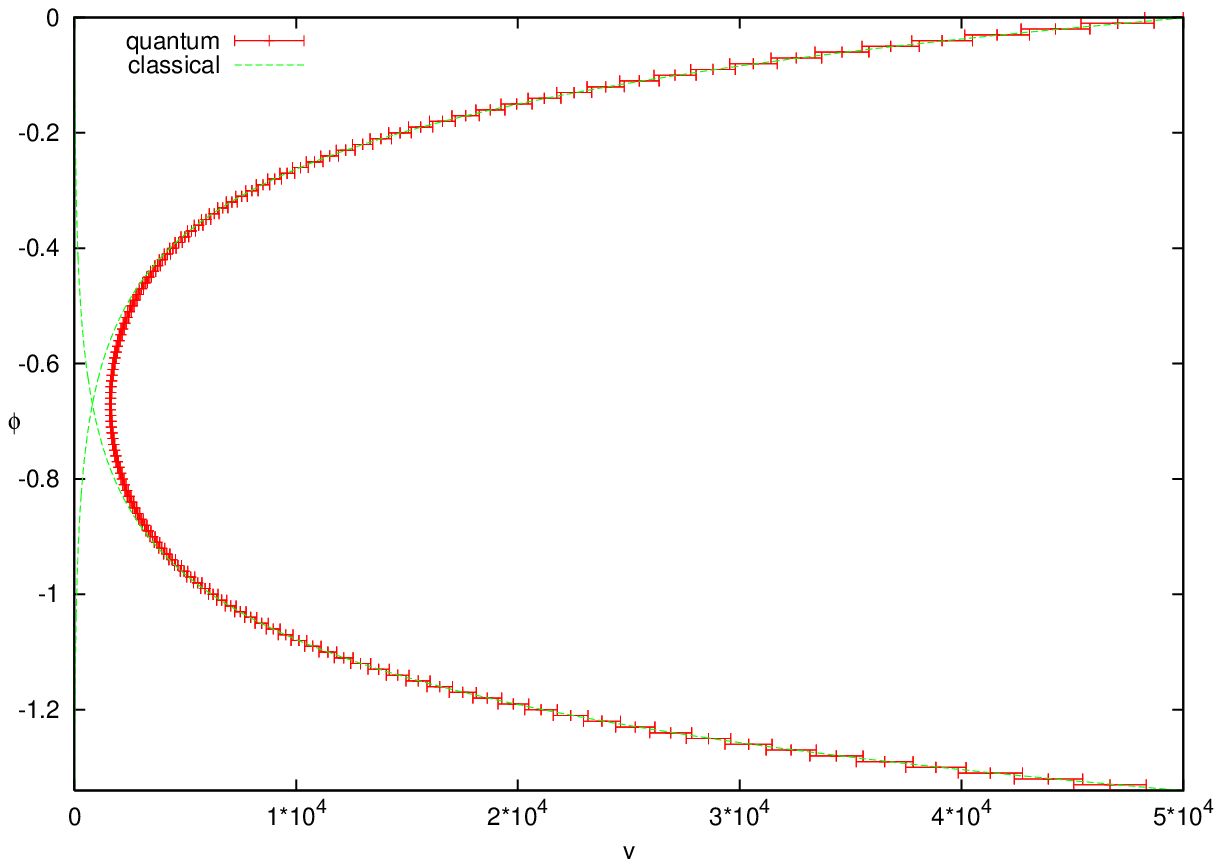}
\caption{$a)$ Classical solutions in k=0, $\Lambda=0$ FRW models
with a massless scalar field. Since $\pphi$ is a constant of motion,
classical trajectories can be plotted in the $\v$-$\phi$ plane. There
are two classes of trajectories. In one the universe begins with a
big-bang and expands and in the other it contracts into a big-crunch.
There is no transition between these two branches. Thus, in a given solution,
the universe is either eternally expanding or eternally contracting.
$b)$ LQC evolution. Expectation values and dispersion of $|\h{\v}|_\phi$,
are compared with the classical trajectory. Initially, the wave function
is sharply peaked at a point on the classical trajectory at which the
density and curvature are very low compared to the Planck scale.
In the backward evolution, the quantum evolution follows the
classical solution at low densities and curvatures but undergoes a
quantum bounce at matter density $\rho \sim 0.41\rho_{\rm Pl}$ and
joins on to the classical trajectory that was contracting to the
future. Thus the big-bang singularity is replaced by a quantum bounce.}
\label{fig:k=0}
\end{center}
\end{figure}

Physics of this quantum dynamics was first studied in detail
using computer simulations \cite{aps3}. Recall that in the
classical theory we have two types of solutions: those that
start with a big-bang and expand forever and their time
reversals which begin with zero density in the distant past but
collapse into a future, big-crunch singularity (see
Fig.\ref{fig:k=0}(a)). The idea, as in the \WDW theory, was to
start with a wave function which is sharply peaked at a point
of an expanding classical solution when the matter density and
curvature are very small and evolve the solution backward and
forward in the internal time defined by the scalar field
$\phi$. As in the \WDW theory, in the forward evolution, the
wave function remains sharply peaked on the classical
trajectory. Thus, as in the \WDW theory, LQC has good infrared
behavior. However, in the backward evolution, the expectation
value of the volume operator deviates from the classical
trajectory once the matter density is in the range $10^{-2}$ to
$10^{-3}$ times the Planck density. Instead of following the
classical solution into singularity, the expectation value of
volume bounces and soon joins a classical solution which is
expanding to the future. Thus, the big-bang is replaced by a
quantum bounce. Furthermore, the matter density at which the
bounce occurs is universal in this model: it given by $\rcr
\approx 0.41 \rho_{\rm Pl}$ \emph{independently of the choice
of state} (so long as it is \emph{initially} semi-classical in
the sense specified above).

The fact that the theory has a good infrared as well as
ultraviolet behaviors is \emph{highly} non-trivial. As we saw,
the \WDW theory fails to have good ultraviolet behavior.
Reciprocally, if in place of $\bar\mu \sim \mu^{-1/2}$ we had
used $\bar\mu = \mu_o$, a constant, the big-bang would again
have been replaced by a quantum bounce but we would not have
recovered general relativity in the infrared regime
\cite{aps2}. Indeed, in that theory, there are perfectly good
semi-classical states at late times which, even evolved
backwards, exhibit a quantum bounce at density of water! (See
section \ref{s7.1}.)

A second avenue to explore the Planck scale physics in this
model was provided by effective equations, discussed in section
\ref{s5}. While assumptions underlying the original derivation
of these equations \cite{jw,vt} seem to break down in the
Planck regime, as is often the case in physics, these effective
equations nonetheless continue to be good approximations to the
quantum dynamics of states under consideration for all times.
Indeed, the effective equations even provide an analytical
expression of the maximum density $\rcr$ whose value is in
complete agreement with the exact numerical simulations
\cite{aps3}! A third avenue was introduced subsequently by
restricting oneself to a `superselection sector' (see below)
and passing to the representation in which states are functions
of $\b$ rather than $\nu$ \cite{acs}. In this representation
one can solve quantum dynamics \emph{analytically} (see section
\ref{s3}). Therefore, one can establish a number of results
without the restriction that states be initially
semi-classical. One can show that \emph{all} quantum states (in
the dense domain of the volume operator) undergo a quantum
bounce in the sense that the expectation value of the volume
operator has a non-zero lower bound in any state. More
importantly, the matter density operator $\hat\rho|_\phi$ has a
\emph{universal upper bound} on $\Hp$ and, again, it coincides
with $\rcr$ \cite{acs}. What is perhaps most interesting is
that this upper bound \emph{is induced dynamically}: $\hat\rho$
is unbounded on the kinematical Hilbert space. The boundedness
implies that in LQC the singularity is resolved in a strong
sense: the key observable which diverges at the big-bang in the
classical theory is tamed and made bounded by quantum dynamics.

Let us summarize. In the simple model under consideration, the
scalar field serves as a good clock, providing us with a
satisfactory notion of relational time. It enables one to go
beyond the `frozen formalism' in which it is difficult to
physically interpret solutions to quantum constraints. Using
relational time, one can define Dirac observables, such as the
density $\h\rho|_{\phi}$ at the instant $\phi$ of internal
time, and discuss dynamics. We then constructed the physical
sector of the \WDW theory and found that the singularity is not
resolved: If one begins with states which are sharply peaked on
a classical solution at late times and evolves them back, one
finds that they remain sharply peaked on the classical
trajectory all the way to the big-bang. In particular, the
expectation value of $\h\rho|_\phi$ increases unboundedly in
this backward evolution. In LQC, by contrast, singularity is
resolved in a strong sense: $\h\rho|_\phi$ is bounded above on
the entire physical Hilbert space! The root of this dramatic
difference can be treated directly to quantum geometry that
underlies LQG: the bound on the spectrum of $\h\rho|_\phi$ is
dictated by the `area gap' and goes to infinity as the area gap
goes to zero. Finally, even though the model is so simple, the
LQC quantization involves a number of conceptual and
mathematical subtleties because one continually mimics the
procedures introduced
in full LQG.\\

\textbf{Remarks:}

1) Note that $\Theta$ in (\ref{qhc4}) is a second order difference
operator and $(\Theta\Psi) (\nu)$ depends only on values of $\Psi$
at $\nu-4\lambda, \nu$ and $\nu+4\lambda$.  Since the physical
states also have to be symmetric, $\Psi(\nu,\phi) = \Psi(-\nu,
\phi)$, we find that the sub-space $\H_\epsilon$ of $\H$, spanned by
wave functions with support just on the `lattice'  $\nu = \pm
\epsilon +4 n \lambda$, is left invariant under dynamics for each
$\epsilon \in [0,4\lambda)$. Furthermore, our complete family of
Dirac observables $\h{p}_{(\phi)}$ and $\h{V}|_\phi$ leaves each
$\H_\epsilon$ invariant. Thus, the physical Hilbert space $\Hp$ is
naturally decomposed into a (continuous) family of separable Hilbert
spaces
\be \Hp = \oplus_\epsilon\,\, \H_\epsilon, \qquad {\rm with}\,\,\,
\epsilon \in [0,4)  \ee
and we can analyze each $\H_\epsilon$ separately. This property will
be used in section \ref{s3} to obtain an analytical solution of the
problem. Numerical simulations \cite{aps3} show that, as one might
expect, qualitative features of physics are insensitive to the
choice of $\epsilon$. The key quantitative prediction ---value of
the maximum density $\rcr$--- is also insensitive.

2) In this discussion for simplicity we used the Schr\"odinger
representation for the scalar field. There is also a `polymer
representation' in which the wave functions are almost periodic in
$\phi$. What would have happened if we had used that representation
to construct $\Hk$? It turns out that we would have obtained the
same quantum constraint and the same physical Hilbert space $\Hp$.

3) It is important to note that in LQC, it is the difference
equation (\ref{qhc4}) that is fundamental, and it is the
continuum limit (\ref{qhc1}) that is physically approximate.
This is a reversal of roles from the standard procedure in
computational physics. Therefore, some of the statements in
\cite{laguna} addressed to computational physicists can be
physically misleading to quantum gravity and cosmology
communities. For example, a sentence in abstract of
\cite{laguna}, ``These bounces can be understood as spurious
reflections'' may be misinterpreted as saying that they are
artifacts of bad numerics. This is certainly not the case
because numerics in \cite{aps3} were performed with all due
care and furthermore the results are in complete agreement with
the analytical solutions found later \cite{acs}. Rather, the
intent of that phrase was to say: `had the physical problem
been to solve a wave equation in the \emph{continuum} and had
one used non-uniform grids, one would also have found bounces
which, from the perspective of continuum physics of this
hypothetical problem, would be interpreted as spurious
reflections in finite difference discretizations'. This is an
illuminating point for computational physicists but is not
physically relevant in LQC where the basic equation is a
difference equation.

\section{Exactly Soluble LQC (sLQC)}
\label{s3}

In this section, we continue with the k=0, $\Lambda$=0 FLRW
model. This model can be solved \emph{exactly} if one uses the
scalar field as an internal clock already in the classical
theory, prior to quantization, and works in a suitable
representation \cite{acs}. This analytical control on the
quantum theory allows to prove further results such as the
generic character of the bounce, obtain an analytical
expression of the upper bound of the energy density operator
and carry out a detailed comparison between the \WDW theory and
sLQC. Furthermore, questions regarding the behavior of
fluctuations and preservation of semi-classicality across the
bounce can be answered in detail.

In section \ref{s3.1} we recast the \WDW quantum constraint in
terms of variables that facilitate the comparison with sLQC. In
section \ref{s3.2} we carry out the loop quantization of this
model following \cite{acs}. Interestingly, the form of the
quantum constraint and the inner product are strikingly similar
in the \WDW theory and sLQC, yet they lead to very different
physical predictions. The reason is that the physical
observables directly relevant to cosmology are represented by
very different operators in the two cases. Physical predictions
of both theories are discussed in section \ref{s3.3}. In
section \ref{s3.4}, we spell out the precise relation between
sLQC and \WDW theory and show that the latter does not follow
as a limit of the former. There is some apparent tension
between the singularity resolution in LQC and the intuition
derived from path integrals on the regime in which quantum
effects become important. This issue and its resolution are
discussed in section \ref{s3.5}

Since cosmologists would not have already read section \ref{s2}, we
have attempted to make this section self-contained. For others we
have included remarks connecting this discussion with that of
section \ref{s2}.

\subsection{The WDW theory}
\label{s3.1}

Let us begin by briefly recalling the Hamiltonian framework and
fixing notation. (For clarifications and details, see section
\ref{s2.1.1})

If the spatial topology is $\R^3$, all spatial integrals
diverge in the Hamiltonian (as well as Lagrangian) framework. A
standard attitude in much of the older literature was just to
ignore this infinity. But then typically quantities on the two
sides of equations have different physical dimensions and there
are hidden inconsistencies. In a systematic treatment, one has
to first introduce an elementary cell $\C$, restrict all
integrals to it, construct the theory and in the final step
remove this infrared regulator by taking the limit $\C \to
\R^3$.

Let us recall that in the canonical framework the \WDW phase space
is coordinatized by $(a,\t{p}_a; \phi, \pphi)$. The scale factor
relates the physical 3-metric $q_{ab}$ to the fiducial one
$\q_{ab}$, associated with the co-moving coordinates, via $q_{ab} =
a^2 \q_{ab}$. It turns out that in LQC it is more convenient to work
with orthonormal triads rather than 3-metrics and with the physical
volume $V$ of $\C$ rather than the scale factor. To facilitate the
comparison between the two quantum theories, we will begin by
writing the \WDW theory in canonical variables which are adapted to
LQC. Thus, the gravitational configuration variable will be $\v =
\varepsilon (V/2\pi G)$ where $\varepsilon =\pm 1$ depending on the
orientation of the physical triad, and $V$ is related to the scale
factor $a$ via $V = a^3 V_o$, where $V_o$ is the volume of $\C$ in
co-moving coordinates. The conjugate momentum $\b = \gamma
\dot{a}/a$ is the Hubble parameter, except for a multiplicative
constant $\gamma$, the Barbero-Immirzi parameter of LQG (whose value
$\gamma \approx 0.2375$ is fixed by the black hole entropy
calculation).

Thus the full phase space is topologically $\R^4$, coordinatized by
$(\v,\b;\, \phi,\pphi)$, and the fundamental Poisson brackets are:
\be \{\b,\, \v\} = 2\gamma \qquad{\rm and} \qquad \{\phi,\,\pphi\} =
1 \ee
Since we wish to use the scalar field $\phi$ as emergent time, it is
natural to consider evolution in a harmonic time coordinate $\tau$
satisfying $\Box \tau =0$. The associated lapse is then $N_\tau =
a^3$ and the Hamiltonian constraint is given by:
\be\label{cl_const} \pphi^2 - 3 \pi G \, \v^2 \b^2 = 0\, . \ee
Let us use quantum states which are diagonal in $\b$. Then the
quantum constraint becomes
\be \label{wdwbcon} \p_\phi^2 \, \chiwb = - \gpi (\b \, \p_b)^2 \,
\chiwb ~. \ee
where on the right side we have used a `covariant' factor ordering
(as in section \ref{s2.2}) and the underbars are again serve to
emphasize that discussing the \WDW theory. The change of orientation
of triads corresponds to a large gauge transformation under which
physics of the model is unchanged. This turns out to imply that the
wave functions $\chiwb$ must satisfy $\chiwb =
-\ul\chi(-\b,\phi)$ \cite{acs}. Therefore, we can incorporate the
invariance under large gauge transformations \emph{simply by
restricting ourselves to the positive} $\b$\emph{-half line.}

The constraint (\ref{wdwbcon}) can be written in a simpler form by
introducing
\be y := \f{1}{(\gpi)^{1/2}} \, \ln \f{\b}{\b_o} \ee
where $\b_o$ is an arbitrarily chosen but fixed constant. Since $\b
\in (0, \infty)$, $y$ is well defined and takes values on the
\emph{full real line.} Then the \WDW constraint takes the form of a
2-dimensional Klein-Gordon equation as in section \ref{s2.2}.
\be \label{wdwqbc} \p_\phi^2 \chiwy = \ul\Theta \, \chiwy,
~~\mathrm{where} ~~ \ul\Theta := \p_y^2  ~. \ee
The idea again is to interpret this equation as providing us with
the evolution of $\chiwy$ in `relational time' $\phi$. Using Fourier
transform, one can naturally decompose solutions to (\ref{wdwqbc})
positive and negative frequency sectors. As explained in section
\ref{s2}, a general `group averaging' procedure \cite{dm,almmt,abc}
leads us to the physical Hilbert space: $\Hpwdw$ consists of
positive frequency solutions to (\ref{wdwqbc}), i.e., solutions
satisfying the positive square root $\-i\p_\phi \chiwy =
\sqrt{\ul\Theta}\, \chiwy$ of (\ref{wdwqbc}), where
$\sqrt{\ul\Theta} = \sqrt{-\p_y^2}$ can be easily defined by making
a Fourier transform. Since we are working with positive frequency
solutions, the physical inner product is given by the standard Klein
Gordon current. In the momentum space, it can be written as
\be \label{ip} (\ul\chi_1, \ul \chi_2)_{\mathrm{phy}} \, = 2
\int_{-\infty}^{\infty} \dd k |k| \, {\bar{\t{\ul{\chi}}}}_1(k)
{\t{\ul{\chi}}}_2(k) 
~ \ee
where $\tilde{\ul\chi}$ is the Fourier transform of $\ul\chi$ and
$k$ is related to the eigenvalues $\omega$ of $\ul \Theta$ as
$\omega = \sqrt{\gpi} |k|$. This expression is just what one would
expect from the 2-dimensional Klein-Gordon theory. Finally, general
initial datum for the physical state at time $\phi = \phi_o$ is of
the form $\ul \chi(y,\phi_o) = ({1}/{\sqrt{2 \pi}})\,
\int_{-\infty}^\infty d k e^{-i k y} \tilde {\ul \chi}(k)$, and
under time evolution one obtains
\be \chiwy = \f{1}{\sqrt{2 \pi}} \big(\int_{-\infty}^0 \dd k e^{-i k
(\phi + y)} e^{i k \phi_o} {\t{\ul \chi}}(k) + \int_0^{\infty} \dd k
e^{i k (\phi - y)} e^{- i k \phi_o} {\t{\ul \chi}}(k)\big)\, , \ee
where the first term is a left moving solution while the second is a
right moving solution. Thus, $\Hpwdw$ can itself be decomposed into
two orthogonal subspaces consisting of right and left moving modes.

As in section \ref{s2.2}, one can introduce a family of Dirac
observables. Since $\h{p}_{(\phi)}$ is a constant of motion, it
is trivially a Dirac observable. On $\Hpwdw$ its action is
given by
\be\label{p_phi} \hat{p}_{(\phi)} \, \chiwy = \hbar \sqrt{\ul
\Theta}\,\, \chiwy ~. \ee
The second Dirac observable is $\hat V|_{\phi_o}$, the volume
of the cell ${\cal C}$ at time $\phi_o$. To define it, we must
first introduce the volume operator: $\h{V} = 2\pi G
\widehat{|\v|} = 2\pi \gamma\lp^2 \widehat{|\nu|}$ where, to
facilitate comparison with section \ref{s2}, we have set $\h\nu
:= \v/\hbar$. In the $\b$ representation, $\hat\nu$ is simply
the self-adjoint-part of $(-2 i \p_\b)$, which, in the $y$
representation becomes:
\be\label{wdw_vol} \hat \nu = -\f{2}{\sqrt{\gpi}b_o} \, \left(P_R
(e^{\sqrt{\gpi} y}i \, \p_y) P_R + P_L (e^{\sqrt{\gpi} y}i \, \p_y)
P_L \right) ~. \ee
where $P_R$ and $P_L$ project on the right and left moving, mutually
orthogonal subspaces. The second Dirac observable, the volume at
internal time $\phi_o$, obtained by `freezing' the given state
$\chiwy$ at time $\phi_o$, operating it by the volume operator, and
then evolving the result to obtain a positive frequency solution to
(\ref{wdwqbc}):
\be \label{vphi} \hat V|_{\phi_o} \, \chiwy = e^{i \sqrt{\ul\Theta}
(\phi - \phi_o)} \,\, (2 \pi\gamma\lp^2 |\hat \nu|) \,\,
\ul\chi(y,\phi_o) \, , \ee
Both the Dirac observables are self-adjoint with respect to
(\ref{ip}) as they must be and, furthermore, \emph{preserve the left
and right moving sectors} of $\Hpwdw$. Therefore, if one describes
physics using this complete set of Dirac observables, it suffices to
work with just one sector at a time. This fact is conceptually
important since, as we will see, the right sector corresponds to
contracting universes and left to expanding \cite{acs}.

\subsection{Loop quantization}
\label{s3.2}

We can now turn to LQC in the $\b$ representation. The underlying
phase space is the same but passage to quantum theory is different
because, by following the procedure used in full LQG \cite{lost,cf}
in this cosmological context, we are led to a theory which is
distinct from the \WDW theory already at the kinematical level. As a
consequence, as we saw in section \ref{s2}, support of the LQC wave
functions $\Psi(\nu,\phi)$ in the volume representation can
restricted to regular `lattices', $\nu = \pm\epsilon + 4n \lambda$
with $\epsilon \in [0,4\lambda)$, and each $\epsilon$-sector is left
invariant by the two Dirac observables \emph{and} evolution. The
step size on these lattices is dictated by $\lambda$ where
$\lambda^2 = 4 \sqrt{3} \pi \gamma \lp^2$ is the `area gap', i.e.,
the lowest eigenvalue of the area operator in LQG. Thus, while
evolution in the \WDW theory is governed by a \emph{differential}
equation, that in LQC by a \emph{difference} equation. The advantage
of the $\b$ representation is that the LQC evolution is again
governed by a differential equation and can be compared more
directly with that in the \WDW theory.

To pass to the $\b$ representation, it is simplest to work with the
$\epsilon=0$ lattice \cite{acs}. (While some technical details
depend on the choice of $\epsilon$, physics is essentially
independent.) Then, since $\Psi(\nu,\phi)$ have support on $\nu =
4n\lambda$, their Fourier transform have support on a \emph{circle}:
While we have $\b\in (-\infty, \infty)$ in the \WDW theory, in LQC
we have $\b \in (0, \pi/\lambda)$. But the quantum Hamiltonian
constraint is a differential equation as in the \WDW theory:
 \be
\p_\phi^2 \, \chilb = \gpi \left(\f{\sin \lam \b}{\lam} \p_{\b}
\right)^2 \, \chilb \, .\ee

As in the case of the \WDW theory, we can make a change of
coordinates to rewrite this constraint as a Klein-Gordon equation:
in terms of the variable $x$,
\be x = \f{1}{\sqrt{\gpi}} \, \ln\left(\tan \f{\lam \b}{2}\right) ~
\ee
the quantum constraint becomes
\be\label{lqc_bcons} \p_\phi^2 \, \chi(x,\phi)  = - {\Theta} \,
\chilx, \qquad {\rm where} \qquad  \Theta := - \p_x^2\, . \ee
Since $\b \in (0,\pi/\lambda)$, it follows that $x\in (-\infty,
\infty)$. The requirement that physics should be invariant under the
change of orientation of the triad now implies $\chi(x,\phi) =
-\chi(-x,\phi)$. This is in striking contrast with the situation in
the \WDW theory, where this requirement was already used in the very
definition of $y$ and left no restriction on $\chiwy$. It is easy to
verify that this restriction implies that the LQC wave functions
$\chi(x,\phi)$ \emph{must have support on both the right and the
left moving sectors}.

Thus, the physical Hilbert space $\Hp$ now consists of all
anti-symmetric, positive frequency solutions to (\ref{lqc_bcons}),
i.e. $\chi(x,\phi)$ satisfying $ \chi(x,\phi) = - \chi(-x,\phi)$ and
$-i\p_\phi\chi(x,\phi) = \sqrt{\Theta}\chi(x,\phi)$, with finite
norm in the inner product (\ref{ip}). The symmetry requirement
implies that every solution $\chilx$ can be written as
\be \label{chilx} \chilx = \f{1}{\sqrt{2}} (F(x_+) - F(x_-))
\ee
where $F$ satisfies (\ref{lqc_bcons}) and $x_\pm = \phi\pm x$; the
right and left moving part of $\chilx$ determines its right moving
part and vice versa. Therefore, the inner product can then be
expressed in terms of the right moving (or left moving) part alone:
 \be
(\chi_1,\chi_2)_{\mathrm{phys}} = - 2 i \int_{-\infty}^\infty \, \dd
x \bar F_1(x_+) \p_x F_2(x_+) ~. \ee

The Dirac observables $\h{p}_{(\phi)}$ and $\h{V}|_{\phi_o}$ have
the same form as in (\ref{p_phi}) and (\ref{vphi}) but, because the
transformation from $\b$ to $y$ in the \WDW theory and to $x$ in LQC
are quite different, the definition of the operator $\h{\nu}$
changes:
\be\label{lqc_vol} \hat \nu = -\f{2 \lam}{\sqrt{\gpi}} \, \left(P_R
(\cosh(\sqrt{\gpi} x)i \, \p_x) P_R + P_L (\cosh(\sqrt{\gpi} x)i \,
\p_x) P_L \right) ~. \ee
This is the second difference that makes the physics of the two
theories profoundly different, in spite of the fact that the states
$\chiwy$ and $\chi(x,\phi)$ satisfy the same dynamical equation.

\subsection{Physical consequences}
\label{s3.3}

\subsubsection{Generic nature of the bounce}
 \label{physicalcons1}

We have discussed that the action of $\hat p_\phi$ is identical in
the \WDW and sLQC. However, crucial differences appear in the case
of the Dirac observable $\hat V|_{\phi}$ corresponding to volume.
Since the right and left moving sectors decouple, let us focus on
the left moving sector. On it, the expectation values of
$\h{\nu}|_\phi$ can be written as
\be (\ul \chi_L, \hat \nu|_\phi \ul \chi_L)_{\rm phy} =
\f{4}{\sqrt{12\pi G}\b_o}\, \int_{-\infty}^\infty \dd y \mid \f{\p
\ul{\chi}_L(y,\phi)}{\p y}\mid^2 \, e^{-\sqrt{12\pi G}\, y}\, .\ee
Using the fact that $\ul{\chi}_L (y,\phi) = \ul{\chi}_L (y_+)$ it
now follows that the expectation value of $\hat V|_{\phi}$ is given
by
\be\label{wdw_vol1} (\ul \chi_L,\, \hat V|_\phi \ul
\chi_L)_{\mathrm{phy}} = 2 \pi \gamma \lp^2 \, (\ul \chi_L,
|\h\nu|_\phi \ul \chi_L)_{\mathrm{phy}} = V_* e^{\sqrt{\gpi} \phi}
\ee
where $V_*$ is a constant determined by the state at any `initial'
time instant and is given by
\be V_* = \f{8 \pi \gamma \lp^2}{\sqrt{\gpi} \b_o}
\int_{-\infty}^\infty \dd y_+ \left|\f{\dd \ul \chi_L}{\dd
y_+}\right|^2 \, e^{-\sqrt{\gpi} y_+} ~. \ee
(An analogous calculation for the right moving modes yields
$(\ul \chi_L, \hat V|_\phi \ul \chi_L)_{\mathrm{phy}} \propto
e^{-\sqrt{\gpi} \phi}$). Hence, for the left moving modes
(which correspond to an expanding universe), the expectation
values $\langle \hat V|_\phi \rangle \rightarrow 0$ as $\phi
\rightarrow -\infty$. Similarly, for the right moving modes,
the expectation values vanish as $\phi \rightarrow \infty$. The
expectation values with the left (right) moving modes, diverge
when $\phi$ approaches positive (negative) infinity. Thus, in
the \WDW theory, a state corresponding to a contracting
universe encounters a big-crunch singularity in the future
evolution, and the state corresponding to an expanding universe
evolves to a big-bang singularity in the backward evolution.
Note that this conclusion holds for any state in the domain of
$\hat V|_{\phi}$. The \WDW quantum cosmology is thus
generically singular.%
\footnote{This conclusion is based on the analysis of the
expectation values of the Dirac observables. The same
conclusion is reached if one considers consistent probabilities
framework a la Hartle \cite{hartle} in this model. A careful
analysis of histories at different times shows that even
arbitrary superpositions of the left and right moving sectors
do not lead to a singularity resolution in the \WDW theory. The
probability that a \WDW universe ever encounters the
singularity is unity \cite{consistent1,consistent2}. A similar
analysis in sLQC, reveals the probability for bounce to be
unity \cite{consistent3}. We discuss these issues in detail in
section \ref{s7.5}. }

Let us now turn to sLQC. Now the expectation value of $\h\nu|_\phi$
is given by
\be (F,\, \h{\nu} F)_{\rm phy} =\f{4\lambda}{\sqrt{12\pi G}}\,
\int_{-\infty}^{\infty} \dd x \mid \f{\p F(x_+,\phi)}{\p x}\mid^2 \,
{\rm cosh}\,(\sqrt{12\pi G} x) \ee
Therefore, Eq (\ref{chilx}) implies the expectation values $\langle
\hat V|_\phi \rangle$ are now given by
\be\label{slqc_vol} (\ul \chi, \, \hat V|_\phi \ul
\chi)_{\mathrm{phy}} \, = \, 2 \pi \gamma \lp^2 \, (\ul \chi,
|\h\nu|_\phi \ul \chi)_{\mathrm{phy}} \, = \, V_+ e^{\sqrt{\gpi}
\phi} + V_- e^{-\sqrt{\gpi} \phi} \ee where \be V_{\pm} = \f{4 \pi
\gamma \lp^2 \lam}{\sqrt{\gpi}} \int_{-\infty}^\infty \dd x_+
\left|\f{\dd F}{\dd x_+}\right|^2 e^{\mp \sqrt{\gpi} x_+} \ee
are positive constants determined by the state at any `initial' time
instant. Unlike the \WDW theory, $\langle \hat V|_\phi \rangle$
diverges in the future ($\phi \rightarrow \infty$) and in the past
evolution ($\phi \rightarrow -\infty$). The minimum of the
expectation values is reached at $\phi = \phi_B$ (the bounce time),
determined by the initial state:
\be\label{bouncetime} V_{\mathrm{min}} = 2 \f{\sqrt{V_+
V_-}}{||\chi||^2}  \qquad {\rm at} \qquad   \phi_B = \f{1}{2
\sqrt{\gpi}} \log \f{V_+}{V_-} \ee
Since $V_+$ and $V_-$ are strictly positive, $\langle \hat
V|_\phi \rangle$ is never zero and the big-bang/big-crunch
singularities are absent. It is important to stress that the
bounce occurs for arbitrary states in sLQC at a positive value
of $\langle \hat V|_{\phi_B}\rangle$ and the resolution of
classical singularity is generic.  Further, the expectation
values $\langle \hat V|_\phi \rangle$ are symmetric across the
bounce time.

The singularity resolution in sLQC can also be understood by
analyzing the expectation value of the time-dependent Dirac
observable corresponding to the matter energy density,
\be \hat \rho|_\phi = \f{1}{2}\, (\hat A|_\phi)^2 ~~~\,
{\mathrm{where}} \, ~~~\hat A|_\phi = (\hat V|_\phi)^{-1/2} \, \hat
p_\phi  (\hat V|_\phi)^{-1/2} ~. \ee
The expectation values $\langle \hat A|_\phi \rangle$ (in any state
$\chi$) can be evaluated as
\be \langle \hat A|_{\phi_o} \rangle \, = \, \f{(\chi, \hat p_\phi
\chi)_{\mathrm{phy}}}{(\chi, \hat V|_{\phi_o} \chi)_{\mathrm{phy}}}
\, = \, \left(\f{3}{4 \pi \gamma^2 G}\right)^{1/2} \, \f{1}{\lam}
\f{\int_{-\infty}^\infty \dd x |\p_x F|^2}{\int_{-\infty}^\infty \dd
x |\p_x F|^2 \cosh(\sqrt{\gpi} x)}\, . \ee
Clearly, these are bounded by $(3/4 \pi \gamma^2 G \lam^2)^{1/2}$.
Since the state is (essentially) arbitrary, this implies that there
is an upper bound on the spectrum of the energy density operator:%
\footnote{Another way to define the expected energy density is:
$\langle \hat \rho|_\phi\rangle = \langle \hat p_\phi\rangle^2/2
\langle V|_\phi\rangle^2$. It leads to the same bound.}
\be \rho_{\mathrm{sup}} = \f{3}{8 \pi \gamma^2 G \lam^2} =
\f{\sqrt{3}}{32 \pi^2 \gamma^3 G^2 \hbar} \approx 0.41
\rho_{\mathrm{Pl}} ~. \ee
The value of the supremum of $\langle \hat \rho|_\phi\rangle$ is
directly determined by the area gap and is in excellent agreement
with the earlier studies based on numerical evolution of
semi-classical states \cite{aps3}. As an example, for semi-classical
states peaked at late times in a macroscopic universe with $\langle
\hat{p}_{(\phi)} \rangle = 5000 \hbar$, the density at the bound
already agrees with $\rho_{\rm sup}$ to 1 part in $10^4$. In the k=1
case, for the universe to reach large macroscopic sizes, $\langle
\h{p}_{(\phi )}\rangle$, has to be far larger. If we use those
values here, then the density at the bounce and $\rho_{\rm sup}$
would be indistinguishable.

\subsubsection{Fluctuations}
\label{s3.3.2}

Because $\h{p}_{(\phi)}$ is a constant of motion, its mean value and
fluctuations are also time independent. For $\hat{V}|_\phi$, on the
hand, both are time dependent. In the last subsection we analyzed
the time dependence of the expectation value. We will now summarize
results on the time dependence of fluctuations $(\Delta \h{V}|_\phi)
= \langle (\h{V}|_\phi)^2 \rangle - \langle \h{V}|_\phi\rangle^2$.
Of particular interest is the issue of whether fluctuations can grow
significantly during the bounce. This issue of potential `cosmic
forgetfulness' has been analyzed from different perspectives and
there has been notable controversy in the literature
\cite{mb2,mb3,cs1,kp2,cm1}. For instance, the case for cosmic
forgetfulness includes statements such as \emph{``It is practically
impossible to draw conclusions about fluctuations of the Universe
before the Big Bang;''} and \emph{ ``in cosmology, fluctuations
before and after the Big Bang are largely independent''} \cite{mb2}.
The case in the opposite direction is summarized in statements such
as \emph{``The universe maintains (an almost) total recall;''} and
\emph{``there is a strong bound on the possible relative dispersion
'on the other side' when the state is known to have, at late times,
small relative dispersion in canonically conjugate variables''}
\cite{cs1}.

We will summarize three mathematical results on this issue, two of
which have been obtained recently \cite{kp2,cm1}. For the question
to be interesting, one has to assume that on one side of the bounce,
say to the past, the states are sharply peaked about a classical
trajectory at early times, in the sense that the relative
fluctuations in both $\h{p}_{(\phi)}$ and $\hat{V}|_\phi$ are small,
and investigate if they remain small and comparable on the other
side of the bounce, i.e., in the distant future.

We will begin with a powerful general result \cite{kp2}. It is based
on a novel scattering theory which extends also to other
cosmological models which are not exactly soluble. There is no
restriction on states%
\footnote{That is, results hold for all states in the domains of
operators considered.}
and the results are analytic, with excellent mathematical control on
estimates. The focus is on the fluctuations of logarithms
$\ln\h{p}_{(\phi)}$ and $\ln\h{V}|_\phi$ of the Dirac observables
under consideration. If a state is sharply peaked so that these
dispersions are very small, then they provide excellent
approximations to the relative fluctuations $ \Delta_R\, \h{V}|_\phi
:= ((\Delta \h{V}|_\phi)^{1/2})/ (\langle \h{V}|_{\phi_o}\rangle)$
and $\Delta_R\, \h{p}_{(\phi)} := ((\Delta \h{p}_{(\phi)})^{1/2})/
(\langle \h{p}_{(\phi)}\rangle)$. The analysis provides an
interesting inequality relating the fluctuations of $\ln\h{V}|_\phi$
in the distant future and in the distant past and the fluctuation in
$\ln\h{p}_{(\phi)}$ (which is constant in time):
\be \label{kpbound} | \sigma_+ - \sigma_- | \leq   2 \sigma_\star
\ee
where
\be \sigma_\pm = \langle \Delta \ln
\f{\h{V}|_\phi}{2\pi\gamma\lambda\lp^2}\, \rangle_\pm, \quad {\rm
and} \quad \sigma_\star = \langle \Delta \ln(\f{\hat
p_{(\phi)}}{\sqrt{G} \hbar}) \rangle \, .\ee
Here the operators have been divided by suitable constants to make
them dimensionless. (Taking logarithms has a conceptual advantage in
that, in the spatially non-compact case, the result is manifestly
independent of the choice of the cell $\C$ made to construct the
theory.) Thus, if we begin with a semi-classical state in the
distant past well before the bounce with, say, $\sigma_- =
\sigma_\star =\epsilon \ll 1$ then we are \emph{guaranteed} that the
dispersion $\sigma_+$ in the distant future after the bounce is less
than $3\epsilon$. In this rather strong sense semi-classicality is
guaranteed to be preserved across the bounce. Furthermore, the bound
is not claimed to be optimal; the preservations could well be even
stronger. However, note also that (\ref{kpbound}) does not rule out
the possibility that $\sigma_+$ is \emph{much smaller} than
$\epsilon$ in which case the relative fluctuation in volume after
the bounce would be \emph{much smaller} than those before. On the
other hand there is no definitive result that, for semi-classical
states of interest, the relative fluctuations before and after the
bounce can in fact be significantly different. We are aware of only
one numerical calculation \cite{mb3} indicating this and it has been
challenged by a more recent result \cite{cm1} summarized below.

For special classes of states, one can obtain stronger results,
ensuring that the fluctuations in volume in the distant past and in
the distant future are comparable. Recall that, given \emph{any}
state, the expectation values $\langle \h{V}|_{\phi_o}\rangle$ of
the volume operator are symmetric around a bounce time $\phi_{B}$.
One can similarly show \cite{cs1} that the expectation values
$\langle \hat V^2|_\phi \rangle$ of the square of the volume
operators are also symmetric about a time $\phi = \phi_B^\prime$
which also depends on the choice of the state. In general, the two
times are not the same. But they have been shown to be the same for
generalized Gaussians \cite{cs1}. From the definition of fluctuation
$\Delta \h{V}|_\phi$ it now follows that the relative fluctuations
$\Delta_R \h{V}|_\phi$ are also symmetric about $\phi=\phi_B$. Thus,
for these states the memory of fluctuations is preserved
\emph{exactly}. How big is this class of states? Recall from
(\ref{chilx}) that each physical state $\chilx$ of LQC is determined
by a function $F(x)$ whose Fourier transform has support just on the
positive half $k$-line. The generalized Gaussians are those states
for which the Fourier transform is the restriction to the positive
half line of functions
\be F(k) = k^n\, e^{-\f{(k-k_o)^2}{\sigma^2} + i p_ok} \ee
where $n$ is a positive integer, the parameters $k_o, \sigma$ are
positive and $p_o$ is any real number. (The factor $k^n$ ensures
that $F(k)$ is sufficiently regular at $k=0$.) This is a `large' set
in the sense that it forms an overcomplete basis. But the result
holds just for these states and \emph{not} their superpositions.

The third result is along the same lines; it allows more general
states but now the dispersions are no more exactly symmetric. In
addition to the multiplicative factors $k^n$, the function $F(k)$ is
now allowed squeezing: the parameter $\sigma$, in particular, can be
complex. In this case, one bound is given by \cite{cm1}
\be \label{cmbound} \Big{|}\,\Delta_R^+\, \h{V}|_\phi  -
\Delta_R^-\,\, \h{V}|_\phi\, \Big{|}\,\, \le \,\,
\big(\f{\kappa}{\langle \h{p}_{(\phi)}\rangle}\big)\,\,
\Delta_R^\pm\,\, \h{V}|_\phi \ee
where the superscripts $\pm$ refer to the distant past and distant
future, and $\kappa$ is a constant, $\sim\, 66$ in Planck units.
Since the factor $\Delta_R^\pm\, \h{V}|_\phi$ appears on the right
side, this bound implies that, for the generalized squeezed states,
the concern that the fluctuation in the future can be much smaller
than that in the past (or vice versa) is not realized. The physical
content of this bound is most transparent in if the spatial sections
are compact with $\mathbb{T}^3$ topology. To make the discussion
more concrete let us suppose that, when the hypothetical universe
under consideration has a radius equal to the observable radius of
our own universe at the CMB time, it has the same density as our
universe then had. For such a universe $\langle \h{p}_{(\phi)}
\rangle \approx 10^{126}$ in Planck units. Thus, the coefficient on
the right side of (\ref{cmbound}) is $\sim\,10^{-124}$!

To summarize, there are now two types of results on fluctuations: a
general result (that holds for arbitrary states) and bounds the
volume dispersion in the distant future of the bounce in terms of
its value in the distant past and the dispersion in the scalar field
momentum \cite{kp2}, and, stronger bounds for special classes of
states \cite{cs1,cm1}.

\subsection{Relation between the WDW theory and sLQC}
\label{s3.4}

As we saw in section \ref{physicalcons1}, in the model under
consideration, all quantum states encounter a singularity in the
\WDW theory but undergo a quantum bounce in sLQC. The occurrence of
bounce is a direct manifestation of the underlying quantum geometry,
captured by the parameter $\lam$, which becomes important when the
space-time curvature approaches Planck regime. On the other hand,
when the space-time curvature is small, i.e. at large volumes for a
given value of $\pphi$, the \WDW theory is an excellent
approximation to sLQC. So a natural question arises: Can \WDW theory
be derived from sLQC in the limit $\lam \rightarrow 0$?

To address this issue, one has to let $\lambda$ vary. Let us call
the resulting theory sLQC$_{(\lam)}$. Now, a necessary condition for
the \WDW theory to be the limit of sLQC${}_{(\lam)}$ is that, given
any state in the \WDW theory, there should exist a state in
sLQC${}_{(\lam)}$ such that the expectation values of the Dirac
observable $\h{V}|\phi$ in the two states remain close to each other
for all $\phi$, so long as $\lambda$ is chosen to be sufficiently
small. Let us suppose we want the predictions of the two theories to
agree within an error $\epsilon$. Then, there should exist a $\delta
>0$ such that, for all $\lambda <\delta$,
\be \label{vbound} |\langle \hat V|_\phi \rangle_{(\lam)} -
\langle \hat V|_\phi \rangle_{(\mathrm{wdw})}| < \epsilon  \ee
for all $\phi$. From (\ref{wdw_vol1}) and (\ref{slqc_vol}), we
have
\be \langle \hat V|_\phi \rangle_{(\mathrm{wdw})} = V_o\,
e^{\sqrt{12\pi G}\, \phi} \qquad {\rm and} \qquad \langle \hat
V|_\phi \rangle_{(\lam)} = V_+ e^{\sqrt{12\pi G} \, \phi} + V_-
e^{-\sqrt{12\pi G}\, \phi} \ee
whence it immediately follows that if the (\ref{vbound}) is to hold
for all positive $\phi$, then we must have $V_* = V_+$ and $V_- <
\epsilon$. But since $V_-$ is necessarily non-zero, irrespective of
the choice of $\delta$, (\ref{vbound}) will be violated for a
sufficiently large negative value of $\phi$. Note however that for
any fixed positive $\phi_o$, one can choose a sufficiently small
$V_-$ so that (\ref{vbound}) holds in the semi-infinite interval
$(-\phi_o, \infty)$. (Similarly, if we use the right moving sector
of the \WDW theory, (\ref{vbound}) can be made to hold in the
semi-infinite intervals $(-\infty, \phi)$.) But this approximation
fails to be uniform in $\phi_o$ whence the \WDW theory cannot arise
in the limit $\lambda \to 0$ of sLQC$_{(\lam)}$. This argument is
rather general and does not require the specification of a precise
map between the two theories. But a useful map can be constructed
using the basic ideas that underlie renormalization group flows and
brings out the relation between the two theories more explicitly
\cite{acs}.

These considerations naturally lead us to another question: Does the
limit $\lambda \to 0$ of sLQC$_{(\lam)}$ yield a well-defined theory
at all? The answer turns out to be in the negative: \emph{LQC is a
fundamentally discrete theory.} This is in striking contrast to the
examples in `polymer quantum mechanics' \cite{afw,cvz,cvz2} and
lattice gauge theories where a continuum limit does exists when
discreteness parameter is sent to zero. Physical reasons behind this
difference are discussed in \cite{acs}.

\subsection{Path integral formulation}
\label{s3.5}

The key difference between the \WDW theory and LQC is that, thanks
to the quantum geometry inherited from LQG, LQC has a novel,
in-built repulsive force. While it is completely negligible when
curvature is less than, say, 1\% of the Planck scale, it grows
dramatically once curvature becomes stronger, overwhelming the
classical gravitational attraction and causing a quantum bounce that
resolves the big-bang singularity. From a path integral viewpoint
(see, e.g., \cite{swh2}), on the other hand, this stark departure
from classical solutions seems rather surprising at first. For, in
the path integral formulation quantum effects usually become
important when the action is small, comparable to the Planck's
constant $\hbar$, while the Einstein-Hilbert action along classical
trajectories that originate in the big-bang is generically very
large. Thus, there is an apparent conceptual tension. In this
sub-section we will summarize the results of a detailed analysis
\cite{ach3} that has resolved this issue in k=0 FLRW model.
(Extension of this analysis to the k=1 case is direct, see
\cite{hmq}.)

\subsubsection{Strategy}
\label{s3.4.1}

Since LQC uses a Hilbert space framework, it is most natural to
return to the original derivation of path integrals, where Feynman
began with the expressions of transition amplitudes in the
Hamiltonian theory and \emph{reformulated} them as an integral over
all kinematically allowed paths \cite{rpf}. But in non-perturbative
quantum gravity, there is a twist: at a fundamental level, one deals
with a constrained system without external time whence the notion of
a transition amplitude does not have an a priori meaning. It is
replaced by an \emph{extraction amplitude} ---a Green's function
which extracts physical quantum states from kinematical ones and
also provides the physical inner product between them (see, e.g.,
\cite{ach1,ach2,hrvw}). If the theory can be deparameterized, it
inherits a relational time variable and then the extraction
amplitude can be re-interpreted as a transition amplitude with
respect to that time \cite{ach2}. In LQC with a massless scalar
field, we saw that a natural deparametrization is indeed available.
However, more generally ---e.g. if one were to introduce a potential
for the scalar field--- it is difficult to find a \emph{global} time
variable. Since the conceptual tension between LQC results and the
path integral intuition is generic, it is best not to have to rely
heavily on deparametrization. Therefore, the comparison was carried
out in the timeless framework. This is also the setting of spin
foams, the path integral approach to full LQG; see e.g.
\cite{eprl,fk,newlook}. The idea then is to start with the
expression of the extraction amplitude in the Hilbert space
framework of LQC, cast it as a path integral, and re-examine the
tension between the path integral intuition and the singularity
resolution.

As we have indicated in section \ref{s2}, solutions to the
constraint equation, as well as inner product between them, can be
obtained through a group averaging procedure \cite{dm,almmt,abc}.
The extraction amplitude $\E(\nu,\phi; \nu',\phi')$ is a Green's
function that results from this averaging:
\be \label{phy1} \E(\nu_f,\phi_f; \nu_i,\phi_i):=
\sint_{-\infty}^{\infty} \dd{\alpha} \,\, \langle \nu_f,\phi_f | \,
e^{\frac{i}{\hbar} \alpha \h{C}} \,| \nu_i, \phi_i \rangle\ , \ee
where $\h\C = -\hbar^2 (\p_\phi^2 + \Theta)$ is the full Hamiltonian
constraint, $\alpha$ is a parameter (with dimensions $[L^{-2}]$),
and the ket and the bra are eigenstates of the operators $\h{V}$ and
$\h\phi$ on the kinematical Hilbert space $\Hk$.%
\footnote{Since in LQC we restrict ourselves to the `positive
frequency part' there is an implicit $\theta(\hat{p}_{(\phi)})$
factor multiplying $e^{(i/\hbar)\alpha \h{C}}$ in (\ref{phy1}) where
$\theta$ is the unit step function. We do not write it explicitly
just to avoid unnecessary proliferation of symbols.}
The integral averages the ket (or the bra) over the group generated
by the constraint. Since the integrated operator is heuristically
`$\delta(\hat{C})$,' the amplitude $\E(\nu_f, \phi_f; \nu_i,\phi_i)$
satisfies the constraint in both of its arguments. Consequently, it
serves as a Green's function that extracts physical states
$\Psi_{\rm phys} (\nu,\phi)$ in $\Hp$ from kinematical states
$\Psi_{\rm kin} (\nu,\phi)$ in $\Hk$ through a convolution
\be \Psi_{\rm phys}(\nu,\phi) = \sum_{\nu^\prime}\, \sint
\dd{\phi}^\prime\,\, \E(\nu,\phi; \nu',\phi') \,\Psi_{\rm
kin}(\nu',\phi'). \ee
This is why $\E$ is referred to as the \emph{extraction amplitude}.
Since it is the group averaging Green's function, $\E$ also enables
us to write the physical inner product in terms of the kinematical:
\be (\Phi_{\rm phys},\, \Psi_{\rm phys})  :=  \sum_{\nu,\, \nu'}\,
\sint \dd{\phi}\, \dd{\phi'}\, \bar{\Phi}_{\rm kin}(\nu,\phi)\,
\E(\nu,\phi; \nu',\phi') \Psi_{\rm kin}(\nu',\  \phi'). \ee
Thus, in the `timeless' framework without any deparametrization, all
the information in the physical sector of the quantum theory is
neatly encoded in the extraction amplitude $\E(\nu,\pphi;
\nu',\pphi')$. Therefore to relate the Hilbert space and path
integral frameworks, it suffices to recast the expression
(\ref{phy1}) of this amplitude as a path integral.

\subsubsection{Path integral for the extraction amplitude}
\label{s3.4.2}

Let us begin by recalling the procedure Feynman used to arrive at
the path integral expression of the transition amplitude in quantum
mechanics. He began with the Hamiltonian framework, wrote the
unitary evolution as a composition of $N$ infinitesimal ones,
inserted a complete basis between these infinitesimal evolution
operators to arrive at a `discrete time' path integral, and finally
took the limit $N \to \infty$. In the timeless framework, we need to
adapt this procedure to the extraction amplitude $\E(\nu_f,\phi_f;
\nu_i,\phi_i)$, using $e^{\frac{i}{\hbar} \alpha \h{C}}$ in the
integrand of Eq. (\ref{phy1}) in place of the `evolution' operator,
and then performing the $\alpha$ integration in a final step.

More precisely, the integrand of (\ref{phy1}) can be thought of as a
matrix element of a \emph{fictitious} evolution operator
$e^{\frac{i}{\hbar} \alpha \h{C}} $. One can regard  $\alpha \h{C}$
as playing the role of a (purely mathematical) Hamiltonian, the
evolution time being unit. We can then decompose this fictitious
evolution into $N$ evolutions of length $\epsilon = 1/N$ and insert
a complete basis at suitable intermediate steps to write the
extraction amplitude as a discrete \emph{phase space} path integral
and finally take the limit $N\to \infty$ (or $\epsilon \to 0$). As
is standard in the path integral literature, the result can be
expressed as a formal infinite dimensional integral,
\be \label{pi} \E(\v_f,\phi_f; \v_i,\phi_i)  = \sint \dd \alpha\,
\sint [\mathcal{D}\nu(T)]\,\, [\mathcal{D}\b(T)]\,\,
[\mathcal{D}\pphi(T)]\, [\mathcal{D}\phi(T)]\,\,\,
e^{\frac{i}{\hbar} S}\, .  \ee
A detailed calculation \cite{ach3} shows that the action $S$ in this
expression is given by
\be \label{qia} S=\int_{0}^{1}\dd T\left(\pphi {\phi}^\prime
-\frac{1}{2}\b {\nu}^\prime-\alpha \left(\pphi^2-3\pi G \nu^2\,
\frac{\sin^2\lambda \b}{\lambda^2}\right) \right) \ee
where the prime denotes derivative with respect to the (fictitious)
time $T$. Note that the final integration is over \emph{all} paths
in the classical phase space, \emph{including the ones which go
through the big-bang singularity}. Therefore, the tension between
the Hamiltonian LQC and the path integral formulation is brought to
forefront: How can we see the singularity resolution in the path
integral setting provided by (\ref{pi})? The answer is that the
paths are not weighted by the standard Einstein-Hilbert action but
by a `polymerized' version (\ref{qia}) of it which still retains the
memory of the quantum geometry underlying the Hamiltonian theory. As
we will see, this action is such that a path going through the
classical singularity has negligible contribution whereas bouncing
trajectories give the dominant contribution.

Note that the new action (\ref{qia}) retains memory of quantum
geometry through the area gap $\lambda^2$. However, since
$\lambda^2$ depends on $\hbar$, this means that the Einstein-Hilbert
action itself has received quantum corrections. This may seem
surprising at first. However, this occurs also in some familiar
examples if one systematically arrives at the path integral starting
from the Hilbert space framework. Perhaps the simplest such example
is that of a non-relativistic particle on a curved Riemannian
manifold for which the standard Hamiltonian operator is simply
$\hat{H} = - (\hbar^2/2m) g^{ab} \nabla_a\nabla_b$. Quantum dynamics
generated by this $\hat{H}$ can be recast in the path integral form
following the Feynman procedure \cite{rpf}. The transition amplitude
is then given by \cite{bdw2}
\be \langle q,T| q',T'\rangle = \sint \mathcal{D}[q(T)]\,\,\,
e^{\f{i}{\hbar} S} \ee
with
\be S = \sint\, \dd T\, (\f{m}{2} g_{ab}\,{q'}^a {q'}^b \,+\,
\f{\hbar^2}{12m} R)\ee
where $R$ is the scalar curvature of the metric $g_{ab}$ and the
prime denotes derivative with respect to time $T$. Thus the
classical action receives a quantum correction. In particular, the
extrema of this action are not the geodesics one obtains in the
classical theory but rather particle trajectories in a
$\hbar$-dependent potential; the two can be qualitative different.

\subsubsection{The steepest descent approximation}
\label{3.5.3}

We can now use the steepest decent approximation to understand the
singularity resolution in LQC from a path integral perspective. As
in more familiar systems, including field theories, one expects that
the extraction amplitude can be approximated as
\begin{equation}\label{spa}
\E(\nu_f,\phi_f; \nu_i,\phi_i) \sim \left(\det{\delta^2 S}|_0
\right)^{-1/2}\,\,  e^{\frac{i}{\hbar} S_0 }\, ,
\end{equation}
where $S_0$ is the action evaluated along the trajectory extremizing
the action, keeping initial and final configuration points fixed.
However there are two subtleties that need to be addressed. First,
the standard WKB analysis refers to unconstrained systems and we
have to adapt it to the constrained one, replacing the Schr\"odinger
equation with the quantum constraint. This is not difficult (see,
e.g., sections 3.2 \& 5.2 of \cite{crbook}, or Appendix A in
\cite{ach3}). However, as in the standard WKB approximation for the
transition amplitude, the procedure assumes that the action that
features in the path integral has no explicit $\hbar$ dependence.
The second subtlety arises because, in our case, the action does
depend on $\hbar$ through $\lambda \sim \sqrt{\gamma^3\hbar}$.
Therefore to explore the correct semi-classical regime of the theory
now one has to take the limit $\hbar\to 0$, keeping $\gamma^3\hbar$
fixed. To highlight the fact that $\gamma^3 \hbar$ is being kept
constant in the limit $\hbar \to 0$, it is customary to use inverted
commas while referring to the resulting `classical' and
`semi-classical' limit. The conceptual meaning of the `classical'
limit is as follows: $\gamma\to \infty$ corresponds just to ignoring
the new term in the Holst action for general relativity in
comparison with the standard Palatini term \cite{alrev}. What about
the `semi-classical' approximation? In this LQC model, eigenvalues
of the volume operator are given by $(8\pi G\gamma\lambda \hbar)n$
where $n$ is a non-negative integer. Therefore, in the
`semi-classical limit' the spacing between consecutive eigenvalues
goes to zero and $\nu$ effectively becomes continuous as one would
expect. Finally, states that are relevant in this limit have large
$n$, just as quantum states of a rigid rotor that are relevant in
the semi-classical limit have large $j$.

\begin{figure}[h]
\begin{center}
\includegraphics[height=3in,width=3in,angle=0]{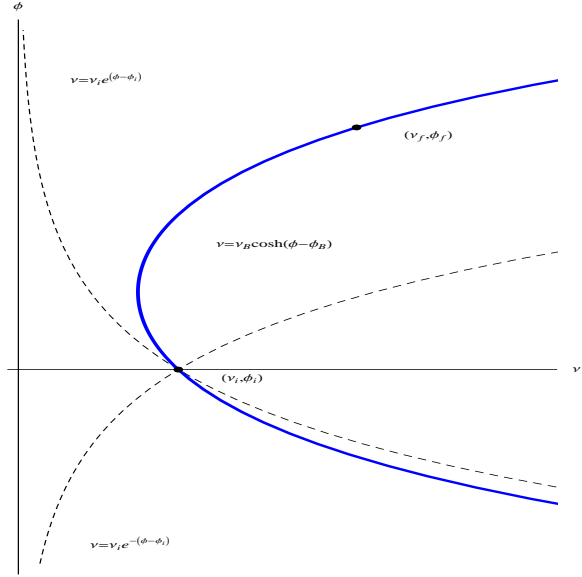}
\caption{For fixed ($\nu_i,\phi_i)$, the (dashed) curves  $\nu_f= \nu_i\,
e^{\pm \sqrt{12 \pi G} (\phi_f-\phi_i)}$ divide the $(\nu_f,\phi_f)$
plane into four regions. For a final point in the
upper or lower quadrant, there always exists a real trajectory joining
the given initial and final points (as exemplified by the thick line).
If the final point lies in the left or right quadrant, there is no real
solution matching the two points. The action becomes imaginary and one
gets an exponentially suppressed amplitude.}
\label{regions}
\end{center}
\end{figure}

We can now evaluate the extraction amplitude in the saddle point
approximation (\ref{spa}). To calculate the Hamilton-Jacobi
functional $S_0$, we first note that the extrema (with positive
`frequency', i.e. positive $\pphi$) are given by
\ba \nu(\phi) & = & \nu_{\B}\cosh(\sqrt{12 \pi G}(\phi-\phi_{\B})),
\label{solns1} \\
\b(\phi) & = & \frac{2\,\, \text{sign}(\nu_{\B})}{\gamma\lambda}
\tan^{-1}(e^{-\sqrt{12 \pi G}(\phi-\phi_{\B})}) \label{solns2}. \ea
where $\nu_{\B}, \phi_{\B}$ are integration constants representing
values of $\nu,\phi$ at the bounce point. As seen from the $\cosh$
dependence of the volume, these trajectories represent bouncing
universes. Since $\nu(\phi)$ can vanish only on the trajectory with
$\nu_{\B}=0$ ---i.e. the trajectory $\nu(\phi) = 0$ for all
$\phi$--- $\nu$ cannot vanish on any `classical, trajectory which
starts out away from the singularity. It is straightforward to
verify that a real `classical' solution exists for given initial and
final points $(\nu_i, \phi_i;\, \nu_f,\phi_f)$ if and only if
\be \label{region} e^{-\sqrt{12 \pi G} |\phi_f-\phi_i|}\, < \,
\f{\nu_f}{\nu_i} \,< \, e^{\sqrt{12 \pi G}|\phi_f-\phi_i|}. \ee
For a fixed $\nu_i, \phi_i$, the `classically' allowed region for
$\nu_f,\phi_f$ consists of the upper and lower quadrants formed by
the dashed lines in Fig.\ref{regions}. For $\nu_f,\phi_f$ in these
two quarters, $S_0$ is real and thus the amplitude (\ref{spa}) has
an oscillatory behavior. Outside these regions the action becomes
imaginary and one gets an exponentially suppressed amplitude. Thus,
the situation is analogous to that in quantum mechanics.

To summarize, in the timeless framework all the physical information
is contained in the extraction amplitude which reduces to the
standard transition amplitude if a global deparametrization can be
found. Following Feynman, one can start with the Hilbert space
expression of the extraction amplitude and recast it as a phase
space path integral. Quantum geometry effects of LQC leave their
trace on the weight associated with each path: The action functional
is modified. This quantum modification of the action governing path
integrals is not an exceptional occurrence; the phenomenon is
encountered already in the transition from Hilbert spaces to path
integrals for particles moving on a curved Riemannian manifold
\cite{bdw2}. It implies that, in the WKB approximation, the
extraction amplitude is dominated by universes that undergo a
bounce. Thus, from the LQC perspective, it would be incorrect to
simply define the theory starting with the Einstein Hilbert action
because this procedure completely ignores the quantum nature of the
underlying Riemannian geometry. For a satisfactory treatment of
ultraviolet issues such as the singularity resolution, it is crucial
that the calculation retains appropriate memory of this quantum
nature. Indeed, this is why in the spin foam models one sums over
quantum geometries, not smooth metrics.

\section{Generalizations}
\label{s4}

In this section we will retain the homogeneity assumption but
consider a number of generalizations of the k=0 FLRW model to
include a cosmological constant, spatial curvature and
anisotropies. In each case we will encounter new conceptual
issues that will require us to extend the strategy developed in
section \ref{s2} to define a satisfactory Hamiltonian
constraint and analyze the dynamics it leads to. While in
retrospect the necessary extensions are very natural, they were
not a priori obvious in the course of the development of the
subject. Indeed, one repeatedly found that the `obvious'
choices can lead to theories which fail to be physically
viable. The fact that even in these simple models one has to
make judicious choices in the intermediate steps in arriving at
the `correct' definition of the Hamiltonian constraint suggests
that the apparent freedom in defining dynamics in full LQG may
in fact be highly constrained once one works out the dynamical
consequences of various choices.

Since the k=0 FLRW model was discussed in the last two sections in
great detail, in this section we will be brief, focusing primarily
on conceptual differences, new technical difficulties and their
resolutions. So far, none of these models could be solved exactly.
Therefore, numerical simulations are now essential. Therefore, by
and large, we will now closely follow the strategies used in
\cite{aps3} rather than those used subsequently to show and exploit
exact solvability of the k=0, $\Lambda$=0 FLRW model discussed in
section \ref{s3}. The numerical results bring out the fact that the
singularity resolution is not tied to exact solubility.

\subsection{Inclusion of spatial curvature: The k=1 FLRW model}
\label{s4.1}

Although the k=1 model is not observationally favored, its
quantization is important for two conceptual reasons. First, it
enables one to test whether the quantum bounce of the k=0 case
survives the inclusion of spatial curvature and, if so, whether the
value of the maximum energy density is robust. The answer to these
questions turns out to be in the affirmative. Second, the k=1 case
provides a sharper test of the infrared viability of LQC. For, if
matter sources satisfy the strong energy condition, the k=1
universes recollapse when its energy density reaches a value
$\rho_{\mathrm{min}} = 3/(8 \pi G a^2_{\mathrm{max}})$. One can now
ask if the recollapse in LQC respects this relation for universes
with large $a_{\rm max}$. This question provides a
\emph{quantitative} criterion to test if LQC agrees with general
relativity when space-time curvature is small. Initially, there was
concern that LQC may not passe this test \cite{gu}. Detailed
analysis showed that it does \cite{apsv}.

There are also two technical reasons that warrant a careful analysis
of k=1 model in LQC. The first concerns the strategy for obtaining
the curvature operator using holonomies round closed loops,
discussed in section \ref{s2.5}. In the k=0 model, one could
construct the necessary closed loops following the integral curves
of the fiducial triads $\e^a_i$. In the k=1 case these triads don't
commute whence their integral curves do not from closed loops. Can
one still define the field strength operator? In the early stage of
LQC it was suggested \cite{bdv} that one should simply add an extra
edge to close the loop. In addition to being quite ad-hoc, this
procedure turns out not to be viable because then the loop does not
enclose a well-defined area. A satisfactory strategy was developed
in \cite{apsv} and independently and more elegantly in
\cite{warsaw1}. The second technical point concerns numerics. In
contrast to the difference operator $\Theta$ used in the k=1 case,
the analogous operator $\Theta_{(k=1)}$ admits a (purely) discrete
spectrum, whence it is much more difficult to find its eigenvalues
and eigenfunctions numerically.

\subsubsection{Classical Theory}
\label{s4.1.1}

The spatial manifold $M$ is now topologically $\mathbb{S}^3$, which
can be identified with the group manifold of SU(2). We will use the
Cartan Killing form $\q_{ab}$ on SU(2) as our fiducial metric. The
fiducial volume $V_o$ of $S^3$ is then given by $V_o =: \ell_o^3 =
16\pi^2$. The fiducial triads $\e^a_i$ on $M$ can be taken to be the
left invariant vector fields of SU(2). As before we will denote the
corresponding co-triads by $\o_a^i$. The three $\e^a_i$ and the
three right invariant vector fields $\ox_i^a$ constitute the six
Killing fields of every metric $q_{ab}$ of the k=1 model. The three
vector fields in each set $\{\ox^i\}$ and $\{\e^a_i\}$ satisfy the
commutation relations of su(2) among themselves and each vector
field from the first set commutes with each vector in the second. 

As in the k=0 case, one can solve and gauge fix the Gauss and
diffeomorphism constraints and coordinatize the gravitational phase
space by $(c,p)$:
\be A_a^i = c \,\, \ell_o^{-1} \o^i_a, \qquad E^a_i = p \,\,
\ell_o^{-2} \, \sqrt{\mathring{q}}\, \e^a_i \ee
where ${\q}$ is the determinant of the fiducial metric ${\q}_{ab}$.
The physical metric $q_{ab}$ is given by $q_{ab} = |p| \ell_o^{-2}\,
{\q}_{ab}$. Computing the extrinsic and intrinsic curvatures on the
homogeneous slices, the curvature of the connection $A_a^i$ turns
out to be
\be F_{ab}^k = \ell_o^{-2} \left(c^2 - c \, \ell_o \right) \,
\epsilon_{ij}^{~~k} \, \o^j_a \, \o^k_b ~. \ee
The gravitational part of the Hamiltonian constraint is given by
\ba C_{\mathrm{grav}} &=& \nonumber -\gamma^{-2} \int_{\cal C} \dd^3
x \, N\, (\det q)^{-1/2} \, \epsilon^{ij}_{~~k} \, E^a_i E^b_j
\bigg[F_{ab}^k - \left(\f{1 + \gamma^2}{4}\right)
\mathring{\epsilon}_{ab}^{~~c} \o_c^k \bigg] \\
&=& - 6 \gamma^{-2} |p|^2 \bigg[\left(c - \f{\ell_o}{2}\right)^2 +
\f{\gamma^2 \ell_o^2}{4}\bigg] \ea
where in the second step, we have chosen the lapse $N = a^3 = V/V_o$
as in the quantization of the k=0 model. The constraint for the
spatially flat model can be obtained if we set $\ell_o = 0$. (If
setting $\ell_o =0$ seems counterintuitive, see \cite{apsv} for
details.)

\subsubsection{Quantum Theory}
\label{s4.1.2}

The  kinematical part of quantization and properties of ${\cal
H}_{\mathrm{kin}}^{\mathrm{grav}}$ are similar to those in the k=0
case, discussed in section \ref{s2.4}. The next step is to express
the Hamiltonian constraint in terms of holonomies
---the elementary connection variables that can be immediate
quantized. In this step, an important technical subtlety
arises: As explained in the beginning of this subsection, the
construction of the required loops $\Box_{ij}$ is more subtle
because the triad vector fields $\e^a_i$ do not commute.
However, since these left invariant vector fields do commute
with the right invariant vector fields $\ox^a_i$, it is now
natural to form closed loops $\Box_{ij}$ by following the left
invariant vector fields along, say, the $i$ direction and the
right
invariant ones in the $j$ direction \cite{apsv,warsaw1}. 
By computing holonomies along such loops and shrinking them, as in
section \ref{s2.5}, so that they enclose the minimum possible
physical area, one obtains the expression of the curvature operator:
\be \h{F}_{ab}^k  = \epsilon_{ij}^{~~k}\, V_o^{-\f{2}{3}}\, \o^i_a
\o^j_b\,\, \left( {\f{\sin^2 \bar \mu (c - \f{\ell_o}{2}
)}{\bar\mu\ell_o}} - {\f{\sin^2 (\f{\bar \mu
\ell_o}{2})}{\bar\mu\ell_o}} \right) \, \ee
where $\bar\mu \sim 1/\sqrt{|p|}$ exactly as in Eq (\ref{mubar}) and
where, as before, we have dropped the hats on trignometric operators
for notational simplicity.

As in the k=0 case, the quantum Hamiltonian constraint simplifies
considerably if one works in the volume representation, i.e., with
wave functions $\Psi(\nu,\phi)$, and uses the equality $\bar \mu c =
\lambda \b$. However, in place of operators of the form $\sin
(\lambda \b)$, we now have to deal with $\sin(\lambda \b -
\ell_o/2)$. To find its action on states $\Psi(\nu)$, we note that
in the kinematical Hilbert space, we have the identity
\be \sin \big(\lambda \b - \f{\ell_o}{2} \big)\, \Psi(\nu) \, = \,
e^{i \ell_o f} \, \sin \lambda \b \, e^{-i \ell_o f}\,\, \Psi(\nu)
\ee
where
\be f = \f{3}{8 \tilde K} \, \mathrm{sgn}(\nu) \, \nu^{2/3} ,\quad
{\rm with}\quad \tilde K := 2 \pi \gamma \lp^2 ~. \ee
Using this identity, it is straightforward to determine the action
of the operator $\h{C}_{\mathrm{grav}}$ on $\Psi(\nu)$. With the
choice of lapse $N = a^3$, for the matter Hamiltonian of the
massless scalar field, the total constraint $\hat{C}_H
\Psi(\nu,\phi) = (\hat{C}_{\mathrm{grav}} + 16 \pi G
\hat{C}_{\mathrm{matt}})\, \Psi(\nu,\phi) = 0$, turns out to be
\ba \label{k1-hc}
\p_\phi^2 \Psi(\nu,\phi) \, &=& \, \nonumber - \Theta_{(k=1)} \, \Psi(\nu,\phi)\\
&=& - \nonumber \Theta \Psi(\nu,\phi) + \f{3 \pi G}{\lambda^2} \,
\nu \bigg[\sin^2\left(\f{\lambda}{\tilde K \nu^{1/3}} \f{\ell_o}{2}
\right) \, \nu \,  - \, (1 + \gamma^2) \left(\f{\lambda}{\tilde K}
\f{\ell_o}{2}\right)^2 \, \nu^{1/3}\bigg] \, \Psi(\nu,\phi)  \\
\ea
where, $\Theta$ is the k=0, second order difference operator  Eq
(\ref{qhc4}). Thus the quantum constraint in k=1 model turns out to
be the one in the k=0 model, with an additional term, due to the
non-vanishing intrinsic curvature, that acts simply by
multiplication. If we set $\ell_o = 0$, we recover the k=0 quantum
constraint. Further, as in the case of the spatially flat model, one
can show that at large volumes and sufficiently smooth wave
functions, the LQC quantum difference equation is well approximated
by the WDW differential equation.
\\

The operator $\Theta_{(k=1)}$ is self-adjoint and positive definite
\cite{warsaw1} and using group averaging one can obtain a physical
Hilbert space ${\cal H}_{\mathrm{phy}}$. As in the k=0 model,
physical states $\Psi(\nu, \phi)$ are `positive frequency' solutions
to (\ref{k1-hc}), i.e. satisfy
\be -i \hbar \p_\phi \Psi(\nu,\phi) = \sqrt{\Theta}
\Psi(\nu,\phi)\ee
and are invariant under the change of orientation of the triad,
i.e., satisfy $\Psi(\nu, \phi) = \Psi(-\nu, \phi)$. The expression
for the inner product turns out to be the same as Eq (\ref{ip3}) in
the k=0 model.

There is however an important difference in properties of operators
$\Theta_{(k=1)}$ and $\Theta$. In contrast to the spectrum of
$\Theta$, the spectrum of $\Theta_{(k=1)}$ is discrete and each
eigenvalue is non-degenerate \cite{warsaw1}. This feature arises
because the extra term in $\Theta_{(k=1)}$ causes its eigenfunctions
to decay exponentially as $|\nu|$ tends to infinity. This decay is a
reflection in the quantum theory of the classical recollapse. Since
$\Theta$ has discrete eigenvalues, considerable care and numerical
accuracy is needed to find these eigenvalues and corresponding
eigenfunctions. This makes the numerical evolution of physical
states much more challenging than in the k=0 model.

\begin{figure}[]
\begin{center}
$a)$\hspace{8cm}$b)$
\includegraphics[width=3.2in,angle=0]{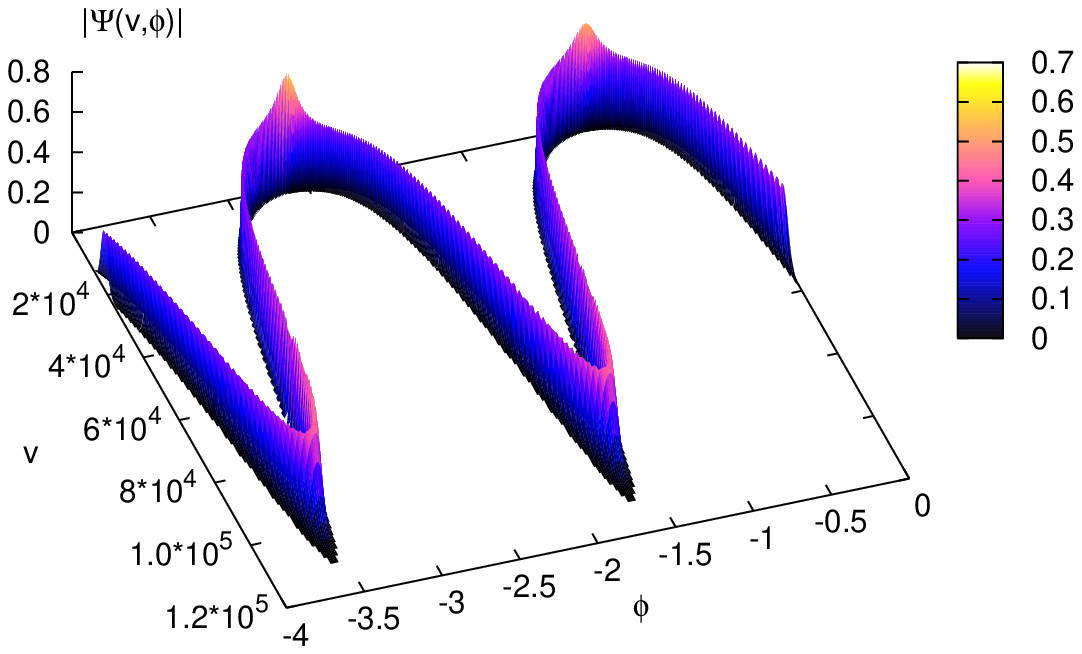}
\includegraphics[width=3.2in,angle=0]{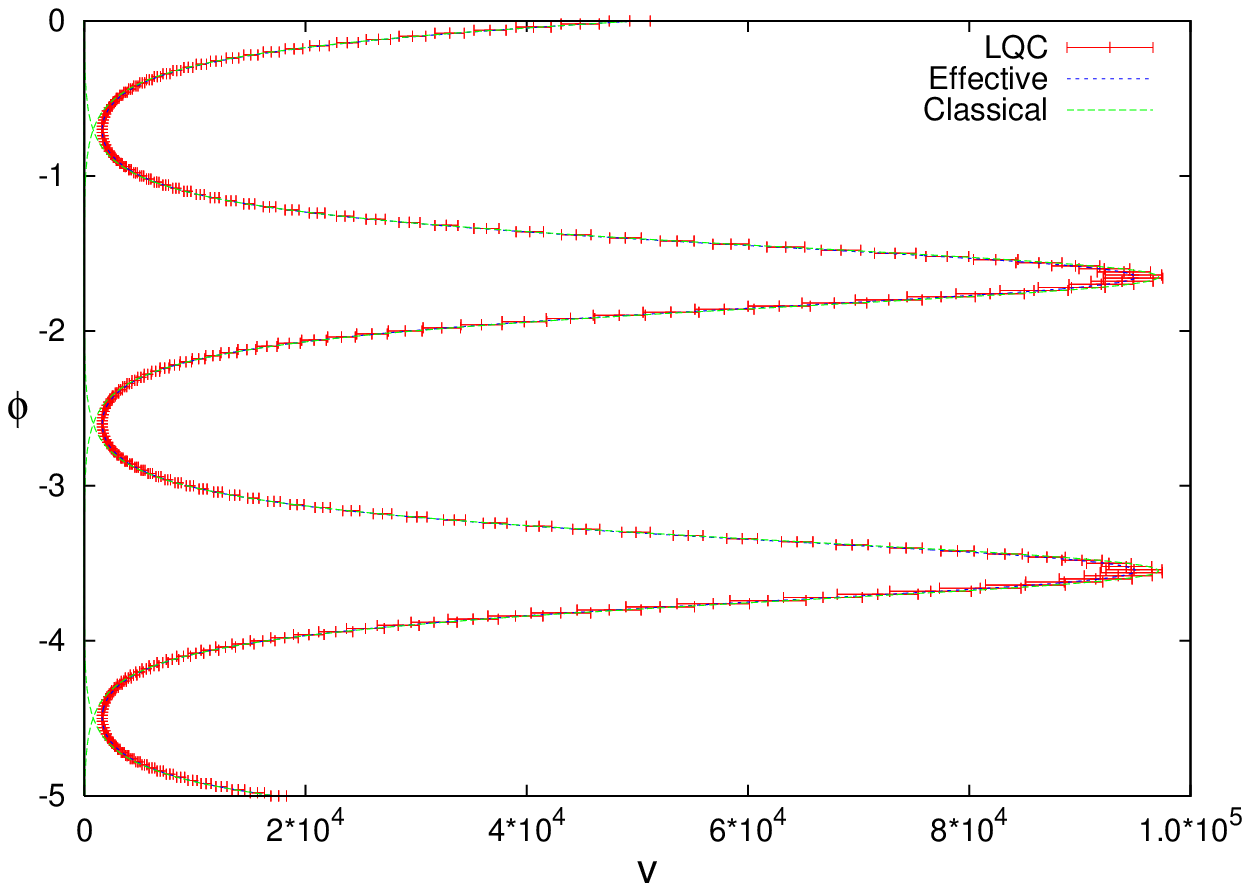}
\caption{Results from a numerical evolution of a state peaked at late
times with the quantum constraint are shown.  $a)$ Plot of the wave function
shows non-singular cycles of expansion and contraction caused by alternating
quantum bounces at $\rho = \rcr$ and the classical recollapse at
$\rho \approx 3/(8 \pi G a^2_{max})$.
It is evident that the peakedness properties of the state do not
significantly change in consecutive cycles. $b)$ Expectation values
of the volume observable are plotted along with the relative fluctuations
and are compared with the classical trajectory and also the trajectory
obtained from the effective Hamiltonian for LQC (see section 5).
The classical trajectory is a good approximation to the quantum dynamics
when space-time curvature is small (large volume regime).
Evolution in LQC shows a recollapse at essentially the same point
as predicted by general relativity. The LQC evolution is non-singular,
whereas classical trajectories undergo a big-bang and a big-crunch.
The effective dynamics trajectory is an excellent approximation to
the quantum dynamics in {\it all} regimes.}
\label{fig:k=1}
\end{center}
\end{figure}

These technical challenges were overcome and extensive numerical
simulations of closed models have been performed \cite{apsv}. In
Fig.\ref{fig:k=1}, we show the results from a typical simulation. As
in the numerical studies with the spatially flat model, one chooses
a state peaked on a classical trajectory when the space-time
curvature is small and evolves it using the quantum constraint.
Using these states and the physical inner product, one then obtains
the expectation values of the Dirac observables: the momentum
$\h{p}_{(\phi)}$, and the volume at a given `time', $\hat V|_\phi$.
The expectation values and the relative fluctuations of
$\h{p}_{(\phi)}$ are constant throughout the evolution. The relative
dispersion of $\hat V|_\phi$ does increase but the increase is
minuscule: For a universe that undergoes a classical recollapse at
$\sim 1$ Mpc, a state that nearly saturates the uncertainty bound
\emph{initially}, with uncertainties in $\h{p}_{(\phi)}$ and
$\h{V}|_\phi$ spread equally, the relative dispersion in
$\h{V}|_\phi$ is still $\sim\,10^{-6}$ after some $10^{50}$ cycles.

The expectation values of volume reveal a quantum bounce which
occurs at $\rho \approx \rcr$ up to the correction terms of the
order of $\lp^2/a_{\mathrm{bounce}}^2$. For universes that grow
to macroscopic sizes, the correction is totally negligible. For
example, for a universe which grows to a maximum volume of $1
Gpc^3$, the volume at the bounce is approximately $10^{117}
\lp^3$! That the bounce occurs at such a large volume may seem
surprising at first. But what matters is curvature and density
and these are \emph{always} of Planck scale at the bounce.
Thus, there is no big-bang singularity; the quantum geometry
effects underlying LQC are again strong enough to cure the
ultraviolet problems of general relativity. What about the
infrared behavior? In this regime there is excellent
quantitative agreement with general relativity. Specifically,
in general relativity $a_{\rm max}$, the scale factor at the
classical recollapse and the density $\rho_{\rm min}$ are
related by $\rho_{\rm min} = 3/(8 \pi G a^2_{max})$. This
relation holds in LQC up to corrections of the order
$O(\lp^4/a^4_{\mathrm{max}} \ell_o^4)$. For a universe that
grows to the size of the observable part of our on universe,
this correction is \emph{completely} negligible and agreement
with general relativity is excellent.

{\bf Remarks:}

1) Using early literature in LQC, Green and Unruh \cite{gu} had
expressed the concern that although the singularity is resolved the
LQC universes may not undergo the recollapse predicted by general
relativity at low densities and curvatures. However the equations
from the early literature they used had several important
limitations. In particular the Hamiltonian constraint was not self
adjoint and physical Hilbert space had not even been constructed.
Therefore, one could not make any reliable physical predictions.
These drawbacks have been overcome and, as we just discussed, LQC
passes the infrared test associated with recollapse with flying
colors. Indeed, even for tiny universes that grow to a maximum size
of only $25 \lp$, general relativity is a good approximation (to one
part in $10^{5}$) in the regime in which the universe has a radius
of about $10 \lp$ to $23\lp$! For universes that grow to a Gpc, the
accuracy is one part in $10^{228}$!

2) Finally, we will comment briefly on the k=-1 model. The procedure
used in \cite{apsv} is not directly applicable because the spin
connection compatible with the triad $\e^a_i$ has off-diagonal
terms. The early works \cite{kv} (and \cite{szulc} where a more
careful treatment was given) suffered from some drawbacks. These can
be overcome by using (non-local) operators corresponding to the
connection itself (introduced in \cite{awe3} and briefly discussed
in section \ref{b2b9}). However, the operator $\Theta_{(k=-1)}$ is
not essentially self-adjoint. To our knowledge, the issue of
possible self-adjoint extensions and robustness of the theory with
respect to this ambiguity have not been studied.

\subsubsection{Inverse volume corrections}
\label{s4.1.3}

For simplicity, we chose to work with harmonic time already at
the classical level, by setting the lapse to be $N_\tau =a^3$.
This removed all the inverse volume factors in the expression
of the classical Hamiltonian constraint, prior to quantization.
However, as we remarked in section \ref{s2.2}, this procedure
does not have a direct analog in full LQG and therefore, in the
first discussions \cite{aps3,apsv} of the FLRW models,
classical analysis was carried out with $N_t=1$ corresponding
to proper time and the scalar field was used as time only to
interpret the final quantum constraint. In that procedure, to
obtain the quantum Hamiltonian constraint it was essential to
define operators corresponding to inverse powers $p^{-n}$ of
`triads', $p$. Even if one works with $N_\tau = a^3$, these
operators are also necessary to define the constraint in the
Bianchi IX model and to define certain physical observables.

Since these operators have to be defined on the kinematical
Hilbert space $\Hkg$, strictly speaking we should define them
on wave functions $\Psi(p)$. However, for brevity, we will
present the main idea using wave functions $\Psi(\nu)$ in terms
of which the final constraint is written. On this $\Hkg$,
$\h\nu$ is a densely defined self-adjoint operator that acts by
multiplication. Now, any measurable function of a self-adjoint
operator is again self-adjoint. Therefore, given a function $f$
of a real variable which is well-defined everywhere on the
spectrum of $\h{\nu}$ except for a set of measure zero,
$f(\h{\nu})$ is also a self-adjoint operator on $\Hkg$. However
since $\{0\}$ is \emph{not} a set of measure zero on the
spectrum of $\h\nu$ ---recall that the corresponding
eigenvector is normalizable in $\Hkg$-- inverse powers of
$\h\nu$ are not a priori well-defined self-adjoint operators.
(Had the Hilbert space been $L^2(\R)$ as in the \WDW theory,
there would have been no such difficulty because zero would be
a point of the continuous spectrum and the set $\{0\}$ then has
zero measure.)

A neat way out of this problem was proposed by Thiemann in full
LQG \cite{tt, ttbook} and his idea has been used widely in LQC.
We will illustrate the essence of this `Thiemann trick' using
the volume representation. Suppose one is interested in
defining the operator corresponding to the classical function
$|\nu|^{-1/2}$. Then one first writes this function using
Poisson brackets involving only \emph{positive} powers of $\nu$
which have well-defined quantum analogs:
\be \label{inv1} |\nu|^{-\f{1}{2}} = \f{i\hbar}{\lambda}\,
({\rm sgn}\, \nu)\,\, e^{i\lambda \b}\,\, \{\,e^{-i\lambda
\b},\, |\nu|^{\f{1}{2}}\,\}\,  \ee
is an exact identity on the phase space. We can now simply
\emph{define} the operator corresponding to the left side by
promoting the right side to an operator, replacing the Poisson
bracket by $1/i\hbar$ times the commutator:
\be\label{inv2} \widehat{|\nu|^{-\f{1}{2}}}\, \Psi(\nu) :=
\f{1}{2\lambda}\, ({\rm sgn}\, \nu)\, \big( e^{i\lambda \b}
\,[\,e^{-i\lambda \b},\, |\h\nu|^{\f{1}{2}}\, ]\, + \,
[\,e^{-i\lambda \b},\, |\h\nu|^{\f{1}{2}}\, ]\,e^{i\lambda \b}
\big)\, \Psi(\nu) \ee
where as usual we have suppressed hats over trignometric
functions of $\b$. (One has to treat the operator ${\rm sgn}\,
\nu$ with due care but this is not difficult.) one can readily
simplify this expression using (\ref{displacement}) to obtain:
\be \label{inv3} \widehat{|\nu|^{-\f{1}{2}}}\, \Psi(\nu) =
\f{1}{2\lambda}\, {\big{|}}\, |\nu+2\lambda|^{\f{1}{2}} - |\nu -
2\lambda|^{\f{1}{2}} \, {\big{|}}\, \Psi(\nu) \, .\ee
This definition has several attractive features. First, the
operator is densely defined and self-adjoint. Second, every
eigenvector of $\h\nu$ is also an eigenvector of this new
operator. Third, for $\nu \gg \lambda$, eigenvalues are
approximately inverses of one another:
\be \label{inv4} |\nu|^{\f{1}{2}}\, \big(\f{1}{2\lambda}\,
\big{|}\, |\nu+2\lambda|^{\f{1}{2}} - |\nu -
2\lambda|^{\f{1}{2}} \, \big{|}\,\big)\,\, \approx \,\, 1 +
\f{\lambda^2}{2\nu^2} + \f{7\lambda^4}{4\nu^4} + \dots \, . \ee
Finally, near $\nu=0$, the left side of (\ref{inv3}) goes as
%
%
$|\nu|/(2\lambda)^{3/2}$ and hence vanishes at $\nu=0$. Thus,
the Thiemann trick provides the desired operator with the
property that, away from the Planck scale, i.e., when $|\nu|\gg
\lambda$, it resembles the naive quantization of the classical
function $1/\sqrt{|\nu|}$. But in the Planck regime, it
provides an automatic regularization, making the operator
well-defined.

Since $\widehat{|\nu|^{-\f{1}{2}}}$ so defined is a
self-adjoint operator on $\Hkg$ its positive powers are again
self-adjoint. Therefore, one can define arbitrary negative
powers of $|\hat \nu|$. But the procedure has an ambiguity: We
could have started out with the classical function $|\nu|^{-n}$
where $n \in ]0, 1[$ in place of $|\nu|^{-1/2}$ and again
constructed operators corresponding to arbitrary inverse powers
of $|\nu|$. The results do depend on $n$ but this is just a factor
ordering ambiguity.%
\footnote{Also, here we have implicitly worked with the $j=1/2$
representation of SU(2). A priori one could have used higher
representations but then they have certain undesirable features
\cite{perez2,kv2}.}
Thus, in the k=1 or Bianchi IX model one can define inverse
volume factors ---and, more generally, inverse scale factor
operators--- up to a familiar ambiguity. Numerical simulations
show that, if one uses states with values of $\pphi$ that
correspond to closed universes that can grow to macroscopic
size, and are sharply peaked at a classical trajectory in the
weak curvature region, the bounce occurs at a sufficiently
large volume that these inverse scale factor corrections are
completely negligible. (Typical numbers are given below). But
these effects are conceptually important to establish results
that hold for all states because the inverse powers of volumes
are tamed in the Planck regime: instead of growing, they die as
one approaches $\nu =0$!

The same considerations hold also in the k=0 case if we use
$\mathbb{T}^3$ spatial topology. However, in the non-compact,
$\R^3$ case, there is a major difficulty. Now $\nu$ refers to
the volume of a fiducial cell $C$. If one rescales the cell via
$\C \to \beta^3 \C$, for the classical function we have
$|\nu|^{-1/2} \to \beta^{-1} |\nu|^{-1/2}$ while the quantum
operator has a complicated rescaling behavior. Consequently,
the inverse volume corrections now acquire a cell dependence
and therefore do not have a direct physical meaning (see
Appendix B.2 of \cite{aps3} for further discussion.) What
happens when we remove the infrared-regulator by taking the
cell to fill all of $\R^3$? Then, the right side of
(\ref{inv4}) goes to 1. Consequently, when the topology is
$\R^3$, while we can construct intermediate quantum theories
tied to a fiducial cell and keep track of cell dependent,
inverse volume corrections at these stages, when the infrared
regulator is removed to obtain the final theory, these
corrections are washed out for states that are semi-classical
at late times.

We will return to these corrections in sections \ref{s5} and
\ref{s6}.

\subsection{Inclusion of the cosmological constant}
\label{s4.2}

In this subsection we will incorporate the cosmological constant.
Basic ideas were laid out in Appendices of \cite{aps3}. However, the
detailed analysis revealed a number of conceptual and mathematical
subtleties both in the $\Lambda <0$ case \cite{bp} and, especially,
in the $\Lambda>0$ case \cite{ap,kp1}. We will explain these
features, the ensuing mathematical difficulties and the strategies
that were developed to systematically address them. In the end, in
all these cases the singularity is resolved, again because of
quantum geometry effects. However, in the $\Lambda >0$ case, the
analysis reveals some surprises indicating that there is probably a
deeper mathematical theory that could account for other aspects of
numerical results more systematically.

For definiteness we will focus on the flat, i.e. k=0, FLRW model,
although these considerations continue to hold in the k=1, closed
case.

\subsubsection{Negative $\Lambda$}
\label{s4.2.1}

In this sub-section we summarize the main results of \cite{bp}. Let
us begin with the classical theory. The phase space is exactly the
same as in section \ref{s2.1} and we are again left with just the
Hamiltonian constraint. However, as one would expect, the form of
the constraint is modified because of the presence of the
cosmological constant $\Lambda$. For lapse $N=1$, in place of
(\ref{hc3}), we now have:
\be \label{hc-lambda} C_H = \f{\pphi^2}{4\pi G |\v|} -
\f{3}{4\gamma^2}\b^2 |\v| + \f{\Lambda}{8}\, |\v| \,\, \approx 0 \ee
One can easily calculate the equations of motion and eliminate
proper time $t$ in favor of the relational time defined by the
scalar field $\phi$. Then, the solution is given by
\be \label{sol-prime-lambda} \v(\phi) = \pm\,\,
\f{\pphi}{\sqrt{\pi|\Lambda| G}}\,\,\f{1}{{\rm cosh}
\sqrt{12\pi\,G\,(\phi-\phi_o)}}  \ee
where we have assumed that the constant of motion $\pphi$ is
positive (so that the relational and proper time have the same
orientation). Since the matter density $\rho = \pphi^2/V^2$
increases as $\v$ decreases, the form of the solution shows
that the universe starts out with infinite density (i.e., a
big-bang) in the distant past and ends with infinite density in
the distant future (i.e., a big-crunch). Since $\v(\phi)$ is
maximum at $\phi= \phi_o$, the matter density is minimum there.
In fact (\ref{sol-prime-lambda}) implies that total energy
density, $\rho_{\rm tot} = \rho + \rho_\Lambda$ vanishes there,
where $\rho_\Lambda = \Lambda/8\pi G$ is the effective energy
density in the cosmological constant term. Consequently, the
Hubble parameter $\dot{\v}/3\v$ vanishes at $\phi=\phi_o$ and
the universe undergoes a classical recollapse. Thus, the
qualitative behavior of the classical solution is analogous to
that in the k=1 case.

\begin{figure}[]
  \begin{center}
    $a)$\hspace{8cm}$b)$
    \includegraphics[width=3.2in,angle=0]{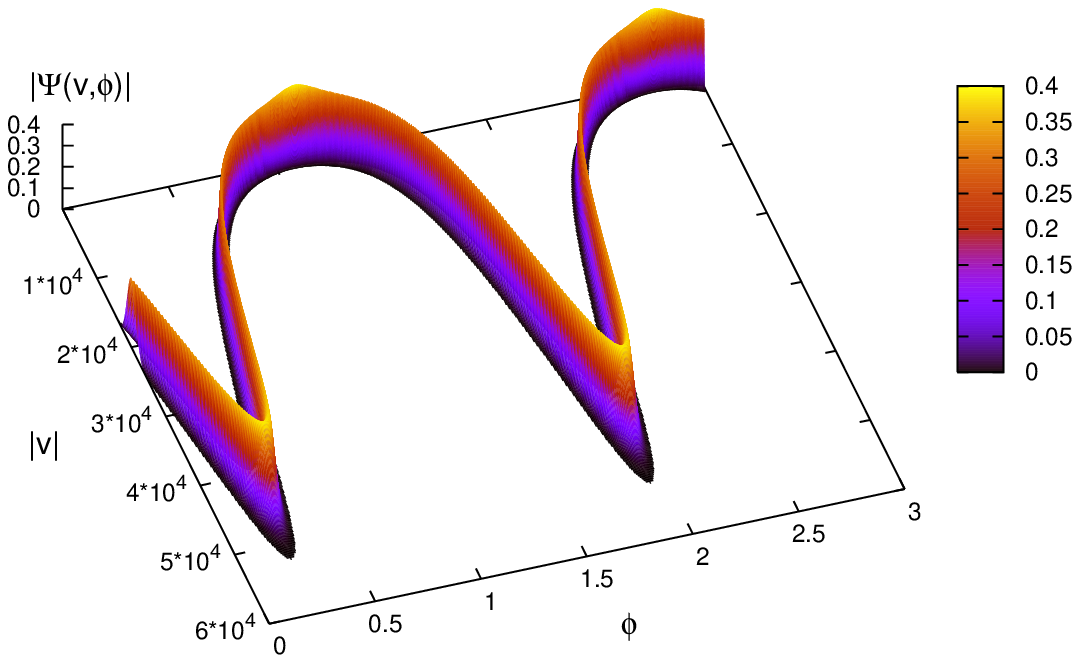}
    \includegraphics[width=3in,angle=0]{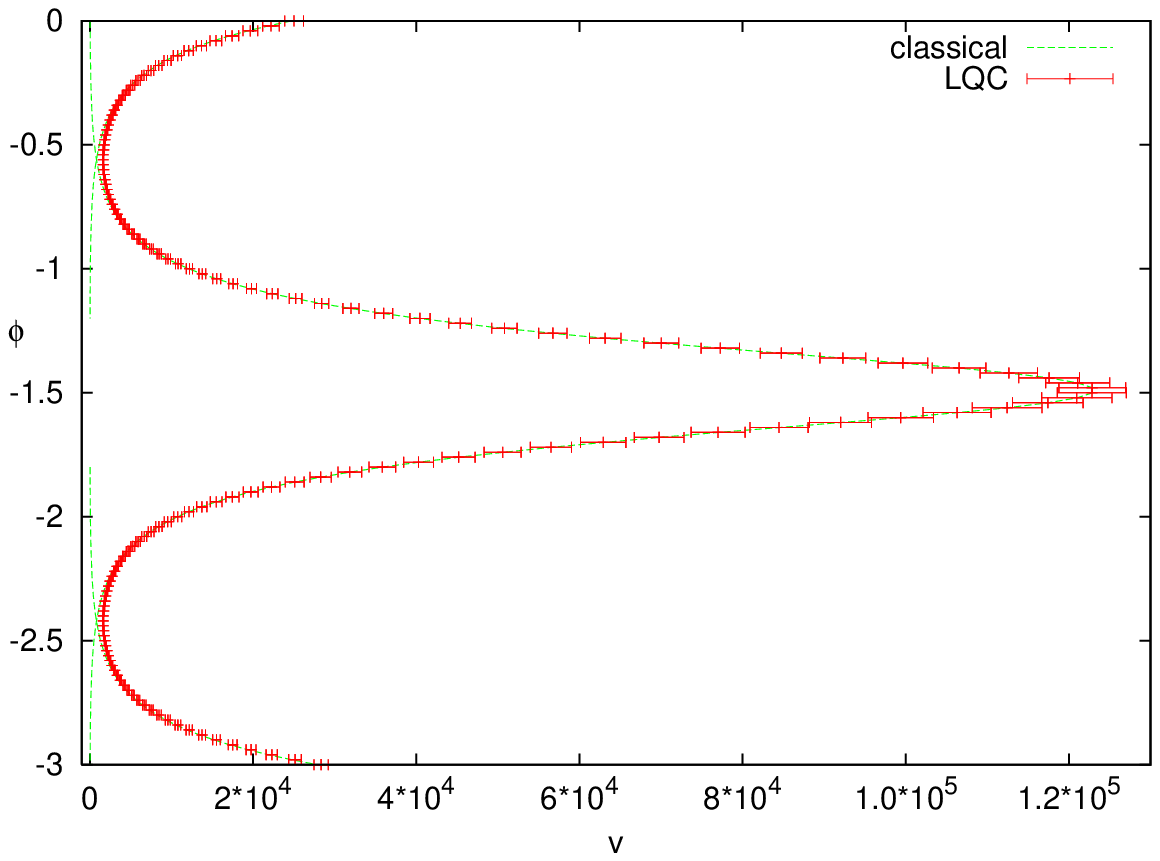}
\caption{Quantum evolution of k=0, $\Lambda<0$ universes is
contrasted with classical evolution. In the classical theory the
universe starts with a big-bang, expands till the total energy
density $\rho_{\rm tot} = \rho +
\Lambda/8\pi G$ vanishes and then recollapses, ending in a
big-crunch singularity. In quantum theory, the big-bang and the
big-crunch are replaced by big bounces and, for large macroscopic
universes, the evolution is nearly periodic. Fig. $a)$ shows the
evolution of a wave function which is sharply peaked
at a point on the classical trajectory with at a pre-specified late time.
Fig. $b)$ shows both the classical trajectory and the evolution of
expectation values of the volume operator. It is clear that the
LQC predictions for recollapse agree very well with those of classical
general relativity but there is a significant difference in Planck regimes.}
\label{fig:negL}
\end{center}
\end{figure}

Quantization is exactly the same as in section \ref{s2}, except that
the final quantum Hamiltonian constraint (\ref{qhc4}) is now
modified by the addition of a cosmological constant term:
\be \label{qhc-prime-lambda} \p_\phi^2\,\Psi(\nu,\phi) =
-\Theta^\prime_{\Lambda}\, \Psi(\nu,\phi) := -\Theta\,
\Psi(\nu,\phi) - \f{\pi G\gamma^2\, |\Lambda|}{2}\,
\nu^2\,\Psi(\nu,\phi)  \ee
where $\Theta$ is the (k=0, $\Lambda=0$) difference operator
defined in (\ref{qhc4}) and $\Lambda$ is negative. As in the
$\Lambda=0$ case, one can introduce an auxiliary inner product
(see (\ref{ip3})) with respect to which $\Theta^\prime_\Lambda$
is a positive definite, symmetric operator on the domain
$\mathcal{D}$ consisting of $\Psi(\nu)$ which have support on a
finite number of points. This operator is essentially
self-adjoint \cite{bp,warsaw3}. The self-adjoint extension,
which we again denote by $\Theta^\prime_{\Lambda}$, is again
positive definite. However, in contrast to $\Theta$, its
spectrum is \emph{discrete}. Furthermore, as eigenvalues
increase, the difference between consecutive eigenvalues
rapidly approaches a \emph{constant, non-zero} value,
determined entirely by $G$ and $\Lambda$. This property has an
important consequence. Let us consider Schr\"odinger states at
a late time which are semi-classical, peaked at a point on a
dynamical trajectory with a macroscopic value of $\pphi$. Such
states are peaked at a large eigenvalue of
$\Theta^\prime_\Lambda$. Since the level spacing is
approximately uniform in the support of such states, their
evolution yields $\Psi(\nu, \phi)$ which are nearly periodic in
$\phi$. They represent eternal, nearly cyclic quantum universes
\emph{even though we are considering k=0, spatially flat
universes.} In each epoch, the universe starts out with a
quantum bounce, expands till the total energy density
$\rho_{\rm tot}$ vanishes, undergoes a classical recollapse and
finally undergoes a second quantum bounce to the next epoch. At
the bounce, $\rho_{\rm tot}$ is well approximated by $\rho_{\mathrm tot}
\approx 0.41\rcr$ as in the $\Lambda=0$ case. For macroscopic
$\pphi$, the agreement between general relativity and LQC is
excellent when $\rho_{\rm tot} = \rho +
\rho_{\Lambda} \ll \rcr$; departures are significant only in
the Planck regimes. In particular, the wave packet faithfully
follows the classical trajectory near the recollapse. Thus,
again, LQC successfully meets the ultraviolet and infrared
challenges discussed in section \ref{s1}. By contrast, as in
the $\Lambda=0$ case, the \WDW theory fails to resolve the
big-bang and the big-crunch singularities.\\

\textbf{Remark:} Because the level spacing between the
eigenvalues of $\Theta^\prime_{\Lambda}$ is not exactly
periodic, there is a slight spread in the wave function from
one epoch to the next. However, this dispersion is
\emph{extremely} small. For a macroscopic universe with
$\Lambda = 10^{-120} \mpl^2$, the initially minute dispersion
doubles only after $10^{70}$ cycles \cite{bp}!

\subsubsection{Positive $\Lambda$}
\label{s4.2.2}

While the change from a continuous spectrum of $\Theta$ to a
discrete one for $\Theta^\prime_{\Lambda}$ had interesting
ramifications, a positive cosmological constant introduces further
novel features which are much more striking \cite{ap,kp1}.%
\footnote{Most of section \ref{s4.2.2} is based on a pre-print
\cite{ap}. We are grateful to Tomasz Pawlowski for his permission to
quote these results here.}
Let us begin with the classical theory. While the Hamiltonian
constraint is again given by (\ref{hc-lambda}), because of the flip
of the sign of $\Lambda$ the solutions
\be {\v}(\phi) = \pm\,\, \f{\pphi}{\sqrt{\pi|\Lambda|
G}}\,\,\f{1}{{\rm sinh} \sqrt{12\pi\,G\,(\phi-\phi_o)}}  \ee
are qualitatively different. First, we have the well known
consequence: because the effective energy density
$\rho_{\Lambda}$ associated with the cosmological constant is
now positive, the universe expands out to infinite volume. The
second and less well known difference is associated with the
use of the scalar field as a relational time variable.  As in
the $\Lambda <0$ case, each solution starts out with a big-bang
in the distant past, expands, but reaches \emph{infinite}
proper time $t$ at a \emph{finite} value $\phi_o$ of the scalar
field. At $\phi=\phi_o$ the physical volume $V$ of the fiducial
cell reaches infinity (whence $\rho_{\rm matt}$ goes to zero).
This means that if we use the lapse tailored to the use of the
scalar field as the time variable, in contrast to the $\Lambda
<0$ case, the Hamiltonian vector field on the phase space is
\emph{incomplete}. This raises the question of whether the
phase space can be extended to continue dynamics beyond
$\phi=\phi_o$. From a physical view point, it is instructive to
consider matter density $\rho$. Along any dynamical trajectory,
its evolution is given by
\be \rho (\phi) = \f{\Lambda}{8\pi G}\, {\rm sinh^2}
(\sqrt{12\pi G}(\phi-\phi_o))\,  \ee
which clearly admits an analytical extension beyond $\phi= \phi_o$.
It suggests that, from the viewpoint of the Hamiltonian theory it is
natural to extend the phase space. While this is not essential in
the classical theory ---one can just stop the evolution at
$\phi=\phi_o$ since proper time and the volume of the fiducial cell
become infinite there--- we will find that it is instructive to
carry out the extension from a quantum perspective.

\begin{figure}[]
  \begin{center}
    $a)$\hspace{8cm}$b)$
    \includegraphics[width=3.2in,angle=0]{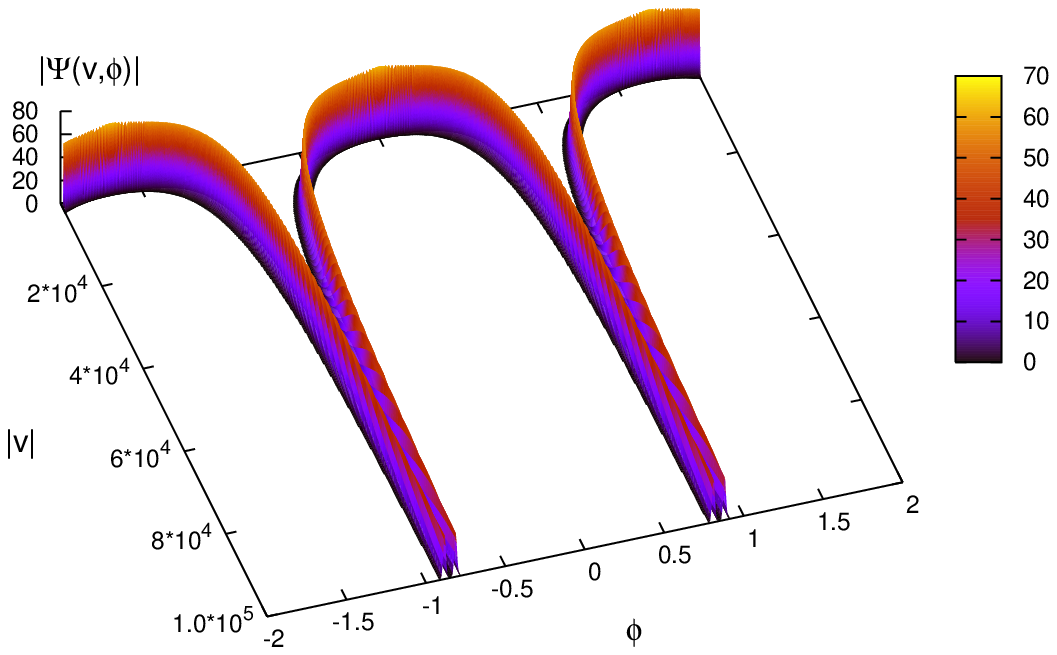}
    \includegraphics[width=3in,angle=0]{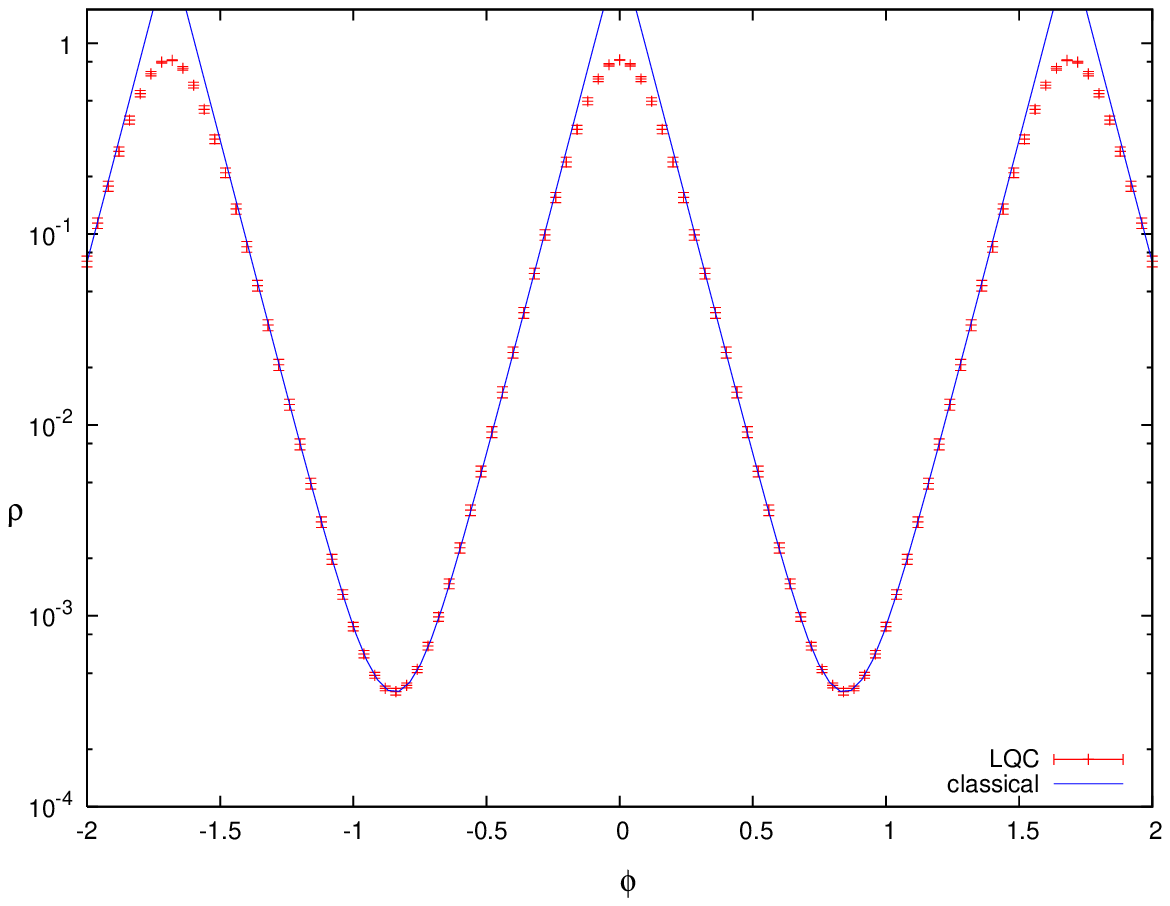}
\caption{Quantum evolution of k=0, $\Lambda>0$ universes is
contrasted with classical evolution. In the classical theory the
universe starts with a big-bang, expands till the matter energy
density $\rho_{\rm matt}$ vanishes. This occurs at a finite value
of the scalar field time and the solution can be analytically
continued. In quantum theory, for \emph{any} choice of the
self-adjoint extension of $\Theta_{\lambda}$, the big-bang and
the big-crunch are replaced by big bounces and, for large
macroscopic universes, the evolution is nearly periodic.
Fig. $a)$ shows the evolution of wave function which
are sharply peaked at a point on the classical trajectory with
$\pphi=5000 \mpl^2$ at a pre-specified late time. Fig. $b)$ shows
both the classical trajectory and the evolution of the expectation
value of the \emph{total} density operator. It is clear that there are
very significant differences in the Planck regime but excellent
agreement away from it.} \label{fig:posL}
\end{center}
\end{figure}

For the analytical extension, let us introduce a new variable
$\theta$ given by
\be \alpha\, \tan\theta = 2\pi\gamma\lp^2\, \nu  \ee
(the right side is just the oriented volume of the fiducial
cell and the constant $\alpha$ has been introduced for
dimensional reasons). To begin with, $\theta\in
]0,\pi/2[$\,\,{}; the big-bang corresponding to $\theta=0$, and
$\phi=\phi_o$ to $\theta=\pi/2$. The Friedmann equation written
in terms of $\theta$ (in place of the scale factor) and $\phi$
(in place of proper time), is analytic in $\phi$:\,\,
$(\p_\phi\,\theta)^2 = \sin^2 \theta\,\,[(4\alpha^2
\Lambda/3\pphi^2)\, \sin^2\theta\, +\, 12\pi G \cos^2 \theta]$
and so is the solution:
\be \label{sol-lambda}\tan\theta(\phi) = \big[ \f{\sqrt{4\pi G}\,
\pphi}{\alpha \sqrt{\Lambda}}\, \f{1}{{\rm sinh} (\sqrt{12\pi G}\,
(\phi-\phi_o))} \,\,\big]\ee
Therefore, it is natural to extend the range of $\theta$ to
$\theta \in ]0, \pi[$. On this extended phase space, dynamical
trajectories start with a big-bang at $\phi=-\infty$ expand out
till $\phi=\phi_o$ where $\rho$ goes to zero and then contract
into a big-crunch singularity at $\phi=\infty$.

Let us now consider the quantum theory. The Hamiltonian constraint
is given by
\be \label{qhc-lambda} \p_\phi^2\,\Psi(\nu,\phi) =
-\Theta_{\Lambda}\, \Psi(\nu,\phi) := -\Theta\, \Psi(\nu,\phi) +
\f{\pi G\gamma^2\,\Lambda}{2}\, \nu^2\,\Psi(\nu,\phi)\, ,  \ee
where $\Lambda$ is now positive. The operator $\Theta_\Lambda$ is
again symmetric on the dense domain $\mathcal{D}$ consisting of
$\Psi(\nu)$ with support on a finite number of points.
\emph{However, unlike the operator $\Theta^\prime_\Lambda$ of
(\ref{qhc-prime-lambda}),\, $\Theta_\Lambda$ fails to be essentially
self-adjoint} \cite{kp1}. This is \emph{not} a peculiarity of LQC;
essential self-adjointness fails also in the \WDW theory. From a
mathematical physics perspective, this is not surprising. Already in
non-relativistic quantum mechanics, if the potential is such that
the particle reaches infinity in finite time, the corresponding
Hamiltonian fails to be essentially self-adjoint in the
Schr\"odinger theory.

The freedom in the choice of self-adjoint extensions has been
discussed in detail in \cite{kp1}. For any choice of extension, the
spectrum of $\Theta_{\Lambda}$ is discrete but the precise
eigenvalues depend on the choice of the extension. However, for
large eigenvalues of $\Theta_{\Lambda}$ the spacing between
consecutive eigenvalues rapidly approaches a constant value which
depends only on the cosmological constant and not on the choice of
the self-adjoint extension. In this respect the situation is the
same as in the $\Lambda <0$ case.

We are most interested in the case when $1/\Lambda \gg \lp^2$ so
that the effective energy density $\rho_\Lambda = \Lambda/8\pi G$ in
the cosmological constant is small compared to the Planck scale,\,
i.e. $\rho_{\Lambda} \ll \rcr$. Furthermore, we are interested in
the quantum evolution of states which are initially sharply peaked
at a point in the classical phase space at which $\pphi^2 \gg \mpl
\lp^3$ and $\rho \ll \rcr$. For these states, numerical simulations
show that the evolution is robust, largely independent of the choice
of the self-adjoint extension. Because these states are sharply
peaked at a large eigenvalue of $\h{p}_{(\phi)}^2$, as in the
$\Lambda<0$ the evolution is nearly periodic. In each epoch, the
universe starts at a quantum bounce, expands till the \emph{matter}
density $\rho$ goes to zero (and the volume of the fiducial cell
goes to infinity), and then `recollapses'. This contracting phase
ends with a quantum bounce to the next epoch. At each bounce, the
\emph{total} energy density $\rho + \rho_{\Lambda}$ is extremely
well approximated by $\rcr\approx 0.41 \rho_{\rm Pl}$ as in sections
\ref{s2} and \ref{s3}. In the region where $\rho \ll \rcr$, the
trajectory is well approximated by the analytical extension
(\ref{sol-lambda}) of the classical solution. In this sense LQC
again successfully resolves the ultraviolet difficulties of the
classical theory without departing from it in the infrared regime.
Again, by contrast, the \WDW theory fails to resolve the
singularities.\\

\textbf{Remarks:}

1) In the classical theory, the volume of the fiducial cell becomes
infinite at a finite value, $\phi =\phi_o$, of the relational time.
In the quantum theory, if a wave-packet were to start in a `tame
regime' well away from the big-bang and follow this trajectory, it
would also reach infinite volume at a finite value of $\phi$.
However, if dynamics is unitary ---as it indeed is for each choice
of the self-adjoint extension of $\Theta_{\Lambda}$--- the wave
function has a well defined evolution all the way to $\phi =\infty$.
If it to remain semi-classical, one can ask what trajectory it would
be peaked on beyond $\phi=\phi_o$. As we have remarked above, the
trajectory is just the analytical extension (\ref{sol-lambda}) of
the classical solution we began with. Thus, if quantum dynamics is
to be unitary, semi-classical considerations imply that an extension
of the classical phase space is inevitable.

2) It is however quite surprising that the evolution of such
semi-classical states is largely independent of the self-adjoint
extension chosen. Part of the reason may be that the classical
trajectories could be extended simply by invoking analyticity. This
`natural' avenue enables one to bypass the complicated issue of
choosing boundary conditions to select the continuation of the
classical solution. But the precise reason behind the numerically
observed robustness of the quantum evolution is far from being clear
and further exploration may well lead one an interesting set of
results on sufficient conditions under which inequivalent
self-adjoint extensions yield nearly equivalent evolutions of
semi-classical states.

3) The Friedmann equation of classical general relativity implies
that, at very late times when the dominant contribution to
$\rho_{\rm tot}$ comes from the cosmological constant term, the
Hubble parameter has the behavior $H = \dot{a}/a \sim
\sqrt{\Lambda}$, whence the connection variable $c \sim \dot{a}
\,\sim \sqrt{\Lambda}V^{1/3}$. Thus, in striking contrast to the
$\Lambda=0$ case, $c$ \emph{grows} at late times even when $\Lambda$
and hence the space-time curvature is very small (in Planck units).
Therefore, had $\bar\mu$ been a constant, say $\mu_o$, as in
\cite{abl}, at late times we would have $\mu_o c \sim \sqrt{\Lambda}
V^{1/3}$. Thus in the $\mu_o$-LQC, $\sin\mu_o c/\mu_o$ does not
approximate $c$ at late times whence the theory has large deviations
from general relativity in the low curvature regime. Thus the
correct infrared limit of LQC is quite non-trivial: the fact that
the field strength operator was obtained using the specific form
$\bar\mu \sim 1/\sqrt{|\mu|} \sim 1/V^{1/3}$ of $\bar\mu$ plays a
key role.

\subsection{Inclusion of an inflaton with a quadratic potential}
\label{s4.3}

At the quantum bounce, the Hubble parameter $H$ necessarily vanishes
and $\dot{H}$ is positive. Therefore, immediately after the bounce,
$\dot{H}$ continues to be positive; i.e., there is a phase of
superinflation. At first there was a hope that this phase of
accelerated expansion may be an adequate substitute for the
inflationary epoch \cite{mb_inflation}. However, this turned out not
to be the case because, in absence of a potential for the scalar
field, the superinflation phase is too short lived. Thus, to compare
predictions of LQC with those of the standard inflationary
paradigms, an inflaton potential appears to be essential at this
stage of our understanding.

Effective equations discussed in section \ref{s5} imply that a
sufficient condition for the quantum geometry corrections to
resolve the big-bang singularity is only that the inflaton
potential be bounded below \cite{as3}. For full LQC, while the
general strategy necessary to construct the theory is clear for
this class of potentials, to handle technical issues and carry
our numerical simulations, it has been necessary to fix a
potential. The simplest choice is a quadratic mass term for the
inflaton. This potential has been widely used in inflationary
scenarios and is compatible with the seven year WMAP data
\cite{wmap,as3}. Therefore in this sub-section we will focus on
the quadratic potential and summarize the status of
incorporating it in full LQC (i.e. beyond the effective
theory).%
\footnote{Most of section \ref{s4.3} is based on a pre-print
\cite{aps4}. We are grateful to Tomasz Pawlowski for his permission
to quote these results here.}

Let us begin with a general remark before entering the detailed
discussion. Recall first that the powerful singularity theorems
\cite{he} of Penrose, Hawking and others strongly suggest that,
if matter satisfies standard energy conditions, the big-bang
singularity is inevitable in general relativistic cosmology.
However, the inflaton with a quadratic potential violates the
strong energy condition. Therefore, in the initial discussions
of inflationary scenarios, one could hope that the presence of
such an inflaton by itself may suffice to avoid the initial
singularity. However Borde, Guth and Vilenkin \cite{bgv}
subsequently proved a new singularity theorem with the novel
feature that it does not use an energy condition and is thus
not tied to general relativity. This important result is
sometimes paraphrased to imply that there is no
escape from the big-bang.%
\footnote{For example, the following remark in \cite{vilenkin}
is sometimes interpreted to imply that the big-bang singularity
is inevitable also in quantum gravity \cite{harper}:  ``With
the proof now in place, cosmologists can no longer hide behind
the possibility of a past-eternal universe. There is no escape:
they have to face the problem of a cosmic beginning.''}
This is not the case: Since the new theorem was motivated by
ideas from eternal inflation, it assumes that the expansion of
the universe is always positive and this assumption is violated
in all bouncing scenarios, including LQC. Thus, the LQC
resolution of the big-bang singularity can evade the original
singularity theorems of general relativity \cite{he} even when
matter satisfies \emph{all} energy conditions because
Einstein's equations are modified due to quantum gravity
effects, \emph{and} it evades the more recent singularity
theorem of \cite{bgv} which is not tied to Einstein's equations
because the LQC universe has a contracting phase in the past.

With these preliminaries out of the way, let us return to the
Hamiltonian constraint in the k=0 FLRW model with an inflaton in a
quadratic potential, $V(\phi) = (1/2) m^2\phi^2${}:
\be \label{hc-inflaton} C_H = \f{\pphi^2}{4\pi G |\v|} -
\f{3}{4}\b^2 {|\v|} + \f{1}{8}\, (\Lambda + 8\pi G m^2\phi^2)\,
{|\v|}\, \, \approx 0 \ee
Note that the potential term is positive and naturally grouped with
the cosmological constant. As a consequence there is considerable
conceptual similarity between the two cases. Phenomenologically, the
contribution due to the cosmological constant term is completely
negligible in the early universe. However, to emphasize the partial
similarity with section \ref{s4.2.2}, we will not ignore it in this
section. In particular, the structure of the quantum Hamiltonian
constraint is similar to that in section \ref{s4.2.2}:
\be \label{qhc-inflaton} \p_\phi^2\,\Psi(\nu,\phi) = -\Theta_{(m)}\,
\Psi(\nu,\phi) := -\big(\Theta -\f{\pi G\gamma^2}{2}\, (\Lambda
+8\pi G m^2\phi^2)\, \nu^2\big)\,\,\Psi(\nu,\phi)\, , \ee
%
There are, however, two qualitative differences between this
constraint and those of other models we have considered so far.
First, since we have a non-trivial potential, $\pphi$ is no longer
conserved in the classical theory and $\phi$ fails to be globally
monotonic in solutions to field equations. Therefore, $\phi$ no
longer serves as a \emph{global} internal time variable.
Nonetheless, we can regard $\phi$ as a \emph{local} internal time
around the putative big bounce because, in the effective theory,
$\phi$ is indeed monotonic in a sufficiently long interval
containing the bounce. The second important difference is that
$\Theta_{(m)}$ carries an explicit `time' dependence because of the
$\phi^2$ term. Therefore, we cannot repeat the procedure of taking
the positive square-root to pass to a first order differential
equation: $-i\partial_\phi\, \Psi = \sqrt{|\Theta|}\, \Psi$ is not
admissible because the square of this equation is no longer
equivalent to the original constraint (\ref{qhc-inflaton}).
Consequently, even locally we do not have a convenient
deparametrization of the Hamiltonian constraint;\, now the group
averaging strategy becomes crucial. For any fixed value of $\phi$,
since the form of $\Theta_{(m)}$ is the same as that of
$\Theta_{\Lambda}$ with a positive cosmological constant (see
(\ref{qhc-lambda})), it again fails to be essentially self-adjoint.
Furthermore, now we have a freedom in choosing a self-adjoint
extension for each value of $\phi$. Given a choice of these
extensions, we can again carry out group averaging and arrive at the
physical sector of the theory. Because we do not have a global
clock, we are forced to use a genuine generalization of ordinary
quantum mechanics: one can make well-defined relational statements
but these have connotations of `time evolution' only locally. Thus,
in any given physical sector, we still have a meaningful generalized
quantum mechanics \cite{hartle} and it is feasible to carry out a
detailed analysis using a consistent histories approach described in
section \ref{s7.5}. However, the question of the detailed relation
between theories that emerge from different choices of self-adjoint
extensions is still quite open. Numerical studies indicate that for
states which are semi-classical, there is again robustness. But, as
we will now discuss, the scope of these studies is much more limited
than that in the case of a positive cosmological constant \cite{ap}.

Suppose we make a choice of self-adjoint extensions and thus fix a
physical sector. The question is whether the expectations based on
effective equations are borne out in this quantum theory. Since the
model is not exactly soluble, we have to take recourse to numerics.
Because physical states must in particular satisfy
(\ref{qhc-inflaton}), we can again use this equation to probe
dynamics. However, the $\phi$-dependence of $\Theta_{(m)}$
introduces a number of technical subtleties.

To discuss these it is simplest to work with states $\Psi(\b,\phi)$
i.e., use the representation in which the variable $\b$ conjugate to
$\nu$ is diagonal. Then, one finds that the Hamiltonian constraint
(\ref{qhc-inflaton}) is hyperbolic in the region $\mathcal{R}$ of
the $(\b,\phi)$ space given by
\be \label{hyperbolic} \f{\sin^2\lambda \b}{\gamma^2\lambda^2} -
\f{8\pi G}{3}\, m^2\phi^2 - \f{\Lambda}{3}\, >\, 0\, .  \ee
(and the choice of a self-adjoint extension of $\Theta_{(m)}$
corresponds to a choice of a boundary condition at the boundary of
$\mathcal{R}$). So far, numerics have been feasible as long as the
wave function remains in the region $\mathcal{R}$. Now, we are
interested in states which are sharply peaked on a general
relativity trajectory in the regime in which general relativity is
an excellent approximation to LQC. The question is: If we evolve
them backward in time using (\ref{qhc-inflaton}), do they remain
sharply peaked on the corresponding solution of effective equations
across the bounce? To answer this question, we need a sufficiently
long `time' interval so that the state can evolve from a density of,
say, $\rho = 10^{-4}\rcr$ where general relativity is a good
approximation, to the putative bounce point and then beyond. For the
wave function to remain in $\mathcal{R}$ during this evolution, the
initial state has to be sharply peaked at a phase space point at
which the kinetic energy is greater than the potential energy
(because in general relativity ${\sin^2\lambda
\b}/{\gamma^2\lambda^2} = H^2 = (8\pi G/3)\, \rho$ and we need
(\ref{hyperbolic}) to hold). Numerical simulations show that such
LQC wave functions do remain sharply peaked on the effective
trajectory. In particular, they exhibit a bounce.

To summarize, to incorporate an inflationary potential in the
effective theory is rather straightforward \cite{as3}. To
incorporate it in full LQC, one has to resolve several technical
problems. So far, the issue of the dependence of the theory on the
choice of self-adjoint extension of $\Theta_{(m)}$ has remained
largely open on the analytical side. On the numerical side,
simulations have been carried out for quantum states which are
initially sharply peaked at a phase space point at which the total
energy density is low enough for general relativity to be an
excellent approximation and, in addition, the kinetic energy density
is greater than potential. In these cases, in a backward evolution,
wave functions continue to remain sharply peaked on effective
trajectories at and beyond the bounce, independently of the choice
of the self-adjoint extension. If we were to allow greater potential
energy initially, with current techniques one can evolve the wave
functions over a more limited range of $\phi$. Therefore in this
case one would have to start the evolution in a regime in which
general relativity fails to be a good approximation. But it should
be possible to adapt the initial state to a trajectory given by
effective equations of LQC and check if, in the backward evolution,
it continues to remain peaked on that trajectory all the way to the
bounce and beyond.

\subsection{Inclusion of anisotropies}
\label{s4.4}

In isotropic models the Weyl curvature vanishes identically
while in presence of anisotropies it does not. Therefore, there
are many more curvature invariants and in general relativity
they diverge at different rates at the big-bang, making the
singularity structure more sophisticated. Indeed, according the
BKL conjecture \cite{bkl1,bkl-ar}, the behavior of the
gravitational field as one approaches generic space-like
singularities can be largely understood using homogeneous but
anisotropic models. This makes the question of singularity
resolution in these models conceptually important.

We will be interested in homogeneous models in which the group of
isometries acts simply and transitively on spatial slices. Then the
spatial slices can be identified with 3-dimensional Lie-groups which
were classified by Bianchi. Let us denote the three Killing fields
by $\ox_i^a$ and write their commutation relations as $[\ox_i,\ox_j]
= \mathring{C}^k_{ij} \ox_k$. The metrics under consideration admit
orthonormal triads $\e^a_i$ such that and $[\ox_i,\e_j] = 0$. As
before the fiducial spatial metric $\q_{ab}$ on the manifold is
determined by the dual co-triad $\o^a_i$ via $\q_{ab} := q_{ij}
\o^i_a \o^j_b$.

Bianchi I, II and IX models have been discussed in LQC. In the
Bianchi I case, the group is Abelian whence the spatial
topology is $\R^3$ or $\mathbb{T}^3$. In the Bianchi II model
the isometry group is that of two translations and a rotation
on a null 2-plane. So we can choose the Killing fields such
that all the structure constants vanish \emph{except}
$\mathring{C}^1_{23} = - \mathring{C}^1_{32} =: \tilde \alpha$.
The spatial topology is now $\R^3$. In the Bianchi IX model,
the isometry group is SU(2) and the spatial topology is
$\mathbb{S}^3$. In the isotropic limit, the Bianchi I model
reduces to the k=0 FLRW model and the Bianchi IX to the k=1
FLRW model.

While the general strategy is the same as that in section
\ref{s2}, its implementation turns out to be surprisingly
non-trivial. In the Bianchi I model for example, first
investigations \cite{chiou_bianchi,cv} attempted to use the
obvious generalization of the expression of the field strength
operator in the isotropic case discussed in section \ref{s2.5}.
However this procedure led to unphysical results ---problems
with the infrared limit and dependence of the final theory on
the initial choice of cell--- reminiscent of what happened in
the $\mu_o$-scheme in the isotropic case
\cite{szulc_bianchi,cs3}. As in the isotropic case, one has to
start anew and implement in the model ideas from full LQG
\cite{awe2} step by step rather than attempting to extend just
the final result. This procedure led to a satisfactory
Hamiltonian constraint which, however, was difficult to use.
Therefore a second new idea was needed to write its expression
in a closed form and then a third input was needed to render
the final expression to a manageable form. The resulting
quantum theory is then free of the limitations of
\cite{chiou_bianchi,cv}. The singularity is indeed resolved in
the quantum theory and, furthermore, there is again an
universal bound on energy density $\rho$ and the shear scalar
in the effective theory. Interestingly, the bound on $\rho$
agrees with the $\rcr$ obtained in the isotropic cases.

In this construction the field strength could be expressed in
terms of the holonomy around a suitable plaquette whose edges
are the integral curves of $\e^a_i$. This procedure turns out
not to be viable in the Bianchi II model because we now have
both, spatial curvature \emph{and} anisotropies. Therefore a
new strategy was introduced in \cite{awe3}: construct non-local
operators corresponding to connections themselves and use them
to define the field strength. This procedure is a
generalization of the one used in Bianchi I case in that in
that case it reproduces the same answer as in \cite{awe2}.
Fortunately, these inputs seem to suffice for more complicated
models. In particular, the LQC of the Bianchi IX model does not
need further conceptual tools. In both Bianchi II and IX models
the singularity is resolved. In the Bianchi II and IX models,
in the effective theory there is again a bound on $\rho$ but
there is no analytical argument that it is optimal (see section
\ref{s5.5}).

In this section, there is no summation over repeated indices which
are all covariant or all contravariant; summation is understood only
if there is a contraction between a covariant and a contravariant
index.

\subsubsection{Bianchi-I model}
 \label{sec_bianchi1}

The space-time metrics of the Bianchi I model can be written as:
\be \dd s^2 = - N^2 \dd \tau^2 + a_1^2 \dd x_1^2 + a_2^2 \dd x_2^2 +
a_3^2 \dd x_3^2, .\ee
If the spatial topology is $\R^3$, we again have to fix a
fiducial cell $\C$. We will take its edges to lie along the
integral curves of the fiducial triads $\e^a_i$ with coordinate
lengths $L_1,\, L_2,\, L_3$ so that the fiducial volume of $\C$
is $V_o = L_1L_2L_3$. In terms of these background structures,
the gravitational connection and conjugate (density weighted)
triad can be written as
\be\label{sym_red} A^i_a \, = \, c^i (L^i)^{-1} \o^i_a, ~~
\mathrm{and} ~~ E^a_i = p_i L_i V_o^{-1} \sqrt{\q}~ \e^a_i ~
\ee
where $\q$ is again the determinant of $\q_{ab}$. Thus, the
gravitational phase space is now six dimensional, parameterized by
$(c^i, p_i)$. The physical volume $V$ of $\C$ is given by $V^2 =
p_1p_2p_3$ and the triad components are related to the directional
scale factors in the metric via
\be\label{triad_sf} p_1 \, = \, \varepsilon_1\, L_2 L_3 \, |a_2
a_3|, ~~ p_2 \,= \, \varepsilon_2 L_1 L_3 \, |a_1 a_3|, ~~ p_3 \, =
\, \varepsilon_3  L_1 L_2 \, |a_1 a_2| ~ , \ee
where $\varepsilon_i \pm 1 $ depending on whether $E^a_i$ is
parallel or antiparallel to $\e^a_i$.

Next, let us discuss the effect of changing the fiducial structures
under which physics must be invariant. As in the case of the
isotropic model (discussed in section \ref{s2.1.2}), $(c^i, p_i)$
are unchanged under  the rescaling of coordinates $x^i$. However, in
contrast to the isotropic case, we now have the freedom of
performing a different rescaling of the cell along each direction,
$L_i \rightarrow \beta_i L_i$, under which the fiducial volume
transforms as $V_o \rightarrow \beta_1 \beta_2 \beta_3 V_o$. The
connection and triad components are not invariant under this
transformation. For example, the $i=1$ components transform as: $c_1
\rightarrow \beta_1 c_1$ and $p_1 \rightarrow \beta_2 \beta_3 p_1$
(and similarly for other components). This fact will be important in
what follows.

As in the isotropic model, our canonical variables $A_a^i, E^a_i;
\,\, \phi,\pphi$ automatically satisfy the Gauss and the
Diffeomorphism constraints and we are only left with the Hamiltonian
constraint:
\be\label{bianchi1_const} \int_{\cal C} N\Big(- \f{1}{16 \pi G
\gamma^2}\, \f{\epsilon^{i j}_{~~k} \, E^a_i E^b_j \,
F^k_{ab}}{\sqrt{q}} +
\f{p_{(\phi)}^2}{2}\f{\sqrt{q}}{V^2}\Big)\, \dd^3 x\,\, \approx
\,\,0\ee
Let us again use the lapse $N = a_1a_2a_3$ adapted to harmonic
time and the expressions (\ref{sym_red}) of $A_a^i, E^a_i$.
Then the Hamiltonian constraint becomes:
\be \label{clH} {C}_H^{(I)} = \f{p_{(\phi)}^2}{2}\, -\, \f{1}{8 \pi
G \gamma^2}{(c_1 p_1 \, c_2 p_2 + c_3 p_3 \, c_1 p_1 + c_2 p_2 \,
c_3 p_3)} \,\, \approx \,\, 0\, . \ee

Hamilton's equations for the triads lead to a generalization of the
Friedman equation
\be\label{gen_fried_class} H^2 = \f{8 \pi G}{3} \rho \, + \,
\f{\Sigma^2}{a^6} ~. \ee
where  $a$ is the mean scale factor defined as $a := (a_1 a_2
a_3)^{1/3}$, $H$ refers to the mean Hubble rate and  $\Sigma^2$
represents shear
\be\label{shearscalar} \Sigma^2 = \f{a^6}{18} \left((H_1 - H_2)^2 +
(H_2 - H_3)^2 + (H_3 - H_1)^2\right) ~. \ee
Eq (\ref{gen_fried_class}) quantifies that the energy density in
gravitational waves captured in the anisotropic shears $(H_i-H_j)$.
From the dynamical equations, it follows that if the matter has
vanishing anisotropic stress ---so $\rho$ depends on
scalar factors $a_i$ only through $a$--- then $\Sigma^2$ is a
constant of motion in general relativity.
\\

Construction of quantum kinematics closely follows the
isotropic case. The elementary variables are the triads $p_i$
and the holonomies $h_i^{(\ell)}$ of the connection $A_a^i$
along edges parallel to the three axis $x_i$ whose edge lengths
with respect to the fiducial metric are $\ell L_i$. The
holonomies can be expressed entirely in terms of almost
periodic functions of the connection, of the form $\exp(i \ell
c_i)$. These, along with the triads, are promoted to operators
in the kinematical Hilbert space $\Hkg$. In the triad
representation, with states represented as $\Psi(p_1,p_2,p_3)$,
the $\h{p}_i$ act by multiplication and the $\exp (i\ell c_i)$
by displacement of the argument of $\Psi$, exactly as in the
isotropic case. 
%
%
Finally, we have to incorporate the fact that physics should be
invariant under a reversal of triad orientation. Therefore the
physical states satisfy $\Psi(p_1,p_2,p_3) =
\Psi(|p_1|,|p_2|,|p_3|)$, whence it suffices to consider the
restrictions of states to the positive octant, $(p_1, p_2, p_3)
\ge 0$. Individual operators in a calculation may not leave
this sector invariant but their physically relevant
combinations do.
\\

To discuss dynamics, we have to construct the Hamiltonian constraint
operator. As in the isotropic case, we can not directly use the
reduced form (\ref{clH}) because it contains $c_i$ while only $\exp
i\ell c_i$ are well-defined operators on $\Hkg$. Therefore we have
to return to (\ref{bianchi1_const}) and seek guidance from full LQG.
As before, the key problem is that of defining the quantum operator
$\h{F}_{ab}{}^i$ and, as in section \ref{s2.5.1}, it can be
expressed as the holonomy operator associated with suitable loops
$\Box_{ij}$. Since the triads $\e^a_i$ commute, we can we can form
the required plaquettes $\Box_{ij}$ by moving along their integral
curves. The question is: What should the lengths $\bar\mu_i$ of
individual segments be? Because of anisotropy, $\bar\mu_i$ can
change from one direction to another. Therefore, we have to return
to the semi-heuristic relation between LQG and LQC and extend the
procedure of section \ref{s2.5.2} to allow for anisotropies. When
this is done, one finds that $\bar\mu$ is replaced by \cite{awe2}
\be \label{bianchi_mub1} \bar \mu_1 = \lam \sqrt{\f{ |p_1|
}{|p_2 p_3|}}, ~~~~ \bar \mu_2 = \lam \sqrt{\f{|p_2|}{|p_1
p_3|}}, ~~~ \mathrm{and} ~~~\bar \mu_3 = \lam
\sqrt{\f{|p_3|}{|p_1 p_2|}} ~ . \ee
In the isotropic case, $p_i =p$ and one recovers Eq
(\ref{mubar}).%
\footnote{In the early work on the Bianchi I model in LQC
\cite{chiou_bianchi,cv}, a short cut was taken by using instead the
`obvious' generalization $\bar{\mu}_i = \lambda/\sqrt{|p_i|}$ of the
isotropic $\bar\mu$. As we explained at the beginning of this
subsection, this led to a theory that is not viable. The lesson
again is that, when new conceptual issues arise, one should not
guess the solution from that in the simpler cases but rather start
ab initio and seek guidance from full LQG.}

The final field strength operator (\ref{F2}) of the isotropic
case is now replaced by
\be \label{F3} \h{F}_{ab}{}^k = \epsilon_{ij}{}^k\,\, \big(\f{\sin
\bar\mu c}{\bar{\mu}L} \o_a\big)^i\,\, \big(\f{\sin \bar\mu c}
{\bar{\mu}L} \o_a\big)^j\quad {\rm where} \quad
 \big(\f{\sin \bar\mu c}{\bar{\mu}L} \o_a\big)^i = \f{\sin
 \bar\mu^ic^i}{\bar\mu^iL^i}\,\, \o_a^i  \ee
(where there is no summation over $i$ in the second expression).
From their definition it follows that the $\bar \mu_i$'s are
invariant with respect to the rescaling of the coordinates but
transform non-trivially under the allowed scalings of the fiducial
edge lengths: $\bar \mu_i \rightarrow \beta_i^{-1} \bar \mu_i$.
Since, the connection components transform as $c_i \rightarrow
\beta_i c_i$, the crucial term $\sin(\bar\mu_i c_i)$ in (\ref{F3})
is invariant under \emph{both} the rescalings. This is key to
ensuring that the physics of the final theory is insensitive to the
initial choice of the infrared regulator, the fiducial cell.

However, because $\bar\mu_1$, say, depends also on $p_2$ and $p_3$,
the operator $\sin(\bar\mu_1 c_1)$ is rather complicated and seems
totally unmanageable at first. However, it becomes
manageable if one carries out two transformations \cite{awe2}:\\
i) Work with the `square-roots' $l_i$ of $p_i$, given by $p_i =
({\rm sgn}\, l_i)\,\,(4 \pi \gamma \lambda \lp^2)^{2/3}\,\, l_i^2$.
Then, one can write the action of $\sin (\bar\mu_i c_i)$ explicitly,
using, e.g.,
\be e^{\pm i \bar \mu_1 c_1}\, \Psi(\lu,\ld,\lt) = \Psi\big(\lu
\pm \f{\mathrm{sgn}(\lu)}{\ld \lt},\, \ld, \lt\big) ~. \ee
But the action of the resulting constraint operator is not
transparent because it mixes all arguments in the wave function
and does not have any resemblance to the
constraint (\ref{qhc4}) we encountered in the isotropic case.\\
ii) A further transformation overcomes these drawbacks. Set
\be v = 2(l_1l_2l_3) \quad \hbox{\rm so that}\quad  V = 2\pi
\gamma\lambda^2 \lp |v|\ee
as in the isotropic case. Since $V$ is the physical volume of the
cell,  $v$ is related to the variable $\nu$ we used in the isotropic
case via $v = \nu\lp/\lambda^2$. This constant rescaling makes $v$
dimensionless and simplifies the constraint. The key idea, now, is
to work with wave functions $\Psi(l_1,l_2,v)$ in place of
$\Psi(l_1,l_2,l_3)$.

Then the final quantum constraint has the familiar form
\be \label{bianchi-qconst} \p_\phi^2 \Psi(\lu,\ld,v;\phi) = -
\Theta_{(I)}\, \Psi(\lu,\ld,v;\phi)\ee
and, because of the reflection symmetry---which is preserved by the
Hamiltonian constraint--- it suffices to specify $(\Theta_{(I)}\,
\Psi)(l_1,l_2,v)$ only for $(l_1,l_2,v)$ in the positive octant
(where $l_1,l_2,v$ are all non-negative). We then have
\ba \Theta_{(I)}\, \Psi(\lu,\ld,v;\phi) &=& \f{\pi G
\hbar^2}{8} \, v^{1/2} \, \big[(v + 2) (v + 4)^{1/2}
\Psi_4^+(\lu,\ld,v) \, - \, (v+2) v^{1/2} \Psi_0^+\lvp\nonumber \\
&& - (v-2) v^{1/2} \Psi_0^-\lvp + (v-2) |v - 4|^{1/2}
\Psi_4^-\lvp\big] \ea
with
\ba \label{Psi4} \Psi_4^\pm\lvp &=& \nonumber \Psi\left(\vfu \lu,
\vfd \ld, v
\pm 4\right) \, + \, \Psi\left(\vfu \lu, \ld, v \pm 4\right) \\
&& \nonumber + ~ \Psi\left(\vfd \lu, \vfu \ld, v \pm 4\right) \,
+ \, \Psi\left(\vfd \lu, \ld, v \pm 4\right) \\
&& + ~ \Psi\left(\lu, \vfd \ld, v \pm 4\right) \, + \,
\Psi\left(\lu, \vfu \ld, v \pm 4\right)  ~, \ea
and
\ba \label{Psi0} \Psi_0^\pm\lvp &=& \nonumber \Psi\left(\vfd \lu,
\vft \ld, v\right) \,
+ \, \Psi\left(\vfd \lu, \ld, v\right) \\
&& \nonumber + ~ \Psi\left(\vft \lu, \vfd \ld, v\right) \, +
\, \Psi\left(\vft \lu, \ld, v\right) \\
&& + ~ \Psi\left(\lu, \vft \ld, v\right) \, + \, \Psi\left(\lu, \vfd
\ld, v\right)  ~. \ea
The physical Hilbert space and the Dirac observables ---the scalar
field momentum $\hat{p}_{(\phi)}$ which is a constant of motion, and
the relational geometrical observables, the volume $\h{V}|_\phi$ and
anisotropies $\h{l}_1|_\phi$ and $\h{l}_2|_\phi$--- can be
introduced on $\Hp$ exactly as in the isotropic case.

Eq (\ref{bianchi-qconst}) is still rather complicated and further
simplifications have been made to make it easier to carry out
numerical simulations \cite{ahtp}. But it is possible to draw some
general conclusions already from this equation. Note first that, it
suffices to specify the wave functions on the positive octant since
their values on the rest of the $l_1,l_3,v)$ space is determined by
the symmetry requirement $\Psi(l_1, l_2, v) = \Psi(|l_1|, |l_2|,
|v|)$. As in the isotropic case, the evolution preserves the space
of wave functions which have support only the lattices $v=4n$ and
$v= \epsilon + 2n$ if $\epsilon \not=0$ in the positive octant. Each
of these sectors is also preserved by a complete set of Dirac
observables, $\hat{p}_{(\phi)}$, volume $\h{v}|_\phi$ and
anisotropies $\h{l}_1|_\phi$ and $\h{l}_2|_\phi$. Therefore we can
focus just on one sector. Let us consider the $\epsilon=0$ sector
since it contains the classical singularity. From the functional
dependence on $v$ of various terms in the quantum constraint, it is
straightforward to deduce that the classical singularity is
decoupled from the quantum evolution: If initially the wave function
vanishes on points with $v=0$, it continues to vanish there. Because
the physical wave functions have support only on points $v =4n$, it
follows that the matter density, for example, can never diverge as
the state evolves in the internal time. In this sense, the
singularity is resolved by the constraint (\ref{bianchi-qconst}). In
the limit, $l_1, l_2, v \gg 1$, and assuming that the wave-function
is slowly varying, one can show that the discrete quantum constraint
(\ref{bianchi-qconst}) approximates the differential quantum
constraint in the \WDW theory. However, in the Planck regime there
is significant difference because the singularity is not resolved in
the \WDW theory. Further details on dynamics are discussed in
section \ref{s5.4} using effective equations.\\

\textbf{Remarks:}

1) As mentioned in section \ref{s1.3}, in the cosmology
circles, there has been a general concern about bouncing models
(see, e.g., \cite{bb}) which may be summarized as follows.
Because the universe contracts prior to the bounce, the shear
anisotropies grow and, in general relativity, this growth leads
to the `Mixmaster chaotic behavior.' Therefore, the singularity
resolution through a bounce realized in isotropic models may
not survive in presence of anisotropies. How does LQC avoid
this potential problem? It does so because of the in-built
corrections to Einstein's equations that result from quantum
geometry. When the dynamics of the model is constructed by
paying full attention to full LQG, as was done in \cite{awe2},
the effective repulsive force created by these quantum
corrections is far more subtle than what a perturbation theory
around an isotropic bouncing model may suggest. There is not a
single bounce but many: Each time a shear potential enters the
Planck regime the new repulsive force dilutes it avoiding the
formation of a singularity. (See section \ref{s5.4}.)

2) The Bianchi I model also enables one to address a conceptual
issue: Does quantization commute with symmetry reduction? There is a
general concern that if one first quantizes a system and then
carries out a symmetry reduction by freezing the unwanted degrees of
freedom, the resulting quantum theory of the symmetry reduced
subsystem will not in general be quite different from the one
obtained by first freezing the unwanted degrees of freedom
classically and then carrying out quantization \cite{kr}. If one
takes the Bianchi I model as the larger system and the isotropic
FLRW model as the symmetry reduced system, this procedure would have
us restrict the wave functions $\Psi(l_1,l_2,v)$ of the Bianchi I
model to configurations $l_1 = l_2 = (v/2)^{1/3}$ and ask if this
quantum theory agrees with that of the FLRW model. The answer, as
suggested in \cite{kr}, is in the negative: the Bianchi I dynamics
(\ref{bianchi-qconst}) does not even leave this subspace invariant!
However, instead of freezing the anisotropies to zero, if one
integrates them out, one finds that the Bianchi I quantum constraint
operator projects down \emph{exactly} to the k=0 FLRW quantum
constraint operator \cite{awe2}. The exact agreement in this
specific calculation is presumably an artifact of the simplicity of
the models. In more general situations one would expect only an
approximate agreement. But the result does bring out the fact that
the issue of comparing a quantum theory with its symmetry reduced
version is somewhat subtle and contrary to one's first instincts the
dynamics of the reduced system may correctly capture appropriate
physics.

3) The treatment we summarized here uses the scalar field as a
relational time variable. LQC of \emph{vacuum} Bianchi models has
also been studied using one of the triad components \emph{or} its
conjugate momentum as internal time \cite{madrid-bianchi,hybrid3}.
This work is conceptually important in that time has emerged from a
geometrical variable \emph{and} one can compare quantum theories
with two different notions of time for the same classical model.
However, technically it used the $\bar\mu_i$ given in the early work
\cite{chiou_bianchi,cv}. For conceptual issues concerning time, this
is not a significant limitation. Nonetheless, for completeness, it
is desirable to redo that analysis with the `correct' $\bar\mu_i$.

\subsubsection{Bianchi II and Bianchi IX models}
\label{b2b9}

In this subsection, we summarize the quantization of Bianchi II and
Bianchi IX space-times \cite{awe3,we}. Since the Bianchi I model was
discussed in some detail here we will focus just on the main
differences from that case.

The Bianchi II model is the simplest example of a space-time with
anisotropies \emph{and} a non-zero spatial curvature. Since the
topology of spatial slices is now $\mathbb{R}^3$, one must introduce
a fiducial cell ${\cal C}$ to construct a Hamiltonian (or
Lagrangian) framework. We will use the same notation for fiducial
structures as in our discussion of the Bianchi I model. By contrast,
the spatial manifold in the Bianchi IX case is compact with topology
$\mathbb{S}^3$. Therefore a cell is \emph{not} needed. In this case
we will use the same notation as in the k=1 model discussed in
section \ref{s4.1}. In particular, the volume $V_o$ of
$\mathbb{S}^3$ defined by the fiducial metric $\q_{ab}$ will be
written as $V_o =: \ell_o^3 (= 16 \pi^2)$.

As before, we will use a lapse $N$ geared to the harmonic time
already in the classical theory. Then the structure of the
Hamiltonian constraint is similar to that in the Bianchi I model but
there are important correction terms. For the Bianchi II model, we
have:
\be\label{bianchi2_cons} {C}_H^{(II)} = {C}_H^{(I)} - \f{1}{8 \pi G
\gamma^2} \big[\alpha \varepsilon p_2 p_3 c_1 - (1 + \gamma^2)
\left(\f{\alpha p_2 p_3}{2 p_1}\right)^2\big] ~, \ee
where $C_H^{(I)}$ is the Bianchi I constraint (\ref{clH}),
$\varepsilon = \pm 1$ is determined by the orientation of the triads
and we have appropriately rescaled $\tilde \alpha :=
\mathring{C}^1_{23}$ and set $\alpha := (L_2 L_3/L_1) \tilde
\alpha$. Note that in the limit, $\alpha \rightarrow 0$, the spatial
curvature in Bianchi II model vanishes and we recover the Bianchi I
constraint (\ref{clH}). Similarly, for the Bianchi IX model one
obtains
\ba\label{bianchi9_cons} {C}_H^{(IX)} &=& \nonumber {C}_H^{(I)} -
\f{1}{8 \pi G \gamma^2} \Big[{\ell_o \varepsilon} (p_1 p_2 c_3 +
p_2 p_3 c_1 + p_3 p_1 c_2) \\
&&  + ~ \f{\ell_o^2}{4} (1 + \gamma^2) \big(2 (p_1^2 + p_2^2 +
p_3^2) - (\f{p_1 p_2}{p_3})^2 - (\f{p_2 p_3}{p_1})^2 - (\f{p_3
p_1}{p_2})^2  \big)\Big] ~ . \ea
The Bianchi-I constraint (\ref{clH}) is now recovered in the limit
taking limit $\alpha \rightarrow 0$. In the isotropic truncation
$c_i =c$ and $p_i =p$, one recovers classical constraint for the k=1
FLRW model. A detailed examination shows that in both cases the
Hamiltonian constraint is left invariant under the change of
orientation of triads, and is simply rescaled by an overall factor
of $(\beta_1\beta_2\beta_3)^2$ under the rescaling of the fiducial
cell $\C$ by $L_i \to \beta_iL_i$. The equations of motion are
unaffected under both these operations.

The kinematics of LQC is identical to that in the Bianchi I model.
However, an important subtlety occurs already in the Bianchi II case
when we try to write a field strength operator using holonomies
around closed loops. First, the triads $\e^a_i$ do not commute.
Therefore, we cannot use them directly to form closed loops. We
already encountered this difficulty in the k=1 FLRW model, where we
could construct the desired loops by moving alternately along
integral curves of $\e^a_i$ and of the Killing vectors $\ox^a_ii$.
One can repeat the same procedure but because the Bianchi II
geometry is \emph{anisotropic}, the resulting holonomy fails to be
an almost periodic function of $c_i$ whence the field strength
operator so constructed is not well-defined on $\Hkg$! One can try
to enlarge $\Hkg$ but then the analysis quickly becomes as
complicated as full LQG. This problem was overcome \cite{awe3} by a
further extension of the strategy which is again motivated by the
procedure used in full LQG \cite{tt,ttbook}: Define a non-local
\emph{connection operator} in terms of holonomies along open
segments
\be \h{A}_a \equiv \h{A}_a^i\tau_i =  \sum_k \, \f{1}{2\ell_k
L_k}\, \big(h_k^{\ell_k} - (h_k^{\ell_k})^{-1} \big)\ee
where the length of the segment along the $k$th direction, as
measured by the fiducial metric, is $\ell_kL_k$ and $\tau_i$ is the
basis in su(2) introduced in section \ref{s1.5}. In full LQG,
because of diffeomorphism invariance, the length of the segment does
not matter. In LQC we have gauge fixed the diffeomorphism constraint
through our parametrization (\ref{sym_red}) of the phase space
variables $A_a^i, E^a_i$ in terms of $(c^i, p_i)$. Therefore we have
to specify the three $\ell_k$. Here one seeks guidance from the
Bianchi I model: the requirement that one should reproduce exactly
the same field strength operator as before fixes $\ell_k$ to be
$\ell_k = 2\bar{\mu}_k$, where $\bar\mu_k$ is given by
(\ref{bianchi_mub1}).%
\footnote{Just as in the older treatment of the Bianchi I model one
fixed $\bar\mu_k$ by using an `obvious' generalization of the
successful FLRW strategy \cite{chiou_bianchi,cv}, here one may be
tempted to use a similar `shortcut' and set $\ell_k = \lambda
\bar\mu_k$, say. But this strategy is not viable because does not
yield the correct field strength operator even in the FLRW model
\cite{ydm}.}
Once this is done, it is straightforward to write the quantum
Hamiltonian constraint both in the Bianchi II and the Bianchi IX
cases in the form $\p_\phi^2 \Psi = \Theta_{(II)}\Psi$ and
$\p_\phi^2 \Psi = \Theta_{(IX)} \Psi$. The operators $\Theta_{(II)}$
and $\Theta_{(IX)}$ contain terms in addition to those in
$\Theta_{I}$. They also have the correct behavior under rescalings
of the fiducial cell $\C$ and under reversal of orientation. They
disappear in the limit $\alpha \to 0$ in the Bianchi I case and
$\ell_o \to 0$ in the Bianchi IX case ensuring that these theories
are viable generalizations of the Bianchi I LQC. The physical
Hilbert space and a complete set of Dirac observables can be
constructed in a straightforward fashion. Finally, in both these
cases the big-bang singularity is resolved in the same sense as in
the Bianchi I model: The singular sector (i.e., states in $\Hk$
which have support on configurations $v=0$) decouples entirely from
the regular sector so that states which start our having no support
on the singular sector cannot acquire a singular component.

There are nonetheless significant open issues. First, the question
of essential self-adjointness of $\Theta_{(I)},\, \Theta_{(II)}$ and
$\Theta_{(IX)}$ is open and it is important to settle it. For
example, if $\Theta_{(IX)}$ turns out to be essentially
self-adjoint, then quantum dynamics would be unambiguous and unitary
in spite of the chaotic behavior in general relativity. This would
not be surprising because that behavior refers to the approach to
singularity which is completely resolved in LQC. Second, while there
have been advances in extracting physics from effective equations in
these models (see section \ref{s5.5}), further work is needed in the
Bianchi II and especially Bianchi IX models. It is even more
important to perform numerical simulations in exact LQC to check
that the effective equations continue to capture the essence of the
quantum physics also in this model. These simulations should also
shed new light on the relevance of the BKL conjecture for LQG.

\section{Effective dynamics and phenomenological implications}
\label{s5}

In this section we summarize the new physics that has emerged
from the modified Friedmann dynamics derived from the effective
Hamiltonian constraint, ${C}_H^{(\mathrm{eff})}$. This
constraint is defined on the classical phase space but
incorporates the leading quantum corrections of LQC. Therefore,
it enables one to extract the salient features of LQC dynamics
using just differential equations, without having to refer to
Hilbert spaces and operators.

The section is organized as follows. In \ref{s5.1}, we outline the
conceptual frameworks that have been used to obtain
$C_H^{(\mathrm{eff})}$. In section \ref{s5.2} we derive the modified
Einstein' s equations using this effective Hamiltonian constraint
and discuss key features of the new physics they lead to. These
include a phase of super-inflation occurring for matter obeying null
energy condition when $\rcr/2 < \rho < \rcr$, and generic resolution
of strong singularities in the LQC for the isotropic model. In
section \ref{s5.3} we consider effective LQC dynamics using a scalar
field in a quadratic potential. A key question is whether an
inflationary phase that is compatible with the 7 year WMAP data will
be seen generically in LQC or if it results only after (perhaps
extreme) fine tuning. We first discuss the subtleties associated
with measures that are needed to answer this question quantitatively
and then summarize the result that `almost all' data at the bounce
surface evolve into solutions that meet this phase at some time
during evolution. In section \ref{s5.4}, we discuss the effective
dynamics for Bianchi-I models and summarize the results on the
viability of ekpyrotic/cyclic models in LQC. We conclude with a
brief summary of various other applications in section \ref{s5.5}.

\subsection{Effective Hamiltonian constraint}
\label{sec_effham} \label{s5.1}

In LQC, the effective Hamiltonian constraint has been derived using
two approaches: the embedding method \cite{jw,vt,psvt} and the
truncation method \cite{mb_as_eff1,mb_as_eff2}. The former is
analogous to the variational technique, often used in
non-relativistic quantum mechanics, where a skillful combination of
science and art often leads to results which approximate the full
answer with surprising accuracy. The latter is in the spirit of the
order by order perturbation theory, which has the merit of being a
more systematic procedure. Both of these approaches are based on a
geometrical formulation of quantum mechanics (see, e.g.
\cite{aats,as}). The basic idea is to cast quantum mechanics in the
language of symplectic geometry, i.e., the same framework that one
uses in the phase space description of classical systems. One starts
by treating the space of quantum states as a phase space,
$\Gamma_{\mathrm{Q}}$, where the symplectic form, $\Omega_{\rm Q}$,
is given by the imaginary part of the Hermitian inner product on the
Hilbert space $\H$. The framework has two surprising features.
First, there is an interesting interplay between the commutators
between operators on $\H$ and the Poisson bracket defined by
$\Omega_{\rm Q}$. Given any Hermitian operator $\h{A}$ on $\H$, we
obtain a smooth real function $\bar{A}$ on $\Gamma_{\rm Q}$ by
taking its (normalized) expectation value on quantum states. Then
one has an \emph{exact} identity on $\Gamma_{\rm Q}$:
\be {\rm If}\quad [\h{A},\, \h{B}] = \h{C}, \qquad {\rm then}
\quad \bar{C} = i\hbar\, \Omega_{\rm Q}^{\alpha\beta}\,
\partial_\alpha \bar{A}\, \partial_\beta \bar{B} \,\, \equiv
\,\, i\hbar\, \{\bar{A}, \bar{B}\}_{\rm Q} \ee
\emph{for all Hermitian operators} $\h{A},\, \h{B}$. Here, to be
explicit, we have chosen an index notation for $\H$, denoting a
quantum state $\Psi$ as a vector $\Psi^\alpha$. Note that this is
not the usual `Dirac quantization prescription' because $\h{A}$ and
$\h{B}$ are arbitrary Hermitian operators and the identity holds on
the infinite dimensional $\Gamma_{\rm Q}$ rather than on the finite
dimensional classical phase space $\Gamma$. The second surprising
fact relates unitary flows generated by arbitrary Hermitian
operators $\h{H}$ on $\H$ and the Hamiltonian flow generated by the
function $\bar{H}$ on $\Gamma_{\rm Q}$. The two flows coincide:
\be (\h{H} \Psi)^\alpha = i\hbar\, (\Omega_{\rm
Q}^{\alpha\beta}\, \p_{\beta} {\bar{H}}){\mid}_{\Psi} \, . \ee
From this perspective, quantum mechanics can be regarded as a
special case of classical mechanics! Special, because $\Gamma_{\rm
Q}$ carries, in addition to the symplectic structure, a Riemannian
metric determined by the real part of the inner product (whence
$\Gamma_{\rm Q}$ is a K\"ahler space). The  metric is needed to
formulate the standard measurement theory but is not be important
for our present considerations.

Because the space of quantum states is now regarded as a phase space
$(\Gamma_{\rm Q}, \Omega_{\rm Q})$, one can hope to relate it to the
classical phase space $(\Gamma, \Omega)$ and transfer information
about quantum dynamics, formulated on $(\Gamma_{\rm Q}, \Omega_{\rm
Q})$, to the classical phase space $(\Gamma, \Omega)$. This
procedure gives the effective equations on $(\Gamma, \Omega)$ that
capture the leading order quantum corrections. We will first
describe the truncation method, and then focus on the embedding
approach which will be used extensively in the rest of this section.

Let us suppose that the classical phase space $\Gamma$ is labeled by
canonical pairs $(q_i,p_i)$. In quantum theory we have the analogous
operators $\h{q}_i, \h{p}_i$ on $\H$. Now, any state $\Psi \in \H$
is completely determined by the specification of the expectation
values of all polynomials constructed from $\h{q}_i, \h{p}_j$.
Therefore, one can introduce a convenient coordinate system on the
infinite dimensional quantum phase space $\Gamma_{\rm Q}$, using the
expectation values $(\bar{q}_i, \bar{p}_i)$ of the basic canonical
pairs, expectation values  $(\overline{q_i^2},\, \overline{p_i^2},\,
\overline{q_i p_j})$ of their quadratic combinations, and so on for
higher order polynomials. The expectation values $(\bar{q}_i,
\bar{p}_i)$ are in 1-1 correspondence with the canonical coordinates
$(q_i, p_i)$ on the classical phase space $\Gamma$ and the rest
represent `higher moments'. The Hamiltonian flow on $\Gamma_{\rm Q}$
determined by the exact unitary dynamics on $\H$ provides evolution
equations for $(\bar{q}_i, \bar{p}_i)$ and all the moments. These
are infinitely many, coupled, non-linear differential equations
dictating the time-evolution on $\bar{\Gamma}_{\rm Q}$ and the full
set is equivalent to the exact quantum dynamics. The idea is to
truncate this set at some suitably low order, solve the equations,
and obtain solutions $\bar{q}_i(t), \bar{p}_i(t)$ which include the
leading order quantum corrections to the classical evolution. The
method is systematic in that, in principle, one can go to any order
one pleases. But in practice it is difficult to go beyond the second
or the third order and there is no a priori control on the
truncation error. This framework makes it possible to calculate the
`back-reaction' of the evolution of higher moments on the desired
evolution of expectation values $\bar{q}_i(t), \bar{p}_i(t)$.
However, in general these effects are sensitive to the initial
choice of state.

In the embedding approach, one seeks an embedding $\Gamma \to
\bar\Gamma_{\rm Q} \subset \Gamma_{\rm Q}$ of the finite dimensional
classical phase space $\Gamma$ into the infinite dimensional quantum
phase space $\Gamma_{\rm Q}$ that is well-suited to capture quantum
dynamics. To define $\bar\Gamma_{\rm Q}$, for any given a point
$\gamma^o \in \Gamma$, where $\gamma^o = (q_i^o, p^o_i)$, one has to
find a quantum state $\Psi_{\gamma^o}$ and these $\Psi_{\gamma^o}$
are to constitute $\bar\Gamma_{\rm Q}$. The first requirement is
kinematic: the embedding should be such that $q_i = \langle
\Psi_{\gamma^o} \h{q}_i \Psi_{\gamma^o}\rangle =: \bar{q}^i$ and
$p_i = \langle \Psi_{\gamma^o} \h{p}_i \Psi_{\gamma^o}\rangle =:
\bar{p}^i$. An elementary example of the required embedding is given
by coherent states: fix a set of parameters $\sigma_i$ let
$\Psi_\gamma^{o}$ be a coherent state peaked at $(q_i^o, p^o_i)$
with uncertainty in $q_i$ given by $\sigma_i$. The second condition
that the embedding captures quantum dynamics is highly non-trivial.
It requires that the \emph{quantum Hamiltonian vector field should
be approximately tangential to} $\bar\Gamma_{\rm Q}$. If this
condition is achieved, one can simply project the exact quantum
Hamiltonian vector field on to $\bar\Gamma_{\rm Q}$ and obtain an
approximate quantum evolution there. But since $\bar\Gamma_{\rm Q}$
is naturally isomorphic to $\Gamma$, this gives us a flow on the
classical phase space \emph{representing the desired quantum
corrected evolution}. For a harmonic oscillator one can make the
Hamiltonian vector field on $\Gamma_{\rm Q}$ \emph{exactly}
tangential to $\bar\Gamma_{\rm Q}$ simply by choosing $\sigma^i$ as
the appropriate function of the frequencies, masses and $\hbar$.
Then there are no quantum corrections to the classical equations;
peaks of the chosen coherent states will follow the exact classical
trajectories. For a general Hamiltonian, the dynamical condition is
difficult to achieve. But if one does succeed in finding an
embedding such that the quantum Hamiltonian vector field is indeed
very nearly tangential to $\bar\Gamma_{\rm Q}$, the resulting
effective equations can approximate the quantum evolution very well.
The error can be estimated by calculating the ratio of the component
of the Hamiltonian vector field orthogonal to $\bar\Gamma_{\rm Q}$
to the component that is tangential. The method is not systematic:
it is an `art' to find a good embedding.

In LQC, existence of such an embedding has been established in
various cases. This method was first used in the context of the
so-called $\mu_o$ quantization \cite{abl} to obtain quantum
corrected effective Hamiltonian constraint for a dust filled
universe \cite{jw}. In the improved $\bar \mu$ quantization
\cite{aps3}, the embedding approach has been used to derive the
quantum corrected Einstein's equations for the FLRW model with a
massless scalar field \cite{vt}, and for arbitrary matter with a
constant equation of state \cite{psvt}.

Let us now discuss the underlying conditions on the choice of
judicious states for the case of a massless scalar field. One starts
with Gaussian states with spreads $\sigma$ and $\sigma_{(\phi)}$ in
the gravitational and matter sectors respectively, peaked at
suitable points of the classical phase space. In terms of the
variables introduced in section \ref{s2}, satisfaction of the
dynamical condition requires that the following inequalities hold
simultaneously: (i) $\nu \gg \lp$, (ii) $\lambda \b \ll 1$, (iii)
$\Delta \b/\b \ll 1$ and $\Delta \nu/\nu \ll 1$, and (iv) $\Delta
\phi/\phi \ll 1$ and $\Delta p_{(\phi)}/p_{(\phi)} \ll 1$. The first
two inequalities are meant to incorporate the idea that the initial
state is peaked at large volumes where the gravitational field is
weak, and the last two inequalities are to ensure that the state is
sharply peaked with small relative fluctuations in both
gravitational and matter degrees of freedom. Some of these
inequalities turn out to be mutually competing and a priori can not
be satisfied without additional constraints of the basic fields
(because, for example, we only have two widths $\sigma$ and
$\sigma_{(\phi)}$, and four uncertainties $\Delta \nu, \Delta b,
\Delta \phi$ and $\Delta\pphi$). But a careful examination shows
that the additional constraint is very weak \cite{vt}: $p_{(\phi)}
\gg 3 \hbar$, which is \emph{very} easily satisfied if the state is
to represent a universe that approximately resembles ours, say in
the CMB epoch.

Using such states, we then compute the expectation values of the
quantum constraint operator $\h{C}_H$, and of the operators
corresponding to the basic phase space variables. The latter define
an the embedding of $\Gamma$ into $\bar\Gamma_{\rm Q}$. One can
check that, within the systematic approximations made, the quantum
Hamiltonian vector field is indeed tangential to the image
$\bar\Gamma_{\rm Q}$ of $\Gamma$ in $\Gamma_{\rm Q}$. The generator
of the Hamiltonian flow provides us with the effective Hamiltonian
constraint:
\be\label{effham1} {C}^{\mathrm{(eff)}}_H \, = \, -\f{3
\hbar}{4 \gamma \lam^2} \nu \, \sin^2(\lam \b) \, + \, {\cal
H}_{\mathrm{matt}} \, + \, O(\sigma^2, \nu^{-2}, \sigma^{-2}
\nu^{-2}) ~, \ee
where ${\cal H}_{\mathrm{matt}}$ denotes the matter Hamiltonian. \\

Approximations used in the derivation of the effective Hamiltonian
constraint seem to break down before the evolution encounters the
Planck regime \cite{vt}. Nonetheless, as often happens in
theoretical physics, the resulting effective equations have turned
out to be applicable well beyond the domain in which they were first
derived.  In particular, they predict a value for the density at the
bounce that is in exact agreement with the supremum of the spectrum
of the density operator in sLQC \cite{acs}, discussed in section
\ref{s3}. Furthermore, extensive numerical simulations have shown
that solutions to the effective equations track the exact quantum
evolution of the peaks of those wave functions that start out being
semi-classical in the low curvature regime. This behavior has been
verified in the k=0 models with or without a cosmological constant,
with an inflaton in a quadratic potential, and in the k=1 model
discussed in sections \ref{s2} - \ref{s4}.

\subsection{Effective dynamics in LQC: Key features}\label{s5.2}

Using the effective Hamiltonian constraint, it is straightforward to
obtain the modified Einstein's equations in LQC. These equations
lead to two important predictions: (i) existence of a phase of
super-inflation following the bounce \cite{ps06}, and (ii) a generic
resolution of strong curvature singularities \cite{ps09}. To analyze
the physical implications of the effective Hamiltonian constraint
(\ref{effham1}), we consider matter with an equation of state
satisfying $P = P(\rho)$, where $P$ denotes the pressure. We further
restrict the attention to the leading part in (\ref{effham1}) and
drop the correction terms proportional to the fluctuations.  For
brevity, we only discuss to case of the spatially flat model.
Extensions to spatially curved models are discussed in
\cite{apsv,kv,sv}.

\subsubsection{Modified Einstein equations and super-inflation}
\label{effham_iso}

The strategy is to consider the classical phase space of the FLRW
model coupled to matter, but now equipped with the quantum corrected
Hamiltonian constraint. This constraint and the Hamilton's equations
of motion it determines provide the full set of effective Einstein's
equation for the FLRW model.

In order to obtain the modified Friedmann equation, one first
computes the Hamilton's equation for $\nu$:
\be\label{eqnudot} \dot \nu = \{\nu, {C}^{\mathrm{(eff)}}_H\} = -
\f{2}{\hbar} \f{\p}{\p \b} {C}^{\mathrm{(eff)}}_H = \f{3}{\gamma
\lam} \, \nu \, \sin(\lam \b) \cos(\lam \b) ~. \ee
Next, since physical solutions also satisfy the constraint
${C}^{\mathrm{(eff)}}_H \approx 0$, we also have
\be \f{3 \hbar}{4 \gamma \lam^2} \, \nu \sin^2(\lam \b) = {\cal
H}_{\mathrm{matt}}\, . \ee
On using the relation (\ref{eq:V_nu}) between $\nu$ and the physical
volume, this equation can be rewritten as
\be\label{eqsinlamrho} \f{\sin^2(\lam \b)}{\gamma^2 \lam^2} = \f{8
\pi G}{3} \rho~, \ee
where, as before, $\rho$ is the energy density of matter. Squaring
(\ref{eqnudot}) and using (\ref{eqsinlamrho}) we obtain
\be \label{fried} H^2 =  \f{\dot \nu^2}{9 \nu^2} = \f{8 \pi G}{3} \, \rho \left(1 -
\f{\rho}{\rcr}\right) \ee
where $H = \dot a/a$ denotes the Hubble rate and $\rcr$ is the
maximum energy density given by $\rcr =  3/(8 \pi G \gamma^2 \lam^2)
\approx 0.41 \rho_{\mathrm{Pl}}$. This is the modified Friedmann
equation we were seeking. Note that, in the expression of the
effective constraint (\ref{eqsinlamrho}), it is the left hand side
that is modified from $\b^2$ to $\sin^2 \lambda \b/\lambda^2$ due to
the underlying quantum geometry. To arrive at the modified Friedmann
equation, we have merely used the equation of motion for $\nu$ and
trignometric identities to shift this modification to the right
side.

Similarly, the modified Raychaudhuri equation can be obtained from
Hamilton\rq{}s equation for $\b$:
\be\label{raych} \f{\ddot a}{a} = - \f{4 \pi G}{3} \, \rho \,
\left(1 - 4 \f{\rho}{\rcr} \right) - 4 \pi G \, P \, \left(1 -
2 \f{\rho}{\rcr} \right) . ~ \ee
It is also useful to obtain the equation for the rate of change
of the Hubble parameter using the modified Friedmann and
Raychaudhuri equations:
\be\label{dotH} \dot H = - 4 \pi G (\rho + P) \left(1 - 2
\f{\rho}{\rcr}\right) ~. \ee
These equations immediately provide a conservation law of
matter,
 \be\label{cl} \dot \rho + 3 H \, (\rho + P) = 0\, .
\ee
Note that although the effective Friedmann and Raychaudhuri
equations are modified due to quantum corrections, the conservation
law is the same as in classical cosmology. Furthermore, as in the
classical theory, Eqs (\ref{fried}),(\ref{raych}) and (\ref{cl})
form an over-complete set: one can alternatively derive the
(\ref{cl}) from the Hamilton's equations for the matter field and
then use (\ref{fried}) to derive the modified Raychaudhuri equation
(\ref{raych}).

In general relativity, the field equations lead to a singularity for
all matter satisfying weak energy condition (WEC), except for the
special case of a cosmological constant. By contrast, one can show
that the modified field equations lead to a non-singular evolution.
In particular, Eq (\ref{fried}) implies that $\dot{a}$ vanishes at
$\rho= \rcr$ and the universe bounces. In the limit $G \hbar
\rightarrow 0$ or equivalently $\lam \rightarrow 0$, as one would
expect, the modified Friedmann and Raychaudhuri equations directly
reduce the classical equations:
\be\label{cfried} H^2 = \f{8 \pi G}{3} \, \rho ~,~~ \f{\ddot
a}{a} = - \f{4 \pi G}{3} \, (\rho  + 3 P) ~, ~~ \dot H = - 4
\pi G (\rho + P) \ee
In this limit $\rcr$ also becomes infinite, the bounce
disappears and the classical singularity reappears.
\\

Let us now consider the time evolution of the Hubble rate. An
interesting phase in the cosmological evolution occurs when $\dot H
> 0$. This is known as the phase of \emph{super-inflation}, as the
acceleration of the universe in this phase is faster than in a
de-Sitter space-time. In the classical theory, for all matter
satisfying the WEC, $\dot H$ is negative. In order to have a
super-inflationary phase in general relativity, one  needs to
violate WEC, e.g., by introducing a phantom matter in the theory
(see, for example, \cite{phantom}). On the other hand, in LQC, the
phase of super-inflation is generic: Since the universe is in a
contracting phase ($H<0$) immediately before the bounce and since it
enters an expanding phase ($H>0$) immediately after the bounce, it
follows that $\dot{H} >0$ at the bounce and hence, by continuity,
also for some time after the bounce. From Eq (\ref{dotH}) one finds
that it occurs when $\rcr/2 < \rho \leq \rcr$, for all matter
satisfying the WEC.%
\footnote{Interestingly, if one considers matter which violates
the WEC, then it does not lead to super-inflation in LQC when
$\rcr/2 < \rho \leq \rcr$ \cite{ps06}.}
Initially there was some hope that this phase may provide the
accelerated expansion invoked in the inflationary scenarios without
having to introduce an inflaton with a suitable potential
\cite{mb_inflation}. However, in general, the phase is extremely
short lived and therefore cannot be a substitute the standard slow
roll. But it has interesting phenomenological ramifications, e.g.,
for production of primordial gravitational waves.
\\
\vfill\break \noindent {\bf Remarks:}

1. Note that the full set of effective equations has two key
properties: i) they are free from infrared problems because they
reduce to the corresponding equations of general relativity when
$\rho \ll \rcr$; and ii) they are all independent of the initial
choice of the fiducial cell. Even though they may seem simple and
obvious, these viability criteria are not met automatically
\cite{dmy} but require due care in arriving at effective equations.

2. The possibility of a phase of super-inflation in LQC was first
noted using a different mechanism than the one discussed here
\cite{mb_inflation}. Inverse volume effects can provide additional
corrections to the modified Friedmann equations which result in
$\dot H > 0$. These were first noted for a scalar field in
\cite{mb_inflation} and were subsequently generalized to matter
satisfying $P = P(\rho)$ \cite{ps05}. With these corrections, the
conservation law (\ref{cl}) also gets modified by additional terms
and the effective equation of state of matter changes \cite{ps05}.
If one assumes an effective Hamiltonian constraint incorporating
purely inverse volume corrections (and neglects all other
modifications originating from quantum geometry), super-inflation
occurs when the effective equation of state is such that the WEC is
violated. Thus, the super-inflationary regime originating from
inverse volume corrections is qualitatively different from the one
discussed above. Also, these corrections are physical only in the
spatially compact case. 

\subsubsection{Absence of strong curvature singularities}

Let us now discuss additional properties of the Hubble rate and
curvature invariants in LQC. In the classical theory of the k=0,
$\Lambda$=0 FLRW models, the Hubble rate diverges for all matter
satisfying WEC when the scale factor vanishes. 
By contrast, the Hubble rate in LQC has a universal upper bound.
From Eq (\ref{fried}), it follows that its maximum allowed value
occurs at $\rho = \rcr/2$ and is equal to
\be |H|_{\mathrm{max}} = \left(\f{1}{\sqrt{3} \, 16 \pi G \hbar
\gamma^3}\right)^{1/2}\,\, \approx\, 0.93 \mpl ~. \ee
It then immediately follows that the expansion of the
congruences of cosmological observers is bounded in LQC.

In the homogeneous and  isotropic model, Ricci scalar, $R$, is
sufficient to capture the behavior of all other space-time
curvature invariants. For the modified Friedmann dynamics of
LQC, the Ricci scalar turns out to be
\be \label{Ricci} R = 6 \left(H^2 + \f{\ddot a}{a} \right) =  8
\pi G \rho \, \left(1 + \f{2 \rho}{\rcr}\right) - 24 \pi G P
\left(1 - \f{2 \rho}{\rcr} \right) ~. \ee
Note that unlike the Hubble rate, $R$ depends both on the energy
density and pressure of the matter. This opens the possibility that
there may exist events where the space-time curvature may diverge,
even though $\rho \leq \rcr$. However, such a divergence occurs for
a very exotic choice of matter with an equation of state: $P(\rho)
\rightarrow \pm \infty$, at a finite value of $\rho$. Interestingly,
it can be shown that it  possible to always extend geodesics across
such events using modified Friedmann equations \cite{ps09}. Thus
unlike in general relativity, even if we allow exotic matter, events
where space-time curvature diverges in LQC are not the boundaries of
space-time.

Let us now examine such events in some detail. From the analysis of
integrals of curvature components for null and particle geodesics,
it is possible to classify the singularities as \emph{strong
curvature} and \emph{weak curvature} types. According to Kr\"{o}lak
\cite{krolak}, a singularity occurring at the value of affine
parameter $\tau = \tau_e$ is strong if and only if
\be\label{krolak} \int^\tau_0 d \tau R_{ab} u^a u^b \,\,\,\,
\to \infty \qquad {\rm as}\qquad \tau \rightarrow \tau_e\, .
\ee
Otherwise the singularity is weak. one can show that this
integral is finite in the isotropic and homogeneous LQC,
\emph{irrespective of the choice of the equations of state}
including the ones which lead to a divergence in the Ricci
scalar \cite{ps09}. Thus, the events where space-time curvature
blows up in LQC are harmless weak singularities.

The finiteness of the integral (\ref{krolak}), can also be used
to show that there exist no strong curvature singularities in
LQC.%
\footnote{The classification of the strength of singularities
is slightly different in Tipler's analysis \cite{krolak}. One can show that
strong curvature singularities are absent in LQC also using
Tipler's criteria.}
This result is in sharp contrast to the situation in general
relativity where strong curvature singularities are generic
features of the theory.

The generic resolution of strong curvature singularities using
the modified Friedmann dynamics implies that not only the
big-bang and the big-crunch, but singularities such as the big
rip studied in various phenomenological models are also
resolved due to underlying quantum geometric effects
\cite{brip_lqc,ps09}. By contrast, such singularities are
difficult to avoid in general relativity unless the parameters
potential of the phantom field are appropriately fine tuned
\cite{phantom}. However, singularities such as the sudden
singularities, which occur due to divergence in pressure with
energy density remaining finite, will not be resolved
\cite{portsmouth,ps09}. It turns out that such singularities
are always  of the weak curvature type, which are harmless and
beyond which geodesics can be extended \cite{ps09}. It is
rather remarkable, that quantum geometric effects in LQC are
able to distinguish between harmful and harmless singularities,
and resolve all those which are harmful.
\\

\noindent {\bf Remark:} Strong and weak singularities have also been
analyzed in the presence of spatial curvature using modified
Friedmann dynamics in LQC, including the  cases where they occur in
the past evolution \cite{sv}. Using a general phenomenological model
of the equation of state, one finds that all strong curvature
singularities in k=1 and k=-1 models are resolved. Furthermore, for
the spatially closed model there exist some cases where even weak
curvature singularities may be resolved. This brings out the
non-trivial role payed by the intrinsic curvature which enriches the
new physics at the Planck scale in LQC.

\subsection{Probability for inflation in LQC}
\label{sec_eff_pro} \label{s5.3}

The inflationary paradigm has been extremely successful in
accounting for the observed inhomogeneities in the CMB which
serve as seeds for the subsequent formation of the large scale
structure in the universe. Consequently, it is widely regarded
as the leading candidate to describe the very early universe.
However, it faces two conceptual issues. First, as we discussed
in section \ref{s4.3}, Borde, Guth and Vilenkin \cite{bgv} have
shown that the inflationary space-times are necessarily past
incomplete; even with eternal inflation one cannot avoid the
initial big-bang. In LQC, on the other hand, thanks to the
quantum geometry effects, the singularity is resolved and
inflationary space-times are past complete.

The second issue is that of `naturalness.' While a given theory
---such as general relativity--- may admit solutions with an
appropriate inflationary phase, is the occurrence of such a
phase generic, or, does it require a careful fine tuning? For
definiteness, let us suppose we have an inflaton $\phi$ with a
quadratic potential, $V(\phi) = (1/2) m^2\phi^2$. Then the WMAP
data \cite{wmap} provides us with a remarkably detailed picture
of the conditions at the onset of inflation \cite{as3}. More
precisely, the data refers to the time $t(k_{\star})$ at which
a reference mode $k_\star$ used by WMAP exited the Hubble
horizon in the early universe. (Today, this mode has a
wave-length about $12\%$ of the radius of the Hubble horizon.)
Within error bars of less than $4.5\%$, the data tells us that
the field configurations were:
\be \label{data} \phi(t(k_{\star})) = \pm 3.15\, \mpl, \qquad
\dot{\phi}(t(k_{\star})) = \mp 1.98\times 10^{-7} \,\mpl^2,
\qquad H(t(k_{\star})) = 7.83 \times 10^{-6}\,\mpl \, , \ee
where, as before, $H = \dot{a}/a$ is the Hubble parameter
\cite{as3}. 
Thus, for the quadratic potential, the 7 year WMAP data
requires the dynamical trajectory of the universe to have
entered a \emph{tiny} neighborhood of the phase-space point
given by (\ref{data}). One's first reaction would be that this
condition would be met only by a \emph{very} small fraction of
all dynamical trajectories. If so, a priori the required
inflationary phase would seem very implausible and the theory
would be left with a heavy burden of `explaining' why it
actually occurred.

Issue of `naturalness' has, of necessity, a subjective element
and one could just say that our universe simply happened to
pass through this tiny region of phase space and, since we have
only one universe, the issue of likelihood is irrelevant.
However, the broader community did not adopt this viewpoint.
Rather, it has sought to sharpen the question of naturalness by
introducing a measure on the space $\mathbb{S}$ of solutions of
the given theory: the required probability is then be given by
the \emph{fractional} volume occupied by those solutions which
do pass through configurations specified by the WMAP data at
some time during their evolution.

To find the measure, the following general strategy was
introduced over twenty years ago \cite{ghs,dp,hp}. Recall that
solutions to the field equations are in 1-1 correspondence with
phase space trajectories, and the natural Liouville measure
$\dd\mu_{\rm L}$ on the phase space is preserved by the
dynamical evolution. On the one hand, this measure is natural
because it is constructed using \emph{just} the phase space
structure. On the other hand, precisely for the same reason, it
does not encode additional information that may be important
for the physics of the specific system. Therefore, it only
provides \emph{a priori} probabilities, i.e. `bare' estimates.
Further physical input can and should be used to provide
sharper probability estimates and a more reliable likelihood.
However, a priori probabilities themselves can be directly
useful if they are very low or very high. In these cases, it
would be an especially heavy burden on the fundamental theory
to come up with the physical input that significantly alters
them.

The question of probability of inflation along these lines has
received a considerable attention in the literature (see for
eg. \cite{hh_inflaton,hw,klm,gt,gns_inflation}). In the general
relativity literature, conclusions on the probability of
inflation have been vastly different, ranging from near unity
\cite{klm} to being exponentially suppressed \cite{gt} (by
factors of $\sim\, e^{-195}$  for the slow roll phase
associated with configurations (\ref{data})). It turns out that
these vast differences arise because procedure is intrinsically
ambiguous within general relativity and different calculations
in fact answer different questions. This is a rather subtle
issue and, although it is explained in detail in
\cite{as3,ck-inflation}, it is still sometimes overlooked (see,
e.g., \cite{turok}).

Let us discuss this point in some detail in the k=0 FLRW model
with a scalar field in a potential $V(\phi)$ (which is bounded
below) either in general relativity or in effective LQC. Let
$\Gamma$ denote the $4$-dimensional phase space, spanned by
$(\v, \b;\, \phi, \pphi)$ (Recall that $\v$ is related to the
scale factor via $\v \sim a^3$ and that, in general relativity,
$\b \sim H$, the Hubble parameter.) Let $\bar\Gamma$ denote the
constraint surface which can be coordinatized by $(\v, \b,
\phi)$ and let $\h\Gamma$ be the 2-dimensional (gauge fixed)
sub-manifold of $\bar\Gamma$ such that each dynamical
trajectory intersects it once and only once. Since $\b$ is
monotonic both in general relativity and LQC, One typically
chooses $\h\Gamma$ to be the sub-manifold of $\bar\Gamma$ with
$\b = \b_o$, a constant, so that $\h\Gamma$ is coordinatized
simply by $(\v, \phi)$. The space $\mathbb{S}$ of solutions is
naturally isomorphic with $\h\Gamma$. The pull-back $\h\Omega$
to $\h\Gamma$ of the symplectic structure $\Omega$ on $\Gamma$
then provides the natural Liouville measure $\dd\h\mu_{\rm L}$
on $\h\Gamma$ and therefore on $\mathbb{S}$. Because the
measure is preserved under dynamical evolution, it is
insensitive to the choice of gauge fixing, i.e., the value of
$\b_o$.

The problem however is that the total Liouville measure of
$\mathbb{S}$ is \emph{infinite} and the volume of the subset of
$\mathbb{S}$ spanned by solutions which enter the phase space region
singled out by the WMAP data is also \emph{infinite}. Therefore,
calculations of probabilities can give any answer one pleases.
However, the infinities are spurious. For, there is a gauge group
${\mathbb G}$ that acts on the solution space,\,\, $\v(t) \to \alpha
\v(t),\, \phi(t) \to \phi(t)$ \,\, under which physics does not
change. (In terms of scale factors, the action is just $a(t) \to
\beta a(t),\, \phi(t) \to \phi(t)$, with $\alpha= \beta^3$.) The
infinity in the Liouville volume arises simply from the fact that
the `length' of this gauge orbit is infinite. Therefore one would
like to factor out by this gauge group and be left with a
1-dimensional space, spanned by $\phi$, with finite total measure.
The problem is that although the group $\mathbb{G}$ does have a
well-defined action on the space $\mathbb{S}$ of solutions, the
symplectic structure $\h\Omega$ and hence the Liouville measure
$\dd\h\mu_{\rm L}$ \emph{fail to be invariant} under the action of
$\mathbb{G}$ \cite{as3}. So, we cannot unambiguously project
$\dd\h\mu_{\rm L}$ down to the space of orbits of $\mathbb{G}$!
Rather than projection, one can simply do a second gauge fixing,
i.e., choose a cross-section $\t{\mathbb{S}}$ of $\mathbb{S}$ (or,
equivalently, of $\h\Gamma$) which is traversed by the orbits of
$\mathbb{G}$ once and only once. Then one can \emph{unambiguously}
define a measure $\dd \tilde\mu_{\rm L}$ on $\tilde{\mathbb{S}}$ (up
to an irrelevant overall constant): In effective LQC one obtains
\cite{as2,as3}
\be \label{measure} \dd\tilde\mu_{\rm L} =
\big(\f{3\pi}{\lambda^2} \sin^2 \lambda \b_o - 8\pi^2\gamma^2
V(\phi)\big)^{\f{1}{2}}\,\,\dd\phi\, , \ee
and the answer in general relativity is given by taking the
limit $\lambda \to 0$ that sends the area gap to zero. The
total measure of $\tilde{\mathbb{S}}$ is \emph{finite} in both
theories because the matter energy density is fixed on the
surface $\b=\b_o$, whence $\phi$ ranges over a finite closed
interval. As a result calculations of probabilities yield
well-defined, finite results. However, the subtlety is that the
measure $\dd \tilde\mu_{\rm L}$ \emph{depends on the initial
choice} $\b=\b_o$ \emph{we made to arrive at} $\h\Gamma$ in a
non-trivial fashion!

To summarize, the Liouville measure $\dd\h{\mu}_{\rm L}$ on the
space $\mathbb{S}$ of solutions is unambiguous, independent of
the gauge fixing choice $\b=\b_o$. But the total Liouville
volume of $\mathbb{S}$ is infinite. Since this infinity can be
directly traced back to the freedom in rescaling $a(t)$ under
which physics does not change, it is natural to get rid of the
spurious infinity by gauge fixing. This procedure does lead us
to a natural measure $\dd\t{\mu}_{\rm L}$ on the space
$\tilde{\mathbb{S}}$ of \emph{physically distinct} solutions.
But this measure now carries a memory of the choice of $\b_o$,
which, in the space-time picture, corresponds to a choice of a
time slice in the solution.

In LQC, we are interested in the evolution to the future of the
bounce and specification of $\phi$ at the bounce surface suffices to
determine a unique \emph{physical} solution in $\t{\mathbb{S}}$. The
measure $\dd\t{\mu}_{\rm L}$ enables us to ask and answer the
following question: What is the \emph{fractional} volume occupied by
the subset of all possible $\{\phi\}_{\rm B}$ at the bounce surface
whose evolution is compatible with the stringent requirements of
WMAP? This fraction is the a priori probability for initial data at
the bounce to lead to a slow roll inflation compatible with the WMAP
data some time in the future.

To carry out the calculation, one has to fix the potential. For the
quadratic potential, $V(\phi) = (1/2) m^2 \phi^2$, the issue was
analyzed in detail in \cite{as3}. The answer turned out to be
\emph{extremely} close to one: The probability that the dynamical
trajectory does \emph{not} pass through the desired configuration
(\ref{data}) within the WMAP error bars is \emph{less than} $3\times
10^{-6}$! Let us amplify this statement. The allowed range of $\phi$
values at the bounce is $|\phi_{\B}| \le 7.44\times 10^5\mpl$ and
\emph{all} trajectories that start out with $|\phi_{\B}| > 5.46
\mpl$ pass through the tiny phase space region selected by WMAP
sometime to the future of the bounce. Note that the matter density
at the bounce is $\rho \approx 0.41 \rho_{\rm Pl}$, while that at
the `onset of the WMAP slow roll' (i.e. corresponding to the
configuration (\ref{data})) is $\rho \approx 7.32 \times 10^{-12}
\rho_{\rm Pl}$. Therefore, at the `onset', the allowed range of
$\phi$ is $|\phi_{\rm onset}| \le 3.19 \mpl$ and the WMAP allowed
region fills only 4.4\% of the full allowed range. Thus,
trajectories from all but a $\sim$\, millionth part of the allowed
values of $\phi$ at the bounce flow to a small region of the allowed
values of $\phi$ at the onset of inflation. In this precise sense,
the small region selected by WMAP is an attractor for the effective
LQC dynamics \cite{ck-inflation,svv}. The details of the effective
LQC dynamics across these 11 orders of magnitude in density and
curvature have a rich structure \cite{as3}. It is quite non-trivial
that this attractor behavior is realized in spite of the fact that
the LQC dynamics is rather intricate and exhibits qualitative
differences from general relativity: for example, certain scaling
relations are now violated and there is a novel super-inflation
phase.

In general relativity, the natural substitute for the big
bounce ---i.e., the natural time to specify initial
conditions--- is the big-bang. But because of the classical
singularity, we cannot carry out this calculation of the a
priori probability there. And there is no other natural instant
of time. One might think of specifying the initial $\phi$ at
the Planck density \cite{klm} and use the Liouville measure
$\dd\t\mu_{\rm L}$ attuned to that time. But there is no reason
to believe that we can trust general relativity in that era.
Alternatively, one can wait till we are in an era where we can
trust general relativity \cite{gt}. But the answer will depend
sensitively on one's choice of $\b_o$ used to carry out the
calculation because there is simply no natural time instant in
this regime. Probabilities calculated in this manner refer to
the fractional volume occupied by the initial data \emph{at
that rather arbitrary time} which is compatible with the WMAP
slow roll. As our discussion above illustrates, later the time,
lower will be the probability.

Let us summarize. In the literature one often asks for the
probability that there are a specified number $N$ of
e-foldings. In general relativity this question is rather loose
because one does not specify when one should start counting and
if one begins at the big-bang the number would be infinite. We
asked a sharper question: given a quadratic potential, what is
the a priori probability of obtaining slow roll inflation that
is compatible with the 7 year WMAP data? But even this question
is ambiguous because of the subtleties associated with the
Liouville measure. One has to fix a time instant, consider all
allowed field configurations at that time, and ask for the
fractional $\dd\t\mu_{\rm L}$-volume of the subset of these
configurations which, upon evolution, enter the small region of
phase space singled out by WMAP. In LQC, there is a natural
choice of the initial time ---the bounce time--- and the
detailed investigation then shows that these initial field
configurations will meet the desired WMAP region some time in
the future with probability that is greater than  $(1\, -\, 3
\times 10^{-6})$. In this precise sense, \emph{assuming} an
inflation in a suitable potential, an inflationary phase
compatible with the WMAP data is almost inevitable in LQC. In
general relativity, because there is no natural `time' for one
to specify initial configurations, and because the probability
depends on one's choice, answers become less interesting.

\subsection{Effective dynamics of Bianchi-I space-times}
\label{s5.4}

As in the case of isotropic models, one can write an effective
Hamiltonian constraint in the Bianchi models to capture the
underlying quantum dynamics and use it to derive modified dynamical
equations incorporating non-perturbative quantum gravity effects. We
will again find that there are striking differences from classical
general relativity; in particular the energy density and the shear
scalar are bounded in effective LQC. We will conclude this
sub-section with an application of this modified Bianchi-I dynamics
to one of the problems faced by the cyclic/ekpyrotic model.

\subsubsection{Modified dynamical equations}\label{sec_bianchi_eff}

The effective Hamiltonian for the Bianchi-I space-time can be
written as
\be\label{effham_bianchi} {C}^{(I,\,\mathrm{eff})}_H =  - ~\f{1}{8
\pi G \gamma^2 (p_1 p_2 p_3)^{1/2}} \left(\f{\sin(\bar \mu_1
c_1)}{\bar \mu_1} \f{\sin(\bar \mu_2 c_2)}{\bar \mu_2} p_1 p_2 +
\mathrm{cyclic}
 ~~ \mathrm{terms}\right) ~~ + ~~ {\cal H}_{\mathrm{matt}}~  \\
\ee where $\bar \mu_i$ are given by Eq (\ref{bianchi_mub1}).
Using Hamilton's equations, one can calculate the time variation of
triads and connections and show that $(p_i c_i - p_j c_j)$ is a
constant of motion (where, as before, there is no summation over
repeated covariant indices). However, unlike in general relativity,
the modified dynamical equations now imply that $\Sigma^2$ of Eq
(\ref{shearscalar}) is not a constant of motion \cite{cv}.

Next, let us consider the matter density $\rho$ and the shear scalar
$\sigma^2 = 6\Sigma^2/a^6$. The density is now given by
\be\label{density_bianchi1_eff} \rho = \f{1}{8 \pi G \gamma^2
\lam^2} \left(\sin(\bar \mu_1 c_1) \, \sin(\bar \mu_2 c_2) +
\mathrm{cyclic} ~~ \mathrm{terms}\right) ~ \ee
whence it is clear that it has an absolute maximum, given by
\be\label{rhomax_bianchi} \rcr = \f{3}{8 \pi G \gamma^2
\lambda^2} \, . \ee
Note that the maximum is identical to the upper bound on the
energy density in the isotropic model. The expression for
$\sigma^2$ can be computed by finding the equations for the
directional Hubble rates $H_i = \dot a_i/a_i$, which after some
computation yields
\ba\label{sigmasqlqc} \sigma^2 &=& \nonumber \f{1}{3 \gamma^2
\lam^2} \Big[\big(\cos(\bar \mu_3 c_3) (\sin(\bar \mu_1 c_1) +
\sin(\bar \mu_2 c_2))  - \cos(\bar \mu_2 c_2) (\sin(\bar \mu_1 c_1)
+ \sin(\bar \mu_3 c_3))\big)^2 \\ && \nonumber + \big(\cos(\bar
\mu_3 c_3) (\sin(\bar \mu_1 c_1) + \sin(\bar \mu_2 c_2)) - \cos(\bar
\mu_1 c_1) (\sin(\bar \mu_2 c_2) + \sin(\bar \mu_3 c_3))\big)^2 \\
&&  + \big(\cos(\bar \mu_2 c_2) (\sin(\bar \mu_1 c_1) + \sin(\bar
\mu_3 c_3))  - \cos(\bar \mu_1 c_1) (\sin(\bar \mu_2 c_2) +
\sin(\bar \mu_3 c_3))\big)^2\Big] ~. \ea
This expression implies that $\sigma^2$ has the maximum value
\cite{gs01}:
\be \sigma^2_{\mathrm{max}} = \f{10.125}{3 \gamma^2 \lam^2} ~.
\ee
These bounds bring out a key difference between the isotropic
and anisotropic cases. In the isotropic case, there is a single
curvature scalar and it ---as well as the energy density---
assumes the maximum value at \emph{the} bounce. In the
anisotropic case, by contrast, the Weyl curvature is non-zero
and shears serve as its (gauge invariant) potentials. Since
both the density and shears have an upper bound, one now has
not just one bounce where the density assumes its maximum
value, but other `bounces' as well, associated with shears.
There are regimes in which the shear grows and assumes its
upper limit. At that time, the energy density is not
necessarily maximum. Still the quantum geometry effects grow
enormously, overwhelm the classical growth of shear, diluting
it, and avoid the singularity. Thus the effective dynamics
leads to the following qualitative picture: Any time a
curvature scalar enters the Planck regime, the `repulsive'
forces due to quantum geometry effects grow and cause a bounce
in that scalar. Thus, even in the homogeneous case, the bounce
structure becomes much more intricate and physically
interesting.

The complicated form of the modified dynamical equations in the
Bianchi-I model, makes the task of writing an analog of the
generalized version of Friedmann-like equation
(\ref{gen_fried_class}), very difficult in LQC. However, if we
neglect terms of the order $(\bar \mu_i c_i)^4$ and higher, then one
can derive such an equation by following the steps as outlined in
the isotropic case, and it turns out to be \cite{cv},
\be H^2 = \f{8 \pi G}{3} \rho \left(1 - \f{\rho}{\rcr}\right) +
\f{\Sigma_{\mathrm{cl}}^2}{a^6} \left(1 - 3
\f{\rho}{\rcr}\right) - \f{9}{8 \pi G}
\f{\Sigma_{\mathrm{cl}}^4}{\rcr a^4} + O((\bar \mu_i c_i)^4) ~.
\ee Here $\Sigma_{\mathrm{cl}}$ the value of $\Sigma$ in
the classical limit of the theory. Unfortunately, this equation only
captures the dynamics faithfully only when anisotropies are small.
It will be an interesting to obtain the generalized Friedmann and
Raychaudhuri equations using the effective Hamiltonian constraint
without such approximations.

\subsubsection{Singularity resolution in the ekpyrotic/cyclic model}

An interesting application of the effective LQC dynamics is to the
ekpyrotic/cyclic model. In this paradigm, one makes the hypothesis
that the universe undergoes cycles of expansion and contraction
driven by  a scalar field in a potential of the form \cite{cyclic}
\be\label{cyclicpot} V(\phi) = V_c\, (1 - e^{-\sigma_1
\phi})\,\, \exp(-e^{-\sigma_2 \phi}) \, , \ee
where $V_c$, $\sigma_1$ and $\sigma_2$ are the parameters of
the model. The model has been proposed as an alternative to
inflation, where the contracting branch plays an important role
in the generation of structures in the universe.

\begin{figure}[tbh!]
\includegraphics[angle=0,width=0.45\textwidth]{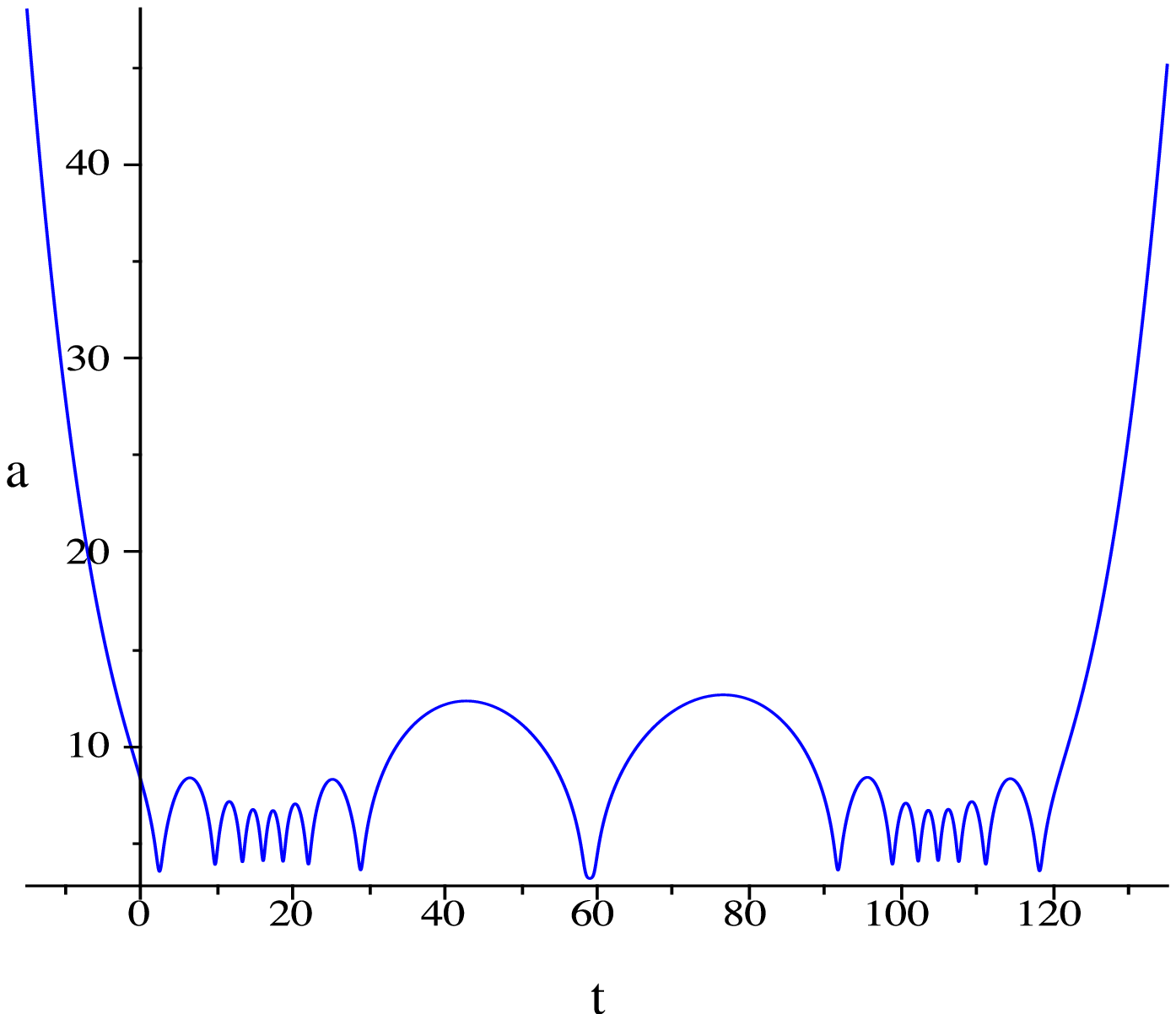}
\includegraphics[angle=0,width=0.45\textwidth]{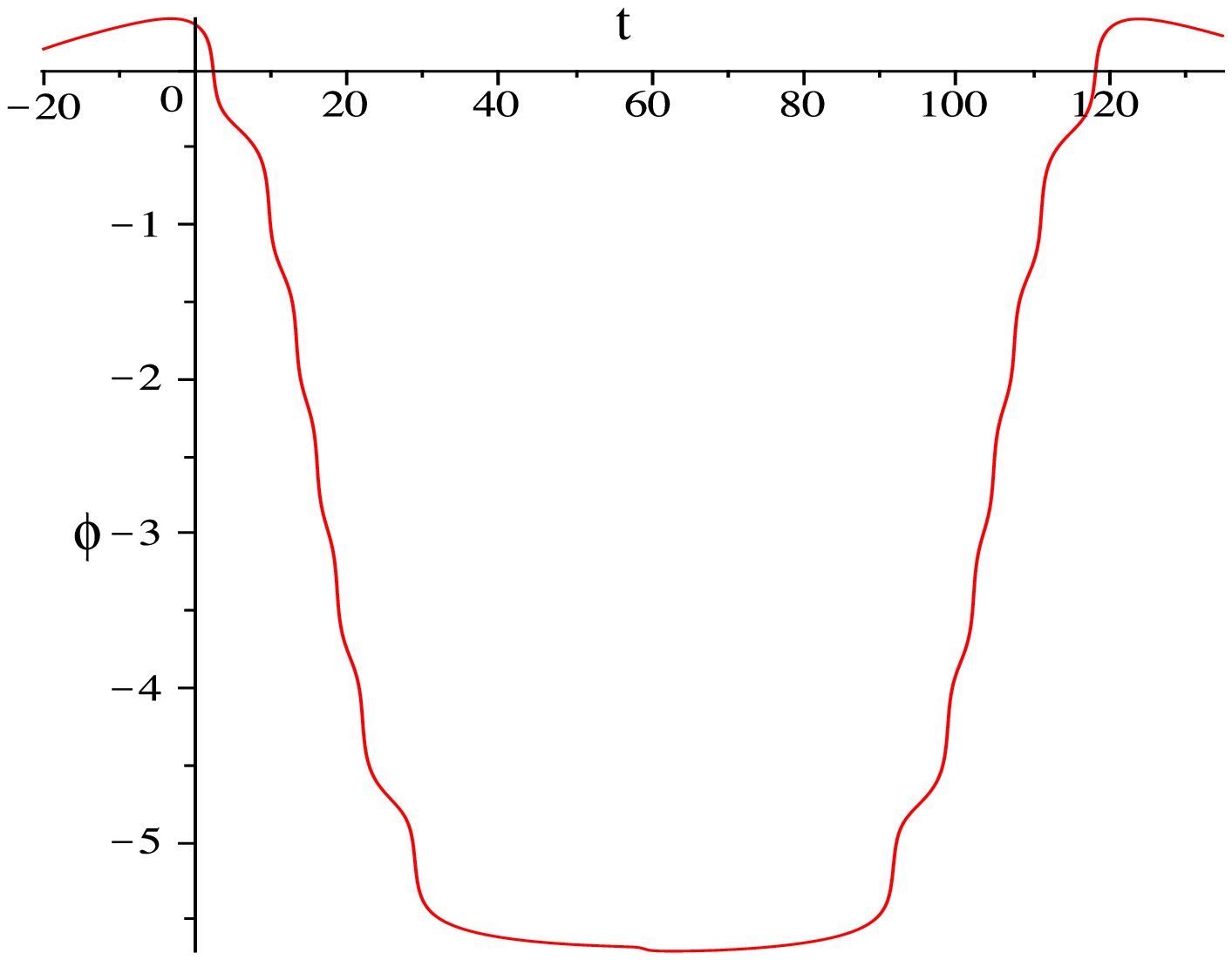}
\caption{Plot of the behavior of the scale factor and the scalar
field in effective loop quantum dynamics of Bianchi-I with potential
(\ref{cyclicpot}). The choice of the
parameters are $V_c = 0.01, \sigma_1 = 0.3 \sqrt{8 \pi}$ and
$\sigma_2 = 0.09 \sqrt{8 \pi}$. The initial values are
$\phi = 0.4$, $\dot \phi = -0.03$, $p_1 = 64$, $p_2 = 72$, $p_3 = 68$,
$c_1 = -0.6$ and $c_2 = -0.5$. (All units are Planckian). The initial value of
$c_3$ was determined using the vanishing of the Hamiltonian constraint.
}\label{cyclic_fig1}
\end{figure}
 \begin{figure}[tbh!]
\includegraphics[angle=0,width=0.40\textwidth]{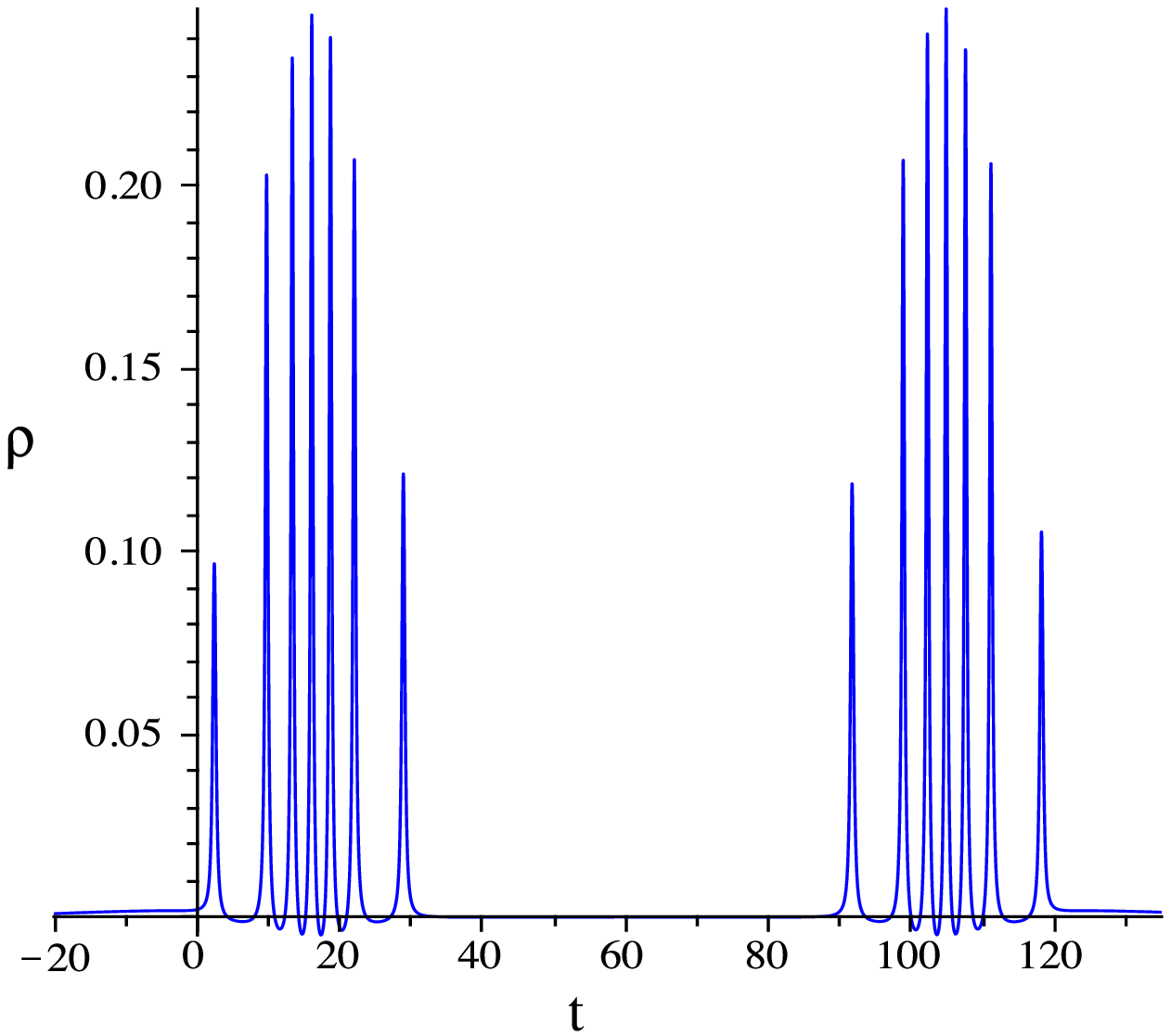}
\includegraphics[angle=0,width=0.45\textwidth]{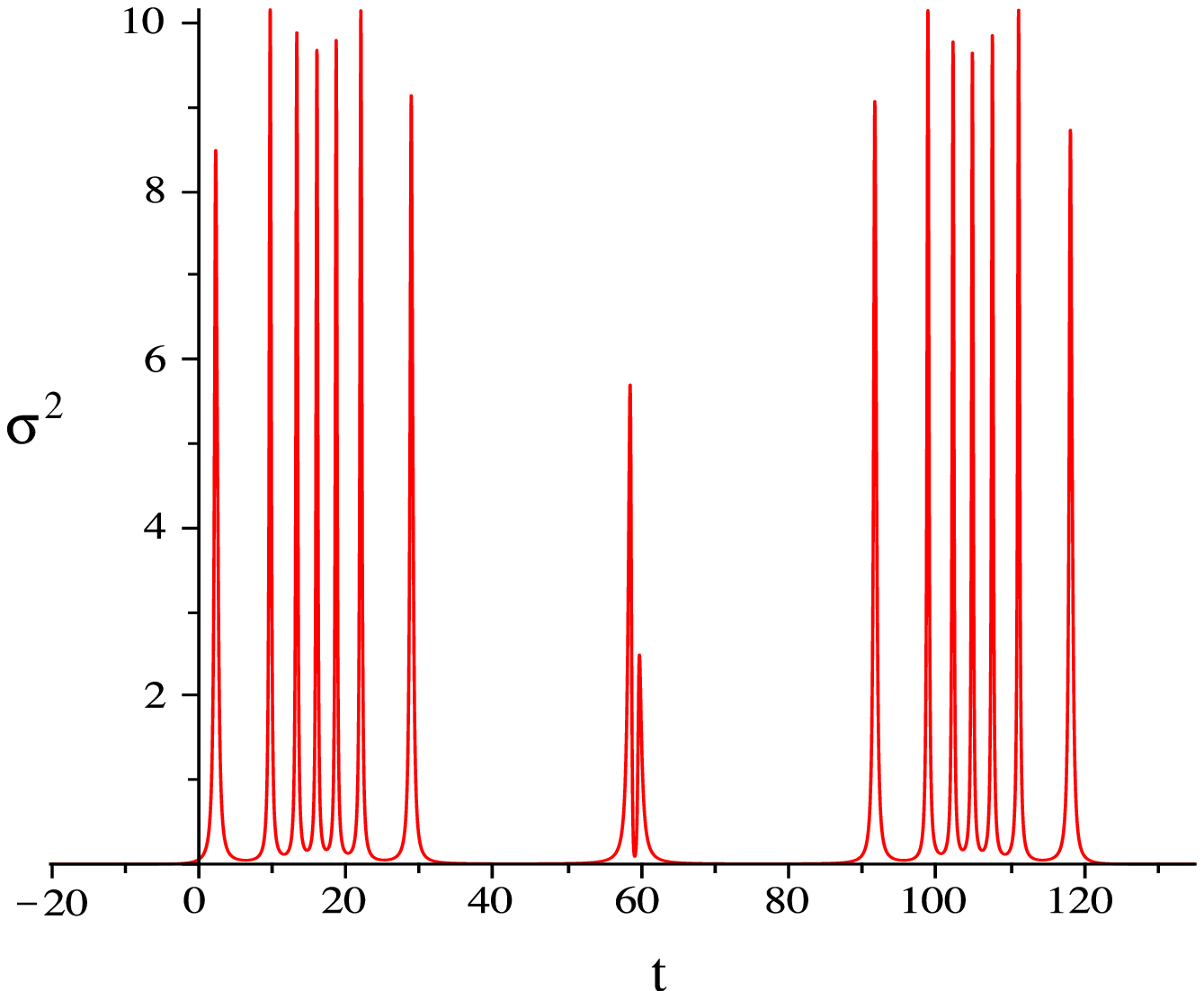}
\caption{Behavior of the energy density and the shear scalar $\sigma^2$.
Parameters in this plot are the same as those in Fig. \ref{cyclic_fig1}. The plot of energy density shows slight negative
values, which occur when the potential is negative. The Hubble rate remains real, and the evolution is well defined in this regime because of the compensation  from the shear scalar term, which is always positive.}
\label{fig2_cyclic}\end{figure}

Various stages of the evolution can be summarized as follows.
In the contracting phase, the field $\phi$ evolves from the
positive part of the potential and enters the steep negative
well where its equation of state becomes ultra-stiff. This
phase of evolution is thought to be responsible for generation
of nearly scale invariant perturbations in the universe. After
passing through the steep negative region of the potential, the
field rolls towards $\phi = -\infty$ which corresponds to the
big-crunch singularity. If one can somehow continue the
evolution through this big-crunch and if the field rolls back
in the expanding branch from $\phi = -\infty$, then at late
times it can lead to an accelerated expansion of the universe.
Eventually, the expansion of the universe stops due to Hubble
friction and the universe enters into a contracting phase to
repeat the cycle.

Note that the model faces a dual challenge: one has to obtain a
non-singular transition of the scale factor of the universe, and
furthermore, one has to achieve a turn-around of $\phi$ in the
vicinity of $\phi = - \infty$. It is relatively straightforward to
see from the classical Friedmann equation that such a turn-around is
not possible when the potential is negative. In any case, the
classical dynamical equations are inadequate to address these key
issues because they can not be trusted in the regime near the
singularity. A natural question then is whether non-perturbative
quantum gravity effects encoded in dynamical equations in LQC, can
provide insights on the singularity resolution in the
ekpyrotic/cyclic model and, if so, whether the scalar field
undergoes the desired turn around in the ensuing dynamics. This
issue has been analyzed both in the isotropic and anisotropic
versions of the model. In the isotropic case, one finds that as the
scalar field rolls towards $\phi = - \infty$, its energy density
rises and approaches $\rcr$. Once it equals $\rcr$, the scale factor
of the universe bounces and the singularity is avoided \cite{svv}.%
\footnote{The singularity resolution in the ekpyrotic/cyclic
model was first considered in the isotropic LQC in \cite{bms}.
However modifications to the effective dynamics originating
from the non-local nature of the field strength tensor were
ignored in that work.}
However, the scalar field does not turn around and it continues
rolling towards $\phi = - \infty$. By contrast, if one
considers the effective dynamics of the Bianchi-I model, one
finds that not only is the singularity resolved but also the
field turns around for a large range of initial conditions
\cite{csv}. The turn around of the field is tied to the
non-trivial role played by the shear scalar in the dynamical
equations, and occurs even if the anisotropies are small.

In Fig.\ref{cyclic_fig1}, we show the results from a numerical
analysis of the effective dynamical equations with the
ekpyrotic/cyclic potential. It depicts a rich structure of multiple
bounces in the Planck regime, and a turn-around of $\phi$ leading to
a viable cycle. Fig.\ref{fig2_cyclic} shows the behavior of the
energy density and the shear scalar. Since the anisotropies are
non-vanishing, the energy density does not reach its maximum value
$\rcr$ when the scale factor bounces in the Planck regime. For a
deeper understanding of the viability of this model, further work is
needed to quantify the constraints on the required magnitude of
anisotropies. However, current studies show that they can be very
small \cite{csv}. Thus, modified dynamics  of Bianchi-I model in LQC
successfully resolves a challenging problem of the ekpyrotic/cyclic
model. Note that attempts to resolve this problem using ghost
condensates, such as in the `New Ekpyrotic Scenario'
\cite{new_ekpy}, encounter problems of instabilities
\cite{new_ekpy_ghost}. In contrast, in effective LQC, one does not
need to introduce exotic matter to resolve the singularity, and
there are no such instability problems.

\subsection{Summary of other applications}
\label{s5.5}

{\it Pre-big-bang models and LQC:} The proposal of pre-big-bang
models \cite{pbb}, based on string cosmology, is an attractive
idea but faces the problem of non-singular evolution from the
contracting to the expanding branch. A phenomenological study
based on exporting the non-perturbative quantum gravitational
effects as understood in LQC to these models reveals that this
problem can be successfully resolved both in the Einstein and
the Jordan frame without any fine tuning of the initial
conditions \cite{pbb_lqc1}. Similar results have been derived
in the presence of a positive potential \cite{pbb_lqc2}. Thus,
modified FLRW dynamics in LQC alleviates a major problem of the
pre-big-bang scenarios and provides a clear insight on the form
of the required non-perturbative terms from higher order
effective actions in the string cosmologies.
\\

{\it Covariant action for LQC:} A natural question in LQC is
whether the modified Friedmann dynamics can be obtained from a
covariant effective action. The issue is subtle because while
isotropic LQC has only two gravitational degrees of freedom,
effective actions in the conventional treatments generally
require additional degrees of freedom. Furthermore, for metric
theories, the requirements of second-order dynamical equations
and covariance uniquely leads to the Einstein-Hilbert action up
to a cosmological constant term. However, if one generalizes to
theories where the metric and the connection are regarded as
independent, one \emph{can} construct a covariant effective
action that reproduces the effective LQC dynamics
\cite{os}.%
\footnote{Attempts have also been made to write an effective action
in a metric theory, but these have not been successful in capturing
the regime in which non-perturbative quantum gravitational effects
become significant, including the bounce \cite{effact_metric}.}
For a simple functional form, $f({\cal R})$ where ${\cal R}$ is the
curvature of the independent connection, the action turns out be an
infinite series in ${\cal R}$.
It is possible that the functional form of the effective action is
not unique and various effective actions involving higher order
curvature invariants such as ${\cal R}_{\mu \nu} {\cal R}^{\mu \nu}$
exist.  Nonetheless, the availability of an effective action
capturing non-perturbative quantum gravity modifications has sparked
considerable interest in the construction of loop inspired
non-singular models in the Palatini theories (see Ref. \cite{os_rev}
for a review).
\\

{\it Correspondence with brane world scenarios and scaling
solutions:} Interestingly, properties of solutions to the modified
Friedmann dynamics in LQC have some similarities and dualities with
properties of solutions in the brane world scenarios (see
\cite{roy_brane} for a comprehensive review of brane-worlds). These
have been investigated in two ways. First, by considering the
modifications to the matter Hamiltonian originating from inverse
scale factor effects \cite{clm} and second by modifications to the
gravitational part of the Hamiltonian constraint \cite{ps06}. The
latter bear a more natural analogy with the brane world cosmologies
where one obtains $\rho^2$ modifications albeit with an opposite
sign. Therefore in the brane world scenarios bounces can occur only
if the brane tension is allowed to be negative \cite{ss}. Finally,
scaling solutions have been obtained in effective LQC both for
equations resulting from the modification of the gravitational part
of the Hamiltonian constraint \cite{cmns,lqc_scale} and from the
modifications in the matter Hamiltonian coming from inverse scale
factor effects \cite{lidsey_scale}. These facilitate qualitative
comparison of the effective LQC dynamics with that in the brane
world scenarios.
\\

{\it Dark energy models in LQC:} Dynamical behavior of various
dark energy models have been studied using the modified
Friedman equations in LQC and comparisons have been made with
general relativity. These include, interacting dark energy
model with dark matter \cite{darkenergy_lqc}; multi-fluid
interacting model in LQC \cite{multi_fluid}; dynamics of
phantom dark energy model \cite{darkenergy_phantom};
quintessence and anti-Chaplygin gas \cite{quint_chap} and
quintom and hessence models \cite{quintom_lqc1}. The latter
have also been used to construct non-singular cyclic models
\cite{quintom_lqc2}.
\\

{\it Tachyonic fields:} Various applications of tachyonic
matter have been studied in LQC. The singularity resolution has
shown to be robust \cite{tachyon_lqc1} and attractors have been
found for a tachyon in an exponential potential
\cite{tachyon_lqc2}. The results on tachyonic inflationary
scenario have been recently extended to the warm inflation
\cite{tachyon_lqc3} (see also \cite{tachyon_lqc4} for another
warm inflationary model in LQC).
\\

{\it Singularity resolution in k=1 model:} In section IV A, we
discussed the loop quantization of the k=1 model with a massless
scalar field as source and demonstrated the singularity resolution.
Using the effective Hamiltonian constraint, resolution of
singularities has been studied for more general forms of matter.
Since the space-time is spatially compact, inverse volume
corrections are well-defined and they lead to some interesting
features. It can be shown that the scale factor in k=1 model
undergoes bounces even if one includes only the inverse volume
modifications to the effective Hamiltonian constraint \cite{st}. For steep
potentials, where the initial conditions to resolve the singularity
have a set of measure zero in general relativity, bounce occurs in
LQC under generic conditions. A
generalization has been recently been investigated in the
Lemaitre-Tolman geometries where, again only the inverse volume
corrections are included and occurrence of bounces has been compared
between marginal and non-marginal cases \cite{mb_ltb}.  In another application
\cite{alex_karami}, bounces have also been studied in an alternative
quantization  which captures the elements of quantization used in
Bianchi-II \cite{awe3} and Bianchi-IX model \cite{we} (see section
\ref{b2b9} for details).
\\

{\it Bianchi-I models and magnetic fields:} An important application
of the effective dynamics in Bianchi models deals with the study of
matter with a non-vanishing anisotropic stress. Such a matter-energy
stress tensor leads to non-conservation of the shear scalar
$(\Sigma^2)$ in the classical theory as well as LQC. It is of interest
to understand how the presence of such matter affects the bounce in
LQC. By including magnetic fields, an investigation on these lines
was carried out in \cite{mv_bianchi}. The result is that the bounds
on energy density and shear scalar in the Bianchi-I model are not
affected, although there is a significant variation in anisotropies
across the bounce.
\\

{\it Physics of Bianchi-II and Bianchi-IX models:}  Effective
Hamiltonian constraints of Bianchi-II and Bianchi-IX models,
obtained in \cite{awe3} and \cite{we} respectively, have been
used to explore some of their physical consequences in detail
\cite{gs01}. It turns out that while one can obtain an upper
bound on the energy density $\rho$ without imposing energy
conditions in the Bianchi II model, in contrast to the Bianchi
I model, these conditions are now necessary to bound the shear
scalar $\sigma^2$. In the Bianchi-IX model, because of the
spatially compact topology, inverse volume corrections to the
effective Hamiltonian constraint are meaningful and have to be
included. They turn out to play an important role in placing an
upper bound on the energy density. In both of these models, the
upper bounds on the energy density and the shear scalar,
obtained by imposition of the WEC, turn out to be higher than
the universal bounds in the Bianchi-I model: They turn out to
be $\rho_{\rm max} \approx 0.54 \rho_{\mathrm{Pl}}$ and
$\sigma^2_{\rm max} \approx 127.03/(3 \gamma^2 \lambda^2)$ in
the Bianchi II model and $\rho \approx 7.91 \rho_{\mathrm{Pl}}$
and $\sigma^2_{\rm max} \approx 1723.64/(3 \gamma^2 \lambda^2)$
in the Bianchi-IX model. However, it is not known that these
bounds are optimal.

\section{Beyond Homogeneity}
\label{s6}

While kinematics of full LQG is well established, there is
still considerable ambiguity in the definition of the full
Hamiltonian constraint. Nonetheless significant progress has
been made in some inhomogeneous contexts by extending the
strategy that has been successful in mini-superspaces. These
explorations are phenomenologically important since they have
the potential to open new avenues to observable consequences of
quantum gravity in the early universe. They are also
significant from a more theoretical angle since they provide
guidance for full LQG.

In this section we will discuss three directions in which
inhomogeneities are being incorporated: i) the one polarization
Gowdy midi-superspaces which admit gravitational waves with or
without a scalar field as the matter source; ii) inhomogeneous
test quantum fields on cosmological quantum geometries; and
iii) cosmological perturbations on FLRW backgrounds. In each
case, one begins with the appropriately truncated (but infinite
dimensional) sector of the classical phase space of full
general relativity (with matter), and uses suitable gauge
fixing to make quantization tractable. The resulting
Hamiltonian theories exhibit a clean separation between
homogeneous and inhomogeneous modes and, mathematically, one
can regard the inhomogeneous modes as an assembly of harmonic
oscillators. Quantum theory is then constructed by treating the
homogeneous modes a la LQC and the inhomogeneous modes as in
the standard Fock theory. It is quite non-trivial that this
`hybrid' procedure provides a coherent, non-perturbative
quantization of an infinite dimensional phase space, consistent
with the underlying truncation of the full theory.

The relation between these theories and the eventual, full LQG
will become clear only once there is a satisfactory candidate
for the full Hamiltonian constraint. Nonetheless, it seems
reasonable to hope that these truncated theories will provide
good approximations to full dynamics for states in which the
`energy' in the inhomogeneities is sufficiently small even in
the Planck era \cite{hybrid5}. In this case, the inhomogeneous
modes will be affected by the quantum geometry effects --such
the bounces--- of the dominant homogeneous modes  but they
would be too weak for their own quantum geometry effects to be
important. These conditions are compatible with some of the
mainstream inflationary scenarios. Therefore this approach may
be useful also in phenomenological studies.

\subsection{The Gowdy Models}
\label{s6.1}

In the Gowdy midi-superspaces the spatial manifold is taken to
be $\mathbb{T}^3$ and all space-times under consideration have
two commuting Killing fields. Thus, inhomogeneities are
restricted just to one spatial direction. Thanks to these
symmetries, one can introduce a geometrically well motivated
gauge fixing procedure, making the model mathematically
tractable. If we further require that the two Killing fields be
hypersurface orthogonal ---so that the gravitational waves have
a single polarization--- the model becomes exactly soluble
classically. Therefore, over the years, the Gowdy
midi-superspace ---and especially the 1-polarization case---
has received considerable attention in the mathematical
literature on cosmology. Long before the advent of LQG, there
were several attempts to quantize this model (see, e.g.
\cite{gowdy-old}). However, in the resulting quantum theories,
the initial, big-bang-type singularity could not be naturally
resolved. By contrast, thanks to quantum geometry underlying
LQC, the singularity has now been resolved both in the vacuum
case and in the case when there is a massless scalar field
\cite{hybrid1,hybrid2,hybrid3,hybrid4,hybrid5}. In what
follows, we will focus on the vacuum case which has drawn most
attention so far.

Let  $(\theta, \sigma, \delta)$ be the coordinates on $\mathbb{T}^3$
with $\partial_\sigma$ and $\partial_\delta$ as the two hypersurface
orthogonal Killing fields. Then the phase space variables depend
only on the angle $\theta$ and one can carry out a Fourier
decomposition with respect to it. This provides a decomposition of
the phase space, $\Gamma = \Gamma_{\rm hom} \times \Gamma_{\rm
inhom}$, where the homogeneous sector is spanned by the zero
(Fourier) modes and the (purely) inhomogeneous sector, by the
non-zero modes. Using underlying symmetries, the inhomogeneous
sector can be fully gauge fixed and one is left with only two
\emph{global} constraints, the diffeomorphism constraint
$(C_\theta)$ and the Hamiltonian constraint $(C_H)$, which Poisson
commute.

The homogeneous sector $\Gamma_{\rm hom}$ can be regarded as
phase space of the vacuum Bianchi I model of section
\ref{s4.4}. However, since the spatial manifold is now
$\mathbb{T}^3$, we do not need to introduce a cell and can set
$L_i = 2\pi$. Quantization of this sector can be performed as
in section \ref{s4.4}.

The inhomogeneous sector $\Gamma_{\rm inhom}$ can be coordinatized
by creation and annihilation variables, $a_m$ and $a_m^\star$
respectively. The global diffeomorphism constraint contains only the
inhomogeneous modes and can be written as
\be \label{gdc} {C}_\theta = \sum_{n = 1}^\infty \, n (a_n^\star
a_n - a^\star_{-n} a_{-n})\, \approx 0 \ee
The Hamiltonian constraint on the other hand involves all
modes:
\be \label{ghc1} C_H^{\rm Gowdy} = C_H^{(I)} + 32 \pi^2 \gamma^2
|p_\theta|\, H_o + \f{(c_\sigma p_\sigma + c_\delta
p_\delta)^2}{|p_\theta|} H_{\rm int} \, \approx 0\ee
where $C_H^{(I)}$ is the Bianchi I Hamiltonian constraint
(\ref{clH}), $H_o$ is the `free part' of the Hamiltonian of
inhomogeneous modes (which represent gravitational waves) and
$H_{\rm int}$ is called the `interaction part' because it mixes the
inhomogeneous modes among themselves. The explicit expressions are
\ba \label{ghc2} C_H^{\rm hom} &=& - \f{1}{8\pi G\gamma^2}
\big(c_\theta p_\theta \, c_\sigma p_\sigma + c_\theta p_\theta
\, c_\delta
p_\delta + c_\sigma p_\sigma \, c_\sigma p_\sigma \big) \nonumber\\
H_o &=& \f{1}{8\pi G \gamma^2}\,\sum_{n \neq 0} \, |n| \, a_n^\star
a_n \quad {\rm and} \quad H_{\mathrm{int}} = \f{1}{8\pi G
\gamma^2}\, \sum_{n \neq 0} \f{1}{2 |n|} \, (2 a_n^\star a_n + a_n
a_{-n} + a_n^\star a^\star_{-n})\nonumber\\ \ea
Note that $H_o$ and $H_{\rm int}$ depend only on the
homogeneous modes and the coupling between the homogeneous and
inhomogeneous modes is made explicit in (\ref{ghc1}).

Quantization of the homogeneous sector follows the procedure
outlined in section \ref{s4.4}. On the inhomogeneous sector one
simply uses the Fock quantization, replacing $a_n$ and $a^\star_n$
by annihilation and creation operators. The kinematical Hilbert
space thus consists of a tensor product $\H_{\rm kin}^{\rm Gowdy} =
\H_{\rm kin}^{(I)} \otimes \mathcal{F}$ of the kinematic Hilbert
space of the Bianchi I model and the Fock space generated by the
inhomogeneous modes. The full quantum constraint operator is given
by \cite{hybrid3}
\be \label{ghc3} \Theta_{\rm Gowdy} = \Theta_{(I)}\, +\, \f{1}{8 \pi
\gamma^2} \Big[32 \pi^2 \gamma^2 \widehat{|p_\theta|} \hat H_o +
\left(\widehat{\f{1}{|p_\theta|^{1/4}}}\right)^2 (\hat \Theta_\sigma
+ \hat \Theta_\delta)^2
\left(\widehat{\f{1}{|p_\theta|^{1/4}}}\right)^2 \, \hat
H_{\mathrm{int}} \Big]\, \ee
where the inverse powers of $|\h{p}_\theta|$ are defined using the
LQC analog of the Thiemann trick \cite{tt,ttbook}. The Bianchi-I
operator $\Theta_{(I)}$ in these works uses a slightly different
---and a more convenient--- factor ordering from that in \cite{awe2}
which we used in section \ref{s4.4}.

Physical states lie in the kernel of the operators $\h{C}_\theta$
and $\Theta_{\rm Gowdy}$. The physical scalar product is chosen by
demanding that a complete set of observables be self-adjoint. The
volume $v = \sqrt{p_\sigma p_\delta p_\theta}$ serves as a local
internal time as in the classical theory with respect to which the
constraint yields a well-posed `initial value problem.' (Had we
introduced a massless scalar field, it would serve as a global
internal time.) One can easily check that, as in the Bianchi models,
states localized on configurations with zero volume decouple form
the rest. In particular, states which initially vanish on zero
volume configurations continue to do so for all time. Thus, the
singularity is again resolved. This resolution is sometimes called
`kinematical' to emphasize that it relies on the properties of the
constraint operator $\Theta_{\rm Gowdy}$ on the kinematical Hilbert
space rather than on a systematic analysis of solutions to the
quantum constraint in the Planck regime. This nomenclature can be
misleading because it may suggest that one only needs the kinematic
set up rather than properties of $\Theta_{\rm Gowdy}$ that generates
dynamics.

Furthermore, details of dynamics have already been explored using
effective equations \cite{hybrid2,hybrid5} and they show that the
behavior observed in the Bianchi I model ---including the bounces--
carries over to the Gowdy models. The analysis also provides
valuable information on the changes in the amplitudes of
gravitational waves in the distant past and distant future,
resulting from the bounce. A limitation of this analysis is that the
effective equations it uses arise from the quantum constraint given
in the older papers \cite{chiou_bianchi,kv} which, as we discussed
in section \ref{s4.4}, has some unsatisfactory features. However, an
examination of the details of the analysis suggests that the
qualitative behavior of the solutions of effective equations derived
from the corrected quantum constraint of \cite{awe2,hybrid3} will be
the same \cite{tp_private}.

In summary, Gowdy models are midi-superspaces which provide an
interesting avenue to incorporate certain simple types of
inhomogeneities in LQC. Although one often refers to the Bianchi I
geometry created by the homogeneous modes as a `background', it is
important to note that this analysis provides an exact ---not
perturbative--- quantization of the one polarization Gowdy model.
However, since the LQG effects created by inhomogeneities are not
incorporated, it is expected to be only an approximation to the full
LQG treatment of the model, but one that is likely to have a large
and useful domain of validity.

\subsection{Quantum field theory in cosmological, quantum space-times}
\label{s6.2}

Singularity resolution in Gowdy models is conceptually important
because, thanks to their inhomogeneity, they carry an infinite
number of degrees of freedom. However, these models do not capture
physically realistic situations because their inhomogeneities are
restricted to a single spatial direction. To confront quantum
cosmology with observations, we must allow inhomogeneities in all
three dimensions. However, an exact treatment of these
inhomogeneities is not essential: to account for the temperature
fluctuations in the cosmic microwave background (CMB), it has
sufficed to consider just the first order perturbations, ignoring
their back reaction on geometry. Indeed, much of the highly
successful analysis in the inflationary paradigm has been carried
out in the framework of quantum theory of \emph{test} fields on a
FLRW background. In LQC, on the other hand, to begin with we have a
\emph{quantum geometry} rather than a smooth curved background. It
is therefore of considerable interest to ask if quantum field theory
on cosmological \emph{quantum} geometries naturally emerges in a
suitable approximation from the Hamiltonian theory underlying LQG
and, if so, what its relation is to quantum field theory on FLRW
space-times, normally used in cosmology.

A priori, it is far from clear that there can be a systematic
relation between the two theories. In quantum field theory on FLRW
backgrounds one typically works with conformal or proper time, makes
a heavy use the causal structure made available by the fixed
background space-time, and discusses dynamics as an unitary
evolution in the chosen time variable. If we are given just a
quantum geometry sharply peaked at a FLRW background, none of these
structures are available. Even in the deparameterized picture, it is
a \emph{scalar field} that plays the role of internal time; proper
and conformal times are at best operators. Even when the quantum
state is sharply peaked on an effective LQC solution, we have only a
probability distribution for various space-time geometries; we do
not have a single, well-defined, classical causal structure.
Finally, in the Hamiltonian framework underlying full LQG, dynamics
is teased out of the constraint. It turns out that, in spite of
these apparently formidable obstacles, the standard quantum field
theory on curved space-times, as practised by cosmologists, does
arise from a quantum field theory on cosmological quantum geometries
\cite{akl}.

In this section we will summarize this reduction. As explained in
section \ref{s6.3.1}, the underlying framework is expected to serve
as a point of departure for a systematic treatment of cosmological
perturbations starting from the big bounce where quantum geometry
effects play a prominent role.

\subsubsection{Setting the stage}
\label{s6.2.1}

While discussing cosmological perturbations, one often uses Fourier
decompositions. Strictly speaking, these operations are well defined
only if the fields have certain fall-offs which are physically
difficult to justify in the homogeneous cosmological settings.
Therefore, to avoid unnecessary detours, in this section we will
work with a spatially compact case; the k=0, $\Lambda=0$ FLRW models
on a 3-torus $\mathbb{T}^3$. To make a smooth transition from
quantum geometry discussed in sections \ref{s2} and \ref{s3}, we
will assume that the FLRW space-time has a (homogeneous) massless
scalar field $\phi$ as the matter source. In addition, there will be
an inhomogeneous, free, \emph{test} quantum scalar field $\varphi$
with mass $m$. It is not difficult to extend the framework to
include spatial curvature or a cosmological constant in the
background and/or more general test fields.

Fix periodic coordinates $x^i$ on $\mathbb{T}^3$, with $x^i \in
(0,\ell)$. The FLRW metric is given by:
\be \label{flrw} g_{ab}\,\dd x^a \dd x^b\, =  - N_{x_0}^2 \dd x_0^2
+ a^2 (\dd x_1^2 +\dd x_2^2 + \dd x_3^2) \ee
where the lapse $N_{x_0}$ depends on the choice of the time
coordinate $x_0$;\,\, $N_{t} = 1$ if $x_o$ is the proper time $t$;
$N_{\tau}= a^3$ if $x_0$ is the harmonic time $\tau$. Since the
solution to the equation of motion for $\phi$ is $\phi =
(\pphi/\ell^3)\, \tau$, if we use $\phi$ as time the lapse becomes
$N_{\phi} = (\ell^3/\pphi) a^3$. Since $\pphi$ is a constant in any
solution, $N_\tau$ and $N_\phi$ are just constant multiples of each
other in any one space-time. However, on the phase space, or in
quantum theory, we have to keep track of the fact that $\pphi$ is a
dynamical variable.

In the cosmological literature, quantum fields on a \emph{given}
FLRW background are generally discussed in terms of their Fourier
modes:
\be \varphi({\vec x}, x_0) = \f{1}{(2\pi)^{3/2}} \sum_{{\vec
k}\in \L} \, \varphi_{\vec k} (x_0)\, e^{i {\vec k} \cdot {\vec
x}}\quad {\rm and} \quad \pi({\vec x}, x_0) =
\f{1}{(2\pi)^{3/2}} \sum_{{\vec k}\in \L} \, \pi_{\vec
k}(x_0)\, e^{i {\vec k}\cdot {\vec x}}\, ,\ee
where, in our case, $\L$ is the 3-dimensional lattice spanned by
$(k_1,k_2,k_3)\in ((2\pi/\ell)\,\, {\Z})^3$,  ${\Z}$ being the set
of integers. One often introduces real variables $q_{\vec k},
p_{\vec k}$ via
\be \varphi_{\vec k} =\f{1}{\sqrt 2}( q_{\vec k} + i q_{-\vec{k}}),
\quad{\rm and}\quad
     \pi_{\vec k} = \f{1}{\sqrt 2} (p_{\vec k} + i p_{-\vec{k}}). \ee
In terms of the canonically conjugate pairs  $(q_{\pm{\vec k}},\,
p_{\pm{\vec k}})$, the Hamiltonian becomes
\be H_\varphi(x_0) = \f{N_{x_0}(x_0)}{2a^3(x_0)}\, \sum_{{\vec k}
\in \L }\, {p}^2_{\vec k} +( {\vec k}^2 a^4(x_0)  + m^2 a^6(x_0))\,
{q}^2_{\vec k}\, . \ee
(Some care is needed to ensure one does not over-count the modes;
see \cite{akl}.) Thus, the Hamiltonian for the test field is the
same as that for an assembly of harmonic oscillators (with time
dependent masses), one for each ${\vec k} \in \L$.

In cosmology, functional analytic issues such as self-adjointness of
$\h{H}_{x_0}$ are generally ignored because
attention is focused on a single or a finite number of modes.%
\footnote{See however some recent developments \cite{unique} that
address the issue of existence and uniqueness of representations of
the canonical commutation relations in which dynamics is unitarily
implemented in spatially compact cosmological models.}
The time coordinate $x_0$ is generally taken to be the conformal
time or the proper time $t$. Thus, in a Schr\"odinger picture,
states are represented by wave functions $\Psi(q_{\vec k}, x_0)$ and
evolve via
\be \label{sch1} i\hbar \partial_{x_0}\psi(q_{\vec k}, x_0)\, =\,
\h{H}_{x_0} \psi(q_{\vec k}, x_0)\, \equiv\,
\frac{N_{x_0}(x_0)}{2a^3(x_0)}\, \Big[p_{\vec k}^2 + ({\vec k}^2
a^{4}(x_0)+ m^2a^6(x_0))q_{\vec k}^2 \Big]\, \psi(q_{\vec k}, x_0).
\ee

\subsubsection{Quantum fields on FLRW quantum geometries}
\label{s6.2.2}

Let us begin with the phase space formulation of the problem. The
phase space consists of three sets of canonically conjugate pairs
$(\nu, b;\, \phi, \pphi;\, q_{\vec k}, p_{\vec k})$. Because we wish
to treat $\varphi$ as a test scalar field whose back reaction on the
homogeneous, isotropic background is to be ignored, only the zero
mode of the Hamiltonian constraint is now relevant. Thus, we have to
smear the Hamiltonian constraint by a constant lapse and we can
ignore the Gauss and the diffeomorphism constraint (see section
\ref{s6.3} for further discussion). As in section \ref{s3}, let us
work with harmonic time so that the lapse is given by $N_\tau =
a^3$. Then, the Hamiltonian constraint becomes:
\be \label{C3}  C_{\tau}\, = \, \f{\pphi^2}{2\ell^3} - \f{3}{8\pi
G}\,\f{\b^2V^2}{\ell^3} \, + H_{{\rm test}, \tau}\ee
where
\be \label{Htest} H_{{\rm test}, \tau} = \f{1}{2}\, \sum_{{\vec
k}}\, \,{p}_{\vec k}^2 + ({\vec k}^2 a^4 + m^2 a^6) q_{\vec k}^2 \ee

As usual, the physical sector of the quantum theory is obtained by
considering states $\Psi(\nu, q_{\vec k}, \phi)$ which are
annihilated by this constraint operator and using the group
averaging technique to endow them with the structure of a Hilbert
space. Thus, the physical Hilbert space $\Hp$ is spanned by
solutions to the quantum constraint
\be \label{hc6} -i\hbar\, \partial_\phi\, \Psi(\nu, q_{\vec k},
\phi) = [\h{H}_o^2 - 2\ell^3\, \h{H}_{{\rm test},
\tau}]^{\f{1}{2}}\, \Psi(\nu,q_{\vec k},\phi )\,=: \h{H} \Psi(\nu,
q_{\vec k}, \phi) .\ee
which have a finite norm with respect to the inner product:
\be\label{ip4} \langle \Psi_1 |\Psi_2\rangle =
\f{\lambda}{\pi}\, \sum_{\v = 4n\lambda}\,
\f{1}{|\v|}\,\int_{-\infty}^{\infty}\!\! \dd q_{\vec k}\,\,
\bar{\Psi}_1(\nu,q_{\vec k}, \phi_0)\, \Psi_2 (\nu,q_{\vec
k},\phi_0)\, , \ee
where the right side is evaluated at \emph{any} fixed instant of
internal time $\phi_0$. In (\ref{hc6}),  $\h{H}_o := \hbar
\sqrt{\Theta}$ governs the dynamics of the background quantum
geometry and $\h{H}_{{\rm test}, \tau}$ of the test field. The
physical observables of this theory are the Dirac observables of the
background geometry ---such as the time dependent density and volume
operators $\h{\rho}|_{\phi}$ and $\h{V}|_{\phi}$--- and observables
associated with the test field, such as the mode operators
$\h{q}_{\vec k}$ and $\h{p}_{\vec k}$. (For the background geometry
we use the same notation as in sections \ref{s2} and \ref{s3}.)

The theory under consideration can be regarded as a truncation of
LQG where one allows only \emph{test} scalar field on FLRW
geometries. Dynamics has been teased out of the Hamiltonian
constraint via deparametrization.  Although we started out with
harmonic time, $\tau$, as in sections \ref{s2} and \ref{s3}, in the
final picture states and observable evolve with respect to the
relational time variable $\phi$. Note that since $\phi$ serves as
the source of the homogeneous, isotropic background, conceptually
this emergent time is rather different from the proper or conformal
time traditionally used in cosmology. In the space-time picture, we
have a quantum metric operator on $\Hp$,
\be \hat{g}_{ab} \dd x^a \dd x^b = - \hat{N}^2_{\phi} \dd \phi^2 +
(\hat{V}|_\phi)^{\f{2}{3}}\, \dd {\vec x}^2 \quad {\rm with} \quad
\hat{N}_\phi = : \h{V}|_\phi\, \h{H}_o^{-1}:\,\, ,\ee
where the double-dots denote a suitable factor ordering (which must
be chosen because $\hat{V}|_\phi$ does not commute with
$\hat{H}_o$). In addition, there is a test quantum field
$\hat\varphi$ that evolves with respect to the internal time
coordinate $\phi$ on the quantum geometry of $\h{g}_{ab}$.

\subsubsection{Reduction}
\label{s6.2.3}

Let us begin by noting the salient differences between quantum field
theory on a classical FLRW background and quantum field theory on
quantum FLRW background. In the first case, the time parameter $x_0$
knows nothing about the matter source that produces the classical,
background FLRW space-time; quantum states depend only on the test
field $\varphi$ (and $x_0$); and in the expression of the
$\varphi$-Hamiltonian the background geometry appears through the
externally specified, $x_0$-dependent parameter $a$. In the second
case, the source $\phi$ of the FLRW background serves as the
relational time parameter; states depend not only on the test field
$\varphi$ but also on the geometry $\nu$ (and the internal time
$\phi$); and in the expression of the $\varphi$-Hamiltonian, the
background geometry appears through \emph{operators} $\h{a}^4$ and
$\hat{a}^6$ which do \emph{not} have any time dependence. Thus,
although dynamics has been expressed in the form of Schr\"odinger
equations in both cases, there are still deep conceptual and
mathematical differences between the two theories. Yet, the second
theory has been shown \cite{akl} to reduce to the first one through
a series of approximations. We will conclude by briefly summarizing
this three step procedure.

\emph{Step 1:} One first uses the fact that $\varphi$ is to be
treated as a test field. Therefore, in the expression (\ref{hc6}) of
the full Hamiltonian $\hat{H} = \h{H}_o + \h{H}_{{\rm test}, \tau}$,
one can regard $\h{H}_o$ as the `main part' and $\h{H}_{\rm
test,\tau}$ as a `perturbation'. After a systematic regularization, the
square-root in (\ref{hc6}) can be approximated as
\be [\h{H}_o^2 - 2\ell^3\, \h{H}_{{\rm test},\tau}]^{\f{1}{2}}
\approx \h{H}_o - (\ell^{-3} \h{H}_o)^{-(1/2)}\, \h{H}_{{\rm test},
\tau}\, (\ell^{-3} \h{H}_o)^{-(1/2)}\ee
Next, since the lapse functions associated with the choices $\tau$
and $\phi$ of time are related by $N_\phi = (\pphi\, \ell^{-3})^{-1}
N_\tau \approx (\ell^{-3} H_o)^{-1} N_\tau$ in the test field
approximation, it follows that $(\ell^{-3} \h{H}_o)^{-(1/2)}\,
\h{H}_{{\rm test}, \tau}\, (\ell^{-3} \h{H}_o)^{-(1/2)}$ is
precisely the matter Hamiltonian $\h{H}_{{\rm test}, \phi}$
associated with time $\phi$. Thus, in the test field approximation,
the LQG evolution equation (\ref{hc6}) is equivalent to
\be \label{hc7} -i\hbar \partial_\phi \Psi(\nu, q_{\vec k}, \phi) =
(\h{H}_o - \h{H}_{{\rm test}, \phi} )\, \Psi(\nu, q_{\vec k}, \phi)
\ee

\emph{Step 2:} The geometry operators $\hat{a}^4$ and $\hat{a}^6$ in
the term $\h{H}_{{\rm test}, \phi}$ do not carry any time
dependence. (This feature descends directly form the classical
Hamiltonian (\ref{Htest}).) The parameters $a^4(x_0)$ and $a^6(x_0)$
in the expression (\ref{sch1}) of the Hamiltonian in quantum field
theory in classical FLRW space-times on the other hand are time
dependent. To bring the two theories closer we can work in the
interaction picture and set
\be \Psi_{\rm int} (\nu, q_{\rm k}, \phi) := e^{-(i/\hbar)\, H_o\,
(\phi-\phi_o)}\,\, \Psi (\nu, q_{\vec k}, \phi)\ee
so that the evolution of $\Psi_{\rm int}$ is governed by $H_{{\rm
test},\phi}$ and of the quantum geometry operators by $\h{H}_o$:
\be i\hbar\p_\phi \,\Psi_{\rm int} (\nu, q_{\vec k},\phi ) =
\h{H}^{\rm int}_{\phi, \vec{k}} \,\,\Psi_{\rm int} (\nu, q_{\vec
k},\phi ) \quad {\rm and} \quad \hat{a}(\phi) = e^{-(i/\hbar)
\h{H}_o(\phi-\phi_o)}\, \h{a}\, e^{(i/\hbar) \h{H}_o(\phi-\phi_o)}\,
. \ee

\emph{Step 3:} The test field approximation also implies that the
total wave function can be factorized: $\Psi_{\rm int}(\nu,q_{\vec
k}, \phi) = \Psi_o(\nu, \phi_o) \otimes \psi(q_{\vec k}, \phi)$.
Next, to make contact with quantum field theory in \emph{classical}
FLRW space-times, one is led to take expectation values of the
evolution equation in a semi-classical quantum geometry state
$\Psi_o(\nu, \tau)$. We know from sections \ref{s2} and \ref{s3}
that these states are sharply peaked on a solution to the effective
equation. Therefore, \emph{if we ignore fluctuations of quantum
geometry}, we can replace $\langle \h{a}^n|_\phi \rangle$ with just
$\bar{a}^n(\phi)$ where $\bar{a}|_\phi$ is the expectation value of
$\hat{a}|_\phi$. Then the evolution equation (\ref{hc6}) reduces to:
\be \label{sch3} i\hbar\partial_\phi\, \psi(q_{\vec k}, \phi) =
\f{\bar{N}_\phi}{2\bar{a}^3(\phi)}\,\, \Big[\, \h{p}_{\vec k}^2 +
({\vec k}^2 \bar{a}^{4}(\phi)+ m^2 \bar{a}^6(\phi)) \h{q}_{\vec
k}^2\, \Big]\, \psi(q_{\vec k}, \phi)\, . \ee
This is exactly the Schr\"odinger equation (\ref{sch1}) governing
the dynamics of the test quantum field on a classical space-time
with scale factor $\bar{a}$ containing a massless scalar field
$\phi$ with momentum $\bar{p}_{(\phi)} = \bar{a}^2\ell^3/
\bar{N}_\phi$. This is the precise sense in which the dynamics of a
test quantum field on a classical background emerges from a more
complete QFT on quantum FLRW backgrounds. Note however that, even
after our simplifications, the classical background is \emph{not} a
FLRW solution of the Einstein-Klein-Gordon equation. Rather, it is
the effective space-time $(M, \bar{g}_{ab})$ a la LQC on which the
quantum geometry $\Psi_o(\nu,\phi)$ is sharply peaked. But as
discussed in sections \ref{s2} and \ref{s3}, away from the Planck
regime, $(M, \bar{g}_{ab})$ is extremely well-approximated by a
classical FLRW space-time.

To summarize, quantum field theory on the (LQC-effective) FLRW
space-time emerges from LQG if one makes two main approximations: i)
$\h\varphi$ can be treated as a \emph{test} quantum field (Steps 1
and 2 above); and ii) the fluctuations of quantum geometry can be
ignored (step 3). This systematic procedure also informs us on how
to incorporate quantum geometry corrections to quantum field theory
on classical FLRW backgrounds.

\subsection{Inflationary perturbation theory in LQC}
\label{s6.3}

In this sub-section we summarize a new framework for the
cosmological perturbation theory that is geared to systematically
take into account the deep Planck regime at and following the bounce
\cite{aan}. This approach provides examples of effects that may have
observational implications in the coming years because their seeds
originate in an epoch where the curvature and matter densities are
of Planck scale.

\subsubsection{Inflation and quantum gravity}
\label{s6.3.1}

The general inflationary scenario involves a rather small set of
assumptions: i) Sometime in its early history, the universe
underwent a phase of rapid expansion during which the Hubble
parameter was nearly constant; ii) During this phase, the universe
was well described by a FLRW solution to Einstein's equations
together with small inhomogeneities which are well approximated by
first order perturbations; iii) Consider the co-moving Fourier mode
$k_o$ of perturbations which has just re-entered the Hubble radius
now. A few e-foldings before the time $t(k_o)$ at which $k_o$ exited
the Hubble radius during inflation, Fourier modes of quantum fields
describing perturbations were in the Bunch-Davis vacuum for
co-moving wave numbers in the range $\sim\, (10k_o,\, 2000k_o)$;
and, iv) Soon after a mode exited the Hubble radius, its quantum
fluctuation can be regarded as a classical perturbation and evolved
via linearized Einstein's equations. Analysis of these perturbations
implies that there must be tiny inhomogeneities at the last
scattering surface whose detailed features have now been seen in the
CMB. Furthermore, time evolution of these tiny inhomogeneities
produces large scale structures which are in excellent qualitative
agreement with observations.  These successes have propelled
inflation to the leading place among theories of the early universe
even though the basic assumptions have some ad-hoc elements.

But as we discussed in section \ref{s6.3.1}, the scenario is
incomplete because it also has an in-built big-bang singularity
\cite{bgv}. Therefore it is natural to ask: What is the
situation in LQC which is free of the initial singularity? Does
the inflationary paradigm persist or is there some inherent
tension with the big bounce and the subsequent LQC dynamics
that in particular exhibits a superinflation phase? If it does
persist, one would have a conceptual closure for the
inflationary paradigm. This by itself would be an important
advantage of using LQC.

But there could be observational pay-offs as well. A standard
viewpoint is that because of the huge expansion --- the Bohr radius
of a hydrogen atom is expanded out to ~95 light years in
65-e-foldings of inflation--- characteristics of the universe that
depend on the pre-inflationary history will be simply washed away
leaving no trace on observations in the foreseeable future. However,
this view is not accurate when one includes quantum effects.
Suppose, as an example, the state of quantum fields representing
cosmological perturbations is not the Bunch-Davis vacuum at the
onset of inflation but contains a small density of particles. Then,
because of the \emph{stimulated emission} that would accompany
inflation, this density does \emph{not} get diluted. Moreover, this
departure from the Bunch-Davis vacuum has potential observable
consequences in non-Gaussianities \cite{agullo-parker}. So, it is
natural to ask: Do natural initial conditions, say at the bounce,
lead to interesting departures from the Bunch-Davis vacuum at the
onset if inflation?

\subsubsection{Strategy}
\label{s6.3.2}

To address such issues, one needs a perturbation theory that is
valid all the way to the big bounce. Now, in the textbook treatment
of cosmological perturbations, one begins by linearizing Einstein's
equations (with suitable matter fields) and then quantizes the
linear perturbations. The result is a quantum field theory on a FLRW
background. As we saw in section \ref{s5}, effective equations capture
the key LQC corrections to general relativity. Therefore, to
incorporate cosmological perturbations one might think of carrying
out the same procedure, substituting the FLRW background by a
corresponding solution of the effective equations. But this strategy
has two drawbacks. First, we do not have reliable effective
equations for full LQG which one can linearize. Second, even if we
had the full equations, it would be \emph{conceptually incorrect} to
first linearize and then quantize them because the effective
equations already contain the key quantum corrections of the full
theory. If one did have full LQG, the task would rather be that of
simply truncating this full theory to the appropriate sector that,
in the classical limit, reduces to FLRW solutions with linear
perturbations.%
\footnote{An `in between' strategy would be to consider quantum
fields satisfying the standard linearized Einstein's equations but
on a background space-time provided by the effective theory.
Unfortunately, this procedure is ambiguous: Because the background
does not satisfy Einstein's equations, sets of linearized equations
which are equivalent on a FLRW background now become inequivalent.
With a judicious choice of a consistent set, this procedure may be
viable. A significant fraction of the literature on cosmological
perturbations in LQC is based on this hope (see e.g. section
\ref{s6.4}).}

In absence of full LQG, an alternative strategy is the one that has
driven LQC so far: \emph{Construct the Hamiltonian framework of the
sector of general relativity of interest and then pass to the
quantum theory using LQG techniques.} This strategy has been
critical, in particular, in the incorporation of anisotropies. Had
we tried to incorporate them perturbatively on the \emph{effective}
isotropic geometry provided by LQC, singularities would not have
been resolved. Instead, we considered the Hamiltonian theory of the
full anisotropic sector ---say the Bianchi I model--- and then
carried out its loop quantization. As we saw in section \ref{s4.4},
this procedure involves new elements beyond those that were used in
the isotropic case and they were critical to obtaining the final,
singularity-free theory. The idea behind the new framework is to use
this philosophy for cosmological perturbations.

The first task then is to identify the appropriate truncation of the
classical phase space. Let us begin by decomposing the full phase
space of general relativity into a homogeneous and a \emph{purely}
inhomogeneous part. For this, it is simplest to assume, as in
section \ref{s6.2}, that the spatial manifold $M$ is topologically
$\mathbb{T}^3$. (The $\R^3$ topology would require us to first
introduce an elementary cell, construct the theory, and then take
the limit as the cell occupies full space.) As in section \ref{s6.2}
let us fix spatial coordinates $x^i$ on $\mathbb{T}^3$ and use them
to introduce a fiducial triad $\e^a_i$ and co-triad $\o_a^i$. Then,
every point $(A_a^i, E^a_i)$ in the gravitational phase space
$\Gamma_{\rm grav}$ can be decomposed as:
\be A_a^i = c \ell^{-1}\, \o_a^i\,+\, \a_a^i, \quad {\rm and} \quad
E^a_i = p \ell^{-2}\, \sqrt{\q}\, \e^a_i\, +\, \ep^a_i \ee
where
\be c = \f{1}{3\ell^2}\, \sint_{M}\,\, A_a^i \e^a_i\, \dd^3 \vec{x}
\quad {\rm and} \quad p= \f{1}{3\ell}\, \sint_{M}\,\, E^a_i \o^a_i\,
\dd^3 \vec{x}\, .\ee
$\a_a^i$ and $\ep^a_i$ are \emph{purely inhomogeneous} in the sense
that the integrals of their contractions with $\e^a_i$ and $\o_a^i$
vanish. Matter fields can be decomposed in the same manner. This
provides us a natural decomposition of the phase space into
homogeneous and inhomogeneous parts: $\Gamma = \Gamma_{\rm H} \times
\Gamma_{\rm IH}$.

We have to single out the appropriate sector of this theory that
captures just those degrees of freedom that are relevant to the
cosmological perturbation theory. For this one allows only a scalar
field $\Phi$ as the matter source, and expands it as $\Phi= \phi
+\varphi$ where $\phi$ is homogeneous and $\varphi$ purely
inhomogeneous, and introduces a potential $V(\Phi)$. For reasons
given in section \ref{s5.3}, we will set $V(\Phi) = (1/2)m^2\Phi^2$
(with $m \approx 1.21 \times 10^{-6} \mpl$). We wish to treat the
homogeneous fields as providing the background and inhomogeneous
fields, linear perturbations. Thus the situation is rather similar
to that in our discussion of quantum fields in quantum space-times
in section \ref{s6.2}. However, there are also some key differences.
First, we now wish to allow inhomogeneities \emph{also in the
gravitational field}, whence the inhomogeneous sector now contains
$\a_a^i, \ep^a_i$ in addition to $\varphi$. Furthermore, these
fields are now coupled. In particular we now have to incorporate the
Gauss, the vector and the scalar constraints to first order:
\be \sint_M \big( N^i C^{(1)}_i + N^aC^{(1)}_a + N C^{(1)}\big)
\dd^3 \vec{x}\, = 0 \quad \forall\,\, N^i,\,N^a,\,N \ee
where $N^i, N^a, N$ are, respectively, generators of the internal
SU(2) rotations, shift and lapse fields on $M$. While they are
arbitrary, because the constraints are linear in the purely
inhomogeneous fields, only the purely inhomogeneous parts of $N^i,
N^a, N$ matter. It is simplest to solve these linearized constraints
on $\a_a^i, \ep^a_i, \varphi$ and factor out by the gauge orbits
generated by these constraints to pass to a reduced phase space as
in \cite{langlois} (see also, \cite{ghtw1,ghtw2,pert_scalar,dt}).
Since we have 10 configuration variables in $(\a_a^i, \varphi)$ and
seven first class constraints, we are left with three true degrees
of freedom: two tensor modes and a scalar mode. These can be
represented as three scalar fields on $M$. As in section \ref{s6.3}
it is simplest to pass to the Fourier space and work with the two
tensor modes $q_{\vec k}^I$ (with $I=1,2$) and the scalar mode
$q_{\vec{k}}$. We will often group them together as $\vec{q}_{\vec
k}$ and denote the three conjugate momenta as  $\vec{p}_{\vec k}$.
Thus, truncated phase space of interest is given by
\be \label{tps} \Gamma_{\rm Trun} \, =\, \Gamma_{\rm H}\,\times\,
\Gamma_{\rm IH}^{\rm Red} \ee
where the truncation manifests itself in the fact that the reduction
of the inhomogeneous part of the phase space has been carried out
using the first order truncated constraints. Dynamics on
$\Gamma_{\rm Trunc}$ is governed by the Hamiltonian constraint
truncated to the second order:
\be \label{th1} C_{N_{\rm H}} = \sint_M N_{\rm H}\,\, \big(C^{(0)} +
C^{(2)}\big)\, \dd^3\vec{x} \ee
where the subscript $H$ on the lapse field emphasizes that it is a
homogeneous (i.e. constant) function on $M$. Note that the integrand
does not contain $C^{(1)}$ because its integral against a
homogeneous lapse vanishes identically. To pass to quantum theory
using LQC techniques, we will use harmonic time (so $N_{\rm H} =
a^3$). Then, as before (see section \ref{s5.3})
\be \label{th2} {\sint} N_{\rm H}\,\, C^{(0)}\, \dd^3\vec{x} =
\f{\pphi^2}{2\ell^3} - \f{3}{8\pi G\ell^3}\, \b^2 V^2 +
\f{m^2}{2\ell^3}\, \phi^2 V^2\ee
while
\ba \label{th3} \sint_{M} N_{\rm H}\,\, C^{(2)} \dd^3\vec{x} =
H_\tau^{(T)} + H_\tau^{(S)} &:=& \f{1}{2}\, \sum_{I=1}^2
\sum_{\vec{k}\in \L}\, \big[(p^I_{\vec k})^2 + a^4 {\vec{k}}^2
(q^I_{\vec k})^2\big]\,\nonumber\\
&+& \, \f{1}{2}\, \sum_{\vec{k}\in\L}\, \big[ p^2_{\vec k} + a^4
{\vec{k}}^2 q_{\vec k}^2 + f(a, \b,\pphi; m) q_{\vec k}^2 \big] \ea
has a form similar to the Hamiltonian of the matter field in section
\ref{s6.2}. The subscript $\tau$ is a reminder that the lapse is
tailored to the harmonic time $\tau$, the superscripts $(T), (S)$
refer to tensor and scalar modes, and the scale factor $a$
determines the physical volume $V$ via $V= a^3\ell^3$. An important
difference is that the `mass term' in the scalar mode depends on the
background fields and is therefore time dependent.

Thus the truncated phase space adapted to cosmological perturbations
is given by (\ref{tps}). It has only one Hamiltonian constraint
(\ref{th1}) that generates dynamics. This is the Hamiltonian theory
one has to quantize using LQG techniques.

\subsubsection{Quantum Perturbations on Quantum Space-times}
\label{s6.3.3}

Recall that for the WMAP data, modes that are directly relevant lie
in the range $\sim \, (10k_o,\, 2000k_o)$.  Thus, from observational
perspective, we have natural ultraviolet and infrared cutoffs,
whence field theoretic issues are avoided and one can hope to
proceed as in section \ref{s6.2}. There is however one subtlety.
Since the homogeneous part (\ref{th2}) of the constraint now
contains a `time dependent' term $m^2\phi^2$ we face issues
discussed in section \ref{s3.3}. As explained there, at an abstract
mathematical level, these can be handled via group averaging.
However, to obtain detailed predictions via numerical simulations,
with the current state of the art, one has to restrict oneself to
quantum states of the background in which the bounce is dominated by
kinetic energy. Now, given the $m^2\phi^2$ potential, WMAP
observations lead to a very narrow window of initial conditions at
the onset of inflation \cite{wmap,as3}. Fortunately we know
\cite{as3,aps4} that there is a wide class states with kinetic
energy domination at the bounce that meet this severe constraint.
Therefore, while it is conceptually important to incorporate more
general states in this analysis, even with kinetic domination one
can obtain results that are directly relevant for observations.

The idea then is to use the quantization of the truncated theory to
analyze dynamics of perturbations starting from the bounce. In this
theory \emph{both the background and perturbations are treated
quantum mechanically} following section \ref{s6.2}. Since this
`pure' quantum regime is rather abstract, it is instructive to use
effective equations to develop some intuition. (For details, see
\cite{as3}). They show that immediately after the bounce the
background undergoes a superinflation phase, which is followed by a
longer phase during which kinetic energy steadily decreases but is
still larger than the potential energy. The total time for which
kinetic energy dominates is of the order of $10^4$ Planck units. At
the end of this phase the energy density has decreased to about
$10^{-11}\rcr$. Therefore, it suffices to use the full quantum
description ---i.e., treat perturbations as quantum fields on a
quantum geometry--- only till the end of this phase. After that, one
can adequately describe perturbations using standard quantum field
theory on a FLRW background.

Fortunately, during the `pure quantum phase' the background inflaton
$\phi$ evolves monotonically. Therefore, it is appropriate to
continue to interpret the quantum version of the Hamiltonian
constraint (\ref{th1})
\be \label{tqh1} -\hbar^2\p^2_\phi \Psi(\nu, \vec{q}_{\vec k}, \phi)
= \left(\hbar^2\Theta_{(m)} - 2\ell^3 (\h{H}_\tau^{(T)} +
\h{H}_\tau^{(S)}) \right) \, \Psi(\nu, \vec{q}_{\vec k}, \phi) \ee
as providing evolution with respect to the `internal time' $\phi$.
Here, $\Theta_{(m)} = \Theta - (2\pi G\gamma m\phi\nu)^2$ where
$\Theta$ is the LQC difference operator (\ref{qhc4}) representing
the gravitational part of the constraint in the FLRW model, and, as
in section \ref{s6.2}, $\h{H}_\tau^{(T)}$ and $\h{H}_\tau^{(S)}$
depend not only of fields representing linear perturbations but also
on the background operators (see (\ref{th3})). Thus, the form of
quantum constraint is the same as that in section \ref{s6.2}. Our
basic assumptions are incorporated by restricting oneself to any
state $\Psi$ representing a semi-classical wave function that is
sharply peaked at a kinetic energy dominated effective trajectory
near the bounce \cite{aps4} and in which the energy in perturbations
is small compared to the background kinetic energy. Then, as in
section \ref{s6.2}, we can make a series of controlled
approximations to simplify the evolution equation: \\
i) Because of kinetic energy domination, the subsequent evolution is
well-approximated by a first order equation analogous to
(\ref{hc7}):
\be \label{thq2} -i\hbar\p_\phi \Psi(\nu, \vec{q}_{\vec k}, \phi) =
\left(\h{H}_o - 2\ell^3 (\h{H}_\phi^{(T)} + \h{H}_\phi^{(S)})
\right)\, \Psi(\nu, \vec{q}_{\vec k}, \phi) \ee
where the Hamiltonians on the right side evolve the wave function in
the scalar field time $\phi$.\\
ii) One then passes to the interaction picture in which operators
(such as $\h{a}^n \equiv \h{V}^{n/3}/\ell^n$)  referring to
background
geometry evolve with respect to the scalar field time. \\
iii) If the evolved state factorizes as $\Psi(\nu, \vec{q}_{\vec k},
\phi) = \Psi(\nu,\phi) \otimes \psi(\vec{q}_{\vec k}, \phi)$, one
can take the expectation value of the evolution equation w.r.t. the
state $\Psi(\nu,\phi)$ of the background geometry. If furthermore
the fluctuations of the background operators that appear in the
definition of $\h{H}_\phi^{(T)}$ and $\h{H}^{(S)}$ are negligible
compared to the expectation values, one obtains the familiar
evolution equation, analogous to (\ref{sch3}), for the evolution of
the quantum state $\psi(\vec{q}_{\vec k}, \phi)$ of perturbations
in the Schr\"odinger picture.\\

The final evolution equations have the same form as the standard
ones one finds in the textbook theory of quantum fields representing
cosmological perturbations. However, the background fields that
feature in these equations naturally incorporate both the holonomy
and inverse volume corrections of LQC. More importantly,
approximations involved in the passage from quantum fields
representing perturbations on quantum space-times to quantum fields
representing perturbations on curved (but LQC corrected) space-times
are spelled out. Therefore, if numerical simulations show that they
are violated, one can work at the `higher level' before the
violation occurs and still work out the consequences of this
evolution (although that task will involve more sophisticated
numerical work). This is a notable strength of the framework.

This candidate framework provides a systematic procedure to evolve
quantum states of the background and perturbations all the way from
the big bounce to the end of the kinetic dominated epoch. This end
point lies well within the domain of validity of general relativity:
as we already observed, the matter density is some  11 orders of
magnitude below the Planck scale. Therefore, the subsequent quantum
gravity corrections, although conceptually still interesting, will
be too small for observations in the foreseeable future. But, as we
emphasized in the beginning of this sub-section, since the quantum
evolution begins at the bounce, quantum corrections to the standard
scenario arising from the early stages could well have consequences
that are observationally significant, in spite of the subsequent
slow roll inflation.

An important example is the issue of what the `correct' quantum
state of perturbations is at the onset of inflation \cite{aan}.
Since the onset of the slow roll phase compatible with the WMAP
data occurs quite far from the Planck scale, from a conceptual
viewpoint, it seems artificial to simply postulate that the
state should be the Bunch-Davis vacuum there. It would be more
satisfactory to select the state at the `beginning' using
physical considerations, evolve it, and show that it agrees
with the Bunch-Davis vacuum at the onset of slow roll to a good
approximation. But in general relativity the beginning is the
big-bang singularity and one does not know how to set initial
conditions there. In bouncing scenarios ---such as the one in
LQC--- it is natural to specify the initial quantum state of
perturbations at the bounce. If the state can be specified in a
compelling fashion and if it does not evolve to a state that is
sufficiently close to the Bunch-Davis vacuum at the onset of
the WMAP slow roll, the viability of the bouncing scenario
would be seriously strained. If on the other hand the evolved
state turns out to be close, but not too close, to the
Bunch-Davis vacuum, there would be observable predictions,
e.g., on non-Gaussianities \cite{agullo-parker}. These
possibilities call for a detailed application of the framework
outlined in this section.

\subsection{LQC corrections to standard paradigms}
\label{s6.4}

There is a large body of work investigating implications of quantum
geometry in the standard cosmological scenarios based on various
generalizations of effective equations. Much of the recent work in
this area is geared to the inflationary and post-inflationary phases
of the evolution of the universe. These investigations are
conceptually important with potential to provide guidance for full
LQG. However, because there are significant variations in the
underlying assumptions and degrees of precision, and because the
subject is still evolving, it will take inordinate amount of space
to provide an exhaustive account of these developments. Also, as one
would expect, because curvature during and after inflation is very
small compared to the Planck scale, the LQG corrections discussed in
many of these works are too small to be measurable in the
foreseeable future. Therefore, in this subsection, we will provide
only a few illustrative examples to give a flavor of ongoing
research in this area.

The possibility for LQG effects to leave indirect observable
signatures was first considered in \cite{tsm} where it was argued
that the quantum geometry induced violations of slow roll conditions
could leave an imprint in the CMB at the largest scales. Since then,
a wide spectrum of calculations of LQG corrections to the standard
cosmological scenario have appeared in the literature and the
subject has evolved significantly. In the early years, emphasis was
on models \emph{inspired} by LQC bounces and the new phase of
superinflation. In more recent studies LQC is used more
systematically and underlying assumptions are also more
stream-lined.

A considerable effort has been made to investigate implications of
the quantum geometry effects of LQC on the standard inflationary
scenarios. As in the analysis of perturbations in the quasi
de-Sitter inflationary paradigm, one considers a Bunch-Davies vacuum
as the initial state, and studies the evolution of perturbations
using effective equations that incorporate quantum gravity
effects. As an example, the Fourier modes $\phi_k$ of tensor
perturbations satisfy a second order evolution equation which is
very similar to the one in the standard inflationary scenario,
except for a modification in the value of frequency resulting from
quantum geometry (see for eg. \cite{barrau3}). One then computes
correlation function for modes in the Bunch-Davies vacuum state,
which yields the power spectrum. Once these modes exit the Hubble
horizon, the quantum correlation function is assumed to become a
classical perturbation as in the conventional scenario, and is
evolved using classical linearized equations. LQC effects are thus
captured in the power spectrum of the perturbations, influencing
both the amplitude and the spectral tilt which measures the
departure from scale invariance. Such effects have been investigated
in various works (see eg.
\cite{barrau1,barrau2,barrau3,als,zl1,pert_tensor1,pert_tensor3,barrau4,
bct,pert_tensor4,wl1,gh_invsf,mn1,cc_invsf,ms_invsf,sh_invsf}).

These and other investigations along such lines have created
interesting frameworks to make observable predictions. However, a
shortcoming of these calculations is that they typically focus just
on one or two aspects of quantum geometry. Definitive predictions
will have to pool together all the relevant effects. Once this is
done, it will become clear which effects are quantitatively
significant and require further detailed analysis to arrive at
reliable predictions that can be tested against observations. In
addition, these explorations are of interest on the theoretical side
even as they stand because they provide concrete illustrations of
quantum effects one can expect in full LQG.

Broadly speaking, here are two kinds of LQG corrections that could
potentially affect the evolution of cosmological perturbations.
These are: (i) modifications originating from expressing field
strength of the gravitational connection in terms of holonomies, and
(ii) corrections due to inverse volume (or scale
factor) expressions in the constraint.\\

\noindent {\it Holonomy corrections and cosmological perturbations}:
As discussed in sections \ref{s2} -- \ref{s5}, in isotropic models
holonomy corrections lead to a $\rho^2/\rcr$ modification of the
Friedmann equation with a bounce of the background scale factor
occurring at $\rho = \rcr$, and a phase of super-inflation for
$\rcr/2 \leq \rho < \rcr$.  In the early attempts to include these
modifications \cite{copeland1}, only the fluctuations in scalar
field were considered on the unperturbed homogeneous background and,
strictly speaking, the analysis was restricted just to a tiny
neighborhood near the bounce. The power spectra of scalar and tensor
perturbations were then computed and the scalar perturbations were
shown to have a nearly scale invariant spectrum for a class of
positive potentials. Scalar perturbations have also been computed
under similar assumptions for multiple fluids \cite{multi_fluid}.

As a next step, perturbations of geometry and matter were
investigated about backgrounds provided by solutions to the
effective LQC equations. Often, it is implicitly assumed is that one
can still use the standard general relativistic perturbation
equations (but on the LQC modified background) because one works in
the era starting from the onset of slow roll inflation where the
matter density and curvature are several orders of magnitude below
the Planck scale. Not surprisingly then it was found that, for
scalar perturbations, the corrections to the standard scenario are
too small to be of interest to observational cosmology \cite{als}.
(Similar conclusions were reached in \cite{zl1}). There are also
more detailed and systematic calculations in which the holonomy
corrections are first incorporated in an effective Hamiltonian
constraint and the constraint is then perturbed to a linear order
\cite{pert_tensor1}. In this framework, tensor perturbations have
been studied in various works \cite{barrau1,pert_tensor3, barrau3,
pert_tensor4} in presence of an inflationary potential.%
\footnote{Recently, holonomy correction on scalar perturbations have
also been computed \cite{wl1}. However it is not clear whether
perturbation equations satisfy all the constraints under evolution.}
These results indicate that the scenario in which perturbations are
generated in the bounce followed with a standard phase of inflation
is consistent with observations \cite{barrau3,pert_tensor4}. They
also point to some new interesting features. The first is a $k^2$
suppression of power in the infrared which is a characteristic of
the bounce. Second, it has been suggested that certain statistical
properties of perturbations are sensitive to the presence of a
contracting phase prior to the bounce and to the bounce itself.
Calculations have been performed to estimate the size of the
imprints of these effects on phenomenological parameters that could
be constrained or measured by the next generation B-mode CMB
experiments \cite{barrau4}. However, it is probably fair to say
these studies are yet to capture a unique signature of LQC which can
not be mimicked by other models.
\\

\noindent {\it Inverse volume corrections and cosmological
perturbations}: Because these corrections are simplest to study in
the context of perturbations in LQC, there is a large body of
literature on the subject for both scalar and tensor perturbations
(see e.g.
\cite{gh_invsf,mn1,cmns,copeland1,sh_invsf,cc_invsf,ms_invsf}).
These investigations have been carried out under a variety of
scenarios: approximating the asymptotic region in deep Planck regime
by a particular variation of the scale factor \cite{mn1};
considering various values of $j$ labeling representations of SU(2)
used to compute the inverse volume corrections \cite{ms_invsf}; etc.
The more conservative and better motivated of these assumptions do
hold away from the Planck regime. But then the inverse volume
corrections are too small to be of interest observationally.
However, they can provide limits on the details of some LQG
scenarios.

In homogeneous models, inverse volume corrections are well defined
when the spatial topology is compact. In the non-compact case, they
depend on the choice of the cell ---the infrared regulator--- used
in constructing the quantum theory and seem to disappear in the
limit as the regulator is removed. Therefore, strictly speaking, in
the k=0 case these corrections are interesting only if the topology
is $\mathbb{T}^3$ (rather than $\R^3$). The topological restriction
is not always spelled out and indeed it would be of interest to
investigate in some detail if there are interesting regimes,
\emph{compatible with the WMAP data}, in which these corrections are
significant. Irrespective of the outcome, the corrections are
conceptually interesting in the $\mathbb{T}^3$ case as well as the
$\mathbb{S}^3$ topology in the k=1 case.

In the non-compact case, introduction of a cell is essential because
of the assumption of spatial homogeneity. Therefore, one may hope
that in an inhomogeneous setting a cell would be unnecessary and the
inverse volume effects would be well-defined also in the non-compact
topology. With this motivation, inverse scale factor modifications
have been introduced by considering certain deformations of the
constraint functions of general relativity that can still yield a
closed algebra \cite{pert_anomaly}, and by postulating an effective
Hamiltonian based on the results found in
homogeneous models. Corrections to the linear order scalar perturbations have been
computed \cite{pert_scalar}. These have been extended to the
proposal of `lattice refinement' \cite{mb_lattice}, and issues of gauge invariant scalar perturbations have been
analyzed \cite{pert_gi}. Linear perturbations have also been
computed for the  tensor modes
\cite{pert_tensor1,pert_tensor2,pert_tensor_fr,barrau2}.  More
recently, inflationary observables have been studied with the goal
of constraining them with the observational data
\cite{pert_inf,bct}.

The notion of `lattice refinement' has evolved over the years. In
recent works, the basic idea is to decompose the spatial manifold in
elementary cells and approximate the inhomogeneous configurations of
physical interest by configurations which are homogeneous within any
one cell but vary from cell to cell. Physically, this is a useful
approximation. However it requires a fresh input, that of the cell
size (or lattice spacing), and the inverse volume effects are now
sensitive to this new scale. The fiducial cell of the homogeneous
model is replaced by a physical cell in which the universe can be
taken to be homogeneous. This new scale may well be ultimately
provided by the \emph{actual}, physical quantum state of the
universe. However, to our knowledge, a procedure to select such a
inhomogeneous state has not been spelled out. Even if one somehow
fixes a state, a systematic framework to arrive at this scale
starting from the given state has not yet been constructed. Finally,
we are also not aware of any observational guidance on what this
scale should be during say the inflationary epoch, i.e., the epoch
when one generates the perturbations in these schemes. Therefore,
although the underlying idea of lattice refinement is attractive, so
far there appears to be an inherent ambiguity in the size and
importance of the inverse volume effects in this setting.

Finally, we will amplify on a remark in the opening paragraph of
this subsection. The matter density and curvature at the onset of
inflation is some 11 orders of magnitude below the Planck scale.
Therefore one would expect that quantum gravity corrections would
also be suppressed by the same order of magnitude. Yet, some of the
works that focus on the inflationary and post-inflationary era
report corrections which are higher by several orders of magnitude.
This would indeed be very interesting for quantum gravity but the
mechanism responsible for these impressive enhancements needs to be
spelled out to have a better understanding of and confidence in the
physical relevance of these results.
\\

\noindent {\it Thermal fluctuations and k=1 model}: So far our
summary of LQC phenomenology has been focused on inflationary
scenarios. A possible alternative to inflation is provided by the
idea that the primordial fluctuations observed in CMB are of thermal
origin. In conventional scenarios this proposal faces two serious
drawbacks. First, there is the horizon problem. A second difficulty
is that in the general relativity based scenarios, the spectral
index of thermal fluctuations is far from being nearly scale
invariant. In LQC, first problem is naturally resolved by the
bounce. In addition, the inverse volume corrections, which are
well-defined for the spatially closed model, lead to important
modifications to the effective equation of state \cite{ps05} and
affect thermodynamic relations in a non-trivial way. Preliminary
investigations that incorporate these modifications indicate that
the thermal spectrum can in fact be approximately scale invariant in
the allowed region of parameter space \cite{joao_ps}. Thus, the
analysis reveals that the LQC corrections could revive this
alternative to inflation. This possibility should be investigated
further.

\section{ Lessons for Full Quantum Gravity}
\label{s7}

LQC has been developed by applying the basic principles and
techniques from LQG to symmetry reduced systems. Although the
symmetry reduction is quite drastic, LQC has the advantage that it
provides us with numerous models in which the general quantization
program underlying LQG could be completed \emph{and, more
importantly,} physics of the Planck regime could be explored in
detail. Consequently, LQC has now begun to provide concrete hints
for full LQG and, in some instances, for any background independent,
non-perturbative approach to quantum gravity.

In recent years, LQG has advanced in two directions. On one front
the Hamiltonian theory has been strengthened by introducing matter
fields which can serve as `internal clocks and rods' (see, e.g.,
\cite{ghtw1,ghtw2,warsaw-full}) and, on the other, significant
progress has been made in the path integral approach through the
Engle, Pereira, Rovelli, Livine (EPRL) and Freidel-Krasnov (FK)
models \cite{eprl,fk,newlook}. LQC sheds new light on a number of
issues on both these fronts. For example, the subtleties that arose
in the definition of the Hamiltonian constraint already in LQC have
begun to provide guidance for the treatment of the Hamiltonian
constraint in full LQG. Similarly, LQC has provided a proving ground
to test the paradigm that underlies the new spin foam models.
Specifically, it has been possible to examine the open issues of LQG
through the LQC lens. The outcome has provided considerable support
for the current strategies and, at the same time, revived or opened
conceptual and technical issues in LQG. More generally, thanks to
ideas that have been introduced over the past decade, any approach
to quantum gravity has to take a stand on the issue of entropy
bounds. Are these bounds universal, or can they be violated in the
deep Planck regime? Should they be essential ingredients in the very
construction of a satisfactory quantum theory of gravity or should
they arise as inequalities that hold in certain regimes where
quantum effects are important but an effective classical geometry
still makes sense? Further, fundamental issues such as a consistent
way to assign probabilities in a quantum universe can be addressed
in the consistent histories paradigm. LQC provides a near ideal
arena to explore such general issues as well.

In this section we will discuss a few examples to illustrate such
applications of LQC.

\subsection{Physical viability of the Hamiltonian constraint}
\label{s7.1}

In full LQG there is still considerable freedom in the definition of
the Hamiltonian constraint. Perhaps the most unsatisfactory feature
of the current status is that we do not know the physical meaning of
these ambiguities. For, once the implications of making different
choices are properly understood, we could rule out many of them on
theoretical grounds and propose experiments to test the viability of
the remaining ones. Now, these ambiguities descend even to the
simple cosmological models. Therefore it is instructive to
re-examine the well-understood k=0, $\Lambda$=0 FLRW and the Bianchi
I models and ask what would have happened if we had used other,
seemingly simpler definitions of the Hamiltonian constraint. Such
investigations have been performed and the concrete results they led
to offer some guidance for creating viability criteria in the full
theory.

The first criterion is to demand internal coherence in the following
sense: physical predictions of the theory must be independent of the
choices of regulators and fiducial structures that may have been
used as mathematical tools in its construction. Let us begin with
the k=0, $\Lambda=0$ FLRW model. In classical general relativity,
spatial topology plays no role in the sense that the reduced field
equations are the same whether spatial slices are compact with
$\mathbb{T}^3$ topology or non-compact with $\R^3$ topology.
However, as we emphasized earlier, in the non-compact case the
symplectic structure, the Hamiltonian and the Lagrangian all diverge
because of spatial inhomogeneity. Since these structures are needed
in canonical and path integral quantization, this divergence impedes
the passage to quantum theory. As we saw in section \ref{s2}, a
natural strategy is to introduce an infrared regulator, i.e., a cell
$\C$, and restrict all integrations to it. But we have the rescaling
freedom ${\cal C} \rightarrow \beta^3 {\cal C}$ where $\beta$ is a
positive real number. How do various structures react to this
change? At the classical level, we found in section \ref{s2.1} that,
although the symplectic structure and the Hamiltonian transform
non-trivially, physics
---the equations of motion for geometry and matter fields--- are all
invariant under the rescaling.

What about the quantum theory? Now the dynamics is governed by the
Hamiltonian constraint. The expression of the Hamiltonian constraint
contains the gravitational connection $c$. However, since only the
holonomies of these connections, rather than the connections
themselves, are well defined operators, the systematic quantization
procedure leads us to replace $c$ by $\sin\bar\mu c/\bar\mu$ where
$\bar\mu$ can be thought of as the `length' of the line segment
along which the holonomy is evaluated. Now, in the older treatments,
$\bar\mu$ was set equal to a constant, say $\mu_o$ (see, e.g.,
\cite{mb1,mb-livrev,abl,bdv,bms,mb_inflation}). With this choice,
the Hamiltonian constraint becomes a difference operator with
uniform steps in $p \sim a^2$. The resulting quantum constraint is
non-singular at $a=0$ \cite{abl} and evolution of states which are
semi-classical at late times leads to a bounce \cite{aps2}.

However, a closer examination shows that this dynamics has several
inadmissible features. First, the energy density at which bounce
occurs scales with the change $\C \to \beta^3 \C$ in the size of the
fiducial cell. As an example, for the massless scalar field the
density at bounce is given by \cite{aps2}
\be\label{rhomaxmu0} \rho_{\mathrm{max}}^{(\mu_o)} =
\left(\f{2^{1/3} ~ 3}{8 \pi G \gamma^2 \lambda^2}\right)^{3/2}
\f{1}{p_{(\phi)}} ~, \ee
where, as before, $\lambda^2$ is the area gap of LQG.  Keeping the
fiducial metric fixed, under the rescaling of the cell ${\cal C}
\rightarrow \beta^3 {\cal C}$, we have $p_{(\phi)} \rightarrow
\beta^3 p_{(\phi)}$ and hence $\rho_{\mathrm{max}}^{(\mu_o)}
\rightarrow \beta^{-3} \rho_{\mathrm{crit}}^{(\mu_o)}$. But the
density at the bounce should have a direct physical meaning that
should not depend on the size of the cell. Moreover, if we were to
remove the infrared regulator in this final result, i.e. take the
limit in which $\C$ occupies the full $\R^3$, we would find
$\rho_{\mathrm{max}}^{(\mu_o)} \rightarrow 0$! This severe drawback
is also shared by more general matter \cite{cs1}. The effective
Hamiltonian corresponding to this $\mu_o$ quantization implies
\be \label{rhomuo} \rho = \f{3 \sin^2(\mu_o c)}{8 \pi \gamma^2 G
\lambda^2 |p|}  
\ee
for general matter sources. Since $|p| \rightarrow \beta^{2} |p|$,
and $c \rightarrow \beta c$, $\rho$ has a complicated rescaling
property. But $\rho$, the physical density, cannot depend on the
choice of the size of the cell $\C$! Thus the $\mu_o$-scheme for
constructing the Hamiltonian constraint, simple as it seems at
first, fails the theoretical criterion of `internal coherence' we
began with.

But it also has other limitations. One can bypass the theoretical
coherence criterion by restricting oneself to the spatially compact
$\mathbb{T}^3$ topology where one does not need a fiducial cell at
all. But now we still have a problem with the quantum gravity scale
predicted by the theory: with a massless scalar field source, from
Eq (\ref{rhomaxmu0}) one finds that for large values of
$p_{(\phi)}$, the density $\rho_{\mathrm{max}}^{(\mu_o)}$ at the
bounce is very small. To see how small, let us recall an example
from section \ref{s3.3.2}. Consider a hypothetical k=0, $\Lambda$=0
universe with $\mathbb{T}^3$ topology and massless scalar field.
Suppose that, when its radius equals the observable radius of our
own universe at the CMB time, it has the same density as our
universe then had. For such a universe $\pphi \approx 10^{126}$ in
Planck units so the density at the bounce would be
$\rho_{\mathrm{max}}^{(\mu_o)} \approx 10^{-32} {\rm gm}/{\rm
cm^3}$!

A third problem with $\mu_o$ quantization is the lack of correct
infrared limit in a regime well away from the bounce in which the
space-time curvature is small. This is most pronounced when matter
violates strong energy condition, as in the inflationary scenario.
For such matter, $\mu_o$ quantization predicts that the expanding
k=0 universe will recollapse at a late (but finite) time, in
qualitative disagreement with general relativity at low densities
\cite{cs1} (see also \cite{npv}) and showing incompatibility with
the inflationary scenario \cite{ns_inflation}.

Thus, irrespective of the choice of spatial topology, $\mu_o$
quantization suffers from severe problems both in the ultraviolet
and infrared regimes. These problems also arise in other proposals
for quantization of the Hamiltonian constraint. One such proposal
results from an attempt to put the constraint operator in LQC in a
more general setting \cite{mb_lattice}. In this proposal, one
effectively works with a more general pair of canonical variables
for the gravitational phase space
\be P_g = c |p|^m \qquad {\rm and}\quad  g = \f{|p|^{1-m}}{1 - m} ~.
\ee
and writes the gravitational part of the constraint as a
difference operator with uniform steps in $g$. The $\mu_o$
scheme now corresponds to the specific choice $m=0$. For the
FLRW model with a massless scalar field, following the
procedure used in LQC, one can derive the maximum density at
the bounce for any $m$ to obtain \cite{cs1}
\be \rho_{\mathrm{max}}^{(\mu_m)} =  \f{3}{8 \pi G \gamma^2
\lambda^2} \left(\f{8 \pi G}{6} \gamma^2 \lambda^2 \,
p_{(\phi)}^2\right)^{(2m + 1)/(2m - 2)} ~. \ee
Thus, except for $m=-1/2$, the energy density at bounce depends on
$p_{(\phi)}$ and hence suffers from the same problems which plague
$\mu_o$ quantization. In addition, analysis of different $m$
parameterizations shows that various quantization ambiguities
disappear in the continuum limit only if $m=-1/2$ \cite{ns_factor}.

The $m=-1/2$ choice corresponds \emph{precisely} to the `improved
dynamics \cite{aps3} discussed at length in sections \ref{s2} and
\ref{s3}. This scheme is free of all the problems discussed above.
The density at the bounce predicted by the effective theory is an
absolute constant $\rcr \approx 0.41 \rho_{\rm Pl}$; not only is it
insensitive to the choice of $\C$ but it is also the absolute upper
bound of the spectrum of the (time dependent) density operator
$\h{\rho}|_\phi$ in the quantum theory. It agrees with general
relativity at weak curvatures even when one includes a cosmological
constant or inflationary potentials violating the strong energy
condition. Finally, it was arrived at not by an ad-hoc prescription
but by using a systematic procedure (discussed in section
\ref{s2.5.2}).
\\

Ambiguities in the quantization of the Hamiltonian constraint have
also been studied for the Bianchi-I model. There is an operator that
meets all the viability criteria discussed above in the FLRW case
\cite{awe2}. It was presented in section \ref{s4.4} and its
effective dynamics was discussed in section \ref{sec_bianchi_eff}.
This choice of the Hamiltonian constraint leads to universal bounds
on energy density and the shear scalar.

This is in striking contrast with the results from an earlier
quantization of Bianchi-I model \cite{chiou_bianchi}, based on an
`obvious' generalization, $\bar\mu_i \propto 1/\sqrt{|p_i|} \propto
1/a_i$, of the successful $\bar\mu$-scheme in the FLRW model
\cite{aps3}. Again, this seemingly straightforward strategy does
lead to a theory but one with severe limitations
\cite{chiou_bianchi_scl,szulc_bianchi,cs3}. First, for the $\R^3$
topology one again needs a cell $\C$ and the rescaling freedom is
now enlarged to $L_i \to \beta_i L_i$ where $L_i$ are the lengths of
$\C$ with respect to a fiducial metric. Under these rescalings we
have $c_1 \to \beta_1 c_i$ and $p_1 \to \beta_2\beta_3 p_1$ etc. If
the three $\beta_i$ are unequal, the `shape' of the cell changes and
even the effective constraint fails to be invariant under such
rescalings \cite{szulc_bianchi,cs3}:
\be\label{effhamb1} {C}^{\mathrm{(eff)}}_{\rm grav} = - \f{1}{8
\pi G \gamma^2} \Big(\f{\sin(\bar \mu_1 c_1) \sin(\bar \mu_2
c_2)}{\bar \mu_1 \bar \mu_2} \, p_1 p_2 + \mathrm{cyclic} \,
\mathrm{terms} \Big) \ee
where $\bar \mu_i = \lambda/\sqrt{p_i}$.%
\footnote{One may be tempted to restrict the fiducial cell to be
`cubical' to avoid this rescaling problem. However, the distinction
between cubical and non-cubical cells is unphysical since for {\it
any} non-cubical fiducial cell, one can always choose a fiducial
flat metric such that the cell is cubical and vice versa.}
Secondly, on the constraint surface of this effective theory, one
obtains
 \be \rho = \f{1}{8 \pi G \gamma^2 \Delta \lp^2}
\Big(\f{\sqrt{|p_1| |p_2|}}{|p_3|} \sin(\bar \mu_1 c_1) \sin(\bar
\mu_2 c_2) + \mathrm{cyclic} \, \mathrm{terms} \Big)\, . \ee
Note that, in comparison to Eq (\ref{density_bianchi1_eff})  ---the
effective equation that follows from the quantum theory summarized
in section \ref{s4.4}---  here the trignometric functions of
connection are multiplied with terms containing $p_i$. As a result
the energy density ---a physical quantity that can not depend on the
choice of a cell---  fails to be invariant under all permissible
rescalings of $\C$. One may attempt to avoid this problem by
restring oneself only to compact topology, $\mathbb{T}^3$. However,
ultraviolet problems still remain because ratios such as
$\sqrt{|p_1| |p_2|}/|p_3|$ grow unboundedly  when one of the scale
factors approaches zero or infinity. For generic initial conditions,
it is possible for evolution in the Bianchi-I model to lead to such
regimes, making the energy density increase without a bound.
Thus, again, an apparently straightforward avenue to the
construction of the Hamiltonian constraint leads to a theory that is
untenable both because it fails the `internal coherence' criterion
in the non-compact case, and because it has an undesirable
ultraviolet behavior irrespective of topology.

These examples illustrate that, even though a priori it may seem
that there is considerable freedom in defining the Hamiltonian
constraint, one can introduce well motivated criteria that can serve
as Occam's razor. The two LQC examples we discussed bring out four
important points: i) Internal coherence, a good ultraviolet behavior
and the requirement that quantum dynamics should not lead to large
deviations from general relativity in tame regimes, already
constitute powerful constraints;  ii) It is important to work out a
few basic consequences of the proposed theory and not be satisfied
only with a mathematically consistent definition of the Hamiltonian
constraint;  iii) Seemingly natural choices in the definition of the
Hamiltonian constraint can lead to theories that are not physically
viable; and, iv) Although the totality of requirements may seem
oppressively large at first, they \emph{can be} met if one follows a
well motivated path that is conceptually well-grounded. Indeed,
recent investigations of parameterized field theories suggest that a
suitable analog of the $\bar\mu$ scheme of \cite{aps3,awe2} is
likely to be necessary also in full LQG \cite{lv}.
\\

\noindent {\bf Remark:} In addition to the above
considerations, another viability criterion arises from
investigating stability of the quantum difference equation and
conditions under which it leads to a semi-classical behavior.
In the Bianchi-I model, for example, the quantization proposed
in \cite{bck} leads to an unstable difference equation and is
therefore problematic \cite{ns_lattice}. These methods provide
a complimentary way to narrow down the quantization
ambiguities.

\subsection{LQG with a scalar field}
\label{s7.2}

The fact that the massless scalar field provides a global relational
time variable in LQC greatly simplifies the task of solving the
constraint and constructing the physical sector of the theory. It
also enables one to introduce convenient Dirac observables and
extract the physics of the deep Planck regime. It is therefore
natural to ask if this scheme can be extended to full LQG. In
fact a proposal along these lines was made already in the nineties
\cite{rs-ham,kr2}. However there was no significant follow-up
because of two reasons. First, the deparametrization procedure in
\cite{rs-ham} simply assumed that the scalar field would serve as a
good time variable. In the classical theory, this amounts to
assuming that every solution of Einstein-Klein Gordon system admits
a foliation by space-like surfaces on each leaf of which the scalar
field $\phi$ is constant. This cannot hold in the spatially
non-compact (e.g. asymptotically flat) context because the total
energy of such a scalar field would have to diverge (for physically
interesting lapse fields). Even in the spatially compact context, it
appears implausible that the assumption will hold in the full
classical theory. The second problem was that, at the time, LQG
kinematics had not been fully developed to have sufficient
confidence in the then treatment of quantum constraints. Advances in
LQG \cite{ghtw1,ghtw2} and especially LQC \cite{aps3} led to a
re-examination of the proposal and recently it was revived in a
somewhat different and sharper form \cite{warsaw-full}. In this
subsection we will summarize the main ideas of this work.

Let us begin by recalling the classical Hamiltonian framework of
this model. Fix a compact 3-manifold $M$. The phase space $\Gamma$
of the Einstein-Klein-Gordon system consists of quadruples $(A_a^i,
E^a_i;\, \phi, \pphi)$. They are subject to the Gauss, the
diffeomorphism and the Hamiltonian constraints. The Gauss constraint
does not play an important role and can be handled both in the
classical and quantum theories in a standard fashion
\cite{alrev,crbook,ttbook}. Let us therefore focus on the other two
constraints. They are usually written as
\ba \label{const1}
C_a &:=& C_a^{\g} + \pphi\, D_a \phi\,\, \approx 0\nonumber\\
C &:=& C_{\g} + \f{1}{2} \f{\pphi^2}{\sqrt{q}} + \f{1}{2} q^{ab}
D_a\phi D_b\phi \sqrt{q}\,\, \approx\,\, 0
 \ea
where the label `$\g$' refers to the gravitational parts of the
constraints whose explicit form will not be needed here. Note that
$C_a, C_a^{\g}, C, C_{\g}$ are all differentiable functions on
$\Gamma$ whence their Hamiltonian vector fields are well-defined
everywhere.

The first observation is that the surface $\bar\Gamma$ in $\Gamma$
where these constraints are satisfied is the same as the surface on
which the following constraints hold \cite{kr2}:
\ba \label{const2}  C_a &:=& C_a^{\g} + \pphi\, D_a \phi\,\,
\approx 0\nonumber\\
C' &:=& C_\g + \f{1}{2} \f{\pphi^2}{\sqrt{q}} \mp \sqrt{C^2_{\g} -
q^{ab} C_a^{\g} C_b^{\g}} \,\, \approx 0 \ea
where the quantity under the square-root is guaranteed to be
non-negative in a neighborhood of $\bar\Gamma$. In what follows, we
restrict to this neighborhood in classical considerations. In this
neighborhood, the new constraint function $C'$ is also
differentiable everywhere on the phase space \emph{except} on the
surface $\Gamma_{\rm sing}$ on which $C^2_{\g} - q^{ab} C_a^{\g}
C_b^{\g}$ vanishes somewhere on $M$. Away from $\Gamma_{\rm sing}$,
one can calculate the Poisson brackets between these constraints.
The Poisson brackets of the Gauss and the diffeomorphism constraints
with all constraints are the standard ones. What about the
Hamiltonian constraints? Recall that the Poisson brackets between
$C(x)$ do not vanish and, moreover, the right side of the brackets
involves \emph{structure functions} rather than structure constants.
In the rich literature on quantum geometrodynamics, this
complication has often been referred to as the principal obstacle in
canonical quantization of the theory. The situation is entirely
different for $C'(x)$ \cite{warsaw-full}: \emph{the Poisson bracket
among the new Hamiltonian constraints} $C'(x)$ \emph{vanish
identically}! This is a major simplification.

The next key observation \cite{kr2,warsaw-full} is that the scalar
constraint $C' \approx 0$ in (\ref{const2}) is well tailored for
deparametrization: It is of the form
\be \label{const3} \pphi^2 = \sqrt{q} \big[- C_\g  \pm \sqrt{C^2_\g
- q^{ab} C_a^\g C_b^\g}\, \big]\ee
where the right side contains phase space functions that depend
\emph{only on the gravitational fields}. Thus, it is exactly of the
form we encountered in homogeneous LQC models with a massless scalar
field. Therefore, as in LQC, one can hope to pass to the quantum
theory by imposing, in addition to the quantum Gauss and the
diffeomorphism constraints which have been studied extensively
\cite{almmt,alrev,crbook,ttbook}, the (infinite family of) quantum
Hamiltonian constraints:
\be \label{const4} \hbar^2 \f{\delta^2\Psi (A, \phi)}{\delta
\phi(x)^2} \, = - \Theta_{\rm LQG}(x)\, \Psi (A, \phi) \ee
where $\Theta_{\rm LQG}$ is to be the quantum operator corresponding
to the right side of (\ref{const3}). After making a key observation
that elements necessary to arrive at a rigorous definition of
$\Theta_{\rm LQG}$ are already in place in the LQG literature, a
concrete candidate is proposed in \cite{warsaw-full}. The next idea
is to construct the physical Hilbert space by taking the positive
square-root of (\ref{const4}) as in LQC and introduce relational
Dirac observables along the lines of the LQC observables
$\h{V}|_\phi$ and $\h{\rho}|_\phi$ (defined in sections \ref{s2} and
\ref{s3}). (Even explicit formulas for such Dirac observables are
available but their physical meaning is often unclear.) This is the
set up necessary to extract physics from the quantum evolution and
check if it satisfies viability criteria such as those outlined in
section \ref{s7.1}. Since there is considerable freedom in defining
$\Theta_{\rm LQG}$, these checks are essential to streamline and
reduce the choices. Developments in cosmology can provide guidance
for this task and also supply an anchor to interpret the resulting
quantum theory.

The potential for a strong interplay between this program and LQC is
illustrated by the following considerations. Let us first briefly
return to the classical theory. Note first that in arriving at
(\ref{const3}) from (\ref{const1}), spatial derivatives of $\phi$
are \emph{not} set equal to zero; one does not even assume that
space-times of interest admit a space-like foliation by $\phi = {\rm
const}$ surfaces. Nonetheless, there is an interesting result. A
large portion $\bar\Gamma_{\rm reg}$ of $\Gamma$ admits a positive
lapse function $N$ such that if one starts at a point on
$\bar\Gamma_{\rm reg}$ at which $\phi$ is constant on $M$ then,
along the Hamiltonian vector field of $C'(N)$, the scalar field
$\phi$ remains constant on $M$ but increases monotonically. Thus,
the solution to the Einstein-Klein-Gordon system determined by the
initial point in $\bar\Gamma_{\rm reg}$ does admit a space-like
foliation by $\phi$ = const surfaces at least locally, i.e., as long
as the dynamical trajectory generated by $C'(N)$ remains in
$\bar\Gamma_{\rm reg}$.%
\footnote{$\bar\Gamma_{\rm reg}$ is the portion of the constraint
surface which excludes points of $\Gamma_{\rm sing}$ \emph{and} and
has the property that $\pphi$ is nowhere vanishing on $M$. The lapse
is given by $N= \sqrt{q}/\pphi$; it is `live', i.e. depends on the
dynamical variables. Of course all these arguments are at the same
level of rigor as is commonly used in the Hamiltonian framework in
general relativity; they are not rigorous in the sense of functional
analysis used in PDEs.}
The homogeneous sector $\bar\Gamma_{\rm hom}$ of $\bar\Gamma$ (on
which all fields are homogeneous) is of course in $\bar\Gamma_{\rm
reg}$ and, if the initial point lies on $\bar\Gamma_{\rm hom}$ then
the dynamical trajectory never leaves it. Therefore, it is
reasonable to assume that in an open neighborhood of homogeneous
solutions to the Einstein-Klein-Gordon system, space-times would
admit a space-like foliation by $\phi$ = const surfaces for a long
time (as measured by $\phi$), and this time would grow as the
strength of inhomogeneities decreases.

Therefore, it would be helpful to apply this LQG framework to the
cosmological perturbation theory discussed in section \ref{s6.3}.
Specifically, it would helpful to expand out the right hand side of
the constraint (\ref{const3}) around a FLRW background and then pass
to the quantum theory, focusing in the first step just on the tensor
modes (thereby avoiding gauge issues and the technically more
complicated form of the perturbed part of the Hamiltonian
constraint). The procedure outlined in section \ref{s6.3} may
provide valuable guidance in removing ambiguities in the definition
of the operator $\Theta_{\rm LQG}$. If the mathematical program
outlined in \cite{warsaw-full} is completed, LQC would also play a
key role in the physical interpretation of that theory. The most
`secure' interpretation is likely to come from states that are
sharply peaked at homogeneous geometries in the regime in which
general relativity is a good approximation. The quantum Hamiltonian
equation would naturally provide a evolution of these states in the
internal time $\phi$. Results from LQC suggest natural, physically
important questions that this theory could address. Is the
singularity still resolved in spite of the presence of an infinite
number of degrees of freedom which are now treated
non-perturbatively? Do LQC results correctly capture the qualitative
features of the Planck scale physics? Thus, the cosmological context
provides a natural home for this sector of LQG which is poised to
make technical advances in the coming years.

\subsection{Spin foams}
\label{s7.3}

The goal of spin foam models (SFM) is to provide a viable path
integral formulation of quantum gravity. Because there is no
background space-time, the framework underlying this program has
certain novel features that are not shared by path integral
formulations of familiar field theories in Minkowski space. Loop
quantum cosmology offers a simple context to test the viability of
these novel elements \cite{ach1,ach2,chn,cgo,hrvw}. Conversely, SFM
offer a possible avenue to arrive at LQC starting from full LQG
\cite{rv,bkrv1,bkrv2}.

\subsubsection{Conceptual setting}
 \label{s7.3.1}

As we discussed in section \ref{s1}, the kinematical framework
of full LQG is well established. A convenient basis in $\Hk$ is
provided by the so-called spin network states
\cite{rs2,jb2,alrev,crbook,ttbook}. The key challenge is to
extract physical states by imposing quantum constraints on spin
networks. Formally this can be accomplished by the group
averaging procedure which also provides the physical inner
product \cite{dm,almmt,alrev,crbook,ttbook}. From the LQG
perspective, the primary goal of SFMs is to construct a path
integral to obtain a Green's function ---called the
\emph{extraction amplitude}--- that captures the result of
group averaging. As explained in section \ref{s3.5}, in the
timeless framework of quantum gravity, the extraction amplitude
determines the full content of quantum dynamics, just as the
transition amplitudes do in the familiar field theories in
Minkowski space-times.

Heuristically, the main idea behind this construction can be
summarized as follows \cite{perezrev}. Consider a 4-manifold
$M$ bounded by two 3-surfaces, $S_1$ and $S_2$, and a
triangulation $\mathcal{T}$ of $M$. One can think of $S_1$ as
an `initial' surface and $S_2$ as a `final' surface. One can
fix a spin network on each of these surfaces to specify an
`initial' and a `final' quantum 3-geometry. A quantum
4-geometry interpolating between the two is captured in a
`colored' 2-complex $\mathcal{T}^\star$ dual to the simplicial
decomposition where the coloring assigns to each 2-surface in
$\mathcal{T}^\star$ a half integer $j$ and to each edge an
intertwiner. The idea is to obtain the extraction amplitude by
summing first over all the colorings for a given $\mathcal{T}$,
and then over triangulations, keeping the boundary states
fixed. The second sum is often referred to as the \emph{vertex
expansion} because the $M$-th term in the series corresponds to
a $\mathcal{T}^\star$ with $M$ vertices (each of which
corresponds to a simplex in $\mathcal{T}$). Since each colored
$\mathcal{T}^\star$ specifies a quantum geometry, the sum is
regarded as a path integral over physically appropriate
4-geometries.

Group field theory (GFT) provides a conceptually distinct
method to obtain the same vertex expansion. The underlying idea
is that gravity is to emerge from a more fundamental theory
based on abstract structures that, to begin with, have nothing
to do with space-time geometry. Examples are matrix models for
2-dimensional gravity and their extension to 3-dimensions
---the Boulatov model \cite{bou}--- where the basic object is a
field on a group manifold rather than a matrix. The Boulatov
model was further generalized to 4-dimensional gravity
\cite{crbook,gft1,gft2}. The resulting theory is again
formulated on a group manifold, rather than space-time.
However, as in familiar field theories, its Lagrangian has a
free and an interaction term, with a coupling constant
$\ul{\lambda}$. In the perturbation expansion, the coefficient
of $\ul\lambda^M$ turns out to be $M$th term in the spin foam
vertex expansion.

Over the last 3-4 years SFM have witnessed significant advances
(see, e.g., \cite{eprl,fk,newlook}). In particular it was shown
that, thanks to the discreteness of eigenvalues of the area
operator, the sum over colorings has no ultraviolet
divergences. More recently, it has been argued that the
presence of a \emph{positive} cosmological constant naturally
leads to an infrared regularization \cite{infrared}. These are
striking results. However, a number of issues still remain
because so far a key ingredient in the spin foam sum over
histories ---the vertex amplitude--- has not been
systematically derived following procedures used in well
established field theories, or, from a well understood
Hamiltonian dynamics. More importantly, because the number of
allowed triangulations grows \emph{very} rapidly with the
number of vertices, and a compelling case for restricting the
sum to a well controlled subset is yet to be made, the issue of
convergence of the vertex expansion is wide open. Finally,
there is also a conceptual tension between the SFM and GFT
philosophies. Do the $\ul{\lambda}^M$ factors in the
perturbative expansion of GFT merely serve as book-keeping
devices, to be set equal to $1$ at the end of the day to
recover the SFM extraction amplitude? Or, is there genuine, new
physics in the GFT at lower values of $\ul\lambda$ which is
missing in the spin foam program? For this to be a viable
possibility, $\ul\lambda$ should have a direct physical
interpretation in the space-time picture. What is it?

These and other open issues suggest that the currently used SFM and
GFT models will evolve significantly over the next few years. Indeed
this seems likely because, thanks to the recent advances, this
community has grown substantially. Therefore, at this stage it seems
appropriate to examine only of the underlying, general paradigms
rather than specifics of the models that are being pursued. Is it
reasonable to anticipate that the `correct' extraction amplitude
will admit a vertex expansion? Are there physical principles that
constrain the histories one should be summing over? Can symmetry
reduced models shed some light on the physical meaning of
$\ul\lambda$ of GFT? LQC has turned out to be an excellent setting
to analyze these general issues.

\subsubsection{Cosmological spin foams}
 \label{s7.3.2}

For definiteness, let us consider the Bianchi I model \cite{chn}
with a massless scalar field as in section \ref{s4.4}. A convenient
basis in the kinematical Hilbert space $\Hk$ is now given by $|\nu,
\vec{l}, \phi\rangle$ where $\vec{l} = l_1,l_2$ are the eigenvalues
of the anisotropy operators and, as before, $\nu$ and $\phi$, of the
volume and scalar field operators. This is the LQC analog of the
spin network basis of LQG. As in section \ref{s3.5}, the group
averaging procedure \cite{dm,almmt,abc} provides us with the
extraction amplitude:
\be \label{e-amp1} \E(\nu_f,\vec{l}_f,\phi_f;
\nu_i,\vec{l}_i,\phi_i):= \sint_{-\infty}^{\infty} \dd{\alpha} \,\,
\langle \nu_f,\vec{l}_f,\phi_f | \, e^{\frac{i}{\hbar} \alpha \h{C}}
\,| \nu_i, \vec{l}_i, \phi_i \rangle\ , \ee
where $\h\C = -\hbar^2 (\p_\phi^2 + \Theta_{(I)})$ is the full
Hamiltonian constraint of the Bianchi I model. (As in section
\ref{s3.5} for simplicity of notation we have chosen not to write
explicitly the $\theta$ function that restricts to the positive part
of the spectrum of $\hat{p}_{(\phi)}$.) Again, since the integrated
operator is heuristically `$\delta(\hat{C})$,' the amplitude
$\E(\nu_f, \vec{l}_f, \phi_f; \nu_i,\vec{l}_i,\phi_i)$ satisfies the
Hamiltonian constraint in both sets of its arguments. Consequently,
it serves as a Green's function that maps states $\Psi_{\rm kin}
(\nu,\vec{l},\phi)$ in the kinematical space $\Hk$ to states
$\Psi_{\rm phys} (\nu,\vec{l},\phi)$ in $\Hp$ through a convolution
\be \Psi_{\rm phys}(\nu,\vec{l},\phi) = \sum_{\nu^\prime,
\vec{l}^\prime}\, \sint \dd{\phi}^\prime\,\,
\E(\nu,\vec{l},\phi; \nu^\prime,\vec{l}^\prime,\phi^\prime)
\,\Psi_{\rm kin}(\nu^\prime,\vec{l}^\prime, \phi^\prime). \ee
and enables us to write the physical inner product in terms of the
kinematical:
\be (\Phi_{\rm phys},\, \Psi_{\rm phys})  :=
\sum_{\nu,\vec{l}\, \nu^\prime,\vec{l}^\prime}\, \sint \dd
{\phi}\, \dd {\phi}^\prime\, \bar{\Phi}_{\rm
kin}(\nu,\vec{l},\phi)\, \E(\nu,\vec{l},\phi;
\nu^\prime,\vec{l}^\prime,\phi^\prime) \Psi_{\rm
kin}(\nu^\prime,\vec{l}^\prime\ \phi'). \ee

As in section \ref{s3.5}, one can just follow the
\emph{mathematical} procedure first given by Feynman \cite{rpf} by
regarding $\E(\nu_f,\vec{l}_f, \nu_f;\, \nu_i, \vec{l}_i, \phi_f)$
as a transition amplitude for the initial state to evolve to the
final one in `unit time interval' during the evolution generated by
a fictitious `Hamiltonian' $\alpha\h{C}$. Again, the `time interval'
and the `Hamiltonian' are fictitious mathematical constructs because
in the physical example under consideration, we are in the timeless
framework and $\h{C}$ is the constraint operator, rather than the
physical Hamiltonian. There is however one difference from section
\ref{s3.5}. To make contact with spin foams one has to remain in the
configuration space with paths representing \emph{discrete quantum
geometries}: Since we are not interested in semi-classical
considerations, there is no need to express the extraction amplitude
as a sum over classical phase space paths.

In the Feynman procedure, one first divides the `unit time
interval' into a large number $N$ of segments, each of length
$\epsilon=1/N$, by introducing $N-1$ decompositions of
identity, $|\nu_{N-1}, \vec{l}_{N-1}\rangle \langle
\vec{l}_{N-1}, \nu_{N-1}|\,\, \ldots\,\, |\nu_{1},
\vec{l}_{1}\rangle \langle \vec{l}_{1}, \nu_{1}|$, between the
final and initial states in the expression of the amplitude
$\E$. This enables one to write $\E$ as a sum over discrete
quantum histories. The key new step is to reorganize this sum
by grouping together all paths that contain \emph{precisely}
$N$ volume transitions \cite{ach1,ach2}. Then, after taking the
limit $N \to \infty$, one obtains
\be \E(\nu_f,\vec{l}_f,\phi_f;\, \nu_i,\vec{l}_i,\phi_i) =
\sum_{M=0}^\infty \Big[\sum_{\nu_{M-1},\ldots, \nu_{1} } \, A(\nu_f,
\nu_{M-1},\ldots , \nu_1, \nu_i;\,\, \vec{l}_f,\vec{l}_i;\,\,\phi_f,
\phi_i)\Big]\, \equiv \, \sum_{M=0}^\infty A(M)\, .
\label{e-amp2}\ee
The partial amplitude $A(\nu_f, \nu_{M-1},\ldots , \nu_1, \nu_i;\,\,
\vec{l}_f,\vec{l}_i;\,\, \phi_f, \phi_i)$ is obtained by summing
over all paths in which the volume changes from $\nu_i$ to
$\nu_1$,\, $\nu_1$ to $\nu_2$,\,\, \ldots \,\, $\nu_{M-1}$ to
$\nu_f$. These ordered sequences of volume transitions can occur at
any values of $\phi$ and can be accompanied by arbitrary changes in
anisotropies $\vec{l}$ subject only to the initial and final values,
$\phi_i, \vec{l}_i$ and $\phi_f, \vec{l}_f$. Still the amplitude
$A(\nu_f, \nu_{M-1},\ldots , \nu_1, \nu_i;\,\,
\vec{l}_f,\vec{l}_i;\,\, \phi_f, \phi_i)$ is a well-defined, finite
expression which can be expressed in terms of the matrix elements of
$\Theta_{(I)}$. In the final expression, $A(M)$ is the contribution
to the extraction amplitude arising from \emph{precisely} $M$ volume
transitions, subject just to the initial and final conditions. Thus,
fixing $M$ is analogous to fixing the number of vertices in SFM and
summing over intermediate volumes and anisotropies is analogous to
summing over `colorings' (spins and intertwiners) \cite{chn}. In
this sense, (\ref{e-amp2}) is the analog of the vertex expansion of
SFM.

There is also a neat analog of the GFT perturbative expansion of the
extraction amplitude. The idea is to split ${\Theta_{(I)}}$ as
$\Theta_{(I)} = \h{D} + \h{K}$ where $\h{D}$ is diagonal in the
volume basis $|\nu\rangle$ and $\h{K}$ is \emph{purely}
off-diagonal, and regard $\h{D}$ as the `main part' of the
constraint and $\h{K}$ as a `perturbation'. Again, this is only a
mathematical step that allows us to use the well-developed framework
of perturbation theory. To highlight our intention, let us introduce
a coupling constant $\ul\lambda \in (0,1)$ and consider, in place of
$\h{C}$, the operator $\h{C}_{\ul\lambda} =
-\hbar^2(\partial_\phi^2+ \h{D} + \ul{\lambda} \h{K})$. Then we can
use the standard perturbation theory in the interaction picture to
calculate the transition amplitude $\E_{\ul\lambda}$. One is
directly led to:
\be \E_{\ul\lambda}(\nu_f,\vec{l}_f,\phi_f;\,
\nu_i,\vec{l}_i,\phi_i) = \sum_{M=0}^\infty
{\ul{\lambda}}^M\,\,A(M)\, \label{e-amp3}\ee
where the coefficients $A(M)$ are \emph{exactly} the same as in
(\ref{e-amp2}). Furthermore, this perturbative treatment enables one
to extract the meaning of truncating the expansion after, say, first
$M_o$ terms as is done in practice in spin foam calculations. One
can show that the resulting truncated series satisfies the
Hamiltonian constraint to order $O(\ul\lambda^{M_o})$. To recover
the full extraction amplitude of direct physical interest, one has
to simply set $\ul\lambda=1$ in (\ref{e-amp3}). Thus, Eq.
(\ref{e-amp3}) of LQC is analogous to the expression of the
transition amplitude obtained in GFT.

GFT considerations naturally lead us to ask if there is a physical
meaning to the generalization $\E_{\ul\lambda}$ of the extraction
amplitude for $\ul{\lambda}\not=1$. This issue as been analyzed in
detail in the isotropic case and, somewhat surprisingly, the answer
turns out to be in the affirmative \cite{ach2}. Suppose, as in our
discussion so far, we are interested in LQC/SFM with $\Lambda=0$ but
let us allow a cosmological constant $\Lambda$ in the GFT-like
perturbation series. Then, one can show that the $\Lambda=0$\,
LQC/SFM theory is unitarily equivalent to LQC/GFT with $\Lambda =
3(1-\ul\lambda)/2\gamma^2 \l_o^2$ \cite{ach2}! This fact leads to an
intriguing possibility. From a GFT perspective the ``correct''
physical interpretation of the $\ul{\lambda}$-theory would be that
it has a non-zero cosmological constant. At weak coupling, i.e.
$\ul{\lambda}\approx 0$, we have $\Lambda \sim 1/\lp^2$. This is
just as one may expect from the `vacuum energy considerations' in
Minkowski space. As $\ul{\lambda}$ increases and approaches the SFM
value $\ul{\lambda} =1$, we have $\Lambda \rightarrow 0$. Now,
$\ul{\lambda}$ is expected to run in GFT. So, $\Lambda$ would seem
to run from its perturbative value $\Lambda \sim 1/\lp^2$ to
$\Lambda \sim 0$ (from above). If the flow reached close to $\ul\lambda
=1$ but not exactly $\ul\lambda =1$, GFT would say that there is a
small positive $\Lambda$. This is only a speculative scenario
because it mixes precise results in the isotropic LQC model with
expectations in the full GFT. It is nonetheless interesting because
it provides an attractive paradigm to `explain' the smallness of the
observed value cosmological constant using GFT.

To summarize, Hamiltonian LQC provides a well-defined, closed
expression of the extraction amplitude. Therefore one can use it to
probe questions raised in sections \ref{s7.3.1}. We found that one
can simply rewrite it as a discrete series (\ref{e-amp2}) mimicking
the vertex expansion of SFM or as a perturbation series
(\ref{e-amp3}) mimicking the expression of the extraction amplitude
in GFT. (The only assumption made in arriving at these expressions
is that the infinite sum $\sum_M$ can be interchanged with the
integration over $\alpha$ in (\ref{e-amp1}).) In this sense,
cosmological spin foams provide considerable support for the general
paradigm underlying SFM and GFT. Finally, here we worked with the
timeless framework. However we can also deparameterize the theory
using the scalar field as a global, relational time variable as in
section \ref{s4.4}. Then the constraint is reduced to a standard
Schr\"odinger equation with $\phi$ as time and one can meaningfully
speak of transition amplitudes as in ordinary quantum mechanics. One
can show that there is an underlying conceptual consistency: the
extraction amplitude in the timeless framework reduces to this
transition amplitude in the deparameterized framework \cite{ach2}.\\

\textbf{Remarks:}

1) It has been suggested that, to obtain the correct extraction
amplitude, one may need to sum over only those quantum
geometries which have a positive time orientation \cite{do}.
However, it has been difficult to incorporate this condition in
the current spin foam models \cite{eprl,fk,newlook}.
Cosmological spin foams can be used to understand the
difference from a physical perspective. The proposed condition
is directly analogous to the LQC restriction to positive (or
negative) frequency solutions to the constraints \cite{ach2}.
Without this condition, the inner product between physical
states obtained by group averaging the basis elements
$|\nu,\vec{l}, \phi\rangle$ would have been real, while the
correct inner product coming from the Hamiltonian theory is
complex. In the current spin foam models, the inner product
between physical states defined by spin network states is also
real.  Thus, to the extent cosmological spin foams provide
hints for the full theory, it would appear that a restriction
to time-oriented quantum histories is indeed necessary to
complete the program.

2) Most of the spin foam literature to date focuses on the vacuum
case. On the other hand, the Bianchi-I model considered in this
section came with a scalar field \cite{chn}. In a timeless framework
a scalar field is unnecessary. What happens if we consider the
vacuum Bianchi-I model? In this case, although the extraction
amplitude is again well defined in the Hamiltonian framework, if one
mimics the Feynman procedure, the vertex expansion has to be
regulated because it contains distributions term by term
\cite{hrvw}. The analysis provides guidance on viable regulators
that may be helpful more generally. However, one of its aspects is
not entirely satisfactory: it is not possible to remove the
regulator at the end of the day, whence the final answer still
carries the memory of the specific regulator used. Thus, in this
model, the inclusion of a matter fields actually simplifies the
vertex expansion. Perhaps there is a lesson here for the full
theory.

3) Because this section is devoted to lessons from LQC, we have
focused on hints and suggestions that LQC has for the SFM and GFT
programs. However, the bridge also goes the other way: In
\cite{rv,bkrv1,bkrv2} proposals have been made to use SFM to go
beyond homogeneous LQC. Although these proposals do not yet
incorporate inhomogeneities of direct physical relevance, the
underlying ideas are conceptually interesting.

\subsection{Entropy bound and loop quantum cosmology}
\label{s7.4}

Bekenstein's seminal work \cite{bek} in which he suggested that the
black hole entropy should be proportional to its area has motivated
several researchers to ask if there is a geometrical upper bound on
the maximum thermodynamic entropy that a system can have. The
heuristic idea is the following. The leading contribution to the
black hole entropy is given by (1/4)th of the area of its horizon in
Planck units. Since black holes are the `densest' objects, one may
be tempted to conjecture that, in a complete quantum gravity theory,
the number of states in any volume $V$ enclosed by a surface of area
$A$ would be bounded by the number of states of a black hole with a
horizon of area $A$, i.e. by $\exp {A/4\lp^2}$. However, this simple
formulation of the idea quickly runs into difficulties. Several
improvements have been proposed. The most developed of these
proposals is Bousso's covariant entropy bound \cite{bousso1}. The
conjecture is as follows: Given a spatial 2-surface $\mathcal{B}$
with an area $A$, if $\L$ is the hypersurface generated by the
non-expanding null geodesics orthogonal to $\mathcal{B}$, then the
total entropy flux $S$ across the `light sheet' $L$ (associated with
matter) has an upper bound given by $S \leq A/4 G \hbar$.

The conjecture has two curious features. First, it is not clear how
to define the entropy flux without there being an entropy current
$s^a$. If there is such a current, the flux across $\L$ can be
defined as the integral of a 3-form: $S = \int_\L\, s^a
\epsilon_{abcd}$ where $\epsilon_{abcd}$ is the volume 4-form. But
`fundamental' matter fields do not have an associated entropy
current. Therefore one can test the conjecture only for
phenomenological matter such as fluids. Thus, there is tension
between the notion that the bound should be fundamental and the
domain in which it can be readily tested. The second curious feature
is related to the fact that the bound makes a crucial use of the
Planck length. Indeed, it trivializes if $\lp \to 0$ and, as we show
below, it is violated in classical general relativity. Yet, it also
requires a smooth classical geometry so that light sheets can be
well defined. Since quantum fluctuations are likely to make the
space-time geometry fuzzy, there is a tension between the classical
formulation of the bound and the quantum world it is meant to
capture.

In spite of these limitations, the bound does have a large domain of
validity \cite{bousso2}: It holds if the entropy current and the
stress-energy tensor satisfy certain inequalities on $\L$ and these
inequalities can be motivated from statistical physics of ordinary
matter so long as one stays away from the Planck regime. Also, the
bound has attracted considerable attention because it has
`holographic flavor' and it has been suggested that `holography'
should be used as a building block of any quantum gravity theory,
much as the equivalence principle was used in general relativity
\cite{holography}. It is thus natural to ask what the status of the
bound is in any putative quantum gravity theory. Specifically, there
are two questions of interest. There are situations in which the
bound fails in classical general relativity if applied in the Planck
regime. The hope is that an appropriate quantum gravity treatment
would restore it. The first question is: Does this happen? The
second question is: Does the bound result only if the theory is
constructed with a fundamental `holographic' input or does it simply
emerge in suitable regimes of the theory where it can be
meaningfully formulated? It turns out that LQC is a near ideal arena
to address both these questions \cite{awe1}.

Let us consider a k=0, $\Lambda=0$ FLRW space-time filled with a
radiation fluid. The space-time metric is given by
\be \dd s^2 = - \dd t^2 + a(t)^2 (\d r^2 + r^2 \d \Omega^2) ~, \ee
and choose the surface $\mathcal{B}$ to be a round 2-sphere in a
homogeneous slice $t=$ const. The past-directed, ingoing null rays
orthogonal to $\mathcal{B}$ provide us with its light sheet $\L$.
Because the space-time is conformally flat, these rays will all
converge on a point $p$ (in which case $\L$ would be the portion of
the future light cone of $p$ bounded by $\mathcal{B}$) or on the
singularity. In what follows we will consider only those
$\mathcal{B}$ for which the first alternative occurs. Then if we
denote by $t_f$ the time defined by $\mathcal{B}$ and $t_i$ is the
time defined by the point $p$, we have $t_i >0$ and $A = 4 \pi^2
a(t_f)^2 r_f^2$ where $r_f = \int_{t_i}^{t_f} \dd t'/a(t')$. For the
null fluid we have $p= 1/3\rho$, and, assuming that the universe is
always instantaneously in equilibrium, by the Stefan-Boltzmann law,
$\rho = (\pi^2/15 \hbar^3) T^4$. The entropy current is therefore
given by $s^a = (4/3) (\rho/T) u^a = (4\pi^2/45\hbar^3) T^3 u^a$,
where $u^a$ is the unit vector field orthogonal to the $t=$ const
slices. Thus, the entropy current is completely determined by the
temperature. Using the classical Friedman equation and the
stress-energy conservation law, it is straightforward to obtain the
temporal behavior of energy density, and therefore of temperature:
\be\label{T_cl} T(t) = \left(\f{45 \, \hbar^3}{32 \pi^3 G
t^2}\right)^{1/4} ~. \ee
Therefore, if we were to move $\mathcal{B}$ to the past, the norm of
the entropy current ---and hence the flux of total entropy across
its light sheet $\L$--- would increase. A simple calculation gives
\cite{awe1}:
\be \f{S}{(A/4\lp^2)}\, =\, \f{2}{3}\, \Big(\f{2G\hbar}{45 \pi
t_f^2}\Big)^{1/4} \Big(1 - \sqrt{\f{t_i}{t_f}} \Big) ~. \ee
Note that $t_i/t_f <1$ and as we move $t_f$ closer to the
singularity at $t=0$, the right hand grows, just as one would expect
from the behavior (\ref{T_cl}) of temperature. An explicit
calculation shows that the right side can exceed $1$ by an arbitrary
amount when $t_f$ is so close to the singularity that $\rho \gtrsim
8.5 \rho_{\rm Pl}$. Thus in a radiation filled FLRW universe, the
bound can be violated by an arbitrary amount, but only in the Planck
regime near the singularity. The question naturally arises: Do
quantum gravity effects restore it?

As explained in the beginning of this subsection, it is not easy to
analyze this issue because of the dual demand on the calculation:
Quantum gravity effects should be so strong as to resolve the
singularity and at the same time one should be able to define
space-like surfaces $\mathcal{B}$ \emph{and} their light sheets $\L$
unambiguously. Fortunately, LQC meets this dual challenge
successfully. As we have seen, non-perturbative loop quantum gravity
effects \emph{are} strong enough to resolve the singularity and yet
we have a smooth effective geometry which accurately tracks the
quantum states of interest across the big bounce. So the first
question we began with can now be sharpened: Does the Bousso bound
hold in the effective space-time of LQC?

The analysis of entropy flux across $\L$ using the effective field
equations can be carried out along the same lines as before. The
modified Friedmann and Raychaudhuri equations again imply the
standard continuity equation whence we again have $\rho \propto
a^{-4}$. However, the modified Friedman and Raychaudhuri equations
change the time dependence of the scale factor, energy density, and
hence the temperature of the photon gas. The temperature now carries
a dependence on the underlying quantum geometry via $\rcr$:
\be\label{T_lqc} T(t) = \left(\f{45 \, \hbar^3}{32 \pi^3 G t^2 +
\f{3 \pi^2}{\rcr}}\right)^{1/4} ~. \ee
As one would expect, the temperature achieves its maximum value at
the bounce, $t=0$. But in stark contrast with the classical theory,
it does not diverge. Therefore the entropy current is also finite
everywhere, including the bounce. As in the classical theory, one
can compute the entropy current through $\L$ and obtain the desired
ratio \cite{awe1}:
\be \f{S}{(A/4\lp^2)} = \f{16}{9} \big(\f{\pi^2 G^4\hbar\rcr}{15}
\big)^{1/4} \, \f{1}{\sqrt{\f{32 \pi G t_f^2}{3} + \f{1}{\rcr}}} \,
\big[t \, \, \, {}_2F_1\big(\f{1}{2}, \f{1}{4}; \f{3}{2} , - \f{32
\pi G \rcr}{3} t^2\big) \bigg]_{t_i}^{t_f} ~ , \ee
where ${}_2F_1$ is a hypergeometric function which can be plotted
numerically. From the fact that the temperature has an upper bound,
it follows that the entropy current also has one and therefore we
know analytically that the right side is finite. But is it less than
1 as conjectured by Bousso? \emph{One finds $S/(4\lp^2) < 0.976$ for
all round $\mathcal{B}$!} In retrospect this is perhaps not all that
surprising because even in general relativity the bound is violated
only at densities $\rho \gtrsim 8.5 \rho_{\rm PL}$ and in LQC we
have $\rho \le \rcr \approx 0.41 \rho_{\rm Pl}$. But note that in
LQC space-time is extended and there are many more `potentially
dangerous' surfaces $\mathcal{B}$ in the Planck regime: one has to
allow $\mathcal{B}$ whose light sheet $\L$ can meet and go past the
$t=0$ slice.

Thus, we have answered the first question: the quantum geometry
effects of LQG do restore the Bousso bound for the round 2-spheres
$\mathcal{B}$ in the radiation filled FLRW universes. We can now
turn to the second question: Is the bound fundamental? In this
calculation, the fundamental ingredient was quantum geometry of LQG.
That construction did not need anything like `holography' as an
input. Yet, the Bousso bound emerged in the effective theory where
it could be properly formulated. Indeed, the conjecture cannot even
be stated in full LQC where the geometry is represented by wave
functions rather than smooth metrics. Thus, the analysis suggests
that the covariant entropy bound —--and its appropriate
generalizations that may eventually encompass quantum field theory
processes even on `quantum corrected' but smooth space-times—-- can
emerge from a fundamental quantum gravity theory in suitable
regimes; they are not necessary \emph{ingredients} in the
construction of such a theory.

\subsection{Consistent histories paradigm}
\label{s7.5}

Quantum theory is incomplete unless it includes a procedure to
assign probabilities to events or histories. In open quantum
systems, which allow an interaction with an environment or a
classical external system, common ways to assign probabilities are
the `Copenhagen' interpretation and the environmental decoherence.
However, for closed quantum systems, such as in quantum cosmology,
these formulations are of little use. A quantum universe, has
neither an environment to enable a decoherence nor an external
system inducing a collapse of its wave function. Therefore one needs
a new strategy. A natural avenue is provided by the consistent
histories approach that stems from the work of Hartle, Halliwell and
others \cite{hartle-halliwell}. (See \cite{hartle} for a detailed
review and an extensive bibliography). This approach relies on
computing a `decoherence functional' between different histories and
assigning consistent probabilities only to those histories whose
interference vanishes.
While the underlying ideas are very general, their application to
full quantum gravity has, of necessity, remained rather formal.
However, recently, this program has been carried out to completion
to answer some key questions in the k=0 WDW theory and sLQC. In the
following, we first provide a brief summary of the paradigm and then
sketch the analysis for the WDW theory following
\cite{consistent1,consistent2}. Analysis for LQC has been performed
in a similar fashion in \cite{consistent3}.

The consistent histories approach uses three main inputs: (i) fine
grained histories, which constitute the most refined description of
observables in a given time interval; (ii) coarse grained histories,
formed by dividing fine grained histories in mutually exclusive sets
governed by a specified range of eigenvalues of observables of
interest, and, (iii) a decoherence functional which is a measure of
the interference between different branch wave functions
corresponding to coarse grained histories.

As an example, for a family of observables $A^{\alpha}$ (labeled by
$\alpha$) with eigenvalues $a_{k_i}^\alpha$ at time $t=t_i$, a
coarse grained history in which the eigenvalues fall in the range
$\Delta a_{k_i}^\alpha$ in a time interval $t \in (t_1,t_n)$  is
denoted by a class operator
\ba C_h &=& \nonumber  P_{\Delta a_{k_1}}^{\alpha_1} (t_1) \,
P_{\Delta a_{k_2}}^{\alpha_2} (t_2) \, ...\,
P_{\Delta a_{k_n}}^{\alpha_n} (t_n) \\
&=& U(t_0 - t_1) P_{\Delta a_{k_1}}^{\alpha_1} U(t_1 - t_2)
P_{\Delta a_{k_2}}^{\alpha_2} \, ...\, U(t_{n-1} - t_n)
P_{\Delta a_{k_n}}^{\alpha_n} U(t_n - t_0) ~. \ea
Here $P^{\alpha_i}_{\Delta a_{k_i}} (t_i)$ denote Heisenberg
projections at time $t_i$ in the range $\Delta_{a_{k_i}}$, and
$U(t)$ is the propagator defined by the Hamiltonian $H$. The branch
wave function $|\Psi_h\rangle$ corresponding to the coarse grained
history $h$ is determined by the action of the class operator:
$|\Psi_h \rangle = C_h^\dagger |\Psi \rangle$. The interference
between two coarse grained histories is measured by the normalized
decoherence functional: $d(h, h') = \langle \Psi_{h'} | \Psi_h
\rangle$ which depends on the two histories $h,h'$ as well as a
pre-specified state $\Psi$. One can unambiguously assign
probabilities to histories $h$ and $h'$ if the histories decohere,
i.e., if $d(h,h') =0$. In this case, the probability of the coarse
grained history $h$ is given by $p(h) = d(h, h)$. In text-book
quantum mechanics, one typically considers a single measurement. In
this case, the class operator $C_h$ is self-adjoint, the
coarse-grained histories automatically decohere and one can assign
probabilities unambiguously. In the more general context of multiple
projectors, the class operator $C_h$ is no longer self-adjoint and
therefore histories do not automatically decohere. In this case, to
assign probabilities, one has to first find coarse-grained histories
that do.

Let us now turn to quantum cosmology and use this framework to
calculate the probability for occurrence of singularities. Answering
such questions, however, requires a reasonably good mathematical
control on the quantum theory, in particular, knowledge of the
physical inner product, families of observables and their properties
and a notion of evolution and dynamics. Fortunately, these
structures are now available in the WDW theory and sLQC of the k=0
model coupled to a massless scalar field \cite{acs}. (The particular
framework we will use in the \WDW case is a direct spin-off of
sLQC). As we will discuss below, consistent histories analysis for
WDW theory shows that the probability that a WDW quantum universe
ever encounters a singularity is unity, independent of the choice of
a state, even if arbitrary superpositions of expanding and
contracting branches are allowed \cite{consistent1,consistent2}. By
contrast, the analysis for sLQC shows  that the probability for the
universe to undergo a non-singular bounce is unity
\cite{consistent3}.

Let us consider the WDW theory. Following section \ref{s2.2} we can
restrict our attention to: i) the positive frequency solutions of
the quantum constraint (\ref{qhc1});  ii) the scalar field momentum
$\hat p_{(\phi)}$, which is a constant of motion, and, iii) the
relational observable  $\hat z|_{\phi_o}$ which measures the
logarithm of the physical volume of the universe at internal time
$\phi_o$. (For a treatment directly in terms of the volume $\hat
V|_{\phi_o}$ itself, see \cite{consistent1,consistent2}). The
propagator for the consistent histories framework is defined as
\be U(\phi - \phi_o) = e^{i \sqrt{\ul \Theta} (\phi - \phi_o)}
~. \ee
We will be interested in coarse grained histories in the range of
eigenvalues $\Delta z$ at internal time $\phi=\phi_o$. The
corresponding projector consistent with the inner product (Eq
(\ref{ip1})) is trivially just
\be P_{\Dz} = \int_{\Dz} \d z \, |z\rangle \langle z| ~. \ee
Using the propagator and the projector, we can define the class
operator corresponding to coarse grained histories with
logarithm of volume in the interval $\Dz$
\be C_{\Dz|_{\phi^*}} = U^\dagger (\phi^* - \phi_o) P^z_{\Dz}
U(\phi^* - \phi_o) = P^z_{\Dz}(\phi^*) \ee
which leads to the branch wave function:
\be |\Psi_{\Dz|_{\phi^*}} (\phi)\rangle = U(\phi - \phi_o)
C^\dagger_{\Dz|_{\phi^*}} |\Psi\rangle ~. \ee
The histories trivially decohere when the ranges of $z$ have no
overlap. Given a normalized state $\Psi(z,\phi)$,  the probability
for the universe to have logarithm of volume in the range $\Dz$ at
$\phi = \phi^*$ is given by the obvious expression
\be p_{\Dz} (\phi^*) = \int_{\Dz} \, \d z \, |\Psi(z,\phi^*)|^2
~. \ee
Thus, because so far we are considering just one measurement, the
framework and the formulas are the familiar ones from ordinary
quantum mechanics.

One can now pose questions about the probability for the occurrence
of a singularity in the \WDW theory. Since $\hat p_{(\phi)}$ is a
constant of motion (and $\rho = \pphi^2/2V^2$), one way to answer
this question is by computing the probability for histories to enter
an arbitrarily small interval of volume, i.e., an interval $\Delta
z^\star = (-\infty, z^\star]$ in the logarithmic volume where
$z^\star$ is an arbitrarily large negative number. Now, in the \WDW
theory, the right (contracting) and the left (expanding) moving
modes belong to superselected sectors. Therefore, one can carry out
the calculation separately in each sector, and find the probability
for left moving states to encounter the singularity in the distant
past (and similarly, for the right moving states in the distant
future). For \emph{arbitrary} left and right moving states (in the
domains of operators under consideration) , one obtains
\be \lim_{\phi \rightarrow - \infty} p^L_{\Delta z^\star} (\phi) = 1
~~~~ \mathrm{and} ~~~ \lim_{\phi \rightarrow \infty} p^R_{\Delta
z^\star}(\phi) = 1 ~ \ee
for any $z^\star$, however large and negative. Thus, the probability
that an expanding WDW universe grows from an arbitrarily small
volume turns out to be unity. Similarly, the probability that a
contracting WDW universe enters an arbitrary small region of volume
in the distant future is unity. In this calculation limits $\phi
\rightarrow \mp \infty$ are necessary because we are allowing
\emph{arbitrary} states. If one restricts the analysis  to a
semi-classical state which has a finite spread in volume,
probability for the histories to enter $\Delta z^\star$ would become
unit at a finite value of $\phi$ (which would depend on the state).
The expectation value calculations in section \ref{s3.1} suggested
this outcome. But that suggestion comes from intuition developed in
quantum mechanics where the expectation values are tied to
ensembles. The consistent histories analysis provides a precise
statement for the single universe now under consideration. Note that
because the decoherence between the mutually exclusive histories is
exact (i.e., $d(h,h') =\delta_{h,h'}$), and the probability is
sharply 1, there is complete certainty that the singularity is not
resolved in either the left or the right moving sector of the \WDW
theory.

So far, we considered left and right moving states independently and
there may be concern that the certainty of our answer is a
consequence of our not allowing superpositions. Indeed, for a
superposed state $|\Psi\rangle = p_L|\Psi^L\rangle + p_R
|\Psi^R\rangle$, with $p_L+p_R =1$, the expectation value of the
volume observable never vanishes. Naively, this could be taken as
evidence that the universe described by this state avoids
singularities. Furthermore, for this state,
\be \lim_{\phi \, \rightarrow \, - \infty} p_{\Delta z^\star} (\phi)
= p_L ~~~~ \mathrm{and} ~~~ \lim_{\phi \, \rightarrow  \, \infty}
p_{\Delta z^\star}(\phi) = p_R ~, \ee
whence one might conclude that there is a non-vanishing probability
for a universe to have a large volume both in the asymptotic past
and the future, i.e. for the universe to bounce in the \WDW theory.
How does this conclusion fare if we let go our intuition about
expectation values that is rooted in ensembles and examine the issue
in the consistent histories framework? Then the conclusion turns out
to be incorrect \cite{consistent1,consistent2}! We will now
summarize why.

To analyze whether $\Psi = p_L|\Psi^L\rangle + p_R |\Psi^R\rangle$
represents a bouncing universe, it is necessary to compute the
projections not just at  one time slice, but at two time slices
---one at a very early time and the other at a very late time---
and ask: What is the probability of occurrence of any history that
has large volume both in the distant past and in the distant future?
Since the question refers to two instants of time, the required
class operator is now more general than the ones considered in
textbook quantum mechanics.

Let us begin by introducing the necessary ingredients. The class
operator corresponding to a history in which the universe \emph{does
not} enter $\Delta z^\star$ at an early time $\phi_1$ and a late
time $\phi_2$ is given by
\be C_{\mathrm{large}}(\phi_1,\phi_2) = P^z_{(\Delta z_1^\star)_{\rm
c}}\,(\phi_1)\,\, P^z_{(\Delta z_2^\star)_{\rm c}}\,(\phi_2) \ee
where $(\Delta z^\star)_{\rm c}$ is the complement of the interval
$\Delta z^\star$. The coarse grained history selected by this
operator has volume larger than that represented by $z^\star$ at the
two times considered. Therefore in the limit at $\phi_1 \to -\infty$
and $\phi_2 \to \infty$,\, $C_{\mathrm{large}}$ can be interpreted
as the class operator $C_{\rm bounce}$ representing bouncing
histories. Similarly, the class operator representing a history that
the universe does enter the interval $\Delta z^\star$ at both
$\phi_1$ and $\phi_2$ is given by
\be C_{\mathrm{small}}(\phi_1,\phi_2) = C_{\Delta z_1^\star} +
C_{\Delta z_2^\star} - C_{\Delta z_1^\star;\Delta  z_2^\star} ~. \ee
Since the coarse grained history selected by this operator has
volume smaller than (or equal to) that represented by $z^\star$ at
the two times considered, in the limit at $\phi_1 \to -\infty$ and
$\phi_2 \to \infty$,\, $C_{\mathrm{small}}$ can be interpreted as
the class operator $C_{\rm sing}$ representing histories that
encounter singularities both in the future and the past.

One can show that the branch wave function for \emph{any} superposed
state $|\Psi\rangle = p_L|\Psi^L\rangle + p_R |\Psi^R\rangle$ to be
in $(\Delta z^\star)_{\rm c}$ both in the asymptotic past and the
future is zero: $|\Psi_{\mathrm{bounce}}\rangle = \lim_{\phi_1
\rightarrow -\infty} \lim_{\phi_2 \rightarrow \infty}
C^\dagger_{\mathrm{large}}(\phi_1,\phi_2) |\Psi\rangle = 0 $. This
shows that the probability associated with any bouncing
coarse-grained history is identically zero. Similarly, the branch
wave function corresponding to the history that is singular both in
the asymptotic past and the asymptotic future turns out to be:
\be |\Psi_{\mathrm{sing}}\rangle = \lim_{\phi_1 \rightarrow -\infty}
\lim_{\phi_2 \rightarrow \infty} C^\dagger_{\mathrm{small}}
(\phi_1,\phi_2) |\Psi\rangle =|\Psi\rangle  \ee
irrespective of how small the volume defined by $z^\star$ is. The
decoherence functional between the histories which bounce and are
singular vanishes identically. Therefore it is meaningful to assign
probabilities to these two coarse grained histories. The probability
that a universe is ever singular, turns out to be unity. Thus,
contrary to what one might have naively expected, arbitrary
superpositions of contracting and expanding solutions do not avoid
singularity in the WDW theory. We emphasize that this result holds
for any superposed state. Again, the prediction is completely
unambiguous because the two coarse grained histories decohere
completely and because the associated probabilities are $1$ and $0$.

A similar calculation has been performed in sLQC \cite{consistent3}.
It turns out that the probability that an arbitrary superposition of
left and right moving states ever encounters a singularity is $0$
and the probability for the bounce to occur is $1$. Again, as in the
WDW theory, these results bring out the precise sense in which the
expectations based on results of section \ref{s3.1} are correct,
without having to rely on the intuition derived from ensembles.

In summary, the consistent histories paradigm can be successfully
realized for the k=0 universes both in the \WDW theory and LQC,
thanks to the complete mathematical control on the quantum theory of
these models. It provides precise answers to questions concerning
probabilities of the occurrence of singularities in a quantum
universe. This analysis provides a road map to use the consistent
histories approach in more general contexts discussed in sections
\ref{s3} and \ref{s6}. These applications could lead to important
insights on some long standing questions such as the quantum to
classical transition in the cosmology of the early universe that is
often evoked to account for the seeds of the large scale structure.

\section{Discussion}
\label{s8}

The field of quantum cosmology was born out of the conviction
that general relativity fails near the big-bang and the
big-crunch singularities and quantum gravity will cure this
blemish. In his 1967 lectures for Battelle Rencontres, John
Wheeler wrote \cite{jw1}:
\begin{quote}
{\sl 
Here, according to classical general relativity, the dimensions
of collapsing system are driven down to indefinitely small
values. ... In a finite proper time the calculated curvature
rises to infinity. At this point the classical theory becomes
incapable of further prediction. In actuality, classical
predictions go wrong before this point. A prediction of
infinity is not a prediction. The wave packet in superspace
does not and cannot follow the classical history when the
geometry becomes smaller in scale than the quantum mechanical
spread of the wave packet. ... The semiclassical treatment of
propagation
is appropriate in most of the domain of superspace 
.... [but] not so in the decisive region.}
\end{quote}

The quote is striking especially because of the certainty it
expresses as to what should happen `in actuality' near
singularities. As we saw in sections \ref{s2} -- \ref{s4}, although
this brilliant vision did not materialize in the \WDW theory, it
\emph{is} realized in all the cosmological models that have been
studied in detail in LQC. However the mechanism is much deeper than
just the `finite width of the wave packet': the key lies in the
quantum effects of geometry that descend from full LQG to the
cosmological settings. These effects produce an unforeseen repulsive
force. Away from the Planck regime the force is completely
negligible. But it rises \emph{very} quickly as curvature approaches
the Planck scale, overwhelms the enormous gravitational attraction
and causes the quantum bounce. Large repulsive forces of quantum
origin are familiar in astrophysics. Indeed, it is the Fermi-Dirac
repulsion between nucleons that prevents the gravitational collapse
in neutron stars. Although this force is rooted in the purely
quantum mechanical properties of matter, it is strong enough to
balance classical gravitational attraction if the mass of the star
is less than, say 5 solar masses. However, in heavier stars,
classical gravity still wins and leads to black holes. \emph{In LQC
the repulsive force has its origin in quantum geometry rather than
quantum matter and it always overwhelms the classical gravitational
attraction.}

Thus, even though we began with just general relativity in 4
dimensions, quantum dynamics contains qualitatively new physics.
This is a vivid illustration of the fact that higher dimensions or
new symmetries are not \emph{essential} for a quantum gravity theory
to open new vistas. Indeed, a general lesson one can draw from LQG
is that, even if a theory is firmly rooted in general relativity
\emph{in the classical domain}, its fundamental degrees of freedom
can be far removed from what the classical continuum suggests. A
coarse grained description can suffice at low energies but the
fundamental degrees of freedom become indispensable at the Planck
scale. They can usher in new physics to overcome the ultraviolet
difficulties of general relativity.\\

Sections \ref{s2} -- \ref{s4} discussed three notable aspects of
this physics beyond general relativity. First, although the exact
solubility of the k=0 FLRW model played a major role in establishing
detailed analytical results \cite{acs} (such as the derivation of
the expression of the maximum density $\rcr$), systematic numerical
studies have shown that the behavior in the Planck regime is much
more general: The bounce persists in the FLRW models with spatial
curvature or cosmological constant which are not exactly soluble
\cite{apsv,bp,ap}. More precisely, one can start with quantum states
which are sharply peaked at late times on a general relativity
trajectory and evolve them towards classical singularities. All wave
functions that initially resemble coherent states undergo a quantum
bounce at $\rho \approx \rcr$ in the FLRW models and the behavior of
these wave functions in the Planck regime is very similar. Since
these models are not exactly soluble, one does not have results for
general states. But the fact that there does exist a large class of
physically interesting states exhibiting this behavior in the Planck
regime is already highly non-trivial. Furthermore, in these models
there are now `S-matrix-type' analytical results that relate the
behavior of generic wave functions well before the bounce to that
well after the bounce \cite{kp2}.%
\footnote{These advances required not only new conceptual ideas
(e.g., relational time and specific Dirac observables) but a careful
handling of hard mathematical issues (such as essential
self-adjointness of the gravitational constraints and control on
their spectra) and the development of accurate numerical methods
(that are attuned to the delicate mathematical properties of various
operators). Because of our intended audience, in this review we
focused on the conceptual issues and physical predictions and could
not do justice to the seminal mathematical contributions, especially
from the Warsaw group, nor to the powerful numerical infrastructure
that was created almost singlehandedly by Tomasz Pawlowski.}

The second notable feature is that this cure of the ultraviolet
limitations of general relativity does \emph{not} come at the cost
of infrared problems. This is surprisingly difficult to achieve
because, on the one hand, quantum dynamics has to unleash huge
effects in the Planck regime and, on the other, it must ascertain
that departures from general relativity at lower curvatures are so
tiny that they do not accumulate over the immense cosmological time
scales to produce measurable deviations at late times. Indeed, the
early treatments of dynamics in LQC \cite{mb1,abl} managed to
resolve the singularity but, as we saw in section \ref{s7.1}, gave
rise to untenable deviations of general relativity at late times
\cite{aps2,cs1}. To obtain good behavior in both the ultraviolet
\emph{and} the infrared requires a great deal of care and sufficient
control on rather subtle conceptual and mathematical issues. The
resulting `improved dynamics' is now providing useful hints in full
LQG \cite{lv}. Finally, it is pleasing to see that even in the
models that are \emph{not} exactly soluble, states that are
semi-classical at a late initial time continue to remain sharply
peaked throughout the low curvature domain. In the closed k=1 model
as well as $\Lambda <0$ FLRW models the universe undergoes a
classical recollapse. For universes that grow to macroscopic sizes,
predictions of LQC for this recollapse reduce to those of general
relativity with impressive accuracy, thereby providing detailed
quantitative tests of the good infrared behavior. Initially this is
surprising because of one's experience with the spread of wave
functions in non-relativistic quantum mechanics. However, this
behavior is precisely what one would expect if at low curvature
quantum gravity evolution is to agree with that in general
relativity over cosmological time scales.

The third notable feature is the powerful role of effective
equations \cite{jw,vt,psvt} discussed in section \ref{s5}. As is not
uncommon in physics, their domain of validity is much larger than
one might have naively expected from the assumptions that go into
their derivations. Specifically, in all models in which detailed
simulations of \emph{quantum} evolution have been carried out, wave
functions which resemble coherent states at late times follow the
dynamical trajectories given by effective equations even in the deep
Planck regime. These equations have two additional noteworthy
features. First, they arise from a (first order) covariant action
\cite{os}. Second, although they introduce non-trivial corrections
to \emph{both} the Friedmann and the Raychaudhuri equations, the two
modified equations continue to imply the correct equation of motion
for matter (the Klein Gordon equation for the scalar field and the
continuity equation for a general perfect fluid). Thanks to all
these features, there is growing confidence that the effective
equations are likely to have a large domain of validity also in more
complicated cosmological models. Therefore, even as a part of the
LQC community is engaged in verifying their validity in, e.g.,
anisotropic models \cite{ahtp}, others are actively engaged in
working out their consequences, assuming they are valid are more
generally. These efforts have provided interesting insights. First,
recall that cosmological singularities are not restricted to be of
the big-bang or big-crunch type: even isotropic models which have
perfect fluid matter (with an equation of state $P=P(\rho)$ admit a
variety of more exotic singularities \cite{brip_lqc,ps09}. Effective
equations imply that all of the strong curvature singularities in
isotropic and homogeneous models are resolved in LQC \cite{ps09,sv}.
Second, they have revealed the richness of the quantum bounces in
more general models. In the isotropic case, there is a single bounce
at which the scalar curvature and matter density reach their maxima.
In the Bianchi models, the structure is much richer because of the
non-triviality of Weyl curvature. Roughly, every time a shear term
--- a potential for the Weyl curvature--- enters the Planck regime,
the quantum geometry effects dilute them. Thus, in contrast to what
was observed in other bouncing models, \emph{anisotropies never
diverge} in LQC \cite{gs01}. This features removes a principal
concern cosmologists have had \cite{bb}. Singularity resolution is
both richer and more subtle in LQC because in anisotropic models
there is not just one bounce; while there is still a `density
bounce' there are also `bounces' associated with other observables
such as the Weyl curvature.

This striking difference arises because the philosophy in LQC is
different from the one implicitly used earlier in bouncing models.
There, one first obtained a bounce in the FLRW model and then added
other effects as corrections to the effective equations governing
the bounce. In LQC, the philosophy is to first restrict oneself to
an appropriate sector of the full phase space of general relativity
with matter, pass to the corresponding truncated quantum gravity
theory by applying principles of LQG, and finally distill effective
equations from this quantum theory. Therefore, in the Planck regime
these equations can contain qualitatively new features, not seen in
the FLRW model. This is exactly what happens in the anisotropic
models discussed in sections \ref{s4.4} and \ref{s5.4}. The very
considerable research on the BKL conjecture \cite{bkl1,bkl-ar} in
general relativity suggests that, as generic space-like
singularities are approached, `terms containing time derivatives in
the dynamical equations dominate over those containing spatial
derivatives' and dynamics of fields at any fixed spatial point is
better and better described by the homogeneous Bianchi models.
Therefore, to handle the Planck regime to an adequate approximation,
it may well suffice to treat just the homogeneous modes using LQG
and regard inhomogeneities as small deviations propagating on the
resulting homogeneous LQC \emph{quantum} geometries. This is the
philosophy underlying `hybrid quantization' which has successfully
led to singularity resolution in the inhomogeneous Gowdy model
\cite{hybrid1, hybrid2,hybrid3,hybrid4,hybrid5}. However, an
important lesson from LQC is that it would not be adequate to treat
just the isotropic degrees of freedom non-perturbatively and
introduce anisotropies as small corrections.

Finally in section \ref{s7} we discussed a few examples of fresh
insights that LQC has provided into some of the long sanding issues
of quantum gravity, beyond cosmology. This is possible because the
conceptual framework underlying LQC is well grounded and because
there is an excellent mathematical control. An example is provided
by the issue of entropy bounds \cite{bousso1, bousso2}: Should a
suitable entropy bound constitute an essential ingredient in the
very construction of a satisfactory quantum gravity theory, or,
would such bounds simply emerge from a quantum gravity theory on
making suitable approximations in appropriate regimes? The issue is
difficult to analyze because, while the bound has its origin in
quantum gravity, its formulation requires a classical geometry. LQC
provides an excellent setting to test these ideas because it can
meet these stringent requirements. The detailed analysis \cite{awe1}
clearly favors the second possibility. Another example is provided
by spin foams and group field theory. Here, LQC could be used to
test general  ideas underlying these paradigms. Specifically, does
the `extraction amplitude'  ---that replaces the more familiar
`transition amplitude' in the timeless quantum gravity framework---
admit a meaningful `vertex expansion' in line with the goals of
these framework? Since the Hamiltonian theory underlying LQC is
fully under control, this and other more detailed questions could be
explored in detail in LQC. The calculations provide a strong support
for the paradigm but also raise specific questions for further work
\cite{ach1,ach2,chn,hrvw}.

A third example is provided by the application of the consistent
histories framework. In full quantum gravity, of necessity, the
application has been only formal. LQC provides a well controlled
setting which has all the major conceptual difficulties of full
quantum gravity that require a generalization of the standard
`Copenhagen' quantum mechanics
\cite{as-book,hartle,crbook,rgjp1,rgjp2}. The fact that this
framework can be used to address in detail concrete questions of
physical interest in quantum cosmology
\cite{consistent1,consistent2} opens doors for more general
applications. Next, there is a recent proposal to construct LQG for
general relativity coupled with massless scalar fields
\cite{warsaw-full} that draws on strategies used in LQC: the use of
a scalar field as relational time, construction of the physical
Hilbert space from `positive frequency' solutions to the quantum
constraints and introduction of Dirac observables analogous to the
$\h{V}|_\phi,\, \h{\rho}_\phi$ in LQC. In all these examples, LQC
provided conceptual and mathematical tools to address in a concrete
fashion some of the issues we face in full quantum gravity. The last
example represents progress also in the other direction. So far
there are only a few partial results on the precise relation between
full LQG and LQC \cite{je1,je2}. If the proposal of
\cite{warsaw-full} can be shown to be physically viable, it would
provide a natural avenue to systematically descend from LQG to LQC.
Finally, in light of the BKL conjecture \cite{bkl1,bkl-ar} and its
recent formulation adapted to the LQG phase space \cite{ahs}, the
LQC singularity resolution in Bianchi models \cite{awe2,awe3,we}
opens up an avenue to study the fate of general space-like
singularities in quantum gravity. Specifically, one can now hope to
prove theorems in support of the idea that strong curvature,
space-like singularities are absent in LQG.

Returning to the more restricted setting of cosmology, it seems fair
to say that LQC provides a coherent and conceptually complete
paradigm that is free of the difficulties associated with the
big-bang and the big-crunch. Therefore, the field is now
sufficiently mature to address observational issues. Indeed, this is
the most fertile and interesting of directions for current and
future research. Not surprisingly, then, it has begun to attract the
attention of main-stream cosmologists (see, e.g.,
\cite{bms,pbb_lqc1,Einstein_static_lqc,emergent_lqc,clm,cmns,
lidsey_scale,mn1,pert_tensor_fr,joao_ps,gns_inflation,barrau1,
barrau2,barrau3,barrau4,joao1,joao2}).

There is already significant literature that has begun to probe in
detail how the novel effects associated with LQC ---the quantum
bounce, the superinflation phase and the holonomy and the inverse
volume corrections--- can affect the current cosmological scenarios
based on general relativity. Much of this work ---\emph{though not
all}--- is based on inflationary scenarios. On the conceptual side,
because the big-bang singularity is replaced by the quantum bounce
where all fields are regular, one can resolve the difficulties
associated with measures on spaces of solutions and carry out a
well-defined calculation of the a priori probability of inflation. A
pleasant surprise was that the probability turned out to be
extremely close to one in spite of the non-trivialities of dynamics
associated with superinflation and the phase that immediately
follows it \cite{as2,as3}. On the phenomenological side, it is
heartening to see detailed calculations on possible modifications of
spectral indices and close connection to WMAP observations. Because
this literature is still evolving, we do not yet have definitive
predictions on which there is general consensus. Therefore, in
section \ref{s6.4} we only provided illustrative examples. A
significant fraction of this work is devoted to the inflationary and
post-inflation phases. Although these quantum gravity corrections
are conceptually interesting, since the curvature and matter
densities at the onset of inflation are some 11 orders of magnitude
smaller than the Planck scale, these effects will not be observable
in the foreseeable future. Effects that could be relevant for such
observations will have to originate in the deep Planck regime and
not dilute away during inflation. As discussed in section
\ref{s6.3}, there are some viable possibilities along these lines
and a systematic framework ---involving quantum fields (representing
linear perturbations) on quantum (FLRW) space-times--- necessary to
exploit them has become available. Therefore there is much scope for
synergistic work from cosmology and LQG communities in this growing
area.

We hope this review will help to attract a broader participation
from the cosmology community to achieve this goal.

\section*{Acknowledgments}

We are grateful to Ivan Agullo, Aurelien Barrau, Miguel Campiglia,
David Craig, Kristina Giesel, Adam Henderson, Wojciech Kaminski,
Alok Laddha, Ian Lawrie, Jerzy Lewandowski, Roy Maartens, Guillermo
Mena, William Nelson, Jorge Pullin, Carlo Rovelli, David Sloan,
Victor Taveras, Thomas Thiemann, Kevin Vandersloot, Madhavan
Varadarajan, Joshua Willis, Edward Wilson-Ewing and especially
Alejandro Corichi and Tomasz Pawlowski for stimulating discussions
and correspondence. In addition, we wish to thank Ivan Agullo,
Alejandro Corichi, Kristina Giesel, Jerzy Lewandowski, William
Nelson, Tomasz Pawlowski, David Sloan and Edward Wilson-Ewing for
their comments on various sub-sections and Carlo Rovelli, on the
whole manuscript. Finally, Miguel Campiglia, Adam Henderson, Tomasz
Pawlowski, and Edward Wilson-Ewing graciously gave their permission
to use figures from joint work. This work was supported in part by
the NSF grants PHY0854743 and PHY1068743 and the Eberly research
funds of Penn State.

\section*{List of Symbols}

\begin{tabular}{ll}

$a$     & scale factor in isotropic models; also mean scale factor in anisotropic models\\
$a_i$   & directional scale factors in anisotropic models\\
$A_a^i$ & gravitational SU(2) connection 1-form on the 3-manifold $M$, used in LQG\\
$a,b,..$ &space-time or space indices (in Penrose's abstract index notation)\\
$\b$ & a gravitational phase space variable, conjugate to $\v$, defined in section \ref{s2.1} \\
$c$ & symmetry reduced connection component in isotropic models \\
$c_i$ & symmetry reduced connection components in Bianchi models\\
${\cal C}$  & fiducial cell on $M$, essential in spatially non-compact homogeneous models \\
$C_{\mathrm{grav}}$ & gravitational part of the Hamiltonian constraint \\
$C_H$ & classical Hamiltonian constraint \\

$\hat C_H$ & operator corresponding to $C_H$ \\
$C_H^{(\mathrm{eff})}$ & effective Hamiltonian constraint in LQC, defined in section \ref{s5}\\
$\Delta$ & ratio of quantum of area to $\lp^2$, defined in section \ref{s2.5}\\

$e^a_i$  &physical ortho-normal triad on $M$ \\
$\mathring{e}^a_i$ & fiducial orthonormal triad on $M$; $e^a_i = a^{-1}\e^a_i$ in isotropic models \\
$E^a_i$ & physical triad with density weight one on $M$ \\
$\epsilon$ & labels super-selected sectors in ${\cal H}_{\mathrm{phy}}$, defined in section \ref{s2.6} \\
$\varepsilon$ & orientation of the triad \\
$\epsilon^i_{~jk}$ & structure constants of SU(2)\\
${\cal E}(\nu,\phi;\nu',\phi')$ & Green's function obtained from group averaging, defined
in section \ref{s3.5}\\
$F_{ab}^i$  & curvature of $A_a^i$ \\
$\hat F_{ab}^i$ & operator corresponding to $F_{ab}^i$ \\
$F(x)$ & a positive frequency solution of the quantum constraint in sLQC \\
$G$ & Newton's constant \\
$g_{ab}$ & space-time metric \\
$\gamma$ & the Barbero-Immirzi parameter of LQG \\

$\hbar$ & Planck's constant divided by $2 \pi$ \\

$h_\ell$ & holonomy of $A_a^i$ along a line segment $\ell$\\
$H$ & Hubble rate \\

$\hat H$ & Hamiltonian operator \\
$\hat H_o$ & Hamiltonian operator corresponding to the background quantum geometry \\
${\cal H}_{\mathrm{matt}}$ & matter Hamiltonian \\
${\cal H}_{\mathrm{kin}}$ & kinematical Hilbert space\\
$\Hkg$ & gravitational part of the kinematical Hilbert space\\
${\cal H}_{\mathrm{phy}}$ & physical Hilbert space \\

$i,j,..$ &internal SU(2) indices \\

$\mathrm{k}$ & spatial curvature index in  isotropic models\\
$k$ & labels Fourier modes. Related to $\omega$ as $|k| = \omega/{\sqrt{12 \pi G}}$
in section \ref{s2}  \\

$l_i$ & proportional to the square root of $p_i$, defined in section \ref{s4.4} \\
$L_i$ & coordinate lengths of ${\cal C}$ in the Bianchi models \\
$\ell_o$ &cube root of the fiducial volume $V_o$ of the cell ${\cal C}$ in k=1 model\\
$\lp$ &Planck length ($\lp = \sqrt{G\hbar/c^3}$)\\

$\lambda$ & square root of the quantum of area, defined in section \ref{s2.5}\\
$\Lambda$ & cosmological constant \\

$m$ & mass parameter \\
$M$     & spatial manifold\\
$\mpl$ & Planck mass $\mpl = \sqrt{\hbar c/G}$\\
\end{tabular}

\newpage

\begin{tabular}{ll}

$\mu$ & dimensionless length of the edge $\ell$, along which $h_\ell$ is computed \\
$\bar \mu$ & dimensionless length of the smallest plaquette, defined in section \ref{s2.5}.  \\
& also used as an adjective to refer to the `improved dynamics' in LQC \\

$\mu_o$ & a constant related to the area-gap.\\
& also used as an adjective to refer to an older dynamics in LQC\\
$N$ & lapse function \\

$\omega$  &eigenvalue of $\Theta$\\
$\mathring{\omega}^i_a$ &fiducial co-triad, dual to $\e^a_i$\\
$\Omega^{\mu \nu}$ & symplectic form on the phase space \\
$p$ & symmetry reduced triad component in the isotropic models \\
$\hat p$ & operator corresponding to $p$ \\
$p^i$ & symmetry reduced triad components in Bianchi models \\
$\tilde p_{(a)}$ & canonical conjugate momentum of the scale factor, defined in
section \ref{s2.1}\\
$p_{(\phi)}$ & canonical conjugate momentum of $\phi$; Dirac observable for a massless $\phi$\\

$\hat p_{(\phi)}$ & operator corresponding to the Dirac observable for a massless $\phi$ \\

$\phi$  & a homogeneous scalar field; serves as a relational time variable \\
$\phi_B$ & value of the relational time $\phi$ at which bounce occurs in isotropic models \\
$\Phi$ & an inhomogeneous scalar field \\
${q}$ & determinant of ${q}_{ab}$\\
$q_{ab}$ & physical spatial metric on $M$\\
$\mathring{q}_{ab}$ & fiducial metric on $M$ \\

$q_{\pm \vec{k}}, p_{\pm \vec{k}}$ & canonically conjugate
      variables in the Fourier space, defined in section \ref{s6.2} \\
$R$ & scalar curvature of the space-time metric \\
$\rho$ & energy density\\
$\hat \rho|_{\phi}$ & operator corresponding to the Dirac observable for energy density \\
$\rcr$ & maximum value of energy density in LQC \\

$\sigma^2$ & shear scalar in Bianchi models, defined in section \ref{s5.4}\\
$\sigma^2_{\mathrm{max}}$ & maximum value of shear scalar \\
$\Sigma^2$ & shear parameter in Bianchi models;  $\Sigma^2 = \f{1}{6} \sigma^2 a^6$\\

$t$ & proper time \\
$\tau$ & harmonic time \\
$\ul \Theta$ &Quantum constraint operator in the Wheeler-DeWitt theory, defined in section \ref{s2.2}\\
$\Theta$  & Quantum constraint operator in LQC for  flat and isotropic model, defined in section \ref{s2.5}\\

$\v$ & phase space variable corresponding to volume, related to $V$ as $\v = V/(2 \pi G)$\\
$\hat \v$ & operator corresponding to \v \\
$v$ & dimensionless volume variable, related to $V$ as $v = V/(2 \pi \gamma \lambda^2 \lp)$
used in section \ref{s4.4}\\
$\nu$ & variable used to define `volume representation', related to $V$ as $\nu = V/(2 \pi \lp^2)$\\
$\hat \nu$ & operator corresponding to $\nu$ \\
$V$ & physical volume of a spatially compact universe,\\
 & or, of the fiducial cell ${\cal C}$ in the non-compact case\\
$\hat V$ & operator corresponding to $V$ \\
$V_o$ & volume of the universe, or the cell ${\cal C}$, with respect to the fiducial metric $\mathring{q}_{ab}$\\
$\hat V|_{\phi}$ & operator corresponding to the Dirac observable for volume at internal time $\phi$ \\
$V(\phi)$ & scalar field potential \\
$x$ &proportional to the logarithm of the tangent of $(\lambda \b/2)$, defined in section \ref{s3.2}\\
$\ox_i^a$ & Killing fields of physical metrics in homogeneous models\\
$y$ & proportional to the logarithm of $\b$, defined in section \ref{s3.1}\\
$z$ &logarithm of volume, defined in section \ref{s2.2} \\

\end{tabular}

\end{document}